Sergei Ivanovich Ipatov

# Formation and evolution of planetary systems

This is the translation of the Russian book.
This file with English translation can be loaded from https://cloud.mail.ru/public/Yeoy/cDwaFQjVV.

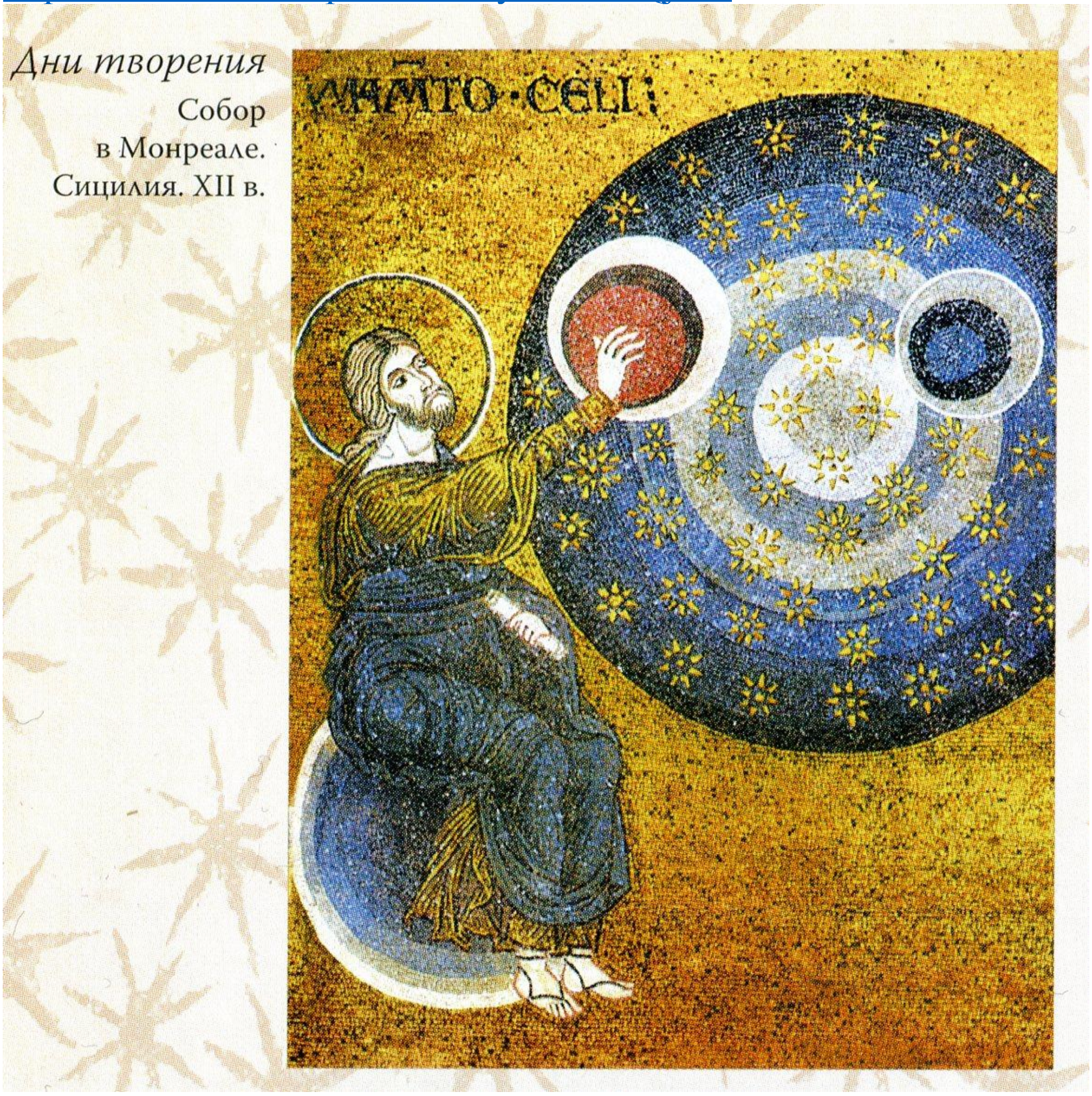

**2025**

**Ipatov Sergei Ivanovich**
Formation and evolution of planetary systems / Sergei Ivanovich Ipatov, 2025.
This is translation of the Russian book

Various problems of the formation and evolution of planetary systems are studied. Most of the studies are devoted to the Solar System. The collapse of the presolar cloud and the accumulation of planets are studied. The author considers the formation of the Earth-Moon system, binary trans-Neptunian objects, and the axial rotations of planets. He discusses the formation of asteroid and trans-Neptunian belts, including the evolution of resonant asteroid orbits and the formation of the Kirkwood gaps in the asteroid belt. The book discusses the migration of bodies during the formation of the Solar System and at present, including a discussion of the delivery of icy bodies from beyond the ice line to the terrestrial planets and the formation of craters on the Moon. The Deep Impact mission is discussed; during this mission the impact module of the spacecraft collided with the comet Tempel 1. The book studies the migration of dust in the Solar System and the formation of the zodiacal belt, as well as the migration of bodies and dust particles ejected from the terrestrial planets and the Moon. The author considers the migration of bodies in some exoplanetary systems (Proxima Centauri, Trappist-1, and Gliese 581), the spectrum of Earth-like exoplanets with different rotation periods, as well as the efficiency of exoplanet searches using microlensing observations with various telescopes.

UDC 523
BBK 22.65

ISBN 978-5-6055159-1-3 - for Russian book

## Introduction

This book, based on the author's publications, examines various problems of the formation and evolution of planetary systems. Much of the research is devoted to the Solar System. It examines the collapse of the presolar cloud, the accumulation of the terrestrial planets, the formation of the Earth-Moon system, binary trans-Neptunian objects, and the axial rotations of the planets. It examines the migration of bodies during the formation of the Solar System and today, including a discussion of the delivery of icy bodies from beyond the ice line to the terrestrial planets and the formation of craters on the Moon. It examines the evolution of resonant orbits of asteroids and the formation of Kirkwood gaps in the asteroid belt. It discusses observations of asteroids and comets and models for determining the probability of detecting near-Earth objects in various regions of the sky. This book also discusses the Deep Impact mission, specifically the removal of cosmic ray signatures from images taken by this spacecraft, and the ejection of material from comet 9P/Tempel 1 following the collision of the Deep Impact impactor with this comet. It also examines dust migration in the Solar System and the formation of the zodiacal belt, as well as the migration of bodies and dust particles ejected from the terrestrial planets and the Moon. It also examines the migration of bodies in some exoplanetary systems (Proxima Centauri, Trappist-1, and Gliese 581), the spectra of Earth-like exoplanets with different axial rotation periods, and the effectiveness of microlensing searches for exoplanets using various telescopes. A brief discussion of some non-astronomical problems in which the author has worked is also provided. The paper discusses studies of the generation of acoustic waves under the influence of fluids on pore walls and their propagation in a porous medium with fluids and gas.

The book is intended for the general reader interested in astronomy, and the material is presented primarily at a popular level. Readers interested in a more detailed presentation of the material can look up the author's publications on the topic of interest. Most of S.I. Ipatov's publications in English and his monograph [78] in Russian, which summarizes Ipatov's 20th-century results, can be downloaded and read free of charge using the internet links in the list of publications provided in the second part of the book. Internet links to S.I. Ipatov's publications can be found on the websites https://sites.google.com/view/siipatov/publications-in-russian, https://siipatov.webnode.ru/publications (rarely upgraded) and in the file https://1drv.ms/w/c/c67d93a65f0a2a17/ERcqCl-mk30ggMYFBAAAAAABye6tCmVUGVWRphNZ_k-WIg?e=DjHucX (the list of publications there is presented in a different order). On these websites, you can also find links to Ipatov's publications that will be published after the publication of this book (the writing of a number of articles on the calculations already performed has been postponed until the completion of this book). A link to the file with this book can also be found there. In Russian this file in docx format can be downloaded from the website https://1drv.ms/w/c/c67d93a65f0a2a17/ETczmLUVNmNIiXcDuG8av_kBO0N_WKVFeZfJeWQ2qbqD1A?e=tOaKMC. I put pdf file with the Russian book on the sites https://dx.doi.org/10.17513/np.649, https://www.elibrary.ru/item.asp?id=88803878, https://www.litres.ru/author/sergeyivanovich-ipatov/, https://doi.org/10.48550/arXiv.2601.08952, https://www.researchgate.net/publication/398948901, https://www.academia.edu/145190350, https://online.fliphtml5.com/Sergei_Ipatov/Ipatov-Formation-planetary-systems/ and https://www.kobo.com/lt/en/ebook/formation-and-evolution-of-planetary-systems. This pdf file with English translation of the whole book can be downloaded from https://cloud.mail.ru/public/Yeoy/cDwaFQjVV. I also plan to put this English translation to reseachgate, academia and kobo. Most of S.I. Ipatov's articles can be downloaded free of charge from the websites https://independent.academia.edu/SergeiIpatov (photographs of article pages that do not exist in digitized form were also posted on this website), https://www.researchgate.net/profile/S_Ipatov/publications and http://arxiv.org. For readers interested in internet links, it is more convenient to have access to the book file or to the file with the list of publications even if they have a printed copy.

References to publications by other authors are given in the text, not at the end of the book. References to a much larger number of publications by other authors can be found in Ipatov's articles. The monograph [78] contained 687 references to publications, and the review article [59] contained 400 references. The total number of references to publications by different authors in all of Ipatov's publications is much greater, and the volume of the book would increase several times if detailed references to the works of different authors were provided. Therefore, the number of references to the works of other authors in the book is limited. The topics of many sections of the book deserve separate books to be written on these topics. A detailed presentation of the material would lead to a large book and would be of interest only to specialists, whereas it is available free of charge using Internet links (the reader can read in more detail what interests him using the links). Ipatov's articles are available free of charge mainly in English, a language that any scientist can speak. Figures and tables are used rarely in the book, and each section has its own numbering, as there are no cross-references between them. A much larger number of figures and tables on the topics discussed can be found in S.I. Ipatov's publications.

This book can serve as a guide to the author's publications. It presents a brief overview of the results obtained by S.I. Ipatov and general information on the topics discussed. The usual reader will receive a general idea of the material presented (and is unlikely to be interested in the details). An interested reader can choose which publications he would like to download free of charge from the Internet, view them and become familiar with the topic of interest in much greater detail, including numerous links to publications by other authors. Reviews of S.I. Ipatov's work on the formation and evolution of planetary systems are given in the monograph [78] (published in 2000), in the articles [59] and [64] and in the popular science article [77]. Chapter 1 of the monograph [78] provides general information about the Solar System (including planets, asteroids, comets, craters, meteorites, and meteor showers). This book was originally written as one of the appendices to a future book about meetings with astronomers. I initially began writing appendices for this future book, but then, seeing the large size of the appendices, decided to publish them separately. The appendix on Russian history was published as a book [537] (the English translation of this book can be downloaded from https://cloud.mail.ru/public/mFnh/tBePebKC2). The appendix on the results obtained by S.I. Ipatov became this book. The book about meetings with astronomers was published in 2026. Files with this Russian book can be found e.g. from https://dx.doi.org/10.17513/np.668, https://www.researchgate.net/publication/406223820, and from several other websites. English translation of the book with photos can be downloaded free from https://cloud.mail.ru/public/DWT8/S6wWoeETJ.

This publication was prepared within the framework of a state assignment for the Vernadsky Geochemical Institute of the Russian Academy of Sciences.

## SOLAR SYSTEM

### 1. The collapse of a presolar cloud and the influx of short-lived radioisotopes into it at the collision of the cloud with the shock wave from a supernova

Under the direction of Alan Boss in 2007, Ipatov and Boss calculated the collapse of a presolar cloud and the injected short-lived radioisotopes into it during the collision of the cloud with the supernova shock wave [36, 38, 97, 377, 382, 392]. Cosmochemical evidence for the existence of short-lived radioisotopes (SLRI), such as 26Al and 60Fe, during the formation of primitive meteorites requires that these isotopes were synthesized in a massive star and then incorporated into chondrites over $\sim 10^6$ years. It was long assumed that the supernova shock wave transferred SLRI to the core of the dense presolar cloud, caused the cloud to collapse, and injected the isotopes. Our simulations using the FLASH2.5 adaptive mesh refinement (AMR) hydrodynamic algorithm show that a 20 km/s shock wave can indeed trigger the collapse of a solar-mass cloud, while simultaneously injecting shock isotopes into the collapsing cloud. These simulations imply that the supernova trigger hypothesis is the most likely mechanism for delivering the SLRIs present during the formation of the Solar System. These models show [36] that, when adding cooling by $H_2O$, $CO_2$, and $H_2$ molecules, a 20 km/s shock wave is capable of triggering the gravitational collapse of a stable, dense solar-mass cloud core, while also injecting corresponding amounts of supernova shock material into the collapsing cloud core. This injected material consists of shock wave gas and dust particles small enough to remain bound to the gas, i.e., submicron-sized particles expected to characterize supernova shock waves and carry SLRI, the decay products of which have been detected in refractory inclusions of chondrite meteorites. These models strongly support the hypothesis that a supernova shock wave carrying SLRI could have triggered the formation of the Solar System.

Similar calculations were also performed [38] with shock wave velocities ranging from 1 km/s to 100 km/s. We found that shock wave velocities in the range of 5 km/s to 70 km/s are capable of causing the collapse of a cloud with a mass equal to 2.2 solar masses, while simultaneously injecting shock wave material. Shock waves with lower velocities do not cause injection, while shock waves with higher velocities do not cause stable collapse. The calculations support the shock wave trigger hypothesis for the formation of the Solar System, although the injection efficiency in existing models is lower than desired. When cooling by appropriate molecules is included, shock waves with velocities in the range of 5 to 70 km/s are capable of causing the gravitational collapse of stable, dense cloud cores, as well as injecting shock wave material into collapsing cloud cores.

## 2. Evolution of two close heliocentric orbits of gravitationally interacting bodies

More details on the evolution of two close heliocentric orbits of gravitationally interacting bodies can be found in Chapter 2 of the monograph [78]. This chapter is based on articles [1, 16-17] and preprints [218-219, 232]. Studies of the interaction of two planetesimals (or a planet and a planetesimal), which is an elementary process in the protoplanetary disk, allow us to better understand the process of evolution of this disk, estimate the feeding zones of growing planets and the influence of planetesimals on the evolution of close orbits of other planetesimals. Planetesimals are the bodies from which planets formed. Planetesimals were formed in the protoplanetary disk by the compression of rarefied gas and dust condensations. The formation of planetesimals was discussed in section 2.3 of the review [59]. The "Greek" and "Trojan" asteroids of Jupiter move in a 1:1 resonance with the motion of Jupiter in the vicinity of the Lagrange points $L_4$ and $L_5$. The "Greeks" move around Jupiter 60° ahead, and the "Trojans" 60° behind.

The mutual gravitational influence of two material points (MPs) moving around a massive central body, the Sun, in the same plane along initially close orbits was taken into account by numerically integrating the equations of motion. Material points have masses but no dimensions. The evolution of the orbits of about a thousand pairs of gravitationally interacting objects-material points (MPs) moving around the Sun was investigated by numerically integrating **the planar three-body problem**. Characteristic orbital changes were studied for the following types of orbital evolution: motions around triangular libration points along tadpole-shaped and horseshoe-shaped synodic orbits (N- and M-types), the case of possible close encounters (up to the radius of the sphere of action) of objects (A-type), and chaotic changes in orbital elements, in which close encounters of MPs are impossible (C-type). The ratio $\mu_1$ of the mass of a larger MP to the mass of the Sun varied from $10^{-9}$ up to $10^{-3}$, not only the restricted three-body problem was considered, but also the case of equal masses of MPs ($\mu_1$=$\mu_2$, $\mu_2$ is the ratio of the mass of the smaller MP to the mass of the Sun). The time interval under consideration in most series of calculations was equal to 2500 or 25,000 revolutions of MPs around the Sun.

In the N-type, the graphs of the changes in the semi-major axes of the MP orbits over time *t* have an N- or I-shape. In this case, in synodic coordinates (rotating around the Sun with an angular velocity equal to that of the first MP), the orbit of the second MP encompasses one triangular libration point and, with near-zero orbital eccentricities *e,* has a tadpole (sickle) shape. The apex of these coordinates lies at the Sun, one of the axes is directed toward the first MP, and the second coordinate axis is perpendicular to the first axis.

In the case of the M-type, the graphs of the dependence of the semi-major axes of the orbits on time *t* have an M- or U-shaped form. In synodic coordinates, the orbit of the second MP covers both triangular libration points. In the case of initially circular sidereal orbits, the synodic orbit has a horseshoe-shaped form. The distance from the Sun to an MP, moving in an eccentric orbit, changes during its revolution around the Sun from $a(1\text{-}e)$ to $a(1+e)$. Therefore, for large eccentricities *e,* synodic orbits have a more complex form than for initially circular sidereal orbits [218]. Even for initial zero orbital eccentricities in the case of $\mu_1=\mu_2=10^{-3}$, in synodic coordinates the horseshoe orbit self-intersected several times during the synodic period of revolution at an angle φ between the directions from the Sun to the MP close to 180° [218]. Analytical formulas for the dependence of the difference in the semi-major axes of the MP orbits on the angle with the vertex in the Sun between the directions to the MP are considered in [16, 218].

Close encounters of bodies (up to the Hill sphere radius $r_s=a(\mu_1/3)^{1/3}$ or the sphere of influence radius $r_s=a(\mu_1)^{2/5}$) are possible only for the A-type. In this case, the changes in the orbital elements are chaotic, and the semi-major axes of the MP orbits are the same at certain moments in time. For the C-type, the orbital elements change chaotically, but there are no close encounters of MPs, and the values of the semi-major axes of the MP orbits cannot become the same. For the P-type, *a* and *e* change periodically, and the synodic orbit of the second MP engulfs the Sun in the same way as with chaotic changes. For this type, a large number of subtypes can be distinguished, each of which is characterized by its own interrelations of changes in the orbital elements. For the vicinity of the 5:2 resonance with the motion of Jupiter, these interrelations are considered in the next section.

Over time, some N orbits may transition to M orbits, and C and M orbits to A orbits. After time $t$>100$T_s$ ($T_s$ is the synodic period of revolution) such transitions between orbital types are rare. In the case of **initially circular orbits** ($e_o=e^o_2=e^o_1=0$, $e^o_1$ and $e^o_2$ are the eccentricities of the MP orbits) for the initial angle with the vertex in the Sun between the directions to the objects $\varphi_o=60^o$, $10^{-9}\le\mu_1\le10^{-4}$ and the considered time interval $T_o\sim10^4t_o$ ($t_o$ is the time of one revolution of the first MT around the Sun), the maximum values of $\varepsilon_o=(a^o_2-a^o_1)/a^o_1$ (where $a^o_1$ and $a^o_2$ are the initial values of the semi-major axes of the MT orbits) for N-, M-, A- and C-types $N_t$ were obtained [16] to be equal to $\alpha=(1.63\text{-}1.64)\mu^{1/2}$, $\beta=(0.77\text{-}0.81)\mu^{1/3}$, $\gamma=(2.1\text{-}2.45)\mu^{1/3}$ and $\delta=(1.45\text{-}1.64)\mu^{2/7}$, where $\mu=\mu_1+\mu_2$. For other values of $\varphi_o$, the values of α, β, and $\delta$ are generally smaller. In the case of $\mu_1=10^{-3}$, M-orbits are unstable, for $\varphi_o=60°$ it was obtained that $\alpha=\beta\approx0.5\mu^{1/3}$, $\gamma=(1.8\text{-}2.1)\mu^{1/3}$ and $\delta=(2.3\text{-}2.4)\mu^{2/7}$, and for other $\varphi_o$ the values of $\delta$ could be smaller by a third. For $\mu_1\leq10^{-5}$ the values of $\gamma$ and $\delta$ with varying $\varphi_o$ generally differed by no more than 10%. In the protoplanetary disk, each planetesimal could closely approach several neighboring planetesimals.

At $10^{-9}\leq\mu_1\leq10^{-4}$ and $\varphi_o=60°$, the fraction of M orbits, i.e. the ratio $(\beta-\alpha)/\delta$, varies between 0.18 and 0.25. The smaller $\mu_1$, the smaller are the fractions of N and A orbits and the greater is the fraction of C orbits. At $\mu_1=\mu_2=10^{-9}$, the fractions of C and A orbits are approximately equal ($\gamma-\beta\approx\delta-\gamma$). In a number of cases, for some $\varepsilon_o$, belonging to a subregion lying within the interval $[\beta, \gamma]$, C orbits were obtained, and for some $\varepsilon_o$, located in the range $[\gamma, \delta]$, the orbital elements changed periodically.

Motion around triangular libration points was studied not only numerically but also analytically. For $\varepsilon_o=0$, the minimum values of $\varphi_o$ corresponding to the N- and M-types were found to be close to 0.4 and $4\mu^{1/3}$ radians, respectively. For these types, the graphs of changes in the semimajor axis $a$ of the orbit over time are N- or M-shaped. The smaller is $\mu$, the closer the M-shaped changes in $a$ are to U-shaped changes. The periods of changes in $a$ over time for N- and M- orbits are maximum for $\varepsilon_o\approx\alpha$ and can be 2-6 times greater than $T_{sf}$, where $T_{sf}$ is the synodic period of revolution for the MPs moving along fixed initial orbits.

In the case of close encounters of objects (A-type), the graphs of changes in orbital elements over time, obtained for different ($\leq10^{-7}$) values of $\varepsilon_r$ integration accuracy per step, are close to each other only until the first very close encounter of the objects. However, the nature and limits of changes in orbital elements for these graphs are approximately the same. Varying $\varepsilon_r$ has the same effect on changes in orbital elements as varying the initial positions of the MPs in orbits.

In the case of $\mu_1\leq10^{-5}$ and $e_o=0$, the maximum eccentricities $e_{max}$ usually do not exceed $(7\text{-}8)\mu_1^{1/3}$ at $\mu_1>>\mu_2$ and $(4\text{-}6)\mu_1^{1/3}$ at $\mu_1=\mu_2$ for the A-type, and do not exceed $(4\text{-}6)\mu_1^{1/3}$ at $\mu_1>>\mu_2$ and $4\mu_1^{1/3}$ at $\mu_1=\mu_2$ for the C-type. When MPs move inside the Hill sphere, the osculating eccentricities of their heliocentric orbits can be significantly greater than these values of $e_{max}$. The maximum distance of MPs from the Sun for the A- and C-types reached $1+1.5\delta$ and $1+2\delta$ at $\mu_1=\mu_2=10^{-9}$ and $1+5.6\delta$ and $1+6.6\delta$ at $\mu_1=10^{-5}>>\mu_2$, respectively.

The data presented above are for initially circular MP orbits. For larger initial orbital eccentricities, the values of $\beta$ are smaller, the values of $\gamma$ and $e_{max}$ are generally larger, and $\alpha$ is almost the same as for smaller initial orbital eccentricities of MPs. For certain orbital orientations for type M, the amplitude of long-period variations in $e$ can exceed $e_o$. The obtained results make it possible, in some cases (e.g., for initially circular orbits), to determine, from the initial data, the nature and limits of changes in the orbital elements of two gravitationally interacting objects.

In a number of cases (especially for $\mu_1\sim10^{-5}>>\mu_2$) for the A-type, MPs or bodies were found to **enter resonant orbits**. Usually, after several hundred revolutions of the bodies around the Sun, these resonant relationships were violated. For most of the resonances considered ($T_2$:$T_1$ = 2:1, 12:5, 13:5, 4:5, 6:5, 7:5, $T_i$ is the period of revolution of the $i$ -th body around the Sun), the semi-major axes and eccentricities of the orbits changed periodically with a small amplitude, and the changes in the longitude of the perihelion were small (Figs. 2.1 and 2.4b in [78]). The motions around the resonances in these cases are oscillations around periodic solutions, which are closed curves in synodic coordinates. In the variant of the calculations presented in Fig. 2.4b in [78], at resonances of 6:5 and 5:4, the semi-major axis of the orbit of the smaller body changed insignificantly, the eccentricity increased almost monotonically, and the longitude of the perihelion decreased. In these cases, in synodic coordinates, the curve around which small oscillations occurred was not closed.

The set of initial data for which, from the initial moment of time, the motion of bodies (or MP) is close to periodic motion is small, but for sufficiently large values of the mass of the larger body, exits to resonant orbits are quite frequent. For $\mu_1=10^{-5}$ and non-zero identical initial eccentricities of the orbits, in six of the ten variants given in [219] during the evolution over more than 200 revolutions of the first body, the average motions of the bodies were resonant. For smaller values of $\mu_1$, exits to resonant orbits were obtained less frequently and the synodic period of revolution was significant. For example, in the case presented in Fig. 2.2g in [78], with the number of revolutions around the Sun (the central body) between 16,000 and 17,500, the bodies were in resonance 19:18. In this case, the values of the eccentricity and longitude of the perihelion remained almost unchanged. Therefore, bodies (for example, comets) whose orbits intersect the orbit of a giant planet can be in resonance with this planet for a significant part of the time if the disturbances from other planets are relatively small.

## 3. Evolution of asteroidal orbits at resonances with Jupiter

Evolution of asteroid-type orbits at resonances of 2:5, 1:3 and 1:2 with the motion of Jupiter was considered by Ipatov in preprints [220, 229-230, 213b], articles [6, 8, 14, 29, 66] and in Chapter 3 of the monograph [78]. These ratios (2:5, 1:3 and 1:2) represent the ratios of the average motions of Jupiter and the asteroid (the ratio of the periods of revolution around the Sun of the asteroid and Jupiter). Such ratios were considered in the USSR and Russia when Ipatov's articles on resonances were published. In foreign literature, as well as in modern publications in Russia, the inverse ratios (5:2, 3:1 and 2:1) are usually considered – the ratios of the periods of revolution of Jupiter and the asteroid. Therefore, unlike my earlier works, below I will consider the ratios 5:2, 3:1 and 2:1, and not 2:5, 1:3 and 1:2. The main focus was on the 5:2 resonance. The results of these calculations were used to explain the origin of the Kirkwood gaps. 5:2 and 3:1. Kirkwood gaps are regions of semi-major axes of orbits that correspond to minima in the asteroid distribution. For these gaps, the ratio of the periods of revolution of Jupiter and the asteroid around the Sun is close to 3:1, 5:2, 2:1, or 7:3. Currently, the most popular hypothesis for the origin of the Kirkwood gaps is that, during the course of their evolution, the eccentricities of the resonant orbits of asteroids reached values at which the perihelia of their orbits lay inside the orbit of Mars. Since the rates of change of the longitude of the perihelion and the longitude of the ascending node of the orbit are different for the asteroid and Mars, such asteroids could leave the gap due to close encounters with Mars. Wisdom in 1982 for the 3:1 gap and Shidlichovsky in 1987 for the 5:2 gap put forward this hypothesis, based on studies of chaotic orbits. Ipatov [6, 14, 230] independently made a similar assumption about the migration of asteroids moving in quasi-periodic orbits from these gap s. His assumption was based on the results of a preprint [220] from 1980, in which, in the case of commensurabilities of 5:2, 3:1, and 2:1, for a number of test asteroids with quasi-periodic changes in eccentricities *e,* an increase in *e* from 0.15 to 0.75, 0.45, and 0.35, respectively, was obtained.

By numerically integrating the full equations of motion of the three-body problem (Sun-Jupiter-asteroid) using the Bulsto method (suggested by Bulirsh and Stoer) I studied the time variations of the orbital elements of more than 500 test asteroids [6, 8, 14, 220, 229-230]. It was noted that the graphs of the changes in the orbital elements over time were the same in calculations with an integration step accuracy of $10^{-8}$, $10^{-9}$ or $10^{-10}$. The initial orbital eccentricities of the asteroids were usually equal to 0 or 0.15, but could reach 0.3. The initial orbital inclinations were usually equal to 0 or 10°, but could reach 30°. For the first time, it was found that in the case of a 5:2 commensurability, the maximum (for various initial orbital orientations) region of initial values of semi-major axes and orbital eccentricities, at which test asteroids reach the orbit of Mars in $10^5$ years, is close to the region free of real asteroids and differs little from a similar region at which the orbit of Earth is reached (Fig. 1). The dependences of these regions (in calculations and observations) on the orbital inclinations of the asteroids are also approximately the same. Many test asteroids that reached the orbit of Mars during the evolution of the orbit also reached the orbit of Earth. The obtained results indicate that close encounters of asteroids with Mars and Earth could be one of the causes of the formation of the 5:2 Kirkwood gap in the asteroid belt. The effective radius of Earth is usually at least twice the effective radius of Mars. The effective radius of a planet is the impact range at which the planet is reached (when a body enters the planet's Hill sphere, it collides with the planet if the minimum distance between the relative velocity line and the planet's center of mass is less than the effective radius). Therefore, meteorites migrating from the 5:2 gap may have fallen on Earth more often than on Mars. Sources of replenishment for the Apollo and Amor groups, as well as of meteorites falling on Earth, could have included bodies from the 3:1, 2:1, 5:2 gaps and from the inner asteroid belt.

The considered test asteroids reached the orbits of Mars and Earth, as a rule, with certain types of relationships between long-period variations in the asteroid's eccentricity and the difference in the longitudes of the perihelion of the asteroid's orbit and Jupiter's. Various types of such relationships were studied in detail. For each of these types, the limits of variations and the periods of long-period variations in the orbital elements were determined. For small eccentricities and orbital inclinations, the relationships between the periods of variations in the eccentricity, inclination, argument of perihelion, and longitude of the ascending node of the asteroid's orbit were considered for a number of test asteroids. Formulas were derived for the transition from rectangular coordinates to orbital coordinates, free from singularities at zero inclinations (see [230] and paragraph 2 of Chapter 3 in [78]).

The article [29] presents the results of calculations (with several values of integration accuracy per step) of the migration of test asteroids at resonances of 3:1 and 5:2 with Jupiter at initial eccentricities $e_o$ orbits equal to 0.05 or 0.15 and initial inclinations $i_o$ equal to 5° or 10°. In Table 4 of the article [29], the values of the probabilities of collisions of such asteroids (during their dynamic life) with Venus, Earth and Mars were, respectively, of the order of $10^{-3}$, $2 \cdot 10^{-3}$ and $5 \cdot 10^{-4}$ at $e_o$=0.15, $i_o$=10° and resonance 3:1, and of the order of $10^{-4}$, $3 \cdot 10^{-4}$ and $10^{-4}$ at $e_o$=0.15, $i_o$=10° and resonance 5:2. At $e_o$=0.05 these probabilities could be several times smaller. Figure 5 of article [29] shows the distribution of migrated asteroids by their semi-major axes and orbital inclinations for $e_o$=0.15, $i_o$=10° and the 3:1 resonance. From this figure, in particular, it follows that the semi-major axes of the orbits were mainly in the range from 2 to 3 AU, but there were also many values between 0.5 and 10 AU. Orbital inclinations could reach 180°.

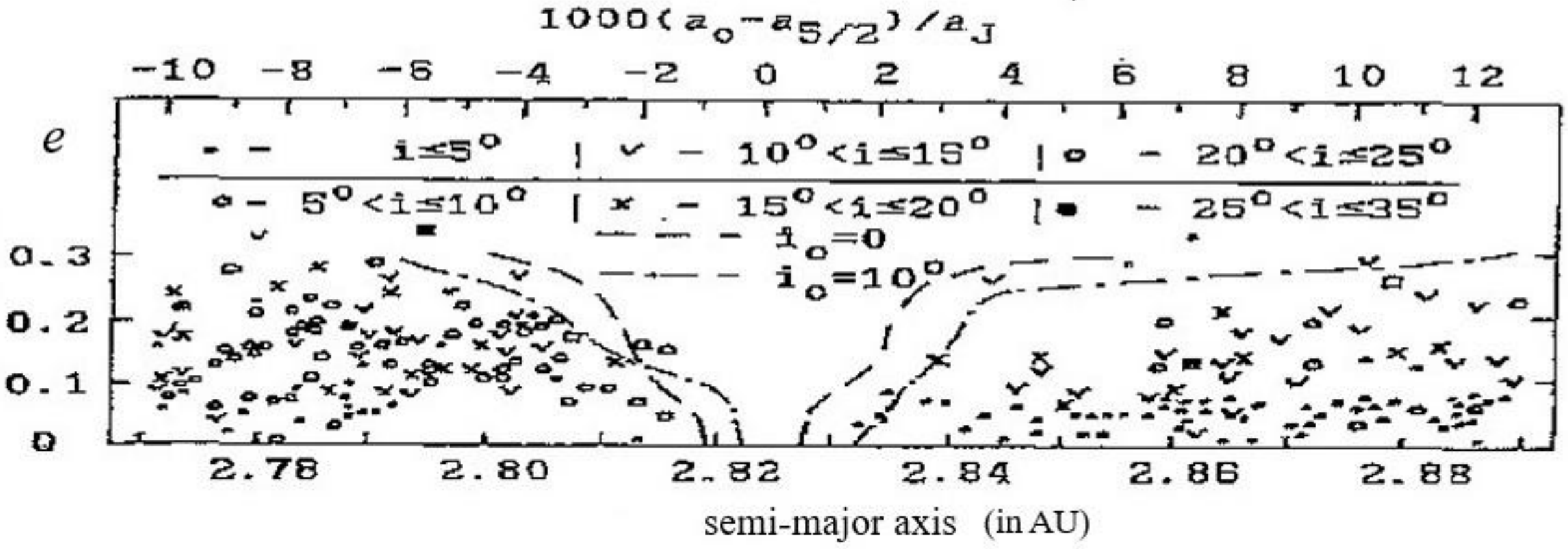

**Figure 1.** Graph of the distribution of asteroids by eccentricities and semi-major axes of their orbits in the vicinity of the Kirkwood hatch 5:2. The dotted and dash-dotted lines mark the boundaries of the maximum range of values of $\Delta a_o = a_o - a_{5/2}$ and $e_o$, at which, for some initial orbital orientations and initial positions of test asteroids in these orbits during the evolution, the maximum eccentricity of the asteroid orbit in the calculations exceeded 0.41, in the flat case and with an initial inclination $i_o$ orbits equal to 10°, respectively. $a_o$ and $e_o$ are the initial values of the semimajor axis $a$ and the orbital eccentricity of the test asteroid, and $a_{5/2}$ – the value of $a$ corresponding to the 5:2 resonance. The upper $x$- axis shows the values of $1000(a_o-a_{5/2})/a_o$ (deviations from resonance), while the lower axis shows $a_o$. Different symbols correspond to different inclinations of the asteroids' orbits. This figure shows that the size of the Kirkwood hatch coincides with the size of the region of initial eccentricities and semi-major axes of the asteroids' orbits, at which the orbit of Mars is achieved during evolution.

## 4. **Modeling the evolution of orbits of celestial bodies using the method of spheres**

Until the 21st century, computer simulation of the evolution of disks consisting of hundreds of gravitating bodies moving around the Sun could not yet be carried out on time intervals of tens to hundreds of millions of years by numerically integrating the equations of motion of the $N$-body problem. In this case, it was possible to use approximate methods for taking into account the gravitational interactions of bodies, and, in particular, the method of spheres. In the method of spheres, it is assumed that outside the sphere under consideration (usually the sphere of action), bodies move around the Sun in unperturbed Keplerian orbits, while inside the sphere, the relative motion is considered within the framework of the two-body problem. In Ipatov's publications [1-5, 7, 9-11, 13, 15, 17-20, 216-217, 221, 225-228] in 1978-1995, the results of calculations in which the mutual gravitational influence of bodies was taken into account using the method of spheres of action were repeatedly used. The radius of the planet's sphere of influence (the region of space in which the planet's relative perturbations are greater than the Sun's relative perturbations) is equal to $r_s = R\cdot\mu^{2/5}$, where $R$ is the planet's distance from the Sun and μ is the ratio of the planet's mass to the Sun's mass.

Below, the algorithm of the sphere method used is briefly described, the results of calculations obtained by this method are compared with the results of numerical integration of the equations of motion, and the characteristic changes in the elements of the orbits of bodies during a single approach to the radius of the sphere of action are investigated. More details on this can be found in [16, 216, 221, 224] and in Chapter 4 of the monograph [78]. Programs for modeling the evolution of disks of bodies merging during collisions using the sphere of action method are given in the IPM RAS report No. O-1211 for 1985. The algorithms given in this report were designed for a PS-2000 multiprocessor computer, but many subroutines were also used for calculations on a single-processor BESM-6 computer.

In contrast to the Öpik's scheme used by other authors, in the algorithm proposed by Ipatov [216] (see also paragraph 2 in Chapter 4 in [78]), **the characteristic time elapsed to approach** (to the radius $r_s$ of the sphere under consideration) of two bodies depended not only on their average relative velocity and the angle $\Delta i$ between the planes of their orbits, but also on the synodic period $T_s$ of the revolution of these bodies and on the sum of the angles $\Delta\varphi$ with the apex in the Sun, within which the distance between the orbit of the "first" body and the projection of the orbit of the "second" body onto the plane of the orbit of the "first" body is less than the radius $r_s$ of the sphere under consideration. Along with the case of a fixed angle $\Delta i$ between the planes of the orbits of the bodies, the case of variable $\Delta i$ was considered for the first time. It was found that the characteristic time before the collision of bodies in the case of variable $\Delta i$ can be several times smaller than for a fixed $\Delta i$ equal to the average value of the variable $\Delta i$. In contrast to Öpik's scheme, the proposed formulas are more convenient for analytical studies of the dependence of the evolution of a disk on the number of bodies comprising it. They also make it possible to take into account some more subtle effects of the interactions of bodies. In [29], a coefficient was added to this probability calculation algorithm that takes into account the difference in velocity at the moment of approach at the point under consideration from the average velocity of motion along the orbit.

Such formulas for the characteristic time to approach to the radius of a sphere (along with the calculation of the probability of collision within the sphere) were also used to estimate the probability of collisions of bodies (dust particles) over a long time interval based on arrays of orbital elements of migrated planetary bodies, obtained by numerical integration of the equations of motion. For example, such estimates were conducted in [28-31, 33, 53, 61, 86-94] when studying the probabilities of collisions of bodies with planets when these probabilities are small and when the numerical integration of the orbital evolution of even hundreds of bodies does not result in their collisions with planets.

The results of calculations using the sphere-of-action method were compared with the results of numerical integration of the equations of motion and demonstrated the success of applying the sphere-of-action method to the problems under consideration. In particular, **a comparison was made between the results** of a study of the evolution of the heliocentric orbits of two gravitationally interacting objects, **obtained by numerical integration, and similar results obtained using the sphere-of-action method** [221]. Based on this comparison, the limits of applicability of the sphere-of-action method were investigated when using spheres of action and spheres of larger radii. For small relative velocities $v_r$ of approaching bodies, the sphere-of-action method poorly approximates the actual motion of bodies within the sphere of action. Moreover, even with a fairly good approximation of motion within the sphere of action, the velocities of bodies leaving the sphere of action may differ significantly from the actual ones. However, in a number of cases, when approximately studying the evolution of the orbits of bodies over a large number of approaches, it is possible to use the sphere-of-action method even for small values of $v_r$.

For a number of initial data, Ipatov [221] compared several dozen graphs of changes over time in the elements of the heliocentric orbits of two gravitating bodies, obtained by modeling the evolution using the method of spheres, with similar graphs obtained by numerically integrating the equations of motion. Analysis of these graphs showed that using the method of spheres it is possible to obtain the limits and main trends in the changes in the elements of the heliocentric orbits. It was found that with an appropriate choice of the radius of the sphere used, the main trends in the evolution and the limits of changes in the elements of the orbits of two gravitating objects moving around the Sun in intersecting orbits are approximately the same for such graphs. In the case when the greatest eccentricity of the heliocentric orbits of the approaching objects $e_m>2\mu^{1/3}$, one should use spheres of action, and for smaller eccentricities, spheres of radius equal to $r_s \approx 2.4R\mu^{1/3}$, where $R$ is the distance of these objects from the Sun, $\mu$ is the ratio of the sum of the masses of the objects to the mass of the Sun.

In a preprint [216] in 1978, the results of calculations of the evolution **of flat disks** of one hundred identical material points (MPs) moving around a central body (the Sun) in initially circular orbits were presented. The total mass of all MPs varied from $10^{-5}$ to 0.1 of the mass of the Sun. In the course of evolution, these disks expanded, the average eccentricity of the orbits of the bodies increased, and the number of MPs with small $e$ was small. The evolution of a disk whose mass was $10^5$ times greater than the mass of the asteroid belt was also studied, and the mass distribution of large MPs was close to the mass distribution of large main-belt asteroids. The average eccentricity $e_{av}$ of the orbits of the bodies was initially equal to 0.15 and increased during the course of evolution. At the same time, the eccentricities of the orbits of the largest MPs decreased. Along with the general expansion of the disk, the concentration of massive MPs by $a$ was obtained. As in the case of disks of equal masses, the orbital eccentricities of the outer edge of the disk (~0.4-0.6) significantly exceeded $e_{av}$ (0.15-0.2). Estimates show that for the average orbital eccentricity of asteroids to increase from zero to 0.15 (the modern average eccentricity) during the existence of the Solar System, the disk mass must significantly exceed the mass of the asteroid belt (taking into account the gravitational influence of only disk bodies) [216]. This, as well as the large (34.8°) orbital inclination of Pallas, one of the largest asteroids, testify in favor of the formation of the asteroid belt under the influence of bodies from the zones of the giant planets.

**The evolution of spatial disks** of bodies [3, 11, 15, 67, 78, 85, 224-225, 227-228, etc.] was investigated by appropriately reducing the spatial evolutionary problem to a flat one based on the results of computer simulations of the evolution of flat disks of high-density test bodies [3, 225]. The density of test bodies was selected in such a way that the changes in the eccentricities of orbits between collisions of test bodies (and hence the eccentricities themselves) were approximately the same as in the evolution of the considered spatial disks of solid bodies (at a density of bodies close to the modern density of the planets). Fine effects of the evolution of spatial disks cannot be obtained using the reduction in question, but for a small number of initial bodies (less than 1000), direct computer simulations of the evolution of spatial disks do not yield them either.

When modeling the evolution of disks of gravitating bodies moving around the Sun using the method of spheres, **the problem arises for selection of pairs of bodies (objects) approaching to the radius of the sphere under consideration**. In the models of the evolution of the protoplanetary circumsolar disk considered in [2-5, 10-11, 13, 15, 20, 216-217, 224, 226-227], during the calculations it was periodically necessary to determine a pair of objects that would be the first to contact (approach to a distance equal to the radius $r_s$ of the sphere under consideration). In [67] and in Appendix 1 in [78], various modifications of the method developed by Ipatov for selecting pairs of contacting objects in the study of the evolution of discrete systems with binary interactions are described. Two approaches to selecting pairs of contacting objects were considered: probabilistic and deterministic.

**In the probabilistic approach,** pairs of approaching objects were selected in proportion to the probabilities of their approaches based on the matrix $p_{ij}$ probabilities of contacts between pairs of objects ([216] and Appendix 1 in the monograph [78]). After modeling each approach, the sum $p_{ij}$ was calculated. For all pairs of current orbits and masses of objects, a random number in the range from zero to this sum was simulated, and the corresponding pair of contacting objects was determined by this number.

**In the deterministic approach**, the pair of objects for which time was selected for which $t_{ij}$ ($p_{ij}$ is inversely proportional to $t_{ij}$) before isolated (from other objects) contact-approach is minimal. A triangular matrix $t_{ij}$ was considered. After contact of $i$ and $j$ objects in this matrix, $i$ and $j$ rows and columns were changed. The task of the algorithm was to select the minimum element in the new matrix $t_{ij}$ by performing the minimum number of operations. The proposed method, called the conditional triangular matrix method, is described in the article [67], in Appendix 1 in [78], in the abstracts [216] and in IPM report No. 1211 for 1985 (IPM report No. 3385 for 1985 discussed the implementation of this method on the PS-2000 multiprocessor computer). It is shown that this method is equivalent to the method of complete enumeration of the elements of the matrix $t_{ij}$. The problem of periodic renumbering of objects due to a decrease in the number of objects under consideration over time, the division of objects into groups, and other algorithm modifications that increase computation speed were also discussed. The calculation results showed that the conditional triangular matrix method (without additional modifications) is 1.5-2 times faster than the virtual contact method used by T.M. Eneev and N.N. Kozlov (IPM preprint, No. 78, 1979).

Initially, Ipatov's calculations used a probabilistic method for selecting pairs of approaching bodies. In publications beginning in 1990 [11, 15, 67, 78, 85, 237, 238, etc.] (as well as in the IPM RAS report No. O-1211 for 1985), calculations made using a deterministic method were considered. The results of the calculations showed that if the number of objects under consideration is large, then with a deterministic selection of pairs of approaching bodies, the characteristic time between successive approaches of the bodies is an order of magnitude shorter than with a probabilistic selection, although in both cases the same formulas for the probability of two bodies approaching to the radius of a sphere were used.

In the article [7] and in paragraphs 5-7 of Chapter 4 in [78], the basic principles of constructing an algorithm for simulating the evolution of a disk consisting of a large number of bodies were discussed. This algorithm allows one to study not only the initial but also the final stages of accumulation of planets from a disk of planetesimals. Based on the calculation results, formulas were obtained that characterize some of the dependencies of the characteristic changes in the orbital elements of bodies and the probability of their collisions during the approach of bodies to the radius of the sphere of action on the masses and orbital elements of the bodies. For the model of the evolution of a disk of approximately identical bodies and the model of the infall of bodies onto a planetary embryo, new formulas were obtained that characterize the number of collisions and approaches (to the radius of the sphere under consideration) of bodies over a certain period of time.

Based on the results of modeling the evolution of disks consisting of hundreds of bodies, Ipatov [3, 7, 228] carried out analytical estimates of the evolution of disks consisting of a large number ($\sim 10^{12}$) of bodies. From these estimates (see also paragraph 2 of Chapter 5 in [78]), it follows, in particular, that the maximum values of the average eccentricity $e_{av}$ of the orbits during the evolution of these disks are no less than during the evolution of disks consisting of hundreds of bodies with the same total mass of bodies. Most of the planetesimals that fell onto the Earth underwent impact evolution. Taking into account the fragmentation of bodies during collisions can increase the formation time of the main mass of the planets several times [7], and a significant increase in the initial number of bodies and taking into account gas drag can reduce the mixing of bodies during the accumulation of planets. The dependences of the evolution time and changes in the average eccentricity on the number of bodies in the disk, their masses and characteristic elements of the orbits are considered.

It was concluded that, during the evolution of disks of various bodies, for a model that takes into account fragmentation and gas drag, the growth times of the largest disk bodies do not exceed (at least by a factor of five) the growth times of bodies for a model of approximately identical initial bodies, provided the disk masses are equal in both cases. Larger bodies grow faster. Therefore, accounting for the differentiation of bodies by mass only strengthens the above statement. Although planetary embryos may be disrupted, melted, and partially vaporized during collisions with large bodies, a significant portion of the fragments may reassemble into a single body under the influence of gravity. However, even with the loss of mass in collisions with large bodies, planetary embryos could still grow overall through the accumulation of smaller bodies.

## 5. Accumulation of the terrestrial planets

It is believed that dust clumps-condensations formed in the solar disk, compressing into planetesimals-solid bodies typically ranging in diameter from kilometers to hundreds of kilometers. These planetesimals participated in the accumulation of terrestrial planets and the embryos of giant planets, which accreted gas. When gas still remained in the protoplanetary disk, the growth of planetary embryos also occurred through the accretion of pebble-like objects accretion. This accretion ended when the planet's embryo reached a certain mass (pebble isolation mass). Accretion of "pebbles" probably played a greater role in the feeding zone of the giant planets than in the feeding zone of the terrestrial planets. The embryos of the giant planets accreted gas. A review of the work on the formation of planetesimals and the accretion of pebble-type objects can be found, for example, in paragraph 14 of Chapter 1 of the monograph [78], in section 2.2 of the article [59] and in the introduction of the article [64]. A discussion of the T.M. Eneev - N.N. Kozlov model, in which planets formed by the collisions of rarefied condensations, was carried out by me in [76]. The actual process of planet formation was very complex and depended on many factors. However, studies based on relatively simple models allow us to make a number of important estimates of the planetary accumulation process.

In studying the accumulation of the terrestrial planets, along with analytical estimates, calculations were made of the evolution of the disks of gravitating bodies that merge during collisions. After the mid-1990s, the gravitational influence of large bodies was taken into account in such calculations by integrating the equations of motion of the bodies. A review of studies on the accumulation of terrestrial planets can be found in Section 3 of the article [59]. In earlier studies, the method of spheres was used to account for the gravitational influence of bodies. In this method (see the previous section for more details), it was assumed that the bodies moved along unperturbed heliocentric orbits outside the spheres under consideration, and the motion of bodies inside the sphere was considered within the framework of the two-body problem. When modeling the evolution of the disks of bodies corresponding to the feeding zones of planets, Ipatov [2, 3, 13, 15, 78, 224] used the method of spheres of action. When selecting pairs of bodies approaching the sphere of action, a "probabilistic" algorithm was initially used, in which pairs of bodies approaching to a distance equal to the radius of the sphere of influence were selected proportionally to the probability of their approach. Later, an efficient "deterministic" algorithm was developed, in which for a selected pair $i$ and $j$ of approaching bodies, the moment $t_{ij}$ the closest isolated approach to the radius of the sphere of action is minimal [67, 78].

Using the method of spheres of action, Ipatov [2, 3, 13, 15, 78, 224] considered the evolution of disks, each of which contained up to 1000 gravitating bodies in the feeding zone of the terrestrial planets, which merged during collisions. The initial distances of the bodies from the Sun varied from 0.36 or 0.4 AU to 1.2 AU, and their total mass was about $1.9m_E$, where $m_E$ is the mass of the Earth. Calculations of the evolution **of flat disks** showed [2, 219] that in the case of nearly circular initial orbits of the bodies, the number of planets formed is greater than four, and the actual number of terrestrial planets is obtained only for initial orbital eccentricities of the bodies $e_0$=0.35. For small initial eccentricities, the actual number of planets can be obtained in a model of the evolution of spatial disks, and in one of the calculation variants, four planets with masses greater than $0.046m_E$ were formed. The actual number of planetesimals was far greater than 1,000, and only a fraction of the material from two colliding bodies formed a new body. However, the computing power of computers in the last century did not allow for more complex models.

Accounting for the fragmentation of colliding bodies could have slightly increased the accumulation time of planets.

In calculations of **the evolution of spatial disks of bodies corresponding to the feeding zone of the terrestrial planets** [3, 13, 15, 78, 224], the initial eccentricities of the orbits of the bodies were taken to be equal to $e_0$=0.02, and it was shown that such eccentricities are achieved fairly quickly when taking into account the mutual gravitational influence of bodies at distances smaller than the radii of their spheres of action. An example of the evolution of such disks is shown in Fig. 1-2. The average eccentricity $e_{av}$ of the orbits of bodies during the course of evolution exceeded 0.2, and in a number of variants at certain moments in time was greater than 0.4. For example, in one of the variants with $e_0$=0.02 and 960 initial bodies, $e_{av}$ was equal to 0.09, 0.20, and 0.35 when the number of bodies in the disk was 500, 250, and 100, respectively. Moreover, bodies located at the disk's edges, with semi-major axes $a$<0.4 AU and $a$>1.2 AU, had large average orbital eccentricities. The orbits of some planets with masses on the order of those of Mercury and Mars acquired eccentricities late in their evolution close to their present-day orbital eccentricities. It should be noted that the increase in the orbital eccentricities of Mercury and Mars (and the orbital inclination of Mercury) could have been caused by the gravitational influence of bodies entering the feeding zones of the terrestrial planets from the feeding zones of the giant planets, and not solely by the influence of large bodies from the feeding zones of the terrestrial planets. Moreover, some of these alien bodies may not have collided with bodies in this feeding zone, but merely gravitationally perturbed their orbits. The high iron content of Mercury's core is usually explained by the loss of most of the mass of its silicate shell through high-speed impacts. At the same time, some of the planetesimals in the vicinity of Mercury's orbit, which passed relatively close to the Sun before their collisions with the embryo of Mercury, could have lost some of the silicate part of their composition during such passages near the Sun and thereby influenced the high iron content in the core of Mercury.

The masses of the embryos of the unformed terrestrial planets could exceed 0.05$m_E$, which helps explain both the tilt of the rotation axis and the period of the axial rotation of the Earth, when such embryos fell onto the growing Earth. According to my estimates, the formation time of 80% of the mass of the largest rocky planet (analogous to the Earth) did not exceed 10 million years, while the total evolution time of the disks of gravitating bodies was on the order of 100 million years. In calculations using the sphere-of-influence method, formation times of the bulk of the planets of the order of 1-10 million years were obtained by considering the "deterministic" method of selecting pairs of approaching bodies, when, when modeling the approach, a pair of bodies with the minimum time before approach was selected [78, 67]. Meanwhile, when using the "probabilistic" method of selecting pairs of approaching bodies, the formation times of the bulk of the planets were almost an order of magnitude longer. That is, the probabilistic approach is less consistent with real evolution. However, even with this method, the maximum values of the average eccentricity and mass and orbit of the resulting planets were approximately the same as with the deterministic method of selecting pairs of bodies.

The time during which the number of bodies in the disk decreased from $N_o$ to $N$, was usually approximately half the time during which the number of bodies in the disk decreased from $N_o$ to $N/2$, and the main time of the evolution of the disks considered fell on the last stages of planet accumulation. Therefore, it was concluded that for $N_o = 10^{12}$ the total time of disk evolution is approximately the same as for $N_o = 10^3$, however, taking into account the fragmentation of bodies during collisions can increase the time of formation of the main mass of the planet several times [7, 78].

As demonstrated by later results of numerical integration of the equations of motion obtained by a number of authors, in which the mutual gravitational influence of bodies was taken into account by integrating the equations of motion, the deterministic approach quite satisfactorily (and better than the probabilistic approach) reflects the actual evolution of the disks of bodies and the formation times of planets. In recent years, the initial number of simulated bodies in disks has been in the thousands, which naturally improves the statistics. The main conclusions about the accumulation of terrestrial planets, based on calculations using the sphere-of-influence method and by numerical integration of the equations of motion, are roughly the same.

The accumulation of the terrestrial planets was also discussed in paragraph 1 of chapter 6 of the monograph [78], in section 3 of the article [59], and in the article [64]. Tables B and C in the article [85] present, for two calculation variants, the composition of the planets that formed from four groups of bodies that were initially located at distances of 0.4–0.6, 0.6–0.8, 0.8–1.0, and 1.0–1.2 AU from the Sun, as well as the semimajor axes and orbital eccentricities of these planets. For planets with masses greater than 0.1 $m_E$, the proportions of bodies that came into the planets from these four distance ranges were in the intervals of 0.11–0.22, 0.19–0.25, 0.19–0.34, and 0.32–0.40; that is, these proportions did not differ significantly for planets that formed at different distances from the Sun. Each of the planets included bodies called planetesimals, which were initially located at the same distance from the Sun. The composition of the large planets was quite close to the composition of the original zone of 0.4-1.2 AU, that is, the planetesimals were quite strongly mixed during the course of evolution, if we take into account the mutual gravitational influence of all the planetesimal bodies. Based on analytical formulas and the results of numerical calculations in [7], it was noted that the main time of accumulation of a planet falls on the final stages of its accumulation. In paragraph 2 of Chapter 6 of the monograph [78], it was noted that when modeling the evolution of disks of bodies in the zone of the giant planets, at some stages of the evolution of these disks, about 1% of the bodies could cross the Earth's orbit.

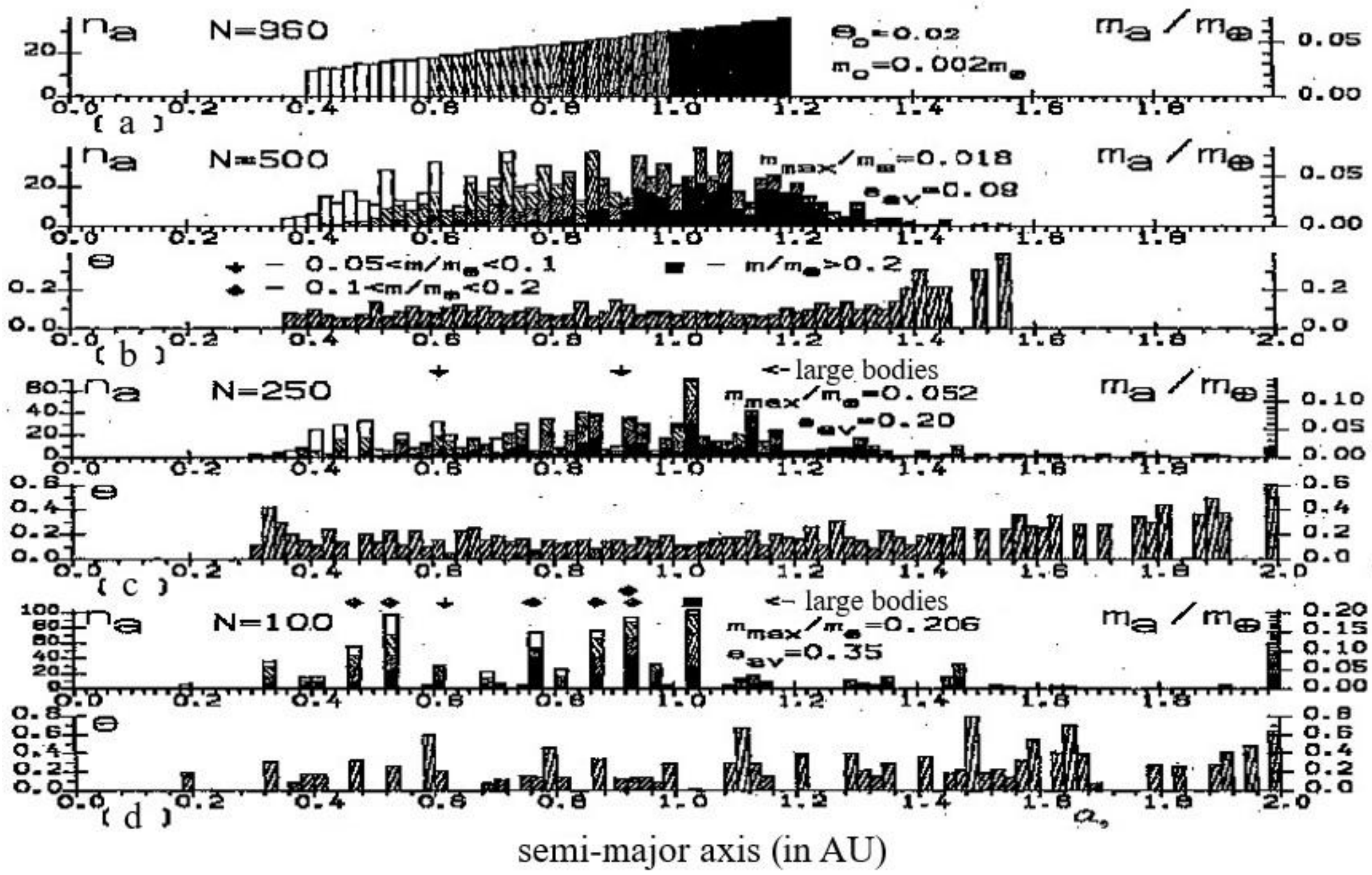

**Figure 1.** Histograms of the distribution of the number of bodies $n_a$ in the initial disk (upper subfigure), as well as the distribution of the masses of bodies and the average eccentricities of the orbits of bodies in the evolving disk (for different numbers N of the remaining bodies) by the semi-major axes of their orbits. The initial disk corresponded to the feeding zone of the terrestrial planets and was located at distances from the Sun from 0.4 to 1.2 AU. The mass of each of the 960 initial bodies was $0.02m_E$, and the eccentricities of their orbits equaled to 0.02. Depending on their initial distance from the Sun, the bodies were divided into four groups. Each group is marked with its own shading in the histograms. The total mass of the bodies in the initial disk was $1.92m_E$. The arrows indicate the semi-major axes of the orbits of bodies whose masses exceeded $0.05m_E$.

Estimates of the time of formation of the main mass of the Earth, on the order of 100 million years, have remained virtually unchanged since the publication of the monograph by V.S. Safronov in 1969 (Safronov V.S., Evolution of the protoplanetary cloud and the formation of the Earth and planets, 1969. Science, 244 p.). Based on the results of modeling the evolution of disks of initially identical bodies that merge during collisions corresponding to the feeding zone of the terrestrial planets [3, 13, 15, 78, 224], as well as on the basis of numerical calculations of the migration of bodies under the influence of planets or their embryos [55] (see also the end of this section), the following **conclusions were made about the accumulation of terrestrial planets:** The inner layers of each terrestrial planet accumulated primarily planetesimals from the vicinity of that planet's orbit. The outer layers of Earth and Venus could have accumulated similar planetesimals from different regions of the terrestrial planets' feeding zones. Earth and Venus could have acquired more than half their mass in 5 million years.

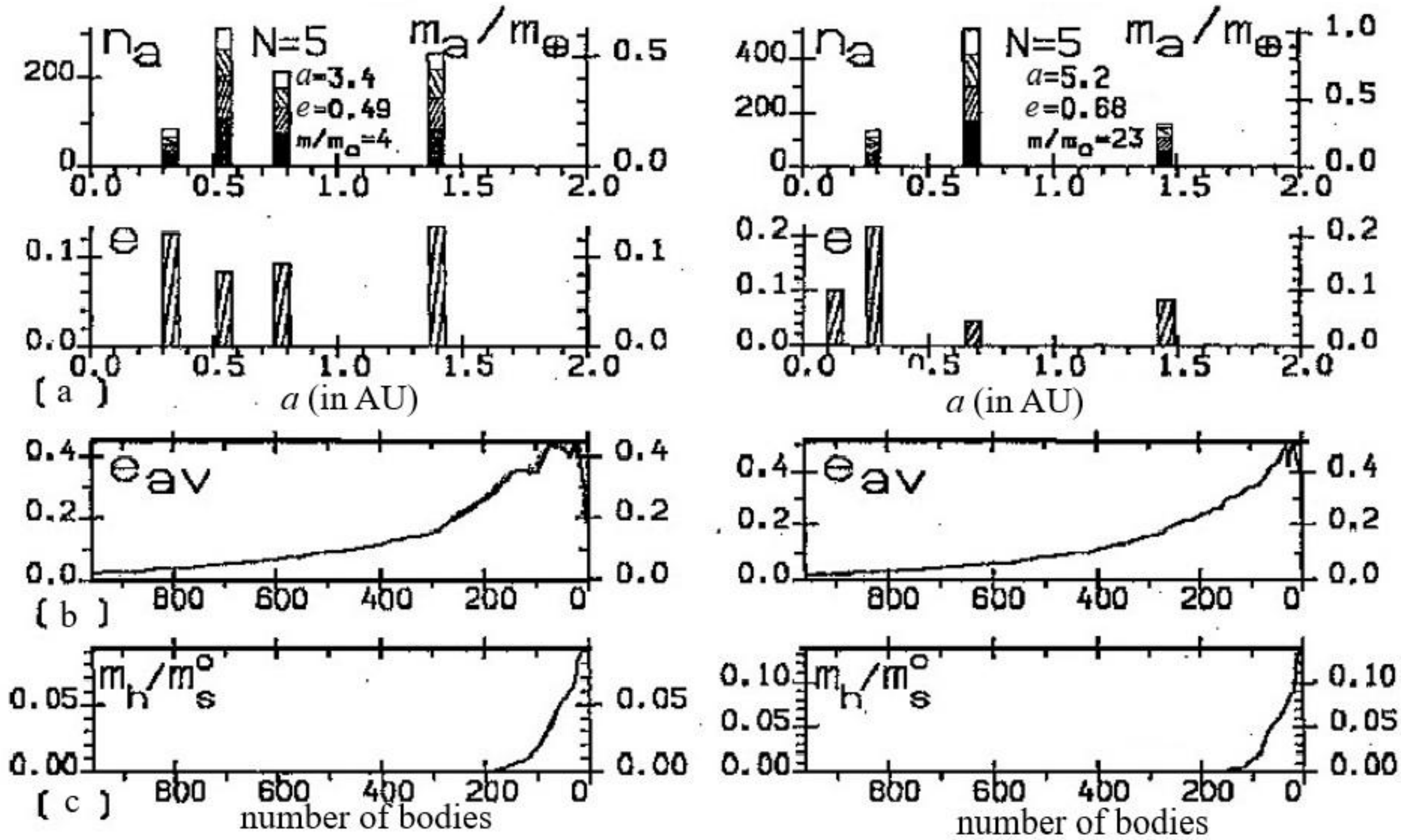

**Figure 2.** Histogram of the characteristics of the formed planets and the dynamics of disk evolution for two calculation variants. The initial data for these disks are given in the previous figure. The upper figures show the masses of the formed planets (on the left - in masses of the initial bodies, on the right - in Earth masses) and the eccentricities of their orbits. The semi-major axis of the orbit of the fifth formed body in both variants exceeded 2 AU. The lower figures show the dependences of the average eccentricity $e_{av}$ of the orbits of the bodies and the ratio $m_h/m_s^o$ of the total mass of the bodies ejected into hyperbolic orbits to the total mass of the initial bodies on the number of bodies in the disk. It is evident that the composition of the formed planets included bodies from different groups of bodies (bodies of different groups were at different distances from the Sun).

It is believed, including results obtained on the basis of the study of Martian meteorites (see review in Section 3.3 in [59]), that the accumulation and solidification of Mars were completed within ~5-10 million years, during which time Mars grew to approximately its present-day size. According to [55], Mars grew more slowly than Earth and Venus, and individual planetesimals in its feeding zone could have remained even after 50 million years. It can therefore be assumed that during the compression of the condensation, a fairly large embryo of Mars with a mass of at least $0.02m_E$ was initially formed, and planetesimals from the feeding zone of Jupiter and Saturn contributed to the more rapid removal of planetesimals from its feeding zone. The formation of a Mars embryo with a mass several times smaller than the mass of Mars as a result of the compression of a rarefied condensation can explain the relatively rapid growth of the bulk of Mars. A similar scenario was proposed in [55] for the formation of the Mercury embryo with the same initial mass. The proportion of planetesimals ejected from the feeding zones of the terrestrial planets into hyperbolic orbits did not exceed 10%. The accumulation of terrestrial planets was also influenced by bodies migrating from the feeding zones of the giant planets and from the asteroid belt.

In section 3.2 of the article [59], the results of modeling **the accumulation of terrestrial planets taking into account the influence of the giant planets, obtained by other authors,** were discussed. The formation and migration of the giant planets are closely linked to the accumulation of terrestrial planets. Planetesimals from the feeding zones of the giant planets, which acquired orbits with small perihelion distances during the evolution of the Solar System, disturbed the orbits of planetesimals and planetary embryos in the feeding zones of the terrestrial planets, as well as of bodies in the asteroid belt, and frequently collided with them. Changes in the orbits of Jupiter and Saturn caused changes in the resonance positions and contributed to the clearing of the asteroid belt, from which some bodies could penetrate into the feeding zones of the terrestrial planets. Therefore, when studying the accumulation of terrestrial planets, the influence of the forming giant planets and bodies in their feeding zones must be taken into account. Various scenarios for this influence have been considered. It has been noted that after 3 million years, the gas disk had decreased in mass by a factor of 20 and had no dynamic influence on the migration of planetesimals.

A number of authors, when studying the formation of the terrestrial planets, considered **the Grand Tack model.** In this model of the early dynamical reorganization of the Solar System, Jupiter, due to interaction with gas in the disk, first migrated toward the Sun to 1.5 AU, and then, after the formation of massive Saturn and the dissipation of gas, began to move together with Saturn back from the Sun, being in a 2:3 resonance with Saturn. As a result of this migration, Jupiter "cleared" the asteroid belt, reduced the amount of material in the feeding zone of Mars, and facilitated the delivery of water to the forming terrestrial planets. For example, it was assumed that Jupiter and Saturn acquired sufficiently large masses of gas over 600,000 years, and during the first 100,000 years they migrated inward from 3.5 and 4.4 AU, respectively, up to 1.5 and 2 AU, and after Saturn's mass increased from $10m_E$ to its present mass, Jupiter and Saturn migrated from the Sun over a period of 500 thousand years to distances of up to 5.25 and 7 AU, respectively. Section 4 of [59] discusses models that consider the subsequent stages of the accumulation of the giant planets and the migration of the embryos of Uranus and Neptune, including **the Nice model**, named after the place where it was created, the Nice Observatory in France. In the Nice model, the trigger for the sharp changes in the orbits of these embryos is believed to be the entry of Jupiter and Saturn into a 1:2 orbital resonance.

The mechanism of migration of the giant planets can explain a number of events in the early history of the solar system, including the presence of numerous trans-Neptunian objects in resonance with Neptune, the formation of the Kuiper belt and the Oort cloud, and a phenomenon known as the late heavy bombardment of the inner solar system (LHB), although this hypothetical mechanism itself is not universally accepted. However, it is used to estimate the contribution of an exogenous source of volatiles to the evolution of the terrestrial planets. It is believed that the delivery of water to Earth by small bodies occurred primarily after these planets had accreted 60-80% of their final mass.

The Nice model was used to study the evolution of asteroid orbits during a drastic change in Jupiter's orbit, which led to a sharp shift in the resonance positions. The probability of asteroid collisions with the Moon was found to be $4\cdot10^{-5}$, while the probability with Earth was 20 times greater. It was concluded that the Nice model provides a good explanation for the formation of Mars and the asteroid belt if such instability occurred within 1-10 million years after the dissipation of the gas disk. In recent years, the formation of the terrestrial planets has been examined primarily based on the aforementioned Nice and Grand Tack models. The first model assumes a sharp change in the orbits of the giant planets when Jupiter and Saturn enter resonance, while the second assumes the migration of Jupiter to the orbit of Mars and back. In [55], it was noted that the features of the formation of the terrestrial planets and the clearing of the asteroid belt can be explained without resorting to the aforementioned models, but on the basis of a relatively smooth decrease in the semi-major axis of Jupiter's orbit and the shift in the positions of resonances due to the ejection of planetesimals by Jupiter into hyperbolic orbits. This model considers the formation of the embryos of Uranus and Neptune near the orbit of Saturn and the migration of these embryos, under the influence of interaction with planetesimals, to the present-day orbits of Uranus and Neptune [11, 15, 78] (see also the next section).

At the beginning of this section, results are presented based on the study of the evolution of disks originally consisting of bodies that merged during collisions. **The migration of planetesimals during the final stages of the formation of the terrestrial planets** was considered by Ipatov in [55]. Calculations were made of the migration of planetesimals within the feeding zone of the terrestrial planets, divided depending on the distance from the Sun into seven regions (0.3-0.5, 0.5-0.7, 0.7-0.9, 0.9-1.1, 1.1-1.3, 1.3-1.5, 1.5-2 AU). The gravitational influence of all the planets was taken into account using the Bulirsh-Stoer method, while in [55] the planetesimals and the planets themselves were considered as material points, and their collisions were not directly taken into account. In a number of variants of the model, instead of the terrestrial planets, their embryos with masses from 0.1 to 0.3 of the masses of modern planets were considered. The arrays of orbital elements of migrated planetesimals obtained in the calculations with a step of 500 years were used in calculating the probabilities of their collisions with the planets, their embryos and with the Moon. Later, calculations similar to [55] were carried out by Ipatov for a model in which planetesimals that collided with planets were excluded from further calculations (the results of these calculations have not yet been published in the article).

In studying the proportions of planetesimals from various zones that entered the planets (or their embryos), in [55] narrower zones from which the planetesimals came were considered than previously, and the change in the composition of the embryos over time was studied, and not only the final composition of the planets. Calculations led to the conclusion that the embryos of the terrestrial planets, whose masses were on the order of one-tenth of the masses of modern planets or less, accumulated mainly planetesimals from the vicinity of their orbits. The inner layers of the terrestrial planets were formed mainly from matter from the vicinity of the planet's orbit. When planetesimals from the feeding zones of Jupiter and Saturn fell onto the embryos of the terrestrial planets when these embryos had not yet acquired the masses of modern planets, matter (including water and volatiles) from this zone could have entered the inner layers of the terrestrial planets and influenced their composition.

With the masses of the Earth and Venus embryos being at least one-third of their present-day masses, the probabilities of impacts of planetesimals formed at distances of 0.7 to 0.9 AU from the Sun onto these embryos differed by no more than a factor of two over the considered time interval $T$>2 million years. Based on the considered model, it was also found that the total mass of planetesimals that migrated from each zone located at distances of 0.7 to 1.5 AU from the Sun and collided with the nearly formed Earth and Venus differed for these planets by no more than a factor of two. The outer layers of Earth and Venus could have accumulated material identical for these planets from different parts of the feeding zone of the terrestrial planets. In the final stages of planetary formation, planetesimals initially located at distances of 1.1 to 2.0 AU from the Sun, could be part of the Earth and Mars in a ratio not very different from the ratio of the masses of these planets.

In the variants in which planetesimals that collided with planets were excluded from the calculations, the evolution time of planetesimal disks was usually equal to several hundred million years. However, in some variants of calculations with small initial eccentricities, individual planetesimals moved in 1:1 resonances with the Earth or Venus and after a time greater than a billion years. In these variants of calculations, with $e_o$=0.05 and $e_o$=0.3, the time interval considered was greater than in [55], and the fraction of bodies that collided with the Sun, with the modern masses and orbits of the planets, was mainly in the range (0.24-0.4) for the initial values $a_o$ of the semi-major axes of the planetesimal orbits between 0.5 and 1.1 AU, and reached 0.75 at $1.5 \le a_o \le 2$ AU. The fractions of planetesimals that collided with Earth and Venus at $a_o \ge 0.7$ AU in the calculation variants excluding planetesimals that collided with planets also differed by no more than a factor of two. For planet masses no greater than half the mass of modern planets and time intervals no greater than 10 million years, no planetesimal collisions with the Sun and their ejection into hyperbolic orbits occurred in most calculation variants.

As in earlier calculations [3, 13, 15, 224] of the evolution of disks of bodies using the spheres of action method, the proportion of planetesimals ejected from the feeding zone of terrestrial planets into hyperbolic orbits did not exceed 10%. At the same time, the probability of a collision with Jupiter of a planetesimal originally located in the feeding zone of terrestrial planets was no more than a few percent of the probability of its collision with Earth, and the probability of a collision with Saturn was an order of magnitude smaller.

The model estimates for the formation of terrestrial planetary embryos presented in [55] were based on a model that took into account the combined gravitational influence of the giant planets and the terrestrial planetary embryos. Additional consideration of the mutual gravitational influence of planetesimals in the feeding zone of the terrestrial planets, as well as planetesimals coming from the feeding zones of the giant planets, leads to increased mixing of matter in the feeding zone of the terrestrial planets, an increase in the probability of collisions of planetesimals with the Sun and their ejection into hyperbolic orbits. With a ratio of the masses of the Earth and Moon embryos equal to 81 and similar to the modern ratio, the ratio of the probabilities of planetesimal fallout onto the Earth and Moon embryos in the variants considered did not exceed 54, and was maximum for embryo masses approximately three times smaller than the modern masses of these celestial bodies.

In the model considered in [55], in which bodies merged with planets during any collisions, the Earth and Venus could have acquired a significant part (more than half) of their mass over 5 million years. In particular, during this time, most of the planetesimals, initially located at a distance of 0.7 to 1.1 AU from the Sun, fell onto the growing Earth and Venus. Taking into account the ejection of matter during collisions of bodies with planets can increase the accumulation time of the planets. The total mass of planetesimals that migrated from each part of the region located at a distance of 0.7 to 1.5 AU from the Sun and collided with the nearly formed Earth and Venus differed for these planets, probably by no more than a factor of two. The outer layers of the Earth and Venus could have accumulated the same material for these planets from different parts of the feeding zone of the terrestrial planets. In the final stages of the formation of the terrestrial planets, planetesimals originally located at distances of 1.1 to 2.0 AU from the Sun may have been incorporated into Earth and Mars in a ratio not much different from the ratio of the masses of these planets.

## 6. Migration of bodies in the zone of the giant planets

Jupiter and Saturn contain a large amount of gas. It is believed that when the mass of the Jupiter embryo reached several Earth masses, a stage of rapid gas accretion began, during which the mass of the Jupiter embryo increased several times over a period of ~0.1 million years. In the model of V.S. Safronov, the total mass of planetesimals in the feeding zone of the giant planets exceeded 100 Earth masses. In some models, the time of increase in Jupiter's mass from zero to approximately its present value was about 2 million years. Calculations of the evolution of gas-free disks of hundreds of gravitating solid bodies, combining during collisions, corresponding to accumulation in the zone of the giant planets, were carried out by Ipatov until the early 1990s [4, 9, 10, 11, 15, 78, 226, 227]. The initial disks considered included planetesimals and embryos of the giant planets (or the planets themselves). The evolution of disks without planetary embryos was also modeled. The initial orbital eccentricities of planetesimals typically equaled to 0.02. Both flat and spatial models were considered.

The results of the evolution of **flat disks of bodies** corresponding to the feeding zones of giant planets are presented in the preprint [226] and more briefly in [4, 9]. The initial disks corresponded to the feeding zone of one or two giant planets. The evolution of a disk corresponding to the feeding zone of Uranus and Neptune (12-37 AU) was most often modeled. Disks consisting of several disks corresponding to all the planets were also considered. For flat disks of initially identical bodies, the mass of the ejected bodies did not exceed one third of the initial mass of the disk of the bodies, and the average orbital eccentricities during evolution could reach 0.2-0.4. Among the formed planetary embryos there were also bodies with negative axial rotation.

The main calculations were carried out for **a spatial model**. The evolution of a spatial disk was modeled by reducing the spatial problem to the evolution of a flat model of test bodies (so that the probabilities of collisions of test bodies were approximately the same as for the spatial problem) [225]. When modeling the evolution of spatial disks without planetary embryos [4] and with planetary embryos [11, 15], it was found that the total mass of planetesimals ejected into hyperbolic orbits amounted to about 80% of their total initial mass when considering only Jupiter and Saturn and about 90% when considering all four giant planets. Let us tell in more detail about the considered evolution of several of the disks.

In [4], when studying **the evolution of spatial disks of identical bodies** merging in collisions, with a total mass from $32m_E$ to $200m_E$, the ratio $k_h$ of the total mass of bodies ejected into hyperbolic orbits to the total mass of bodies remaining in elliptical orbits (including those in the composition of planets) was equal to 2.5-8 for initial disks at a distance of 3.4-12 AU from the Sun and 5-30 for initial disks at a distance of 12-37 AU. In [10], the distribution of ejected bodies by the eccentricities of their orbits was considered. In Fig. 5a of [10], a significant portion of the ejected bodies have orbital eccentricities $e<1.1$, but for individual bodies $e>2$.

In variant B3 of the monograph [78] and the article [4], a case was considered when, at the initial moment of time, in addition to a disk of identical bodies divided into two groups ($8<a\leq16$ AU and $16<a\leq32$ AU) with a total mass $M_o=150m_E$ ($m_E$ is the mass of the Earth), there were also small bodies - "asteroids" ($2<a\leq4$ AU) and almost formed terrestrial planets, as well as "Jupiter" and "Saturn", with the semi-major axis of the orbit of "Jupiter" being 0.3 AU larger, and that of "Saturn" being 2.5 AU smaller than the modern values. The dynamics of the evolution of this disk is presented in Fig. 6.3 in [78] and in Figs. 3-4 in [4]. During the evolution, the average eccentricity $e_{av}$ of the orbits of the bodies reached 0.4. Small bodies were swept out of the asteroid belt, and individual massive bodies completely penetrated the feeding zone of the terrestrial planets. In the modeling, it was assumed that in variant B3, the average inclinations $i_{av}$ (in radians) were several times smaller than $e_{av}/2$. However, even with this these assumptions, the total mass of objects formed beyond the orbit of Saturn was only $0.03M_o$, with the mass of the largest object not exceeded $3m_E$. The obtained results indicate that the formation of Uranus and Neptune in the feeding zones of these planets only from identical solid bodies – planetesimals requires a large (up to $10^3m_E$) mass of solid bodies in these feeding zones. In this case, the mass of the gas and dust cloud in the feeding zones of Uranus and Neptune would be no less than 0.15 solar masses.

In a number of calculations, planets with masses reaching several Earth masses formed within Jupiter's orbit. The formation of such planets was due to the fact that the model under consideration did not account for the fragmentation of colliding bodies at high impact velocities. Some bodies formed during such fragmentation could have remained in the zone of the terrestrial planets and in the asteroid belt, and thus their material could have been incorporated into the terrestrial planets and some asteroids. The orbital eccentricities of bodies colliding within Jupiter's orbit typically exceeded 0.5. Therefore, in reality, no coagulation of such bodies during their collisions should have occurred. Those fragments of colliding bodies that did not intersect Jupiter's orbit were gradually swept out by bodies from the feeding zones of the giant planets that migrated inward. The strong changes in the orbits of the terrestrial planets in the considered calculation variants are due to the fact that in the calculations the masses of the bodies significantly exceeded the masses of the real planetesimals.

The idea that **the embryos of Uranus and Neptune formed near the orbit of Saturn** was first put forward by V.N. Zharkov and A.V. Kozenko in 1990 (Letters to the Astron. J., 1990, Vol. 16, pp. 169-173). Studying the composition of Uranus and Neptune, these authors concluded that the embryos of these planets acquired hydrogen envelopes with a mass of about (1-1.5)$m_E$ in the feeding zones of Jupiter and Saturn before the dissipation of gas from the protoplanetary disk.

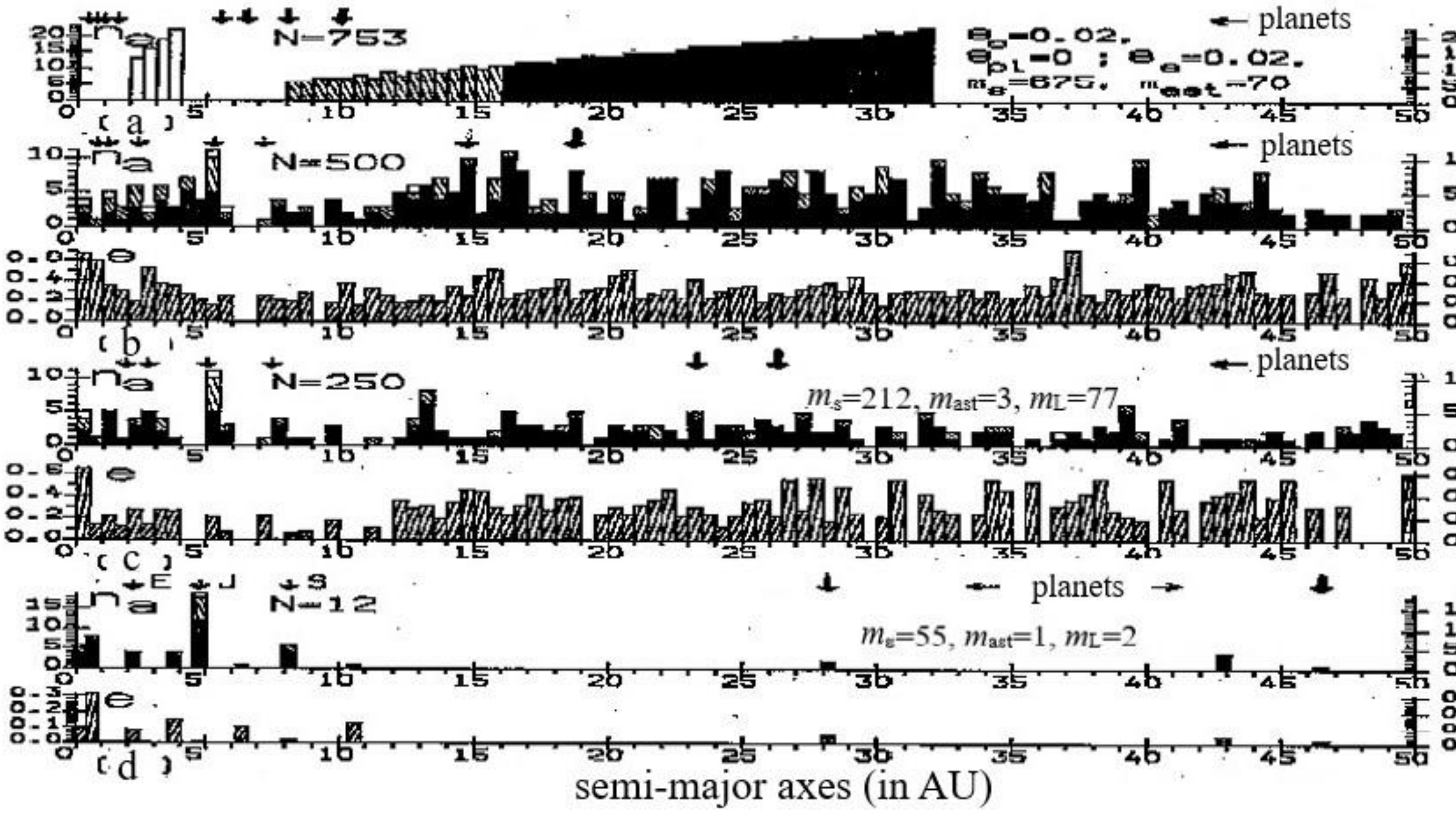

**Figure 1.** Histograms of the distribution of the number of bodies $n_a$ over the semi-major axes of their orbits for the initial disk corresponding to the feeding zone of the embryos of Uranus and Neptune, as well as the distributions of the masses and average orbital eccentricities of the bodies over the semi-major axes of their orbits for several stages of evolution (with the number of bodies equal to N). The disk initially consisted of the embryos of all eight planets, 675 identical bodies with a total mass equal to $m_{so}$=135$m_E$ and of $m_{ast}$=70 “asteroids” (out of a total of 753 celestial objects). The initial eccentricities of the orbits of the bodies and embryos of Uranus and Neptune (with masses equal to 10$m_E$) are equal to 0.02, and the initial orbits of the remaining planets (with almost present masses) are circular. The semi-major axes of the initial orbits of these embryos are equal to 8 and 10 AU. The positions of the planets are indicated by arrows. The evolution of such disks is described in [15] and is also briefly shown in the right-hand graphs of Fig. 2.

Ipatov studied in [11, 15, 239] the evolution of a number of disks that initially consisted of nearly formed planets (except Uranus and Neptune), of the embryos of Uranus and Neptune, and of several hundred identical bodies for which 8<$a$≤32 AU (Figs. 1-2). The initial values of the semi-major axes of the orbits of Jupiter and Saturn were taken to be 5.5 and 6.5 AU, respectively. In the main series of calculations, the initial **values of the semi-major axes of the orbits of the embryos of Uranus and Neptune were equal to 8 and 10 AU**, respectively. The total mass $M_o$ of the initial bodies was taken to be in the range from 135$m_E$ to 180$m_E$; the initial eccentricities of the orbits of the planets were $e_{pl}$=0, and the initial eccentricities of the orbits of the bodies and embryos were $e_o$=0.02. The considered model took into account gravitational interactions and coagulations of bodies not only with planets or their embryos, but also with each other. Gravitational interactions between bodies and planets were accounted using the spheres of action method. Calculations showed that if the orbit of the initial embryo crossed Saturn's orbit, the probability that this embryo would remain in an elliptical orbit at the end of its evolution was an order of magnitude less than the probability of its ejection into a hyperbolic orbit. Therefore, below we will discuss the case in which the perihelia of the initial orbits of the Uranus and Neptune embryos were located beyond

Saturn's orbit. Calculations have shown [11, 15, 239] that, under the influence of planetesimals migrating from the feeding zones of Uranus and Neptune to Jupiter, the nearly formed Uranus and Neptune could have migrated from Saturn's orbit to their present-day orbits over 10 million years, moving throughout this time along slightly eccentric orbits.

Calculations of such migration of Uranus and Neptune were published by S.I. Ipatov [11, 15, 239] in 1991-1993, long before the first works on the above-mentioned Nice model. Moreover, in contrast to the Nice model, the migration of Uranus and Neptune embryos occurred without the giant planets entering resonances, and the total initial mass of planetesimals in the feeding zones of Uranus and Neptune was greater. In the considered variants, it varied from $135m_E$ to $180m_E$. In [15, 78], it was concluded that a disk of planetesimals with a total mass equal to $100m_E$ was sufficient for the migration of Uranus and Neptune embryos to modern orbits. This mass is smaller if we consider larger (than those calculated) semi-major axes of the initial orbits of the Uranus and Neptune embryos (their calculations assumed 8 and 10 AU, respectively). The major changes in the orbital elements of the giant planet embryos occurred over a period of no more than 10 million years, although individual bodies could have fallen onto these embryos billions of years later. If the majority of the disk mass was accounted for small bodies, the migration times of the planetary embryos could have been longer than those assumed for the initial masses of $0.2m_E$, as calculated.

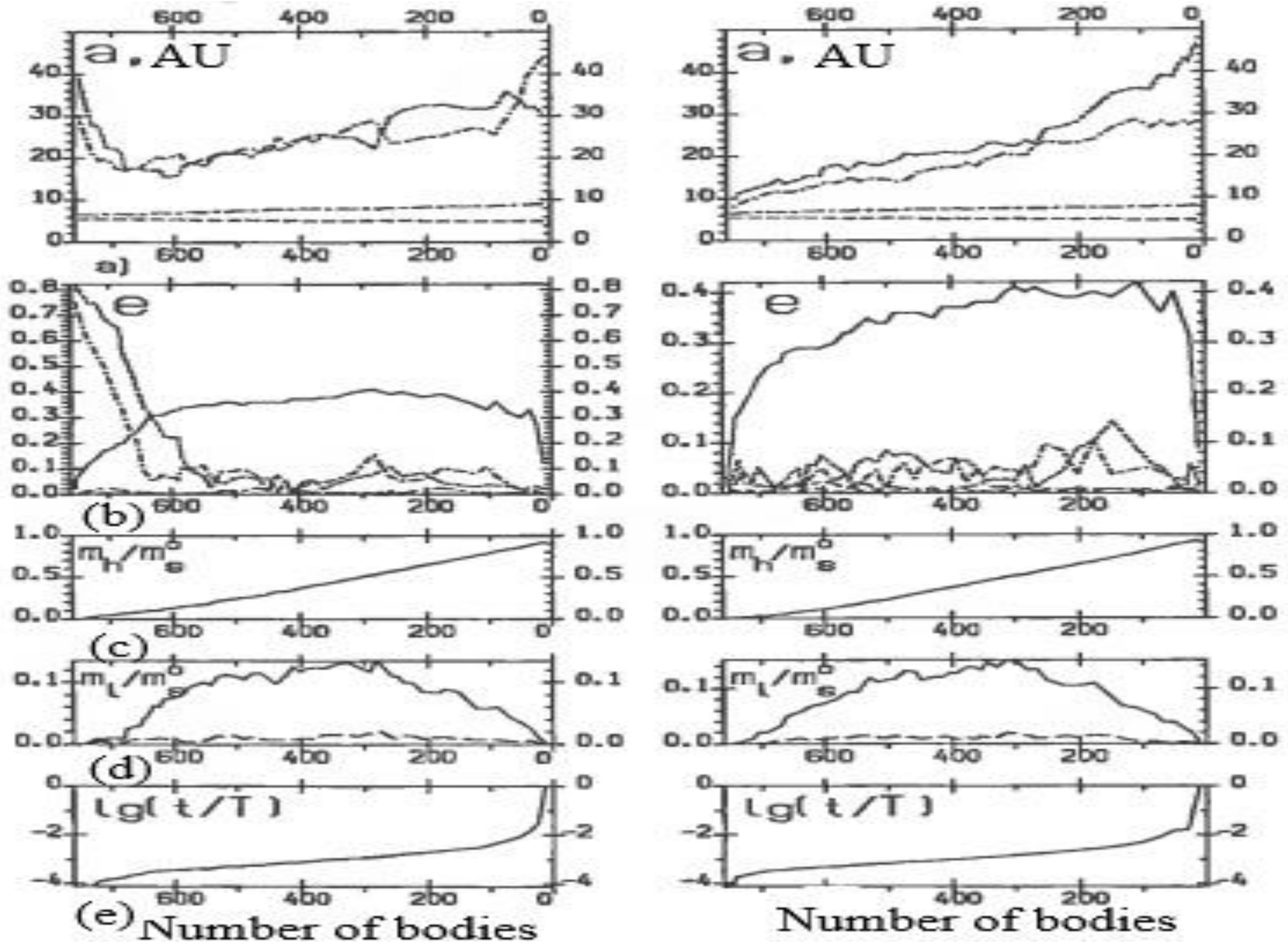

**Figure 2.** Dependences of the semi-major axes (Fig. a) and orbital eccentricities of the embryos of the giant planets (dashed lines in Fig. b), the average orbital eccentricity of bodies (solid line in Fig. b), the ratio of the total mass of bodies ejected into hyperbolic orbits to their initial total mass (Fig. c), the ratio of the total mass of bodies with semi-major axes of orbits greater than 49.5 AU to their initial total mass (Fig. d) and the ratio of the current time to the total evolution time of the disk (Fig. d) depending on the number of bodies in the disk. (According to [15].)

In addition to calculating the migration of giant planets that were initially in circular or nearly circular orbits, the case of large (0.75-0.82) initial eccentricities of the orbits of massive ($10m_E$) Uranus and Neptune embryos was also considered (variant B4 in [78]; [11, 15]). In this model, during interactions of these embryos with planetesimals, the eccentricities of the embryos' orbits decreased, and the embryos could acquire the modern orbits of Uranus and Neptune if the initial perihelia of their orbits lay beyond the orbit of Saturn, while at smaller perihelion distances these embryos were in most cases ejected into hyperbolic orbits. However, the acquisition of such eccentric orbits with perihelia lying beyond the orbit of Saturn by the Uranus and Neptune embryos is unlikely. In [11], based on an additional study of the model, which did not take into account the gravitational interactions of bodies and planets, it was noted that the contribution of the coagulations to the reduction of the initially large eccentricities of the orbits of the embryos of Uranus and Neptune is significantly less than the contribution of the gravitational interactions of these embryos with bodies.

The total mass of the bodies that penetrated beyond the orbit of Neptune could have reached tens of Earth masses. The fact that Neptune has the smallest eccentricity among the giant planets may be related to the fact that the masses of the largest planetesimals in its feeding zone were smaller than in the feeding zones of the other giant planets. The dynamic lifetimes of most planetesimals located inside the orbit of Neptune are less than 100 million years. Therefore, the end of the intense bombardment of the Earth, which occurred 4.5-3.8 billion years ago, could have occurred primarily with objects that came from eccentric orbits located mainly beyond the orbit of Neptune, or from the zone of the outer asteroid belt. In [7], based on an analysis of analytical formulas and the results of numerical calculations, it was noted that for the zone of Neptune, some planetesimals moving in highly inclined and highly eccentric orbits could have survived to the present day.

Let us briefly touch on the growth rate of the giant planets. In his study of the migration of planetesimals initially located at various distances from the Sun, Ipatov also obtained estimates of the fraction of planetesimals that collided with the giant planets. In these calculations, the evolution of orbits of planetesimals under the influence of the planets was modeled by numerically integrating the equations of motion. The probability of a collision of a planetesimal initially located beyond the orbit of Jupiter with Uranus or Neptune did not exceed 0.015, and in most calculations was no more than a few thousandths. Therefore, with a total mass of planetesimals beyond the orbit of Saturn of less than $200m_E$, the massive embryos of Uranus and Neptune, migrating from the zone in the vicinity of Saturn's orbit to their present-day orbits, increased their masses by no more than $2m_E$. Moreover, the probability of a planetesimal collision with Jupiter in most calculations did not exceed 0.05, and with Saturn it was several times lower. It is believed that the mass of the silicate component in Jupiter is $(15\text{-}20)m_E$, and the mass of the solid component in Saturn was greater than that of Jupiter. Planetesimals in the feeding zones of Uranus and Neptune, along with silicates, also contained ices. Therefore, with a total mass of planetesimals beyond the orbit of Saturn of less than $200m_E$, the increase in the mass of Jupiter's silicate component due to such planetesimals probably did not exceed 10 Earth masses. From this, it can be concluded that the total initial mass of planetesimals beyond the orbit of Saturn within $200m_E$ is consistent with the composition of the giant planets. With the ratio of the mass of dust, consisting of rocky matter and ices, to the mass of gas equal to 0.015, the mass of planetesimals, equal to $200m_E$, corresponds to a mass of matter in the disk equal to $0.04M_S$, with the mass of the entire disk being $0.06M_S$, where $M_S$ is the mass of the Sun. Approximately the same values for the mass of the protoplanetary disk in the range of $(0.04\text{-}0.1)M_S$ were obtained in other studies.

In variants B8-B9 of the monograph [78] (these are also the variants in Figs. 4-5 in [4]) the evolution of spatial disks was considered, initially consisting of **eight planets and** 500 identical **bodies in the zones of Uranus and Neptune**. The masses of Uranus and Neptune were taken to be $14m_E$ and $16m_E$, and the semi-major axes of their orbits were 19 and 30 AU, respectively. The semi-major axes of the orbits and the masses of the remaining planets were equal to their modern values. The initial orbits of the planets were circular. The density of the bodies was considered to be 2 g/cm$^3$. This model corresponds to the time when all **the planets had almost formed, but there were still planetesimals in the zone of Uranus and Neptune**. In variant B8 of the monograph [78], the initial eccentricities of the orbits of the planetesimals were equal to $e_0$ =0.02, and the mutual gravitational influence of the planets and bodies was taken into account using the method of spheres of action. As noted in [4], in the zones of Uranus and Neptune for the gasless model the average eccentricity of the orbits of the bodies was most of the time on the order of 0.3-0.4. Therefore, in variant B9 of the monograph [78], the initial eccentricities $e_0$=0.35, and the remaining initial data were the same as in variant B8 of the same work. In this variant, the mutual gravitational influence of the bodies was not taken into account. The total initial mass $M_o$ of bodies in variants B8-B9 of the monograph [78] was equal to $10m_E$.

In calculations with initial bodies in the feeding zones of Uranus and Neptune, the semi-major axis of Jupiter's orbit $a_j$ decreased by $0.005M_o/m_E$ AU, the semi-major axis of Saturn's orbit $a_s$ increased by $(0.01\text{-}0.03)M_o/m_E$ AU, and the semi-major axes of Uranus and Neptune's orbits increased by approximately на $(0.2\text{-}0.3)M_o/m_E$ AU. In variants B8-B9 of the monograph [78], with $M_o$=$10m_E$ and initial body masses of $0.02m_E$ or $0.1m_E$, the orbital eccentricities of Mercury and Mars reached values of approximately 0.3. The actual orbital eccentricities of these planets are 0.2 and 0.09. In reality, most bodies from the zones of the giant planets had smaller masses than in these calculation variants, but the total mass of bodies in the feeding zones of the giant planets could have been on the order of $100m_E$. In principle, the influence of bodies from the feeding zones of the giant planets penetrating into the zone of the terrestrial planets could have been sufficient for Mercury and Mars to acquire their sufficiently large orbital eccentricities.

**Summarizing the results** of studies of migration of bodies in the feeding zones of the giant planets, we note that the total mass of bodies ejected into hyperbolic orbits from the zone of all giant planets was an order of magnitude greater than the mass of solid material that formed the planets. These results are consistent with the values of the initial mass of the protoplanetary cloud of approximately 0.04-0.1 solar masses, accepted by many authors. The results of numerical modeling of planetary accumulation indicate that Jupiter's core and shell may have contained more ices and rocky material than any other planet. The total mass of bodies from the zones of the giant planets penetrating the asteroid belt could have reached tens of Earth masses. The system of giant planets expanded during the accumulation of these planets. In all calculations of the evolution of disks of bodies in the zone of the giant planets with planets [4, 10, 11, 15, 78], the semi-major axis of Jupiter's orbit decreased over time, while the semi-major axes of the orbits of the other giant planets generally increased. It was first found in [11, 15] that the embryos of Uranus and Neptune, with initial masses equal to several Earth masses, could have migrated to the present-day orbits of these planets from a zone located just beyond the orbit of Saturn, moving in slightly eccentric orbits. A large amount of water could have been delivered to the Earth during the accumulation of Uranus and Neptune. The total mass of the bodies penetrating beyond the orbit of Neptune could have reached tens of Earth masses.

Bodies from the feeding zones of the giant planets influenced the accumulation of bodies in the zone of the terrestrial planets. As numerical studies of the migration of planetesimals from the feeding zones of Jupiter and Saturn have shown, the bulk of these planetesimals left this zone within a few million years. This indicates the time of influence of planetesimals coming from the zone of Jupiter and Saturn on the formation of the terrestrial planets. It is important to emphasize that, according to calculations performed within the framework of a model taking into account the interaction of all planets of the Solar System (section 4.6 in [53]), the maximum times of orbital evolution of some planetesimals starting from the zone of Jupiter and Saturn can be much greater than in similar calculations in the absence of Uranus and Neptune (50 and 4 million years, respectively). However, even in this case, the main contribution to collisions with the embryos of the terrestrial planets was made by planetesimals from the feeding zones of Jupiter and Saturn in the first million years after the formation of a significant mass of Jupiter. The duration of this formation is estimated at 1-2 million years from the origin of the Solar System. Meanwhile, some planetesimals from the feeding zones of Uranus and Neptune fell to the Earth hundreds of millions of years later and could even remain in the Solar System to this day. When planetesimals fell from the feeding zones of Jupiter and Saturn onto the embryos of the terrestrial planets, as well as from the outer asteroid belt in the "Grand Tack" model, these embryos had not yet acquired the present masses of the planets, and material (including water and volatiles) from these zones could have accumulated in the inner layers of the forming terrestrial planets and the Moon.

## 7. Formation of the asteroid and trans-Neptunian belts

Bodies from the zone of the giant planets had a great influence on the formation of the asteroid and trans-Neptunian belts. The eccentricities and semi-major axes of the asteroid orbits are shown in Fig. 1. The clearing of the asteroid belt was influenced by the gravitational influence of bodies penetrating this belt from the zone of the giant planets and by collisions of asteroids with these bodies. With changes in the semi-major axes of the orbits of Jupiter and Saturn, the positions of the orbital and secular ($\upsilon_5$, $\upsilon_6$, $\upsilon_{15}$ and $\upsilon_{16}$) resonances with Jupiter and Saturn changed, at which the orbital eccentricities of planetesimals could greatly increase. In [8, 14], it was shown, for example, that the boundaries of the Kirkwood 5:2 gap coincide with the boundaries of the region of initial values of the semi-major axes and orbital eccentricities of asteroids, at which, in the course of evolution, the orbit of Mars is reached and bodies can leave the gap. In the " Grand Tack " model, Jupiter crossed the entire asteroid belt. It was believed that over a billion years, the asteroid belt had been cleared by more than 99%.

In [15] (Fig. 8) and [78] (variant B7 in Fig. 6.8), the results of modeling the evolution of a disk initially consisting of the terrestrial planets, Jupiter, Saturn, 250 planetesimals with a total mass of $M_o$~$10m_E$ and semi-major axes of their initial orbits from 5 to 10 AU, as well as 250 "asteroids" with semi-major axes of the initial orbits from 2 to 4 AU are presented. The initial eccentricities of the orbits of the planetesimals and asteroids were equal to 0.02, and the initial orbits of the planets were circular. The gravitational influence of the planets and planetesimals was taken into account. This model characterizes **the influence** **of bodies from the feeding zones of Jupiter and Saturn on the formation of the asteroid belt**. In these calculations, the semi-major axes of the orbits of Jupiter $a_j$ and Saturn $a_s$ decreased by $0.005M_o/m_E$ and $0.01M_o/m_E$ AU, respectively. Note that when considering bodies in the feeding zones of Uranus and Neptune, the semi-major axis of Saturn's orbit, conversely, increased, as Saturn brought many such bodies closer to the Sun. During the course of evolution, "asteroids" were swept out of the asteroid belt. Jupiter, Venus, and Earth received 8%, 5%, and 2.5% of the asteroids, respectively.

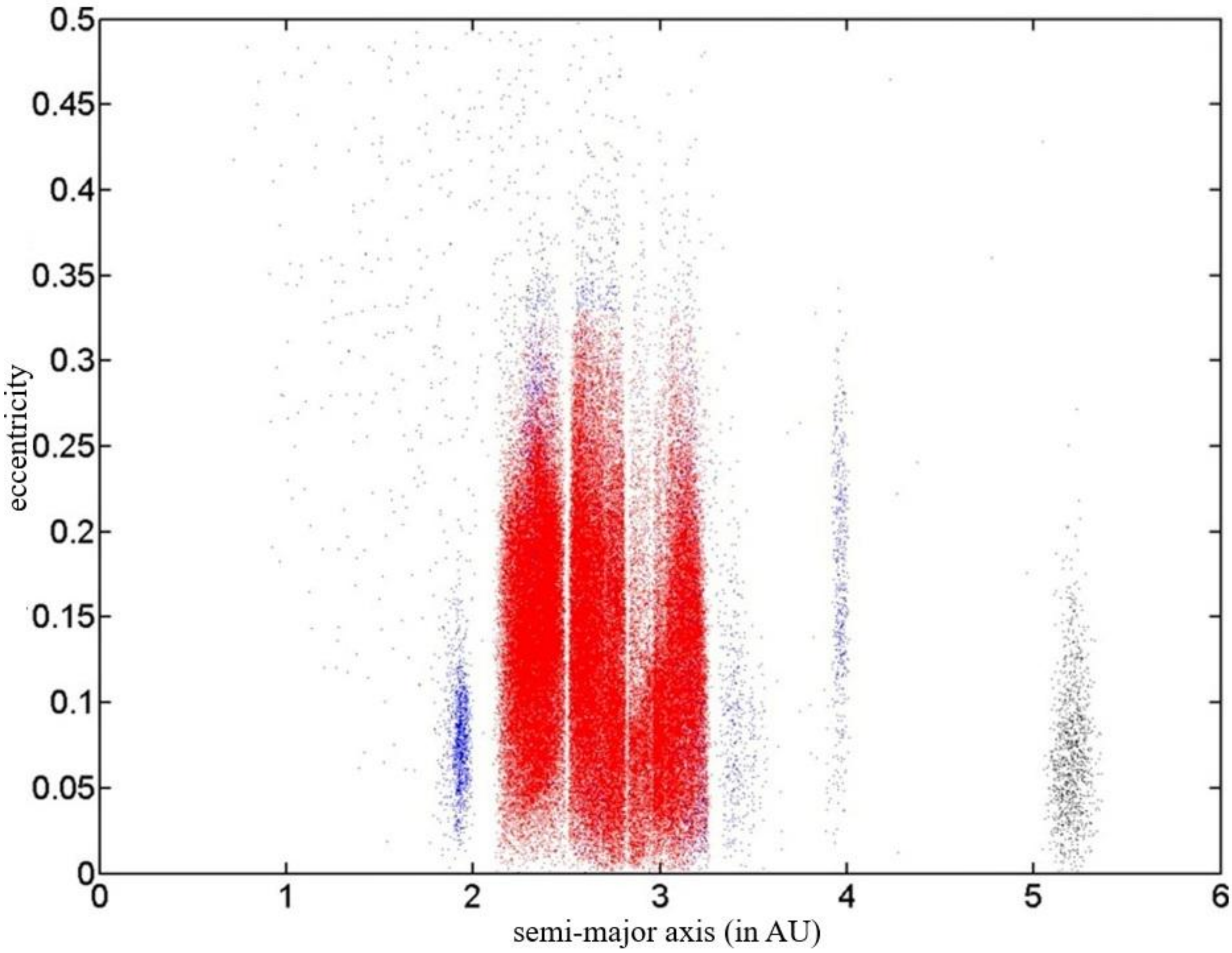

**Figure 1.** Eccentricities and semi-major axes of asteroid orbits.

In the variant presented in Fig. 6.3 of the monograph [78], Jupiter, Saturn and the terrestrial planets were initially considered, as well as 750 gravitating bodies (with a total mass of $150m_E$) with major semi-axes of orbits from 8 to 32 AU and 100 asteroids. In the course of evolution, the number of asteroids decreased by approximately 4 times with a decrease in the number of bodies by 20%. Such a large difference is due to the fact that during this time the orbits of many of the considered bodies had not yet acquired eccentricities close to 1, and some of these bodies had already entered the asteroid belt zone, where evolution proceeded more rapidly due to their proximity to the Sun and shorter orbital periods.

It is believed that some **bodies of the trans-Neptunian belt** were formed directly in this zone, and some bodies came from the zone of the giant planets during the formation of these planets. In 1987, in the article [4], 5 years before the discovery in 1992 of the first (after Pluto) trans-Neptunian object, based on the results of calculations of the formation of giant planets, S.I. Ipatov suggested that in addition to trans-Neptunian objects formed further than 30 AU from the Sun and moving in slightly eccentric orbits, objects formed in the zone of giant planets move in this zone in highly eccentric orbits. These objects are now called " scattered disk objects" (Fig. 2). At present, trans-Neptunian objects (TNOs) have been discovered moving in highly eccentric orbits with perihelion distances $q$ exceeding 40 AU, ("extended scattered disk" objects), while typical scattered disk objects (SDOs) have perihelion distances $q$~35-38 AU. Many of these TNOs with large $q$ could be former planetesimals from the feeding zone of the giant planets.

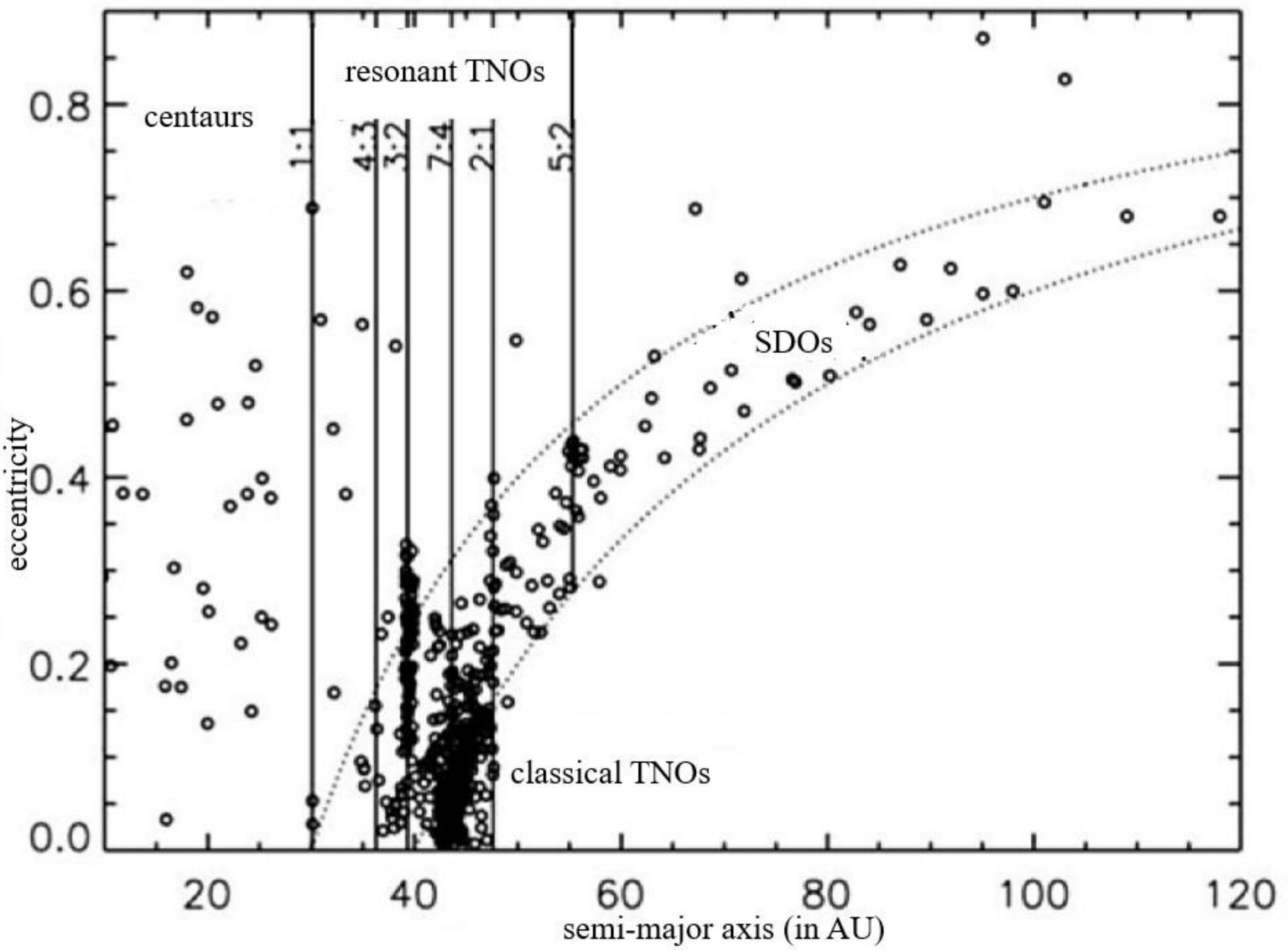

**Figure 2.** Semi-major axes and orbital eccentricities of trans-Neptunian objects (Source: Scott Sheppard, Carnegie Institution for Science).

During the accumulation of the giant planets, the total mass of planetesimals that started from the feeding zone of these planets and moved for some time beyond the orbit of Neptune was equal to tens of Earth masses. These planetesimals increased the eccentricities of the orbits of local trans-Neptunian objects (TNOs), the total initial mass of which could exceed $10m_E$, and swept out most of such bodies from this zone. A small portion of these planetesimals could remain beyond the orbit of Neptune in highly eccentric orbits. This mechanism of formation of TNOs in highly eccentric orbits and the formation of "local" TNOs in weakly eccentric orbits from matter located beyond the orbit of Neptune [4], as well as the first estimates of gravitational interactions of TNOs [5, 221] were considered by Ipatov in 1980-1988. It was shown that over the past 4 billion years, several percent of TNOs could have changed their semi-major axes $a$ by more than 1 AU due to gravitational interactions with other TNOs. Note that even small changes in the orbital elements of TNOs, occurring due to their mutual gravitational influence and collisions, can lead to large changes in the orbital elements of TNOs under the gravitational influence of planets [25].

At some stages in the considered evolution of the disks of bodies, corresponding to the zone of the giant planets (with the number of remaining bodies $N\approx$300-400, comprising about a third to a half of the number of initial bodies), the total mass of bodies with semi-major axes $a$>50 AU exceeded $0.1M_o$, and it was significantly less the end of their evolution. The perihelia of the orbits of many of these bodies lay near the orbit of Neptune, and the average eccentricities of their orbits were close to 0.6. These bodies, along with bodies formed beyond the orbit of Neptune, participated in the formation of the trans-Neptunian belt.

Since some of the bodies in the trans-Neptunian belt can migrate into the solar system, it was noted in [78] that, thanks to the mechanism of exchange of angular momentum with such bodies, some large bodies in the trans-Neptunian belt could increase the semi-major axes of their orbits and move from the inner part of the belt to its middle part.

Trans-Neptunian objects with a diameter $d\geq$100 km, moving in moderately eccentric orbits, could have formed by the compression of large, rarefied dust condensations (with $a$>30 AU), rather than by the accretion of smaller planetesimals. For solid-state accumulation of trans-Neptunian objects from small planetesimals, it is necessary that the accumulation occur at small (~0.001) orbital eccentricities and a massive belt (tens of Earth masses). According to my calculations [22, 26], due to the gravitational influence of forming giant planets, trans-Neptunian objects (TNOs), and migrating planetesimals, such small eccentricities could not exist for the time required for the solid-state accumulation of trans-Neptunian objects with a diameter $d$>100 km. It's likely that some planetesimals with diameters of $d$~100–1000 km in the feeding zones of the giant planets, and even some large main-belt asteroids, could also have formed directly through the compression of rarefied dust condensations. Some smaller objects (TNOs, planetesimals, asteroids) could have been fragments of larger objects, while other similar objects could have formed through the compression of condensations.

The population of trans-Neptunian bodies with diameters $d$>100 km likely has not changed significantly due to interactions between these bodies over the lifetime of the Solar System, as catastrophic collisions of large bodies with diameters $d$~100-150 km are unlikely. A body with a mass ~100 times smaller than a 100-km object is incapable of destroying it, but can change its orbital velocity by several m/s and its semi-major axis by ~0.2% (~0.1 AU). Such events may, generally speaking, be frequent enough to provide a certain flux of Chiron-sized bodies. The semi-major axes of the orbits of the fragments formed in the collisions may differ by 0.1-1 AU from the semi-major axes of the orbits of the parent bodies.

In [22, 25, 266, 267, 269, 274, 275, 278-285, 289-291, 293-295, 299, 300, 301, 302] and in paragraphs 4-5 of Chapter 7 in [78], changes in the orbital elements **of trans-Neptunian belt bodies under the influence of giant planets** were studied for different initial eccentricities and orbital inclinations. The evolution of the orbits of these bodies was also studied after they began to cross the orbit of Neptune. The initial values of the semi-major axes of the orbits of TNOs varied from 35 to 50 AU.

In modeling the evolution of TNO orbits at the 2:3 resonance with the motion of Neptune (in Russia they previously wrote "3:2 resonance") in [25] and in paragraph 5 of Chapter 7 in [78], various types of changes in the elements of TNO orbits were considered. The evolution of the orbits of about one hundred TNOs was studied. Usually, the considered time interval was equal to 20 million years, but sometimes it reached 100-150 million years. For many of these calculation variants, the initial value of the semi-major axis of the TNO orbit $a_0$=39.3 AU. With such $a_0$, the resonant value of $a$ (39.5 AU) was always reached during the evolution. Various types of interrelations between variations in the difference in the longitude of the ascending node for TNOs and Neptune and the argument of perihelion of TNOs were considered. When integrating into the past, the evolution of orbits close to the orbit of the object P/1996 R2, which crossed the orbit of Jupiter, in a number of calculation variants [21] it was found that previously the semi-major axes of the orbits of such objects exceeded 30 AU (Fig. 7.6a-b in [78]), that is, this object came from the trans-Neptunian belt zone.

Based on simple formulas and the results of numerical calculations, estimates were made ([22, 27, 89, 143, 144], paragraph 5 of Chapter 7 in the monograph [78]) of the number of bodies that migrated from the trans-Neptunian belt to the orbits of Jupiter and the Earth. It was noted that about 6% of the bodies that crossed the orbit of Neptune could reach the orbit of the Earth. If we consider the number of TNOs with a diameter $d$>1 km and semi-major axes of orbits from 30 to 50 AU to be equal to $10^{10}$, then according to our estimates [22, 78], about 30,000 former TNOs with a diameter $d$>1 km are moving in orbits that cross the orbit of Jupiter, and about 170 of them also cross the orbit of the Earth. Thus, it was believed that approximately 20% of Earth-crossing objects (ECOs) may be former TNOs crossing the orbits of both Earth and Jupiter. It is possible that an even a greater number of objects originating from beyond Neptune are in Earth-crossing orbits that lie entirely within Jupiter's orbit. The perihelions or aphelions of Earth-impacting bodies generally lie near Earth's orbit.

In 1980, in the preprint [221], it was noted that the limits of variations in the semi-major axes and orbital eccentricities of gravitationally interacting bodies with Pluto's masses for three or four bodies can be several times greater than for two such bodies moving around the Sun. The variations in the semi-major axes of the orbits of TNOs in this case could amount to several astronomical units [5, 221]. These results support the hypothesis of T.M. Eneev that the semi-major axis of Pluto's orbit was previously greater than its current value and that it decreased during gravitational interactions between Pluto and large bodies of the trans-Neptunian belt. It was also noted that two bodies of equal masses can exchange the values of their orbital semi-major axes during close encounters. Some trans-Neptunian objects could migrate from the middle of the belt to its inner part. Moreover, some of these bodies could move into orbits whose eccentricities increased significantly over time under the influence of the giant planets, and such bodies could migrate into the inner Solar System.

## 8. Formation of binary trans-Neptunian objects

It is known that the fraction of small bodies with satellites is 0.1 among main-belt asteroids, 0.15 for near-Earth asteroids, about 0.3 for classical trans-Neptunian objects, and 0.1 for other trans-Neptunian objects. In the paper [51], it is shown that the angular velocities of the condensations used in calculations in (Nesvorný et al., Astron. J., 2010, v. 140, pp. 785–793) as initial data in modeling the compression of rarefied condensations (consisting of dust and/or objects with a diameter less than 1 m), leading to the formation of trans-Neptunian satellite systems, could have been obtained in collisions of condensations whose radii are comparable to their Hill radii. The heliocentric orbits of the colliding condensations could have been close to circular. The model of satellite system formation, considering the collision of two condensations, is consistent with observations according to which about 40% of binary objects discovered in the trans-Neptunian belt have negative angular momentum relative to their centers of mass [52]. The ratio of the tangential component of the collision velocity of condensations to the parabolic velocity on the surface of the condensation is inversely proportional to the distance from the Sun and inversely proportional to the cube root of the mass of the condensation. Therefore, the merger of condensations-preplanetesimals is more likely for more massive condensations and for condensations that were at a greater distance from the Sun (in particular, more likely for condensations in the trans-Neptunian belt than in the asteroid belt) [39, 51].

## 9. Formation of the Earth-Moon system

Various models for the origin of the Moon have been proposed by various scientists, but to date this issue has not been definitively resolved. We will touch upon several of the most significant models here (a more detailed review is given in [54, 59, 82]). The coaccretion theory considered the formation of the Moon from a near-Earth swarm of small bodies. The main source of the swarm, according to the Schmidt-Ruskol-Safronov model, was considered to be the capture of particles from the protoplanetary disk during collisions ("free with free" and "free with bound"). **The mega-impact model** proposed by a number of authors has gained great popularity. It is assumed that the Moon was formed as a result of a catastrophic collision during the formation stage with a Mars-sized body (named Theia), which led to the ejection of molten silicate mantle of the Earth upon impact into close near-Earth orbits. The merged fragments of the ejecta formed the proto-moon, which gradually moved away to the modern orbit due to tidal interaction with the Earth during the process of evolution. The appeal of this hypothesis lay primarily in its explanation of the Moon's average density, which is equal to that of the Earth's mantle. Several modifications of the mega-impact model have been proposed.

Canonical the mega-impact model faces certain difficulties, primarily of a geochemical nature, and is currently subject to serious criticism. It fails to explain the similar isotopic compositions of a number of elements in the Earth and Moon, primarily the isotopes of oxygen, iron, hydrogen, silicon, magnesium, titanium, potassium, tungsten, and chromium. It is difficult to accept that the body that formed the Moon, even from a relatively close vicinity of the forming terrestrial planets, had a composition similar to that of the Earth, since in this model, most of the Moon's matter was formed from the material of the impactor, not the proto-Earth. These results deprive the mega-impact model of its geochemical justification. Furthermore, the mega-impact hypothesis fails to explain the absence of isotopic shifts in lunar and terrestrial matter, since the material ejected by a giant impact should consist of 80-90% vapor, and the isotopic compositions of K, Mg, and Si should change significantly during the evaporation of the melt. The giant impact hypothesis suggests that a magma ocean would form on the planet's surface following the impact. However, the connection between this event and Earth's ancient magma ocean has been questioned.

An alternative to the mega-impact model is **the multi-impact model** and a model of the formation of the Earth and Moon embryos from a single initial rarefied gas-dust condensation in the protoplanetary disk, with the subsequent formation and compression of two fragments. The multi-impact model is based on the hypothesis of multiple collisions of planetesimals with the Earth embryo. In contrast to the mega-impact model, there are several impactors, each with a smaller mass. **The model of the formation of the Earth and Moon embryos from a single initial rarefied gas-dust condensation** in the protoplanetary disk, with the subsequent formation and compression of two fragments, in addition to satisfying geochemical constraints, also allows us to explain the known differences in the chemical composition of the Moon and Earth, including iron deficiency, depletion of volatiles, and enrichment in refractory oxides of Al, Ca, and Ti of the Moon's substance compared to the Earth's. In E.M. Galimov's model, the formation of the Earth's and Moon's cores from the condensation began no earlier than 50 million years after the origin of the Solar System, dated by CAI (calcium-aluminum inclusions, the study of which allows us to determine the age of the Solar System). The model by Galimov et al. is also not free from shortcomings; its criticism is given by Ipatov in [54]. In particular, this model does not indicate where 95% of the iron-depleted substance from the inner part of the condensation went before the moment when the nuclei began to grow at the expense of the substance of its outer part. At zero (relative to the center of the condensation) velocities, the particles of the condensation would have fallen very quickly (the free-fall time is 25 years) into the center of the condensation before the nuclei formed in its inner hot part, where the evaporation of particles alone would have taken tens of thousands of years. Therefore, the existence of the condensation that gave birth to the Earth and the Moon over a period of 50 million years seems unlikely to us, which is supported by the review of publications on the lifetimes of condensations given in section 2.3 of the article [59]. As noted above in the study of the formation of the Earth, the main time of the formation of the Earth probably did not exceed 5 million years.

In contrast to the model of Galimov et al., in which the mass of the initial condensation was equal to the combined mass of the Earth and the Moon, in the model of Ipatov [54, 82], the embryos of the Earth and the Moon formed from a common rarefied condensation with a mass of the order of (0.01-0.1)$m_E$ and an angular momentum sufficient to form the embryo of the Moon. The angular momentum of the parent condensation, necessary for the formation of the embryos of the Earth and the Moon, could have been acquired mainly in the collision of two rarefied condensations, which formed the parent condensation. The angular momentum of the Earth-Moon system could have been acquired in the collision of two condensations (moving in circular heliocentric orbits before the collision) with a total mass no less than the mass of Mars [54]. Taking into account the subsequent growth of the masses of the Earth and Moon embryos to achieve the present angular momentum of the Earth-Moon system, the total mass of the embryos formed during the compression of the parent condensation could not have exceeded 0.01 Earth masses. The growth of the Earth and Moon embryos was considered within the framework of a multi-impact model. Most of the matter that entered the Moon embryo was ejected from the Earth during numerous collisions of planetesimals and smaller bodies with it. Some small bodies also have satellites. A mechanism similar to this model was proposed by Ipatov for the formation of trans-Neptunian satellite systems [51-52].

Generally speaking, the angular momentum of condensation necessary for the formation of the Earth-Moon system could be obtained by accumulation of only small objects A condensation with a final mass $m_f$>0.15$m_E$. However, it is more likely that the main contribution to the angular momentum of the parent condensation was made by the collision of two large condensations. Otherwise, Venus and Mars could also have formed with large satellites (which is not the case) if their parent condensations had acquired sufficient angular momentum. It is likely that, unlike Earth, the condensations that formed the embryos of the other terrestrial planets during compression did not collide with massive condensations at this stage. If this scenario is correct, then the condensation that gave rise to the embryo of Mars did not have high angular momentum, and only the small satellites Phobos and Deimos could have formed during its compression, although another mechanism for their formation is possible. The angular momentum of the condensations that gave rise to the embryos of Mercury and Venus was insufficient even for the formation of small satellites.

If the initially iron-depleted embryos of the Earth and Moon grew solely through the accumulation of planetesimals, to explain the small fraction of iron in the Moon, the mass of the Moon's embryo would have increased by no more than 1.3 times. The mass of the Earth's embryo, with such growth of the Moon's embryo, would have increased by no more than 3 times due to the accumulation of planetesimals in a gas-free environment. In this case, the Earth would not have acquired its current iron content. If the embryos grew solely through the accumulation of solid planetesimals (without the ejection of matter from the embryos), it is problematic to obtain the current iron content of the Earth and Moon for any values of the initial iron content in the initial embryos. To obtain the current iron content of the Earth and Moon for any fraction of iron in the initial embryos, the increase $dm$ in the mass of the embryo must be proportional to $m^p$, where $p$>2. For the motion of solid bodies in a gas-free environment, $p$ does not exceed 4/3. To explain the present-day iron content of the Moon, the fraction of matter ejected from the Earth's mantle and deposited on the Moon's embryo must have been an order of magnitude greater than the sum of the total mass of planetesimals deposited directly on the Moon's embryo and the initial mass of the Moon's embryo formed from the parent condensation, assuming the initial embryo contained the same proportion of iron as the planetesimals. Most of the matter that entered the Moon's embryo could have been ejected from the Earth during numerous collisions of planetesimals (and smaller bodies) with the Earth.

In the problem of the origin of the Earth-Moon system, the question of the cause of the difference in the composition of the matter of these bodies of iron and the role of migration processes in this difference is fundamentally important. Assuming that the proportion of iron in the initial embryo of the Moon and in planetesimals, was 0.33, and the proportion of iron in the Earth's crust and in the Moon, according to modern data, is equal to 0.05 and 0.08, respectively, and using the ratio $0.05k_E+0.33(1-k_E)=0.08$, we can estimate the proportion $k_E$ of the Earth's crustal substance in the Moon as ~0.9. Therefore, to explain such an iron content in the Moon, it should be assumed that the proportion of matter ejected from the Earth's embryo and falling on the Moon's embryo should have been almost an order of magnitude greater than the sum of the total mass of planetesimals that fell directly on the Moon's embryo and the initial mass of the Moon's embryo formed from the parent condensation, provided that this lunar embryo contained the same proportion of iron as the planetesimals. As we can see, the lower the estimate of the proportion of Earth's crustal material in the Moon, the more depleted in iron the lunar embryo formed during compression and the greater its mass. As already noted, most of the material that formed the lunar embryo could have been ejected from Earth during numerous collisions of planetesimals and other bodies with Earth.

It should be emphasized that objects ejected from the Earth's nucleus during collisions with other objects had a greater chance of becoming part of the larger Moon than of merging with similar low-mass objects. This facilitated the formation and growth of a larger satellite than would have been possible had it accumulated solely from material ejected from the Earth. The presence of a satellite formed by the compression of the condensation may explain the absence of a satellite around Venus. Venus, like Earth, was impacted by various planetesimals with roughly the same mass and velocity distributions for both planets. These collisions also ejected material from the surface of Venus, but no satellite formed from this material.

The indicated approach [54], according to which the initial embryos of the Moon and Earth could have formed from a common parental condensation, differs significantly from works on the classical multi-impact model, which considered the formation and growth of the Moon embryo mainly due to the matter of the Earth's crust ejected from the Earth embryo during its numerous collisions with bodies of the protoplanetary disk. The main difference from Ipatov's model [54] is that the initial embryo of the Moon was formed not from the matter ejected from the Earth embryo, but from the same condensation as the Earth embryo, and the subsequent growth of the Earth and Moon embryos, formed during the compression of the parental condensation, was similar to the multi-impact model. The matter that entered the Moon embryo could have been ejected from the Earth during numerous collisions of planetesimals and other smaller bodies with the Earth, and not only during a small number of large collisions.

Ipatov's proposed models for the formation of satellite systems impose certain constraints on the lifetimes of rarefied condensations. For circular heliocentric orbits, the difference between the semi-major axes of which is equal to the radius $r_{Ho}$ of the Hill sphere, the ratio of the periods of motion of two condensations around the Sun is approximately $1+1.5r_{Ha}$, where $r_{Ha}=r_{Ho}/a$. In this case, the angle with the vertex in the Sun between the directions of the two condensations changes by $2\pi\cdot1.5\cdot r_{Ha}\cdot n_r$ radians per $n_r$ revolutions of the condensations around the Sun. Let us assume that the collision of the condensations under consideration occurs when the semi-major axes of their orbits differ by $r_{Ho}$, and the initial angle with the vertex at the Sun between the directions to the condensations is equal to $\pi$ radians. Then the collision will occur after approximately $(3r_{Ha})^{-1}$ revolutions. For a mass equal to $0.01m_E$, the corresponding time between collisions is approximately equal to $10^{8/3}/3\approx155$ revolutions. That is, the collision of the condensations that gave rise to the parent condensation of the embryos of the Earth-Moon system could have occurred within a time of about 100 years after their formation. Similar estimates [59] have shown that to explain the proportion of formed satellite systems for trans-Neptunian objects, the lifetime of the condensations that gave rise to these objects should be on the order of $10^3$ revolutions around the Sun. In other words, the lifetime of the condensation in this zone of the Solar System did not exceed hundreds of thousands of years.

The article [Da Wang et al., 2025, Nature Geoscience, https://doi.org/10.1038/s41561-025-01811-3] showed that the proportion of potassium-40 on Earth as a whole is greater than in the proto-Earth (in the deep regions of the Earth's mantle), and less than in meteorites. In this work, this difference is explained by the mega-impact model. However, in my opinion, these studies of potassium-40 are consistent with the usual model of Earth's growth (without mega-impact), when, over time, its composition included bodies originating at increasingly greater distances from Earth's orbit (the greater the distance from the Sun from which bodies approached Earth, the more potassium-40 they contained).

**The growth of the Moon due to bodies ejected from the Earth** was discussed in [63]. The evolution of the orbits of bodies ejected from the Earth during its accumulation stage and early evolution due to impacts of large planetesimals was also studied. In the considered variants of calculations of the motion of bodies ejected from the Earth, most of the bodies left the Earth's Hill sphere and moved in heliocentric orbits. Their dynamic lifetime reached several hundred million years. At higher ejection velocities $v_{ej}$, the probabilities of collisions of bodies with the Earth and the Moon were generally lower. The probability of a collision of a body ejected from the Earth with the Moon moving in its present orbit was approximately 15–35 times lower than with the Earth at an ejection velocity $v_{ej} \geq 11.5$ km/s. The probability of such bodies colliding with the Moon was generally around 0.004–0.008 for ejection velocities of at least 14 km/s and around 0.006–0.01 for $v_{ej}$ =12 km/s. It was higher for lower ejection velocities and ranged from 0.01–0.02 for $v_{ej}$=11.3 km/s. The Moon may contain material ejected from Earth during the accumulation of the Earth and the Late Heavy Bombardment. Calculations show that the number of bodies ejected from Earth and impacting the lunar embryo would not have been sufficient for the Moon to grow to its present-day mass from a small embryo moving in the Moon's present-day orbit. This result supports the formation of a lunar embryo and its subsequent growth to form most of the Moon's present-day mass near Earth. It seems more likely that the initial lunar embryo, with a mass no greater than 0.1 times that of the Moon, formed simultaneously with the Earth's embryo from a common, rarefied condensation. For more efficient growth of the lunar embryo, it would be desirable for some impacts of impactors with Earth to not simply eject the ejected bodies from the crater, but for some of the material to enter orbit around the Earth, as in the multiple-impact model. The average collision velocities of ejected bodies with the Earth are greater, the greater the ejection velocity. These collision velocities were approximately 13, 14-15, 14-16, 14-20, and 14-25 km/s, with ejection velocities of 11.3, 11.5, 12, 14, and 16.4 km/s, respectively. The collision velocities of bodies with the Moon were also higher at high ejection velocities and were generally within the range of 7–8, 10–12, 10–16, and 11–20 km/s with $v_{ej}$ equal to 11.3, 12, 14, and 16.4 km/s, respectively.

## 10. Formation of axial rotations of planets

The axial rotations of most planets in the Solar System (with the exception of Venus and Uranus) are prograde, i.e., they coincide with the direction of their orbital motion. Many authors believe that most planets acquired their current axial rotations during the formation of planets from the protoplanetary circumsolar cloud. In modeling the evolution **of flat disks of hundreds of solid bodies** orbiting the Sun and merging during collisions using the method of spheres (primarily spheres of action), Ipatov [2-4, 125, 217, 224, 226] was the first to study the axial rotation periods of the resulting planetary bodies. Analysis of the obtained results showed that, as a result of disk evolution, some of the resulting bodies usually acquired positive axial rotations, while others acquired negative axial rotations. Moreover, if the orbital eccentricities (more precisely the values $e_m \mu^{1/3}$, where $e_m$ is the greatest eccentricity of the heliocentric orbits of the two merging bodies, $\mu$ is the mass of the larger of these bodies in solar masses) were relatively small during the course of evolution, then among the resulting bodies, those with positive axial rotations predominated. If, however, evolution proceeded with large orbital eccentricities, then the proportions of the resulting bodies with positive and negative axial rotations were approximately equal. The small number of initial bodies (no more than 1000) in the considered calculation variants, and, consequently, the small ratios of the masses of the colliding bodies, are possibly the reason that the periods of axial rotation of the resulting bodies in the considered variants were generally somewhat shorter than the periods of axial rotation of modern planets. A review of works on the formation of axial rotation of planets is given in paragraph 4 of Chapter 5 of the monograph [78].

**By simulating multiple (hundreds of thousands in each variant) isolated approaches** (up to the radius $r_s$ of the sphere under consideration) of two bodies moving around the Sun in the same plane, calculations were made of the values of $\Delta K = K^+ - K^-$ (where $K^+$ and $K^-$ are the fractions of positive and negative moments of angular momentum of the colliding bodies relative to their common centers of mass among the sum of the moduli of moments of both signs) for different initial data (masses $m_1 = \mu_1 M_S$ and $m_2 = \mu_2 M_S$ ($M_S$ is the mass of the Sun), densities $\rho$ ($k_\rho$ is the ratio of the density of bodies to the density of the Earth), eccentricities $e_1$ and $e_2$ and the semi-major axes $a_1$ and $a_2$ of the heliocentric orbits). The mutual gravitational influence of the bodies was taken into account by the method of spheres (mainly spheres of action). The results of these calculations are presented in preprints [222-223] and summarized in paragraph 4 of Chapter 5 of the monograph [78]. In each of the calculation variants, the eccentricities and masses of the bodies, as well as the direction to the pericenter of the orbit of the first body, were fixed. The direction to the pericenter of the orbit of the second body varied from 0 to $2\pi$ radians. In the calculation variants of $\Sigma$, the ratio $a_2/a_1$ of the semi-major axes of the orbits took on different values, at which the approach of the bodies to $r_s$ was possible. The results of such calculations [222] made it possible to estimate the average values of $\Delta K$ for different orbital eccentricities and densities of bodies and planets. It was shown that the values of $K^+$ and $\Delta K$ (especially in the $\Sigma$ variants) tend to decrease with increasing $a_1 k_\rho^{1/3}$ and $e_m = \max\{e_1, e_2\}$. In some variants for which $a_2/a_1$=const, the dependence of $K^+$ on $a_1 k_\rho^{1/3}$ is not entirely monotonic. In the variants of calculations carried out by the sphere-of-action method, as well as in the numerical integration of the equations of motion of the three-body problem, it was found that the values of $\Delta K$ are maximum at $e_H=e_m(\mu/3)^{-1/3}\sim1$.

From the obtained results, it is evident, in particular, that for the same values of $a_1k_p^{1/3}$, $e_m$=max{$e_1$, $e_2$} and μ=max{$\mu_1$, $\mu_2$}, the values of $\Delta K$ may differ by several times for different values of $e_1$, $e_2$, $\mu_1$ and $\mu_2$. The obtained results show that the values of $\Delta K$ may depend significantly on the ratio $a_2/a_1$. Although the values of $\Delta K$ and the probabilities of collisions of bodies during their approaches to the radius of the sphere of action are greater in the case of nearly tangent orbits than for other orbits, when studying the formation of axial rotations of planets, one should not limit oneself to the study of only nearly tangent orbits (especially when the eccentricities of the orbits of the planet and bodies are considerable). It is desirable to take into account all permissible orbits of bodies.

It is believed that Jupiter and Saturn were formed primarily by gas accretion, while the axial rotation rates of Mercury, Venus, and Pluto were strongly modified by tidal forces. Comparing the $\Delta K$ values corresponding to the modern periods of planetary axial rotation with the $\Delta K$ values obtained using different initial data, we find that, for the axial moments of Earth and Neptune to have been formed by the accretion of small bodies, the majority of the bodies that fell onto these planets must have had weakly eccentric orbits ($e<0.05$). The numerical results obtained by Ipatov [222-223] indicate that the regular component of the axial moments of some planets (e.g., Earth and Neptune) was significantly affected by the fall of large bodies onto the planets.

Within the framework of the solid body accumulation model, the modern **tilts of the planets' rotation axes** are usually explained by the fall of large bodies onto the embryos. Ipatov obtained [223] that the most massive bodies (with a mass of up to several Earth masses) should have fallen onto Saturn and Uranus. If the initial period of the Earth's axial rotation is considered equal to 6 hours, then, according to [223], the tilt of the Earth's rotation axis during a single impact is obtained from its collision with a body with a mass of $0.03m_E$. With the current period of the Earth's axial rotation (24 hours), the mass of the impactor is 4 times smaller ($\sim0.01m_E$). Large bodies that fell onto the planets made a significant contribution not only to the tilts of the rotation axes, but also to the regular components of their axial rotation moments.

Studies of the formation of protoplanet axial rotations through the collisions of **rarefied gas-dust condensations** has shown [223] that their axial moments significantly exceed those of modern planets. The results indicate that if the protoplanet and condensations are homogeneous spheres, coinciding with their Hill spheres, then the axial moment of the rarefied proto-Earth is almost two orders of magnitude greater than that of the modern Earth. However, due to tidal interactions of protoplanets with the Sun, their axial moment decreases significantly during contraction of protoplanets.

## 11. Migration of small bodies from various regions of the Solar System to the terrestrial planets

Water and volatiles are important for the origin and evolution of life in the Solar System and in extrasolar systems. Earth's ocean water and its deuterium-to-hydrogen D/H ratio may be the result of mixing water from several exogenous and endogenous sources with high and low D/H ratios. The problem of celestial body migration in the Solar System is also important for understanding the formation and evolution of the Solar System. Endogenous sources of water could include direct adsorption of hydrogen from nebular gas into magma melts, followed by hydrogen reaction. with iron oxide FeO and accumulation of water by particles in the protoplanetary disk before the beginning of gas dissipation in the inner part of the early Solar System, which could have increased the D/H ratio of deuterium to hydrogen in the Earth's oceans. They also include accumulation of water by particles in the protoplanetary disk before the beginning of gas dissipation in the inner part of the early Solar System. Exogenous sources of water are discussed below. Exogenous sources included the migration of bodies from beyond the ice line. In a number of studies, the outer part of the main asteroid belt was considered as the main source of water on the Earth. Many bodies collided with the terrestrial planets and the Moon during the Late Heavy Bombardment (LHB). Various estimates of this period lie within the range from 4.5 to 3.5 billion years ago.

Ipatov studied the migration of bodies to the terrestrial planets from the main and outer asteroid belts, from the zone of giant planets, and from the trans-Neptunian belt. The migration of bodies to the terrestrial planets from the trans-Neptunian belt is discussed above at the end of Section 7, named "Formation of the asteroid and trans-Neptunian Belts." In particular, in [22], it was estimated that up to 20% of near-Earth objects with a diameter $d>1$ km could originate from the trans-Neptunian belt. Studying the probabilities of collisions of bodies with the Earth is important when considering the problem of asteroid-comet hazard (ACH), which is attracting growing attention as a potential source of threat to terrestrial civilization. Among the near-Earth objects (NEOs), the greatest danger is presented by three main groups of near-Earth asteroids: the Amor, Apollo, and Aten groups, of which bodies from the Amor group come close to the Earth's orbit ($1.017<q=a(1-e)<1.3$ AU), while bodies from the Apollo and Aten groups cross it, with the values of their semi-major axes, respectively, greater than or less than 1 AU. There is also a small group of asteroids of the Atira group with orbits lying inside the Earth's orbit ($Q=a(1+e)<0.983$ AU). Some NEOs reach kilometers in size, and their collision with the Earth is capable of causing a global catastrophe, as has happened more than once in the geological history of our planet. Many researchers believe that asteroids are the main source of NEOs, while comets pose a greater danger due to the suddenness of their appearance.

Until the mid-1990s, in addition to numerical integration, I also used **the sphere-of-action method when modeling the migration of bodies to the Earth's orbit under the influence of planets**. In [19] and in paragraphs 7-8 of Chapter 7 of the monograph [78], in particular, I considered the evolution of disks that initially consisted of 500 bodies with identical semi-major axes $a_o$ (≥1.7 AU) and large (≥0.4) initial eccentricities $e_o$ of their orbits. The results of calculations of such spatial (with an mean inclination of about 15 degrees) disks, presented in Table 1, showed that the fraction of bodies ejected into hyperbolic orbits is $k_h$~5-30 times greater than the fraction of bodies that collided with planets, most of the eccentricities of the hyperbolic orbits of ejected bodies lie between 1.01 and 2, and the fraction of nearly parabolic orbits is small. The number of bodies colliding with Venus was close to the number of bodies colliding with Earth. The fraction of bodies colliding with Mercury and Mars depended significantly on $a_o$ and $e_o$. At $a_o$=2.5 AU and $e_o$=0.4 (for bodies coming from the 3:1 resonance with Jupiter), the fractions of bodies colliding with Mars and Mercury were, respectively, three and half those for Earth. At $a_o$=1.7 AU and $e_o$ =0.5 (typical orbits of the Apollo asteroids) and at $a_o$ =2.82 AU and $e_o$=0.7 (for bodies coming from a 5:2 resonance with Jupiter), the time during which the number of bodies in the disk decreased by a factor of two was two orders of magnitude shorter than the average time before the collision of bodies with Earth. Moreover, more than 10% of the bodies from the 5:2 resonance that collided with the Earth did so less than 1 million years ago, although the times before the last collisions can exceed 100 million years.

**Table 1.** The fraction $p_{E500}$ of bodies colliding with the Earth and the fraction $p_h$ of bodies ejected into hyperbolic orbits with eccentricities $e_h$ in a number of ranges. $k_h$ – the ratio of the number of bodies ejected into hyperbolic orbits to the number of bodies colliding with planets, $T_h$ is the time it takes for the number of bodies in the disk to halve. Since there are 500 bodies in the initial disk, the probability of a collision for one body is 500 times smaller than $p_{E500}$.

| $a_o$, AU | 1.7 | 2.5 | 2.82 | 7 | 25 |
|---|---|---|---|---|---|
| $e_o$ | 0.5 | 0.4 | 0.7 | 0.5 | 0.5 |
| $p_{E500}$ | 0.04 | 0.03 | 0.14 | 0.04 | 0.014 |
| $k_h$ | 8 | 5 | 20 | 30 | 20 |
| $T_h$, million years | 1 | 10 | 0.2 | 0.1 | 1 |
| $p_h$ at $e_h$>2 | 0.004 | 0.008 | 0.0 14 | 0.002 | 0.004 |
| $p_h$ at 1.1<$e_h$<2 | 0.452 | 0.453 | 0.398 | 0.406 | 0.360 |
| $p_h$ at 1.01<$e_h$<1.1 | 0.396 | 0.381 | 0.396 | 0.464 | 0.474 |
| $p_h$ at 1.001<$e_h$<1.01 | 0.108 | 0.1 32 | 0.1 4 8 | 0.1 1 8 | 0.1 44 |
| $p_h$ at 1.0001<$e_h$<1.001 | 0.032 | 0.026 | 0.044 | 0.008 | 0.01 6 |
| $p_h$ at 1.00001<$e_h$<1.0001 | 0.00 8 | 0 | 0 | 0.002 | 0.002 |

Migration of bodies **from the main asteroid belt** was discussed in the article [19] and paragraphs 2 and 8 of Chapter 7 of the monograph [78]. The following conclusions were made. Asteroids that entered the Kirkwood gaps due to the gravitational influence of the largest asteroids may constitute several percent of NEOs. A larger number of bodies may enter the gaps as a result of mutual collisions of asteroids. Therefore, NEOs delivered from the asteroid belt are largely products of high-speed collisions. It was found that the average time until the first collision of an asteroid with a diameter $d$=1 km with an asteroid with a diameter $d'$≥0.1$d$ is less than 10 thousand years. The average lifetime of small ($d$<100 km) main belt asteroids does not exceed the lifetime of the Solar System, and the lifetime of a body with a diameter of 1 m in the asteroid belt can be of the order of a million years. The typical time before an NEO impacts Earth is approximately 100 million years. The probability of an NEO being ejected into a hyperbolic orbit is an order of magnitude greater than the probability of its impact with the Earth. For a half of the bodies that collide with the Earth, the time interval between entering an Earth-crossing orbit and the impact with the Earth does not exceed 10 million years. An impact of NEO with a diameter of $d$>1 km with the Earth can occur on average once every 100,000 years.

In the papers [27-29, 93-94] the evolution of the orbits of ~30,000 **initially Jupiter-crossing objects** (JCOs) with an initial period $P$<20 years was studied. In particular, the comets 2P, 10P, 44P and 113P were considered. The gravitational influence of seven planets (Venus–Neptune) was taken into account. Resonant asteroids and trans-Neptunian objects were also considered. For the integration of the equations of motion, the symplectic algorithm from the SWIFT package was mainly used (Levison, Duncan, Icarus, 1994, v. 108, 18-36). For the problems considered, the results of calculations by this method were close to the results of calculations by the method (Bulirsh, Stoer, Numerishe Mathematik, 1996, v. 8, 1-13). The average dynamic lifetime of JCOs was obtained to be about 100 thousand years, and the average time of motion of an object initially crossing the orbit of Jupiter in an orbit crossing the orbit of the Earth was about 30 thousand years. However, among the 30 thousand JCOs considered, there were several objects that migrated to orbits lying inside the orbit of Jupiter and moved along them for tens of millions of years. The probability of a collision of such an object with a terrestrial planet may be greater than that of the 10 thousand other objects that crossed the orbit of Jupiter. Although only a small fraction of the migrated objects had semi-major axes of orbits $a$ <2 AU, the average time spent by objects initially crossing the orbit of Jupiter in orbits with $a$<2 AU was comparable to the time for $a$=3 AU. The probability of a collision of an object crossing Jupiter's orbit with the Earth was of the order of $4\times10^{-6}$-$4\times10^{-5}$ (interval for different groups of objects).

**The migration of planetesimals from the feeding zone of Jupiter and Saturn to the forming terrestrial planets** was studied using the SWIFT symplectic integrator. In the "JS" series of calculations, the modern orbits and masses of the terrestrial planets, Jupiter, and Saturn were considered. In the $JS_{01}$ series the masses of the terrestrial planets were 10 times smaller than their present values, and Jupiter and Saturn had present masses and orbits (in a number of cosmogonic models, it is believed that Jupiter and Saturn were almost formed when the masses of the terrestrial planets were still far from their present values). In the JN and $JN_{01}$ series, in addition to the initial data for JS and $JS_{01}$, Uranus and Neptune in their present orbits were considered. In the four series of calculations JS, $JS_{01}$, JN and $JN_{01}$, the semi-major axes of the initial orbits of the planetesimals varied **from $a_{min}$=4.5 to $a_{max}$=12 AU**, and the number of planetesimals with a semi-major axis of the orbit close to $a$ was proportional to $a^{1/2}$. The eccentricities of the initial orbits of the planetesimals were 0.3 (such average eccentricities of the orbits of planetesimals were at the final stages of accumulation of the terrestrial planets), and the inclinations of their orbits were 0.15 rad. The probabilities of collisions of the considered planetesimals (from the zones of Jupiter and Saturn) with the Earth were of the order of $2\times10^{-6}$ [53]. In [53] it was noted that about 30% of the water delivered from the feeding zone of Jupiter and Saturn could have been delivered during the growth of the Earth embryo to $0.5m_E$.

In a number of calculations [59, 122, 484, 496], the initial values $a_o$ of the semi-major axes of the orbits of planetesimals-bodies varied from $a_{min}$ to $a_{min}+d_a$, their initial eccentricities were equal to $e_o$, and the initial inclinations were equal to $e_o/2$ rad. In some calculations, the values of $a_{min}$ varied in 2.5 AU increments **from 5 to 40 AU** and $d_a$=2.5 AU, and $e_o$=0.05 or $e_o$=0.3. In other calculations, the values of $a_{min}$ varied in 0.1 AU increments from 3 to 4.9 AU and $d_a$=0.1 AU, and $e_o$=0.02 or $e_o$=0.15. The gravitational influence of the planets (from Venus to Neptune) and the Sun was taken into account. The SWIFT integration package was used to integrate the equations of motion. In the calculations presented in Figs. 1–6, the planets were considered as material points. The probabilities of collisions of bodies with planets and the Moon were calculated based on arrays of orbital elements of migrated bodies. In later versions of the calculations, bodies colliding with planets were excluded from further calculations, and it was taken into account when calculating the probabilities of collisions of bodies with planets. In these new variants of the calculations, the probabilities of collisions of bodies with planets and the Moon were also additionally calculated based on arrays of orbital elements of migrated bodies, since these probabilities are sometimes small, and for some planets (especially for the terrestrial planets during the migration of bodies from the zone of the giant planets) sometimes there were no collisions of bodies with planets during the integration. The results of calculations with the removal of colliding bodies correspond better to the actual evolution, since the colliding bodies no longer appear in further calculations and no longer contribute to the collision probabilities. The results of the calculations on the migration of bodies discussed in this paragraph are planned to be published in more detail later in journal articles. For now, they are briefly presented (and only for calculations with material points) in abstracts [122, 484, 496] and in a review article [59].

Below are Figures 1-2 from [496]. They show the probabilities of collisions of bodies with the Earth, multiplied by a million, for initial values of the semi-major axes of the orbits of bodies from 3.2 to 5 AU (with $a_{min}$ from 3.2 to 4.9 AU) with $e_o$=0.02 and $e_o$=0.15 for various time intervals (from 1 million years to 1000 million years, and for a finite time interval).

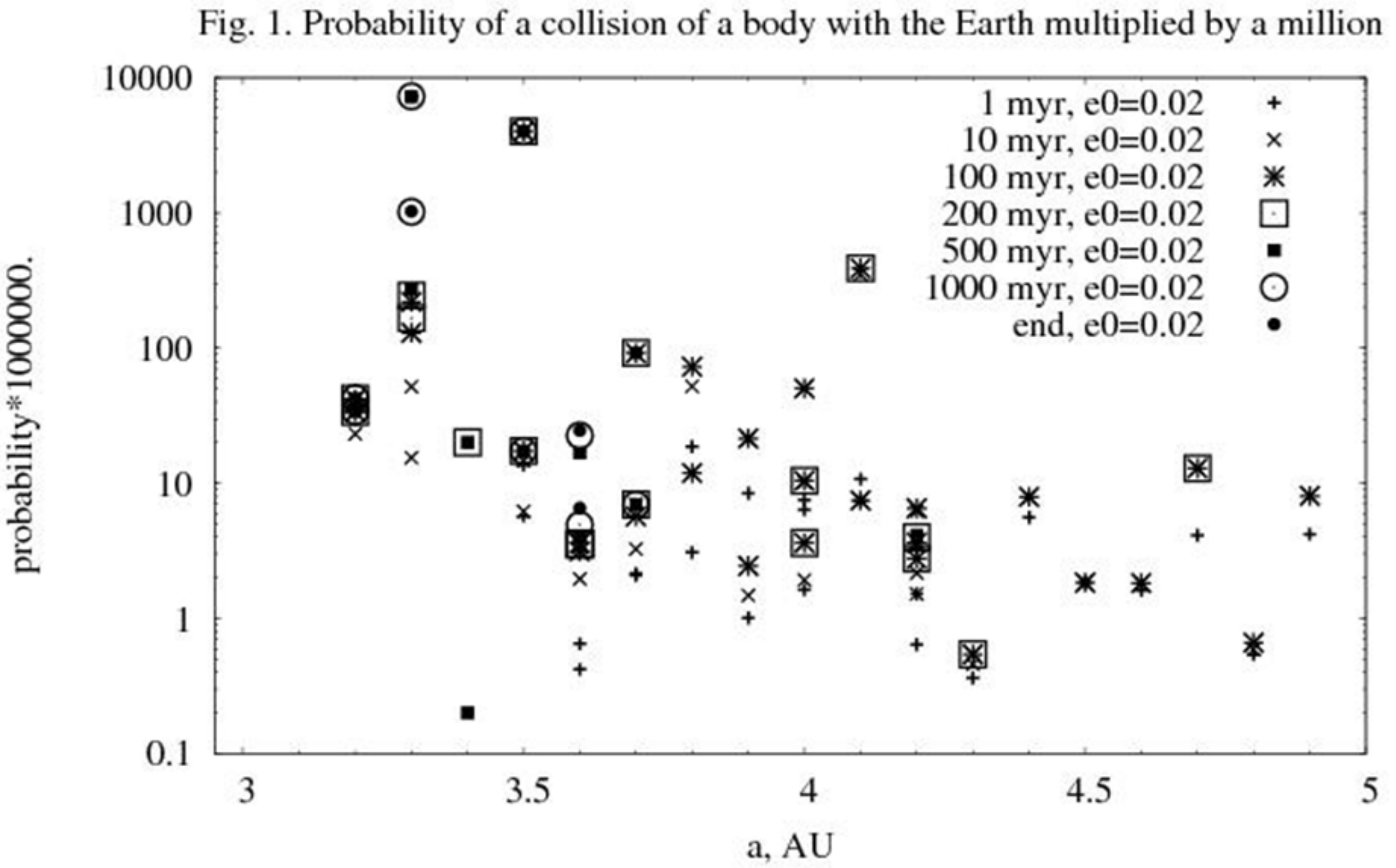

**Figure 1.** Probabilities of collisions of bodies with the Earth, multiplied by a million, for $a_{min}$ from 3.2 to 4.9 AU and $e_o$=0.02 for different time intervals.

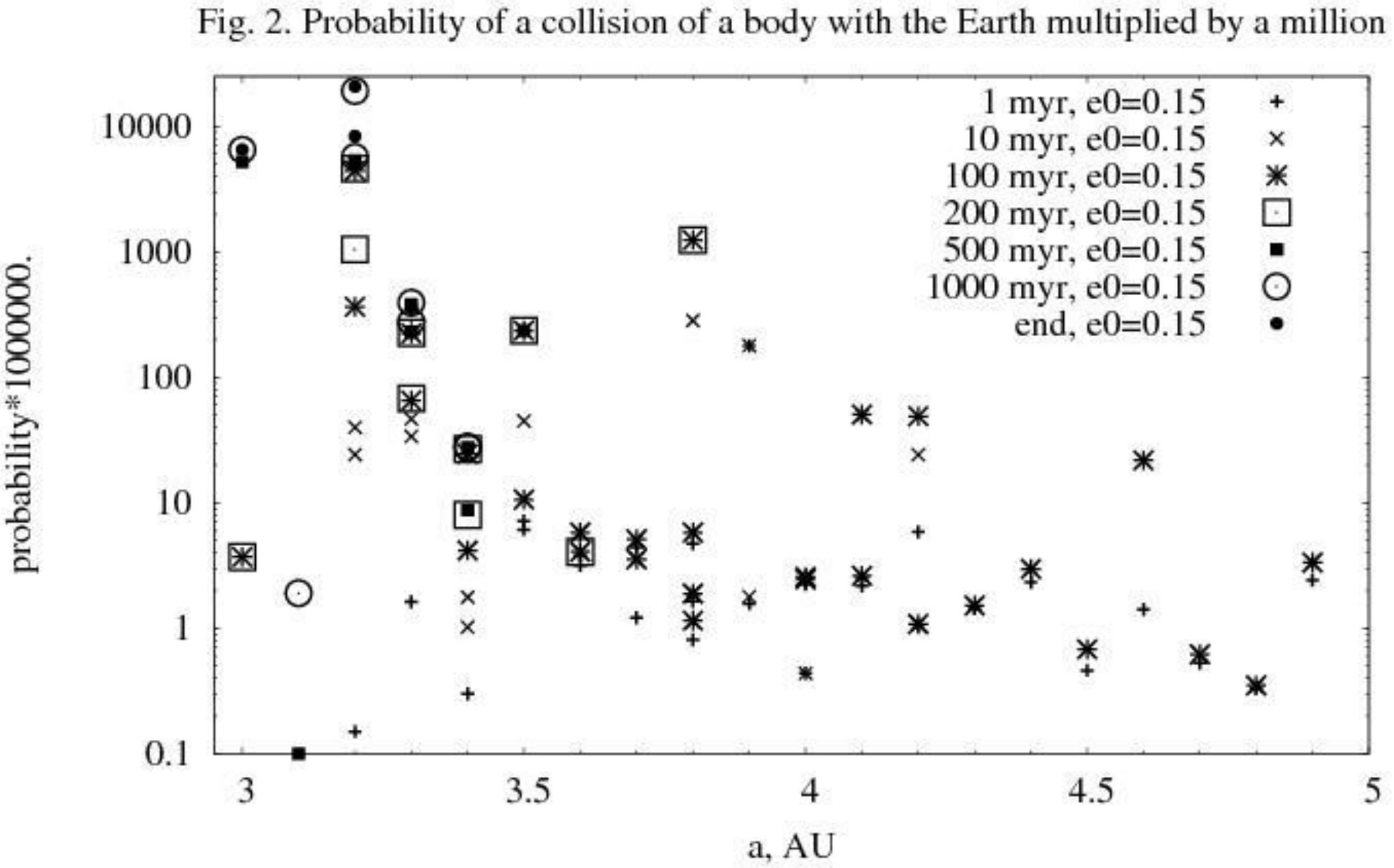

**Figure 2.** Probabilities of collisions of bodies with the Earth, multiplied by a million, for $a_{min}$ from 3.2 to 4.9 AU and $e_o$=0.15 for different time intervals.

Figures 3-6, taken from [484], show the probabilities of collisions of bodies with the Earth, multiplied by a million (the + sign in the upper right corner refers to the description of the notations, and not to the value of the probability for a given $a_{min}$). Figures 3-4 examines the results of calculations for $a_{min}$ from 5 to 40 AU, $d_a$=2.5 AU and $e_o$=0.05 or $e_o$=0.3. Figure 5 shows the case when Uranus and Neptune are excluded from the calculations, and bodies are considered only in the zone from 5 to 10 AU. Figure 6 shows the results of calculations for $a_{min}$ from 2.5 to 40 AU, $d_a$=0 and $e_o$=0.3.

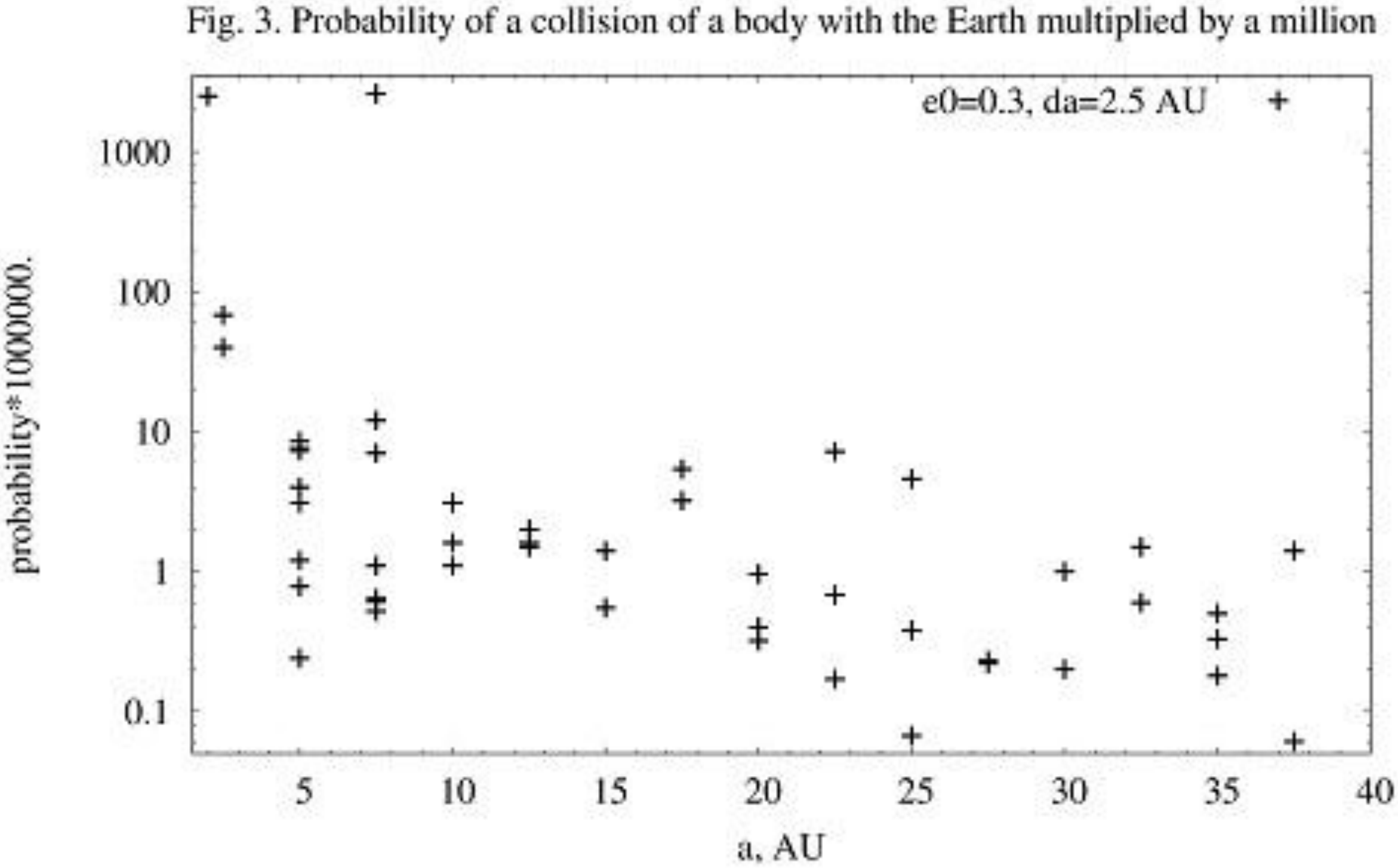

**Figure 3.** Probabilities of collisions of bodies with the Earth, multiplied by a million, for $a_{min}$ from 2.5 to 37.5 AU, $d_a$=2.5 AU and $e_o$=0.3 for the entire time interval considered.

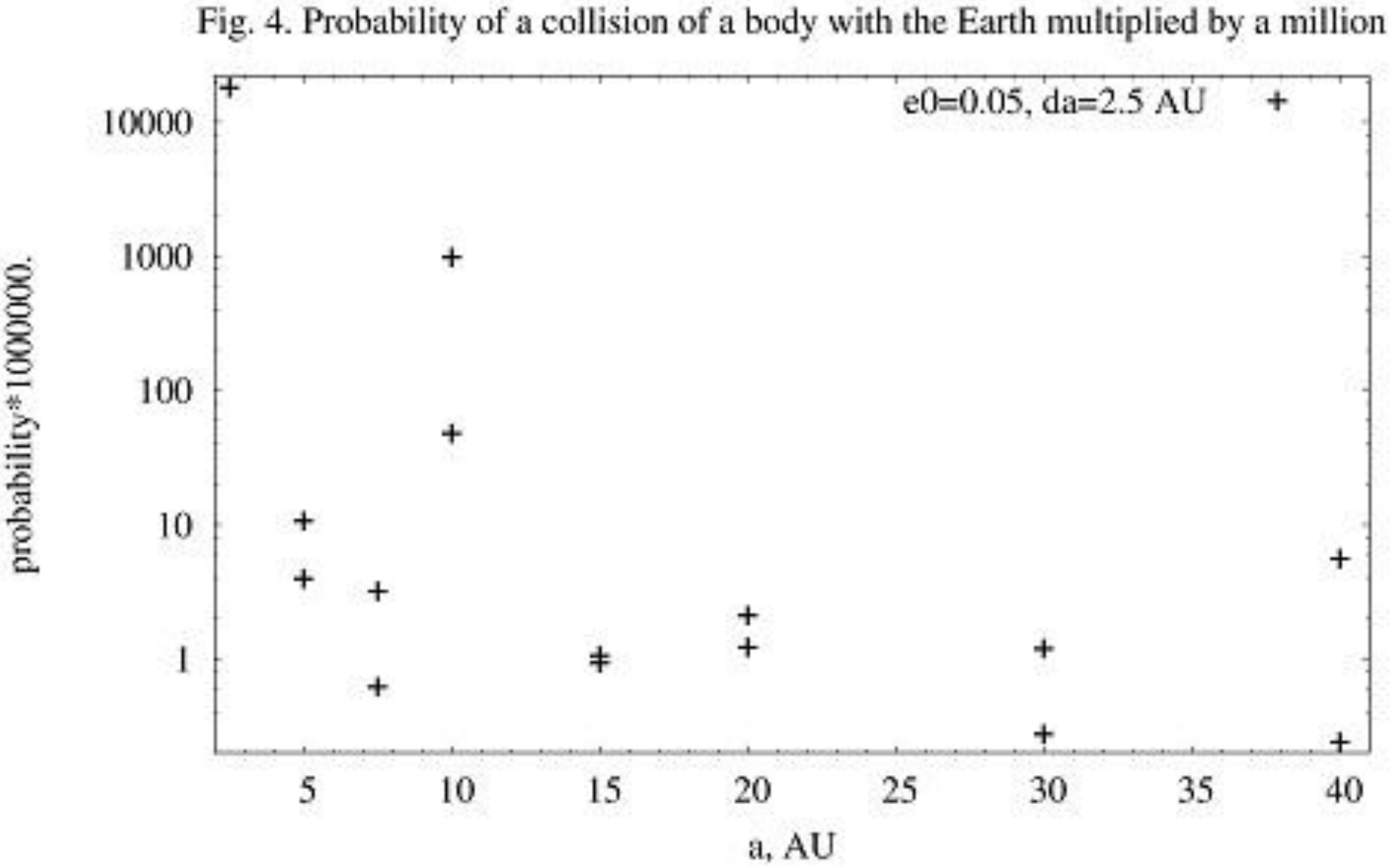

**Figure 4.** Probabilities of collisions of bodies with the Earth, multiplied by a million, for $a_{min}$ from 2.5 to 40 AU, $d_a$=2.5 AU and $e_o$=0.05 for the entire time interval considered.

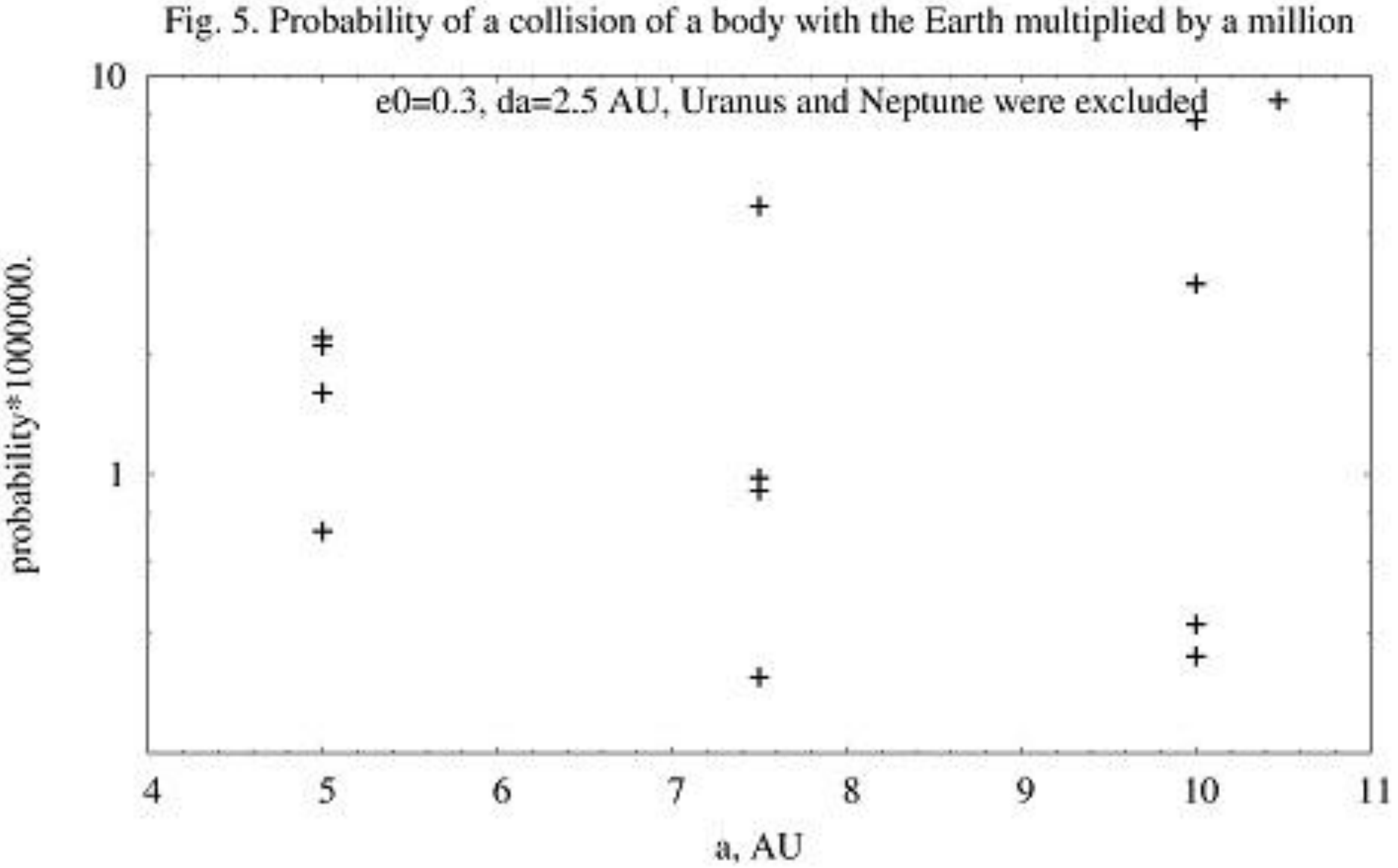

**Figure 5.** Probabilities of collisions of bodies with the Earth, multiplied by a million, for $a_{min}$ from 5 to 10 AU, $d_a$=2.5 AU and $e_o$=0.3 for the entire time interval considered.

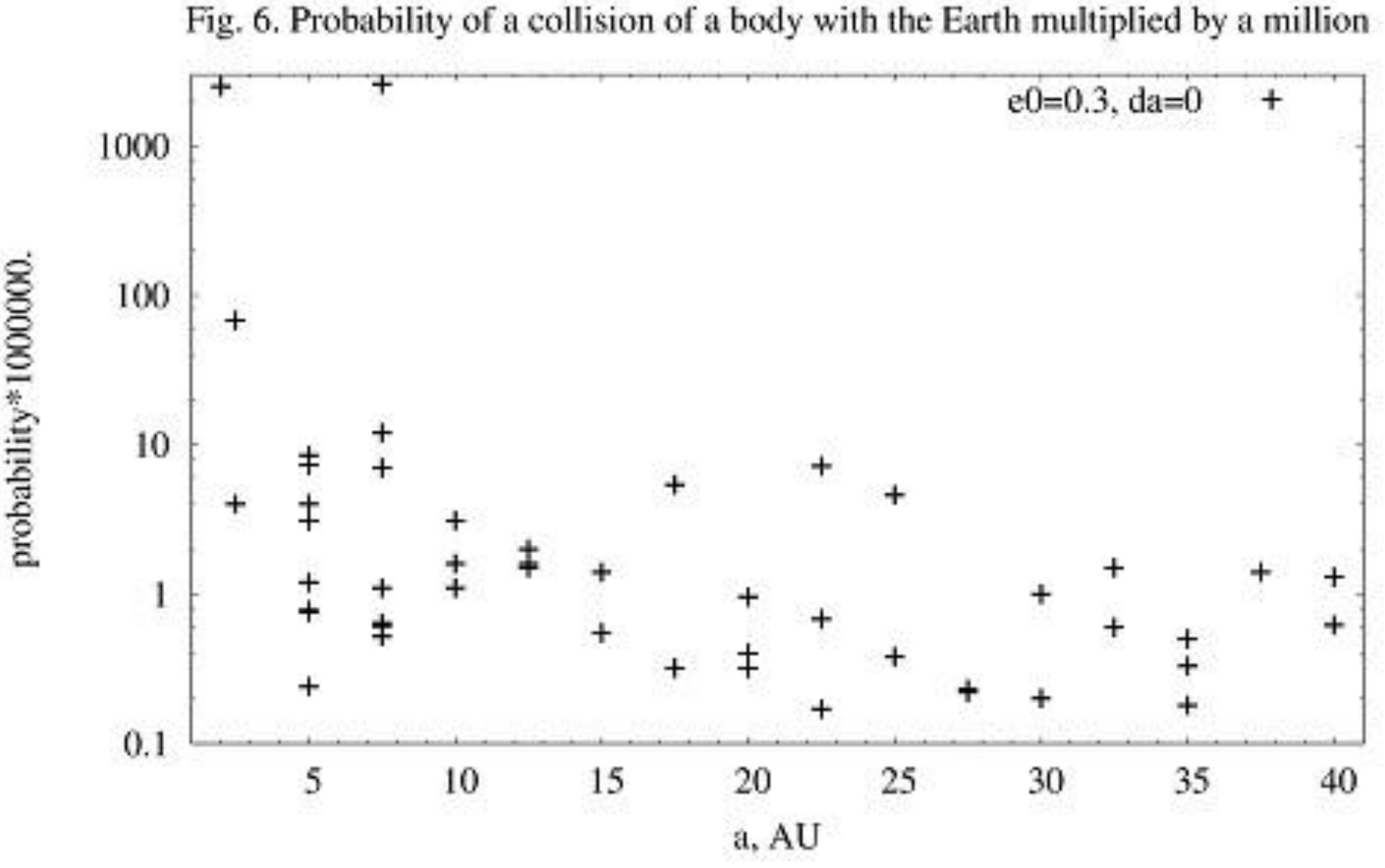

**Figure 6.** Probabilities of collisions of bodies with the Earth, multiplied by a million, for $a_{min}$ from 2.5 to 40 AU, $d_a$=0 AU and $e_o$=0.3 for the entire time interval considered.

The values $p_E$ of probabilities of a body's collision with the Earth were, on average, lower for larger initial values $a_o$ of the semi-major axes of the orbits at $5 \le a_o \le 40$ AU [484]. The values of the probability $p_E$ of a planetesimal's collision with the Earth are about $10^{-6}$ for $a_o$ of about 15-40 AU and about $10^{-5}$ at $a_o$ of about 4-10 AU. For some values of $a_{min}$ and $e_o$, the $p_E$ values calculated for 250 planetesimals may differ by hundreds of times for calculations with nearly identical initial orbits. This difference is caused by the fact that one in a thousand planetesimals could move in an orbit crossing the Earth's orbit for millions of years and have a much higher probability of colliding with the Earth than for other planetesimals. In the variants presented in Fig. 5, when Uranus and Neptune are excluded from the calculations, and the bodies are considered in the zone from 5 to 10 AU, in all variants the probability $p_E$ of a collision of bodies with the Earth did not exceed $10^{-5}$. In some variants with Uranus and Neptune, for the same values of $a_{min}$ and $e_o$, the values of $p_E$ exceeded $10^{-5}$.

Bodies initially located at different distances from the Sun **reached the Earth after different times** *t*. At $3 \leq a_{min} \leq 3.5$ AU and $e_o \leq 0.15$, individual bodies could have fallen on the Earth and the Moon after several hundred million years. For example, for $a_{min}$=3.3 AU and $e_o$=0.02, the value of $p_E = 4 \times 10^{-5}$ for $0.5 \leq t \leq 0.8$ Myr (the time of the Late Heavy Bombardment) and $p_E = 6 \times 10^{-6}$ for $2 \leq t \leq 2.5$ Myr. For $a_{min}$=3.2 AU and $e_o$=0.15, it was obtained that $p_E$=0.015 for $0.5 \leq t \leq 1$ Myr, and $p_E = 6 \times 10^{4}$ for $1 \leq t \leq 2$ Myr. The outer asteroid belt zone could have been one of the sources of the "late heavy bombardment". Most of the fallouts on the Earth of bodies that were initially at a distance of 4 to 5 AU from the Sun, occurred during the first 10 million years. Bodies that initially crossed Jupiter's orbit could arrive at Earth's orbit mainly during the first million years (after the formation of massive Jupiter). The time of migration of bodies to Earth's orbit from the feeding zone of Uranus and Neptune depended on when large embryos of these planets appeared in this zone. According to [15], the main changes in the orbital elements of the embryos of the giant planets occurred in no more than 10 million years. Most of the bodies with $5 < a_{min} < 30$ AU that fell onto the Earth, fell on it within 20 million years. For $a_{min} > 20$ AU, the values of $p_E$ could increase slightly even after hundreds of millions of years, and individual bodies could remain in elliptical orbits even after a time equal to the age of the Solar System.

The total mass of the Earth's oceans is about $2 \times 10^{-4} m_E$. At $p_E = 4 \times 10^{-6}$ and with a total mass of planetesimals $m_\Sigma$ at an initial distance of 5 to 10 AU from the Sun equal to $100 m_E$, the total mass $m_{\Sigma E}$ of planetesimals that fell onto the Earth would be $4 \times 10^{-4} m_E$. For the 10–40 AU zone, at $p_E = 1.5 \times 10^{-6}$ and $m_\Sigma = 100 m_E$, we have $m_{\Sigma E} = 1.5 \times 10^{-4} m_E$. For the 3–4 AU zone, at $p_E = 10^{-3}$ and $m_\Sigma = 10 m_E$, we have $m_{\Sigma E} = 0.01 m_E$. The estimated amount of ice in comets does not exceed 33%. However, some authors believe that the rimordial planetesimals could have contained more ice (~50%) than modern comets. The above estimates show that **the total mass of water ice delivered to Earth** from beyond Jupiter's orbit could be comparable to the mass of the Earth's oceans. Although bodies in the outer asteroid belt had a smaller total mass and likely contained less ice (~10%) than bodies beyond Jupiter, due to a much higher probability of collisions with Earth, bodies from the outer asteroid belt could have delivered even more water to the Earth than bodies from the zone of the giant planets. Due to the decrease in the semi-major axis of Jupiter's orbit, which ejected planetesimals into hyperbolic orbits, the zone in the asteroid belt with a higher probability of delivering planetesimals to the Earth shifted and, over time, encompassed a larger number of planetesimals. Some of this water was lost in collisions of planetesimals with planets, especially with the Moon. Consequently, the amount of water incorporated into the terrestrial planets and the Moon could have been less than the amount of water delivered to these celestial objects.

The total mass of water delivered **to Venus and Mercury** from beyond Jupiter's orbit, calculated per planetary mass, was approximately the same as for Earth, and the mass of water delivered to Mars, calculated per unit of planetary mass, was approximately 2–3 times greater than for Earth [53]. In absolute terms, the mass of water delivered to Mars was 3–5 times less than the mass of water delivered to Earth. The results support the existence of ancient oceans on Mars and Venus, likely partially preserved at depth (Mars) or lost during evolution (Venus).

The mass of water delivered **to the Moon** from beyond Jupiter's orbit could have been less than that for the Earth by no more than a factor of 20. For planetesimals in the feeding zone of the terrestrial planets, the ratio $r_{EM}$ of the number of planetesimals that collided with Earth to the number of collisions with the Moon generally varied from 20 to 40. Planetesimals initially more distant from Earth's orbit arrived at the Moon from more eccentric orbits, and for them, the ratio $r_{EM}$ was smaller than in the case of closer orbits of planetesimals with small eccentricities. In 80% of the variants of body migration calculations for $3 \leq a_o \leq 5$ AU, it was obtained that $16.4 \leq r_{EM} \leq 17.4$. In other variants of calculations for $a_o \geq 3$ AU, the ratio $r_{EM}$ could be in the range from 14.6 to 17.9. The fraction of water evaporated during impacts with the Moon is higher at higher impact velocities and, accordingly, at lower values of $r_{EM}$.

The square of the effective radius $r_{ef}$ of a celestial body of radius $r$ is equal to

$$r_{ef}^2 = r^2[1 + (v_{par}/v_{rel})^2], \quad (1)$$

where $v_{par}$ is a parabolic velocity on the surface of this celestial body, and $v_{rel}$ is the relative velocity of a small body upon its entry into the sphere of action of a celestial object (the exact formula is valid for the relative velocity at infinity). Based on the ratio $r_{EM}$ of the number of planetesimals colliding with the Earth and the Moon, one can estimate the characteristic velocity $v_{relE}$ (relative to the Earth) of planetesimals upon their entry into the sphere of action of the Earth. Considering that $r_{EM}$ is equal to the ratio of the squares of the effective radii of the Earth and the Moon, and using formula (1), for the ratio $v_{relE}/v_{parE}$ of the relative velocity of the body's entry into the sphere of action of the Earth to the parabolic velocity $v_{parE}$=11.186 km/s on the surface of the Earth, we obtain

$$(v_{relE}/v_{parE})^2 = [r_{EM}(v_{parM}/v_{parE})^2 (r_M/r_E)^2 - 1] / [1 - r_{EM}(r_M/r_E)^2] , \quad (2)$$

where $r_M$ and $r_E$ are the radii of the Moon and the Earth, $v_{parM}$ and $v_{parE}$ are the parabolic velocities on the surface of the Moon and Earth, respectively. Considering that for the velocities $v_{colE}$ and $v_{colM}$ of collisions of bodies with the Earth and the Moon the following relationships are valid: $v_{colE}^2=v_{relE}^2+v_{parE}^2$ and $v_{colM}^2=v_{relM}^2+v_{parM}^2$, and setting $v_{relE}=v_{relM}$, for some values *of* $r_{EM}$ in [58] estimates were made of the values of $v_{relE}$ and the characteristic **velocities** $v_{colM}$ and $v_{colE}$ **of collisions of bodies with the Earth and the Moon** for some cases.

For planetesimals near the Earth's orbit (with $0.9 \le a_o \le 1.1$ AU), the characteristic velocities of their collisions with the Moon ranged from 8 to 10 km/s, and with the Earth – from 13 to 15 km/s. For planetesimals coming from other parts of the feeding zone of the terrestrial planets (with $20 \le r_{EM} \le 40$), the spread of characteristic velocities of planetesimals' collisions with the Earth was mainly in the range from 13 to 19 km/s, and with the Moon from 8 to 16 km/s. For most bodies with $a_o \ge 3$ AU (with $16.4 \le r_{EM} \le 17.4$), a similar spread of characteristic collision velocities was 23–26 km/s for the Earth and 20–23 km/s for the Moon. However, the range of collision velocities for all bodies in this zone (at $14.6 \le r_{EM} \le 17.9$) was wider: from 22 to 39 km/s for Earth and from 19 to 38 km/s for the Moon. These velocities are greater than the parabolic velocities on the surfaces of these celestial bodies, allowing some of the bodies ejected during collisions to enter heliocentric orbits.

The characteristic velocities of collisions of planetesimals, initially located relatively close to the orbit of the Earth embryo, with the Earth and Moon embryos, whose masses were ten times smaller than the present-day masses of these celestial objects, generally ranged from 7–8 km/s for the Earth embryo and from 5 to 6 km/s for the Moon embryo. For planetesimals originating from more distant (from Earth's orbit) regions of the feeding zone of the terrestrial planets, the characteristic velocities ranged from 9 to 11 km/s for collisions with the Earth embryo and from 7 to 10 km/s for collisions with the Moon embryo.

## 12. Variations in orbital elements of planets

When modeling the evolution of the orbits of zero-mass bodies under the influence of planets, graphs of changes in the elements of the planets' orbits were also constructed, given in Appendix 2 of the monograph [78] and in the article [24] (which also provides a review of publications on this item). Two tables from these publications are given below. Table 1 presents for all planets the limits $\Delta a=a_{max}-a_{min}$ of changes in the semi-major axes of the orbits, the minimum and maximum distances ($R_{min}$ and $R_{max}$) from the Sun, the minimum and maximum values of the eccentricity and inclination of the orbit ($e_{min}$, $e_{max}$, $i_{min}$, $i_{max}$), $\Delta e=e_{max}-e_{min}$ and $\Delta i=i_{max}-i_{min}$. With the exception of Mercury ($T$=200 million years), the data for the remaining planets are given for $T$=20 million years (integrated into the past). Changes in the orbital elements of the planets were also discussed in the abstract [277].

**Table 1.** Limits of changes in the orbital elements of the planets

| Planet | Mercury | Venus | Earth | Mars | Jupiter | Saturn | Uranus | Neptune | Pluto |
|---|---|---|---|---|---|---|---|---|---|
| $\Delta a$, a.u. | 0.000030 | 0.000033 | 0.0000 62 | 0.000 29 | 0.00 37 | 0.080 | 0.0231 | 0.407 | 1.004 |
| $R_{min}$, a.u. | 0.281 | 0.673 | 0.939 | 1.330 | 4,880 | 8.69 | 17.70 | 29.34 | 28.39 |
| $R_{max}$, a.u. | 0.493 | 0.774 | 1.061 | 1.717 | 5.524 | 10:45 | 20.78 | 31.00 | 51.15 |
| $e_{min}$ | 0.080 | 0.0002 | 0.0002 | 0.0018 | 0.0252 | 0.0074 | 0.0009 | 0.00005 | 0.2065 |
| $e_{max}$ | 0.273 | 0.0697 | 0.0608 | 0.1268 | 0.0618 | 0.0894 | 0.0757 | 0.0222 | 0.2806 |
| $\Delta e$ | 0.193 | 0.0695 | 0.0606 | 0.125 | 0.0366 | 0.082 | 0.0748 | 0.0222 | 0.0741 |
| $i_{min}$, deg. | 1.80 | 0.013 | 0.000 | 0.061 | 1.096 | 0.567 | 0.438 | 0.783 | 13.414 |
| $i_{max}$, deg. | 11.86 | 4.731 | 0.429 | 8.640 | 2.063 | 2.594 | 2.710 | 2.365 | 18.446 |
| $\Delta i$, deg. | 10.06 | 4.718 | 0.429 | 8.579 | 0.967 | 2.027 | 2.272 | 1.582 | 5.032 |

Table 2 shows the average values of the periods $T_e$, $T_i$, $T_\omega$, $T_{\Delta\Omega J}$ and $T_{\Delta\pi J}$ of the changes in the eccentricity, inclination and argument ω of the perihelion of the planet's orbit, the difference $\Delta\Omega_J=\Omega-\Omega_J$ between the longitudes of the ascending nodes of the planet and Jupiter and the difference $\Delta\pi_J=\pi_J-\pi$ between the longitudes of the perihelion of the planet and Jupiter. These data are obtained by analyzing graphs of changes in the elements of the planets' orbits over time. If the angles increase or decrease during evolution, the period is understood as the time it takes for them to change by 360°. The letters *I, D* and *L* in the lines $k_\omega$, $k_{\Delta\Omega J}$ and $k_{\Delta\pi J}$ indicate, respectively, the increase, decrease and libration of the angles. The combination *Il* is given in the case where the angle generally increases, but may vary around 0 for some time. $T_I$ denotes the period of changes in the inclination *i* of the orbit, greater than $T_i$ (the values of $T_I$ are not given in Table 2 if such changes in *i* are not clearly expressed). During evolution, the values of the periods can change several times. Table 2 presents the average values of the periods (mainly for intervals $T$=1 million years or $T$=20 million years).

Note that the periods $T_{\Delta\pi J}$ for Saturn and Uranus are the same and equal to the periods $T_e$ of changes in the eccentricities of the orbits of Jupiter and Saturn (0.054 million years). For Jupiter, the value of $\pi_J$ increases by 360° in 0.3 million years, and the ratio $d\pi_J/dt=\nu_5\approx4.20''$/year. For all planets (except Jupiter) it was found that $T_{\Delta\Omega J}\approx T_i$. The periods $T_{\Delta\Omega J}$ for Venus, Earth, and Mars are approximately the same (equal to 0.07 million years). The values of $\Delta\Omega_J$ for Saturn, Uranus, and Neptune librated near 0, and the value of $\Delta\pi_J$ for Uranus librated near 180°.

**Table 2.** Periods of variations in planetary orbital elements (in million years). Designations are given in the text.

| Planet | Mercury | Venus | Earth | Mars | Jupiter | Saturn | Uranus | Neptune | Pluto |
|---|---|---|---|---|---|---|---|---|---|
| $T_\omega$ | 0.12 | 0.1 | 0.1 | 0.036 | 0.30 | 0.046 | 0.30 | 1.91 | 3.8 |
| $k_\omega$ | *I* | *I* | *I* | *I* | *I* | *I* | *I* | *I* | *L* |
| $T_{\Delta\Omega J}$ | 0.2 | 0.07 | 0.07 | 0.07 | - | 0.049 | 0.44 | 1.87 | 3.7 |
| $k_{\Delta\Omega J}$ | *D* | *L, D* | *L.D* | *D* | - | *L* | *L* | *L* | *D* |
| $T_i$ | 0.2 | 0.07 | 0.07 | 0.07 | 0.049 | 0.049 | 0.043 | 1.87 | 3.7 |
| $T_e$ | 1 | 0.1 | 0.1 | 0.1 | 0.054 | 0.054 | 1.12 | 0.53 | 3.8 |
| $T_{\Delta\pi J}$ | 1.4 | 0.3 | 0.2 | 0.1 | - | 0.054 | 0.054 | 0.36 | 0.28 |
| $k_{\Delta\pi J}$ | *I* | *Il* | *Il* | *I* | - | *I* | *L* | *D* | *D* |
| $T_I$ | | 0.25 | 0.34 | 0.25 | 1.9 | 1.9 | | | 130 |

## 13. Probabilities of collisions of near-Earth objects with the Earth and the Moon and lunar craters

Due to collisions with other celestial bodies and ejection into hyperbolic orbits, the dynamic lifetime of Earth-crossing objects (ECOs) is shorter than the characteristic time before the impact of an ECO with the Earth for the model in which ECOs remain in their orbits forever, do not move anywhere, and do not collide with anything other than the Earth. However, such a model can be used to estimate the characteristic time before the impacts of ECOs with the Earth and the Moon under the assumption that the number and characteristic orbits of ECOs do not change with time. The probabilities of collisions of near-Earth objects with the Earth were discussed in paragraph 8 of Chapter 7 in [78].

For 417 Earth-crossing objects (ECOs) discovered before mid-1999, the mean times $T_{3s}$ before their collisions with the Earth were calculated (using formula (4.4) from paragraph 2 of Chapter 4 in [78]) for fixed values of the semi-major axes, eccentricities, and inclinations of orbits. The mean inclination and mean eccentricity of the orbits of these ECOs were 14.3° and 0.52, respectively. For 11 ECOs, $T_{3s}\leq10$ million years was obtained. In most of these cases, the inclination $\Delta i<3°$. The orbital elements of real asteroids can vary significantly before their collisions with planets, and the times before these collisions can differ greatly from the $T_{3c}$ values obtained for fixed orbital elements. In [19, 26, 28-29, 78], when calculating $T_E=N/\Sigma(T_{3c}^{(k)})$. all ECOs known at the time of calculation were considered ($N$ is the number of ECOs). In [78], the value of $T_E$ was found to be equal to 100 million years for 417 ECOs, equal to 105 million years for 363 Apollo objects, and equal to 80 million years for 55 known Aton asteroids. Aton asteroids are near-Earth asteroids whose orbits intersect the Earth's orbit from the inner side (their distance from the Sun at aphelion is greater than the Earth's perihelion distance, $Q$>0.983 AU, but the semi-major axis is smaller than the Earth's, $a$<1 AU). Apollo asteroids have orbits that intersect the Earth's orbit from the outer side (their distance from the Sun at perihelion is less than the Earth's aphelion distance, $q$<1.017 AU, but the semi-major axis of the orbit is greater than the Earth's, $a$ > 1 AU).

$T_E$=76 million years was obtained for 93 Earth-crossing objects known in 1991. In [28–29], $T_E$ values of 15, 164, and 67 million years were obtained for 110 Aton group asteroids, 643 Apollo group asteroids, and all ECOs, respectively. Amor group asteroids are asteroids with perihelion distances greater than 1.017 AU. Therefore, virtually all such asteroids (until they change their perihelion distances) cannot approach the Earth to within the Earth's sphere of action. Changes in the semi-major axes, eccentricities, and orbital inclinations of asteroids over time were not considered in these calculations. Although the Apollo asteroids accounted for 85% of all ECOs in these calculations, the $T_E$ values for them were 2.4 times greater than for all ECOs. At $T_E$=67 million years, the probability $p_E=1/T_E$ of a collision of an ECO with the Earth in 1 year is $1.5\times10^{-8}$. The number of known ECOs increases every month. As of January 2025, more than 21 thousand of Apolos were known (of which 1,742 received numbers, meaning their orbits became well known).

The differences in the $T_E$ values for the different publications cited above were due to the different number of asteroids considered. The smaller $T_E$ values for all ECOs in [28–29] than in [78] were due to several Aton asteroids with small orbital inclinations discovered in the early 2000s. When the orbital inclination of the Aton asteroid 2000 SG 344 was increased from its current value of 0.1° to 1°, the $T_E$ values in [29] were obtained to be 28 million years and 97 million years for Aton asteroids and all ECOs, respectively. These times are larger than those obtained when considering the Aton asteroid 2000 SG 344 with its current orbital inclination. The above differences in values of $T_E$ illustrate the importance of taking into account the small number of asteroids with a high probability of their collisions with the Earth (the role of small inclinations is also discussed below in the next paragraph). Ipatov [19, 78] noted that when using the same formulas, the value of $T_E$ calculated for all ECOs is several times smaller than for the average values of inclinations and eccentricities for the same ECOs. Studies of the evolution of the orbits of objects that initially crossed the orbit of Jupiter and of resonant asteroids have shown [27-29, 94] that the probability of a collision of one such object with the Earth may be greater than for thousands of other objects that initially had close orbits.

In [28-29], $T_E$ calculations were performed for elements of modern orbits of ECOs. However, similar values of $T_E$=67 million years could also be obtained for an interval of 1 billion years, since some ECOs with a fairly high probability could also have had similar small inclinations earlier. The orbital inclinations of ECOs change over time. In [28-29], 110 Aton asteroids were considered. If the orbital inclinations are randomly distributed in the range from 0 to 11°, then one of them will be less than 0.1° (in the previous paragraph we discussed the contribution to $T_E$ of Aton asteroid 2000 SG 344 with an orbital inclination of 0.1°). If the inclinations are randomly distributed in the range from 0 to 22°, then one of them will have a value in the range from 0 to 0.2°, that is, on average, it will be equal to 0.1°.

We also calculated the values of $T_{3c}$ for various values of $a$, $e$ and $\Delta i$. It was found that the values of $T_{3c}$ increase with increasing $e$ and $\Delta i$ and with decreasing $|a$-1$|$ AU. For $a$=1.25 AU, $\Delta i$=80° and $e\geq0.7$, and for $e$=0.9 and $\Delta i\geq50°$, the values of $T_{3c}$ exceed the age of the Solar System, i.e., the probability of collisions with the Earth of bodies constantly moving along highly inclined and eccentric orbits is small. However, due to changes in the orbital elements of bodies during evolution and the ejection of most of them into hyperbolic orbits, it is very unlikely that even with large initial values of $e$ and $\Delta i$, bodies could have remained near the Earth's orbit since its accumulation. For the model in which $\Delta i$ takes values from 0 to $\Delta i_{max}$, collisions occur more frequently than with $\Delta i=\Delta_{max}/2$, and for the model in which $e$ takes values from $e_{min}$ up to $e_{max}$, collisions occur more frequently than at fixed $e=(e_{min}+e_{max})/2$.

Slightly more than half of near-Earth objects (NEOs) (those with a minimum distance to the Sun (perihelion) of less than 1.3 AU) cross Earth's orbit (i.e., are ECOs). The probability of an NEO being ejected into a hyperbolic orbit is an order of magnitude greater than the probability of its collision with the Earth, and most NEOs leave the Solar System within $t\leq10$ million years. These asteroids can collide not only with the Earth but also with Venus and other planets. Therefore, for most bodies that collide with the Earth, the time interval between entering an Earth-crossing orbit and the impact with Earth is several times shorter than $T_E$ and likely does not exceed 10 million years.

I considered the following model. Let us consider $N_o$ ECOs at the initial moment of time. We study the change in the number $N$ of these ECOs with time $t$. Assuming that the number of ECOs that collided with the Earth during time $\Delta t$ is equal to $\Delta tN/T_E$, and the number of ECOs ejected into hyperbolic orbits or collided with other planets or the Sun is $k$ times greater than the number of collisions with the Earth, we obtain $N=N_o\exp(-(k+1)t/T_E)$. Denoting $\beta_r=N/N_o$, we obtain получаем $t=-T_E\ln\beta_r/(k+1)=-T_d\ln\beta_r$, where $T_d=T_E/(k+1)$ is the dynamic lifetime of the ECOs. For $1+k=10$, $\beta_r=0.5$, $\beta_r=0.1$, $\beta_r=0.01$ and $\beta_r\approx4.54\cdot10^{-5}$, we have $t\approx0.07T_*$, $t\approx0.23T_*$, $t\approx0.46T_E$ and $t=T_E$, respectively. For $\beta_r=0.5$, $1+k=10$, and $T_E$=100 million years, the value of $t=-T_d\ln\beta_r$ is equal to $T_{E*}\approx7$ million years. Thus, a half of all ECOs falling on the Earth may fall on it during a time $t\leq7$ million years after these asteroids began to cross the Earth's orbit. After that, the bodies that fell onto the Earth will mainly be those that became ECOs later. The number of NEOs will be constant and equal to $N_o$ if, in addition to taking into account the above-mentioned collisions and ejections into hyperbolic orbits, we assume that during the time $\Delta t$, $\delta=N_o\cdot\Delta t(k+1)/T_E$ of new ECOs arrive from other regions of the Solar System. The number of 1-km NEOs is estimated to be about 1000. Assuming the number of 1-km ECOs to be equal to about 500, with $\Delta t$=1 million years, $T_E$=100 million years, and 1+$k$=10, we obtain $\delta$=50.

In Chapter 7 of the monograph [78], it was noted that the collisional lifetime of a body with a diameter of 1 m in the asteroid belt is $T_{cN}$~$10^6$ years. Bodies crossing the Earth's orbit often penetrate quite far into the asteroid belt. Therefore, the times $T_{cN}$ before the destruction of such bodies are not much shorter than for bodies in the asteroid belt, and they are several times shorter than $T_{E*}$. These estimates are consistent with the fact that stony meteorites are usually the result of several destructions. The ancient age of iron meteorites (>$10^8$ years) may be due to the fact that the last collision of large asteroids occurred more than $10^8$ years ago, and the iron fragments formed during this collision were almost not fragmented before they fell on the Earth. In paragraph 8.3 of Chapter 7 in [78], the frequency of collisions of bodies of different masses with the Earth was discussed, and in paragraph 8.4, the times before the fall of bodies on various planets were discussed.

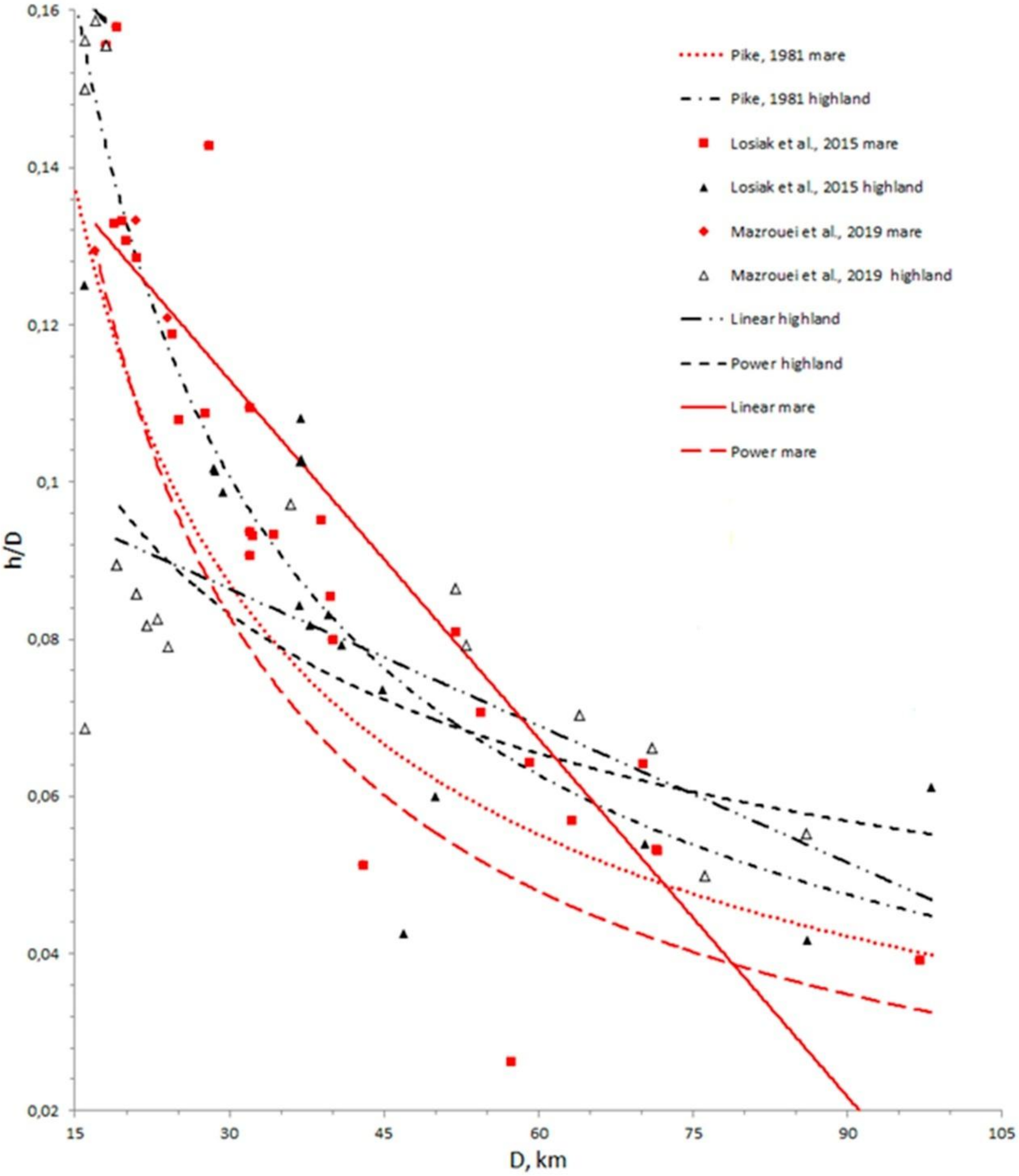

**Figure 1. The** *h/D* ratio of the lunar crater depth to diameter for Copernican craters with a diameter $D \geq 20$ km on mares and highlands. The lines show our approximations of this relationship.

Problems of asteroid-comet hazard and **crater formation** are associated with approaches of small bodies to the Earth. In [56], a comparison was made of the number of lunar craters with a diameter greater than 15 km and an age of less than 1.1 billion years with estimates of the number of craters of this size that could have formed in 1.1 billion years if the number of objects approaching the Earth and the elements of their orbits during this time were close to their modern values. The comparison was made for craters on the entire surface of the Moon and for the region in the Oceanus Procellarum and the maria of the visible side of the Moon. These estimates used the values of the near-Earth object collision probabilities, with the Moon, as well as the dependence of crater diameters on impactor diameters. The number of known Copernican craters with diameters $D \geq 15$ km per unit area on the maria, according to various authors, is at least twice the corresponding number for the rest of the Moon's surface. Our estimates are consistent with the increasing number of near-Earth objects, after possible catastrophic disruptions of large main-belt asteroids that could have occurred over the past 300 million years, but they also do not prove this increase. In particular, they do not contradict the conclusion of the work (Mazrouei et al., Science. 2019. v. 363. 253–255) that the number of near-Earth asteroid impacts on the Moon per unit time increased by 2.6 times 290 million years ago. The number of Copernican lunar craters with a diameter of at least 15 km is possibly greater than that estimated by this work. Taking into account the destruction of some older craters and the changing orbital distribution of near-Earth objects over time may lead to the conclusion that the average number of near-Earth objects over the past billion years could have been close to the current value. With an Earth-orbit-crossing object (ECO) impact probability of $10^{-8}$ per year, our crater number estimates are consistent with a model in which the number of 15-km Copernican craters per unit area for the entire lunar surface would be the same as for the maria region, if the data (Losiak et al., 2015, https://www.lpi.usra.edu/scientific-databases/) for $D$<30 km were as complete as for $D$>30 km. With this ECOs impact probability and this model, the cratering rate over the past 1.1 billion years could have been constant.

In the article [57], the dependence **of the ratio of the crater depth to its diameter** on the diameter for lunar craters on the maria and on the highlands for craters with an age of less than 1.1 billion years was studied. In particular, it was noted that with the same diameter, craters on the lunar maria are deeper than on the highlands, with a crater diameter smaller than 40-50 km (Fig. 1). For larger diameters, craters on the highlands are deeper.

## 14. Dust migration in the Solar System and the formation of the zodiacal belt

The main source of interplanetary dust are particles formed during asteroid collisions and ejected by comets during the sublimation of the icy matrix of their nuclei. Dust particle sizes range from nanometers to millimeters, with the lower limit being the upper limit of the size of molecular clusters. The millimeter-centimeter boundary is used as a conventional boundary between dust particles and meteoroids. Dust particles accompanying the sublimation process from the comet nucleus form dust tori along their orbits, periodically crossed by the Earth and causing meteor showers. Known meteor showers are directly related to the radiants of their parent comets, projected onto specific constellations, after which these meteor showers are named. The amount of material contained in dust particles and small meteoroids, falling daily onto the Earth's atmosphere, ranges from 30 to 180 tons.

Migration of dust particles has been considered by a number of authors. In our numerical models [30-31, 33, 37, 93, 95, 96, 99, 106, 146, 149, 152, 171, 173, 312, 330, 331, 335, 338, 342, 344, 345, 348, 350, 353-355, 358, 367, 374, 389, 400, 403-404, 411] in addition to the gravitational influence of the planets, other factors were also taken into account (radiation pressure, the Poynting-Robertson effect, solar wind). The relative error at the integration step by the Bulirsh-Stoer method was less than $10^{-8}$. The initial orbits of the particles were taken to be the same as those of 500 real asteroids, as well as various comets and planetesimals from the feeding zones of Jupiter and Saturn. The evolution of the orbits of 20 thousand dust particles was studied until the particles left the Solar System or collided with the Sun. Calculations were performed for β values (the ratio of the radiation pressure to the gravitational force) equal to 0.0001, 0.0002, 0.0004, 0.001, 0.002, 0.004, 0.005, 0.01, 0.05, 0.1, 0.2, 0.25 and 0.4. For silicate particles with a density of 2.5 g/cm$^3$, such values of β (β~1/$d$) correspond to particle diameters $d$ of 4700, 2400, 1200, 470, 240, 120, 47, 9.4, 4.7, 2.4, 1.9, and 1.2 μm, respectively. For the same β value for water ice, the diameter $d$ is 2.5 times greater than for silicate particles. In the above-cited works on dust migration, planets were considered as material points, but based on orbital elements obtained with a step of 20 years, we calculated (similar to calculations for small bodies) the average probabilities $P$ of collisions of particles during their lifetime with planets and the average times $T$ during which the perihelia of the particle orbits were less than the semi-major axis of the planet's orbit. A brief review of the work on dust migration is given in Section 8 of the article [59].

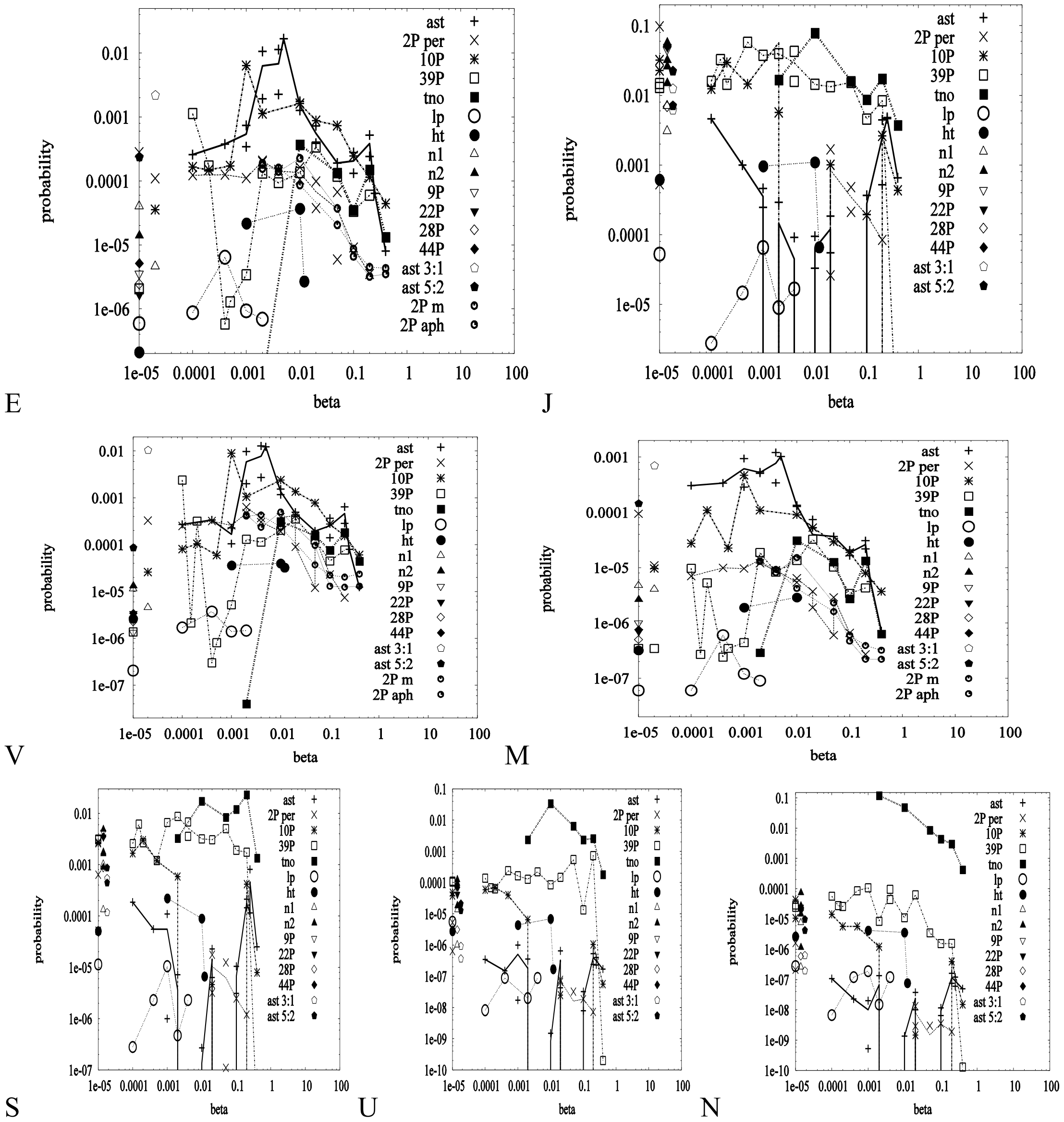

**Figure 1.** Probabilities of collisions of various particles and bodies with the Earth (E), Jupiter (J), Venus (V), Mars (M), Saturn (S), Uranus (U) and Neptune (N) for different values of the parameter β for particles launched from asteroids (ast), trans-Neptunian objects (tno), comet 2P/Encke at perihelion (2P per), comet 2P in aphelion (2P aph ), comet 2P in the middle of the orbit between perihelion and aphelion (2P m), comet 10P/Tempel 2 (10P ), comet 39P/Oterma (39P ) and long-period comets (*lp*) with eccentricity $e$=0.995 and perihelion distance $q$=0.9 AU, and Halley-type comets (*ht*) with $e$=0.975 and $q$=0.5 AU (for the *lp* and *ht* series the initial inclinations were in the range from 0 to 180°). If two collision probabilities (obtained for different calculations) are presented on the graph for one β value, the graph is drawn through the mean value. The probabilities for bodies (β=0) are presented for $\beta \sim 10^{-5}$. The probabilities presented only for bodies were obtained for initial orbits close to the orbits of comets 9P/Tempel 1 (9P), 22P/Kopff (22P), 28P/Neujmin (28P), 44P/Reinmuth 2 (44P), and for test asteroids at the 3:1 and 5:2 resonances with Jupiter at $e$=0.15 and $i$=10° ('ast 3:1' and 'ast 5:2'). For the 'n1' and 'n2' series, the initial particle orbits were close to the orbits of 10-20 different Jupiter-family comets.

For cometary and asteroidal dust particles, the probability $p_E$ of collisions of particles with the Earth had a maximum (~0.001–0.01) at $0.002 \leq \beta \leq 0.01$, i.e. at $d \sim 100$ μm (this value of $d$ is in agreement with observational data). These values of probability $p_E$ for particles usually (except for comet 2P) were an order of magnitude greater than for parent comets. Figures characterizing the values of the probabilities of collisions of dust particles of various origins with planets for different values of β are given in [99] (see also Fig. 1). In [106, 173] the probabilities of collisions of such particles with the Moon, as well as with the embryos of the Earth and the Moon, are given for masses of the embryos 10 times smaller than the modern masses. The probabilities of collisions of dust particles with the Moon were 16-33 times smaller than with the Earth for particles launched from asteroids and comets C10 and C39. For particles launched from Halley-type comets and from long-period comets, this ratio was in the ranges of 10-18 and 9-14, respectively [173]. It has been shown that asteroid particles do not dominate at distances from the Sun $R$>3 AU, while a significant portion of particles at distances of 3–7 AU are of cometary origin. The smaller the particles (and, accordingly, the larger β), the greater is the fraction of particles carried away by the solar wind.

Peaks in the distribution of asteroid dust particles along the semi-major axes of their orbits, corresponding to $n$:($n$ +1) resonances with Earth and Venus, and gaps associated with 1:1 resonances with these planets, are more pronounced for larger particles [30]. The probability of a collision of a trans-Neptunian particle with a diameter $d$<10 microns with the Earth is only a few times smaller than that for an asteroid particle of the same size. Maximum values (of the order of 0.1) of the probability of a collision of particles with Jupiter were obtained for particles formed beyond the orbit of Jupiter and for bodies initially moving in orbits close to the orbit of Comet Encke. The probabilities of collision of trans-Neptunian particles with Uranus and Neptune could reach 0.1 at β~0.002-0.01. In all other cases, the probabilities of collisions of bodies and particles with a giant planet were smaller.

Large quantities of matter, including water and volatiles, could have been delivered by dust particles to the feeding zones of the terrestrial planets immediately after the zone near the orbits of Jupiter and Saturn, where large amounts of dust remained, was cleared of gas. Dust could have been more effective than bodies in delivering organic matter to the planets. This is because dust particles were not subjected to intense heating as they passed through the atmosphere (since they had a higher surface-to-mass ratio). Therefore, dust particles are considered a more likely source of complex organic molecules, including biogenic elements and compounds found in microorganisms, in the interplanetary and interstellar environment than larger bodies. These concepts are consistent with the panspermia hypothesis, in which dust particles could have played an important role in the origin of life on the Earth, as they experience significantly less heating upon atmospheric entry at low angles of attack.

Ipatov also calculated the migration of dust particles originating at various distances from the Sun. In each calculation, **$a_o$ varied from $a_{omin}$ to $a_{omin}$+2.5 AU.** In [171, 173] the results of calculations of dust particle migration are presented for $a_{omin}$ from 5 to 15 AU, initial eccentricities and orbital inclinations equal to 0.3 and 15°, respectively. At β=0.04 the values of $p_{Sun}$ of the fraction of particles that collided with the Sun were in the range from 0.20 to 0.33 and, as a rule, were slightly smaller for large $a_{omin}$. At β=0.0004 (in [171, 173] erroneously indicated 0.004) the values of $p_{Sun}$ did not exceed 0.03. At β=0.04 the values of $p_E$ and $p_{Moon}$ of the probabilities of particle collisions with the Earth and the Moon were two orders of magnitude higher than the values for bodies in similar initial orbits (for such bodies, $p_E$ is on average close to (2-4)·$10^{-6}$). As was obtained for bodies, at β=0.0004, the values of $p_E$ (and $p_{Moon}$) can differ by two orders of magnitude for different calculations with the same $a_{omin}$. In particular, at β=0.0004, $p_E$ took values from $5.7\cdot10^{-7}$ to $4.9\cdot10^{-4}$. For β=0.04, the values of $p_E$ (and $p_{Moon}$) for different calculations differed less than for β=0.0004. The values of $p_E/p_{Moon}$ were in the range from 16 to 31 for β=0.04 and in the range from 14 to 22.5 for β=0.0004.

Calculations of particle migration were also performed for $a_{omin}$ from 2.5 to 37.5 AU and initial eccentricities equaled to 0.05. The values of β were equal to 0.4, 0.04, 0.004, and 0.0004. Dust particles that collided with the Sun or planets or were ejected into hyperbolic orbits were excluded from further integration. The results of these calculations have not yet been published. Most of these particles were ejected from the Solar System, the remaining particles mainly fell on the Sun. The fraction of particles that collided with planets usually did not exceed 1%. For $a_o$ greater than 17 AU, the probabilities of collisions of bodies with the Earth were obtained to be of the order of $10^{-6}$. For dust particles with a diameter of 10-100 microns at $a_o$>17 AU, the probabilities of their collisions with the Earth in most of the calculations considered were on the order of $10^{-5}$-$10^{-4}$. The probabilities of collisions of such particles with Venus typically differed from the probabilities of their collisions with the Earth by no more than two times. The probabilities of collisions of such particles with Mercury were several times (sometimes an order of magnitude) greater than with Mars, and several times less than with the Earth.

The nature of the zodiacal light is directly related to the migration of dust particles. **The zodiacal light** is visible from Earth as a bright, diffuse glow in the west after dusk and in the east before dawn. It has a triangular shape, the brightness of which decreases with increasing elongation (elongation is the angular distance between a planet and the Sun from the perspective of an observer on the Earth). It is caused by a cloud of interplanetary dust lying along the plane of the ecliptic and reflecting sunlight.

Based on the calculated positions and velocities of migrating dust particles that started from various small bodies (asteroids, comets, trans-Neptunian objects) in [37, 155, 378-379, 414] the change in the spectrum of zodiacal light, the spectrum of which generally coincides with the solar Fraunhofer spectrum, was considered in the case of scattering by dust particles at different values of the angle with the vertex in the Earth between the directions to the Sun and the dust particle. The Doppler shift and the width of the characteristic line of Mg I were determined. Graphs of the dependence of the characteristic velocity of this line (determined by the Doppler shift) on the solar elongation were constructed for a number of light scattering functions. The calculated data [37] were compared with the data of WHAM (Wisconsin H-Alpha Mapper) observations of the Doppler shifts and the width of this line in zodiacal light. It was shown that the contribution to the zodiacal light of cometary particles formed within the orbit of Jupiter and particles formed outside the orbit of Jupiter, taking into account trans-Neptunian particles, are approximately the same, and the contribution of each of these two components to the zodiacal light is equal to approximately 1/3 with a possible deviation from 0.3 to 0.1–0.2. The contribution of asteroidal dust (the third component) is estimated at ~0.3–0.5. The contribution of particles generated by long-period and Galeian -type comets does not exceed 0.1–0.15. The same conclusion can be drawn for particles ejected by comets such as Encke 2P (with $e$~0.8–0.9). The average orbital eccentricities of zodiacal particles located at distances of 1–2 AU from the Sun, which better correspond to WHAM observations, range from 0.2 to 0.5, with a more probable value of approximately 0.3.

## 15. Migration of bodies and dust particles ejected from the terrestrial planets and the Moon

During the formation of planets, large bodies fell on them, and collisions with these bodies could have resulted in the ejection of material from the planets. Currently, over 140 lunar meteorites have been reliably identified. As of 2020, 266 Martian meteorites have been reliably identified out of all those found on the Earth. Two meteorites found in the Sahara Desert may be fragments of Mercury. Ejection angles for planetary bodies are known to generally fall between 20° and 55°, with most values between 40° and 50°.

In the calculations performed, the motion of bodies ejected from the terrestrial planets was studied over the dynamic lifetime $T_{end}$ of all bodies. During $T_{end}$, all bodies collided with planets or the Sun or were ejected from the Solar System (reached 2000 AU from the Sun). In cases of such collisions and ejection, they were excluded from further integration. At ejection velocities close to a parabolic velocity, almost all ejected bodies quickly fell back onto the planet. At high ejection velocities, $T_{end}$ was typically around 200–400 million years (could exceed a billion years). In each variant of the calculations, the motion of 250 bodies ejected from a celestial object was studied at fixed values of the ejection angle $i_{ej}$, ejection velocity $v_{ej}$, and ejection point. The ejection velocity $v_{ej}$ varied from the parabolic velocity on the surface of the planet (Moon) to 20 km/s, and the values of the ejection angles $i_{ej}$ varied from 15° to 90°.

Six opposite ejection points on the planet's surface were considered (Fig. 1). For ejection points F and C, the bodies' motion began from the planet's surface points most distant from and closest to the Sun in the direction from the Sun to the Earth, respectively. For ejection points U and D, the bodies' motion began from the planet's surface points with the maximum and minimum z values (with the 0z axis perpendicular to the planet's orbital plane), respectively. Points F, C, U, and D are referred to below as mean points. For ejection points W and B, the bodies' motion began from the planet's surface points in the direction of the planet's motion and from the opposite side of the planet, respectively. When studying the ejection of bodies from the Moon, only point F, at an altitude $h$ from Earth, was considered (primarily for $h$ corresponding to the radius of the Moon's orbit). Note that when determining the velocity of a body relative to the Sun, the velocity vector of the body's ejection from the planet is summed with the velocity vector of the planet's motion relative to the Sun. The orbital velocities of the planets are 47.36, 35.02, 29.78, and 24.13 km/s for Mercury, Venus, Earth, and Mars, respectively. The average orbital velocity of the Moon relative to the Earth is 1.02 km/s. If we consider the Moon's orbit closer to Earth rather than its present-day orbit, we must keep in mind that the orbital velocity is proportional to $r_{EM}^{-1/2}$, where $r_{EM}$ is the distance between the Earth and the Moon. For example, the orbital velocity would be 3 km/s if the radius of the lunar embryo's orbit were 9 times smaller than the semimajor axis of the Moon's present-day orbit.

The gravitational influence of the Sun and all eight planets was taken into account. The symplectic algorithm from the SWIFT package was used to integrate the equations of motion (Levison H.F., Duncan M.J. Icarus. 1994. V. 108. P. 18-36). The considered time step $t_s$ was equal to 1, 2, 5, or 10 days, and the calculation results for different $t_s$ were compared. In most calculation variants, the step $t_s$ was equal to 5 days when studying the ejection of bodies from Earth and Mars and equal to 1 and 2 days for the ejection of bodies from Mercury and Venus, respectively. The probabilities of collisions of bodies with the Moon were calculated based on arrays of orbital elements of migrated bodies (stored with a step of 500 years) similarly to [28, 55].

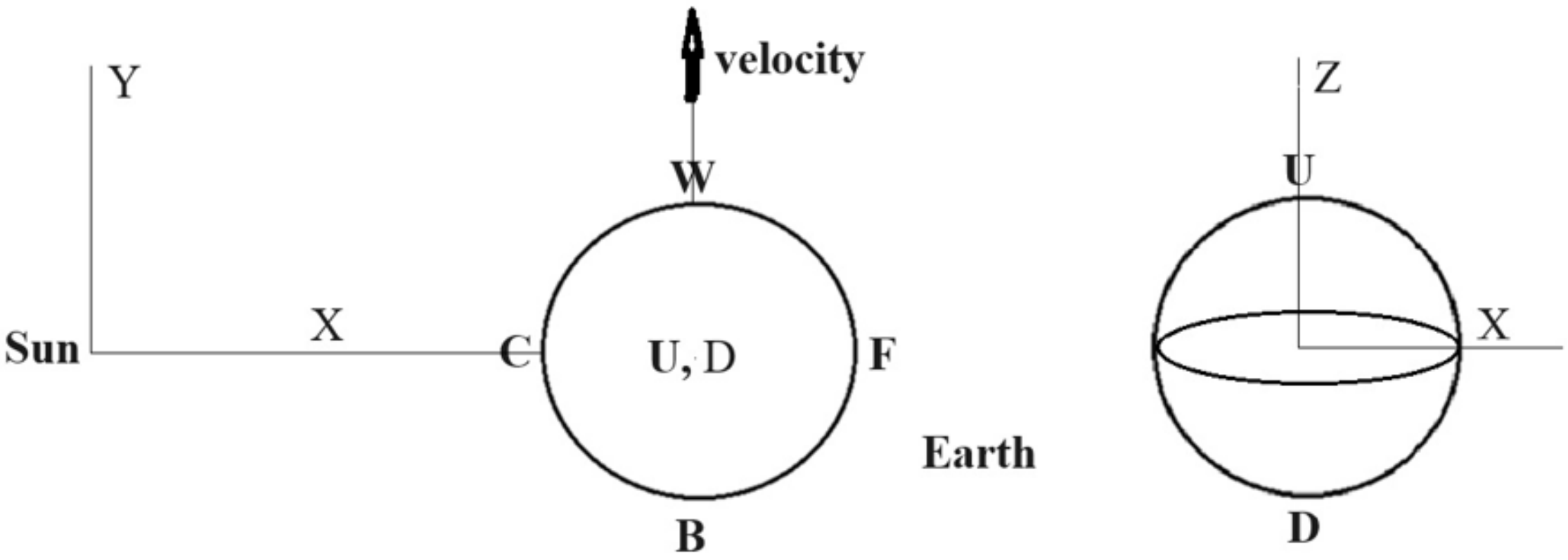

**Figure 1.** Location of ejection points on the planet.

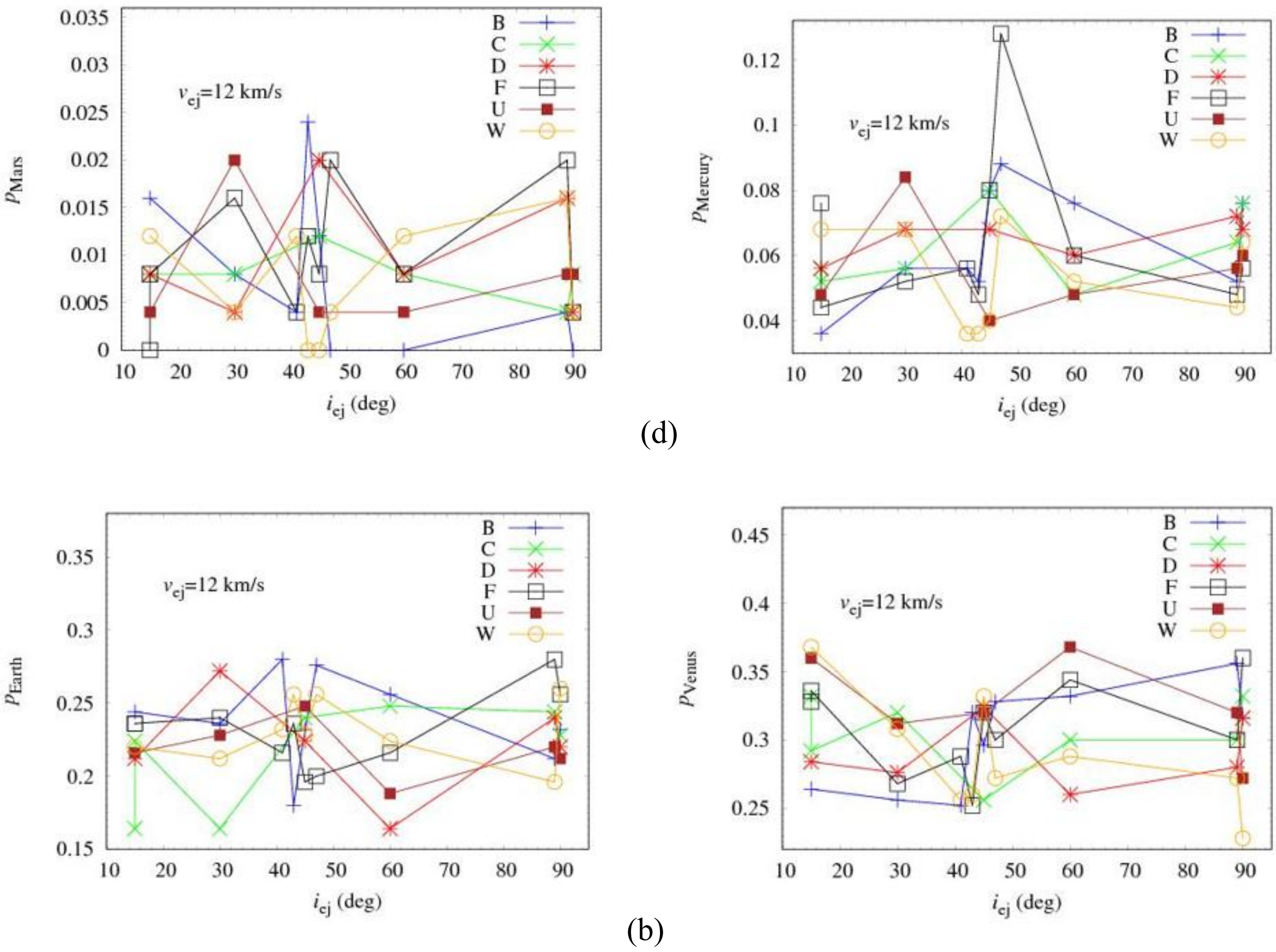

**Figure 2.** Plots of collision probabilities for bodies ejected from Earth and falling on Earth (a), Venus (b), Mars (c), and Mercury (d) at ejection velocity $v_{ej}$=12 km/s for ejection points B, C, D, F, U, and W, depending on the ejection angle $i_{ej}$ (15–90°). The probability values given for $i_{ej}$ near 45° were obtained for $i_{ej}$=45°, but with different integration accuracies per step. The differences in these values indicate a random scattering of probability values due to slight variations in the initial data and the calculation model.

**Ejection of bodies from the Earth**: The motion of bodies ejected from the Earth and the probabilities of their collisions with planets were studied in [63, 65, 118, 124, 512, 514, 516, 518] during the dynamic lifetime $T_{end}$ of all bodies in the calculation variant. In most calculation variants, the bodies started from the Earth's surface. In different variants, the values $i_{ej}$ of the ejection angle were equal to 15°, 30°, 45°, 60°, 89° or 90°. The ejection velocity $v_{ej}$ of bodies from the Earth varied from 11.22 to 20 km/s (mostly equaled to 11.22, 11.5, 12, 14, 16.4 and 20 km/s). The articles [63, 65] present graphs and tables characterizing the probabilities of collisions of bodies ejected from the Earth with planets and the Moon at various velocities and angles of ejection, various ejection points, and various time intervals after ejection. Below, as an example, Fig. 2 presents the probabilities of collisions with planets for bodies ejected from the Earth at a velocity of 12 km/s with various ejection angles. A much larger number of similar figures are presented in [65] for $T$=$T_{\rm end}$ and $T$=10 million years.

The average value of the probability $p_{\rm E}$ of the collision velocity of an ejected body with the Earth depends on the distribution of bodies by $v_{\rm ej}$ and $i_{\rm ej}$. At ejection velocities $v_{\rm ej}$≤11.3 km/s, i.e. slightly greater than the parabolic velocity, most ejected bodies fell onto the Earth. At ejection velocities $v_{\rm ej}$ equal to 11.5 and 12 km/s, the values of $p_{\rm E}$ did not differ significantly for different ejection points on the Earth's surface and over the entire considered time interval were in the ranges of 0.25-0.35 and 0.16-0.27, respectively. For ejection from point C on the Earth's surface, the values of $p_{\rm E}$ could be equal to, greater than, or less than (less than 1.7 times) than for ejection from point F, but on average, across the various calculation options, they were approximately the same. For all ejection points at $T$=10 million years and $i_{\rm ej}$=45°, the values of $p_{\rm E}$ were in the ranges of 0.2-0.28 and 0.12-0.16 at ejection velocities $v_{\rm ej}$, equal to 11.5 and 12 km/s, respectively. The only significant difference was during ejection from the front point W of the Earth's motion at $v_{\rm ej}$=16.4 km/s and $i_{\rm ej}$=45°. In this case, 80% of the bodies were ejected into hyperbolic orbits, and 17% collided with the Sun. The average value of $p_{\rm E}$ depends on the distribution of bodies over $v_{\rm ej}$ and $i_{\rm ej}$. At $T$=10 million years, at ejection from point F and 30≤$i_{\rm ej}$≤60 deg, the value of $p_{\rm E}$ was about 0.28, 0.15-0.17 and 0.06 at $v_{\rm ej}$ equal to 11.5, 12 and 16.4 km/s, respectively. At $i_{\rm ej}$=90° in some variants $p_{\rm E}$ was less than at $i_{\rm ej}$≤60°, and in other variants it was greater. During ejection from point F and $i_{\rm ej}$=90°, the value of $p_{\rm E}$ is half that of $i_{\rm ej}$≤60° for $v_{\rm ej}$=11.5 km/s, but slightly (approximately 1.3-1.4 times) greater for $v_{\rm ej}$≥12 km/s.

Below in Table 1 are given the probabilities of collisions of bodies ejected from the Earth with the Earth, Venus, Mercury, Mars and the Sun ($p_{\rm E}$, $p_{\rm v}$, $p_{\rm me}$, $p_{\rm ma}$, $p_{\rm sun}$) and the probabilities of ejection $p_{\rm ej}$ into hyperbolic orbits over the entire considered time interval (at $T$=$T_{end}$) for several values $v_{\rm ej}$ of ejection velocity and four ejection points C, D, F and U. A range of values of probability was obtained for different ejection angles. For the other two ejection points, the probabilities could differ from the values given in Table 1.

**Table 1.** Probabilities of collisions of bodies ejected from the Earth with the Earth, Venus, Mercury, Mars and the Sun ($p_{\rm E}$, $p_{\rm v}$, $p_{\rm me}$, $p_{\rm ma}$, $p_{\rm sun}$) and the probabilities of their ejection $p_{\rm ej}$ into hyperbolic orbits.

| $v_{\rm ej}$ | $p_{\rm E}$ | $p_{\rm v}$ | $p_{\rm me}$ | $p_{\rm ma}$ | $p_{\rm sun}$ | $p_{\rm ej}$ |
|---|---|---|---|---|---|---|
| 11.3 | 0.45-0.55 | 0.18-0.25 | 0.02-0.06 | 0-0.016 | 0.1-0.25 | 0.01-0.04 |
| 11.4 | 0.3-0.45 | 0.23-0.33 | 0.02-0.08 | 0-0.02 | 0.2-0.32 | 0.01-0.05 |
| 11.5 | 0.25-0.35 | 0.22-0.33 | 0.02-0.08 | 0-0.016 | 0.22-0.33 | 0.02-0.06 |
| 12 | 0.16-0.27 | 0.25-0.37 | 0.04-0.08 | 0-0.02 | 0.3-0.45 | 0.02-0.08 |
| 14 | 0.1-0.2 | 0.2-0.3 | 0.04-0.1 | 0-0.02 | 0.35-0.5 | 0.03-0.15 |
| 16.4 | 0.1-0.15 | 0.18-0.25 | 0.03-0.1 | 0-0.025 | 0.2-0.5 | 0.05-0.33 |
| 20 | 0.08-0.15 | 0.15-0.25 | 0.05-0.1 | 0-0.012 | 0.2-0.6 | 0-0.4 |

In the considered variants at $T$=10 million years and $i_{\rm ej}$ =45°, the values of the probability $p_{\rm V}$ of a collision of a body ejected from the Earth **with Venus** were in the ranges of 0.16-0.22 and 0.13-0.18 at $v_{\rm ej}$ equal to 11.5 and 12 km/s, respectively. At $T$=$T_{end}$ and 11.5≤$v_{\rm ej}$≤12 km/s, the values of $p_{\rm V}$ were in the range of 0.22-0.37. The total number of bodies delivered to the Earth and Venus probably did not differ significantly. In general, the ratio $p_{\rm V}/p_{\rm E}$ of the probability $p_{\rm V}$ of a body's collision with Venus to its probability $p_{\rm E}$ of collision with the Earth was smaller at lower ejection velocities.

At $T$=$T_{\rm end}$ (over the entire time interval considered), the probabilities of collisions of bodies **with Mercury and Mars** were in the ranges of 0.02–0.08 and 0–0.025, respectively. In most calculations, the probability of a body colliding **with the Sun** was approximately 0.05–0.2 at $T$=10 million years, and at $T$=$T_{\rm end}$ it could reach 0.5.

The values $p_{\rm ej}$ of the probability **of the ejection of bodies into hyperbolic orbits** at $T$=$T_{\rm end}$ usually exceeded the values at $T$=10 million years by no more than 2 times. The values of $p_{\rm ej}$ in different series of calculations were in the range of 0.016–0.064 at $T$=$T_{\rm end}$, 30° ≤$i_{\rm ej}$≤60° and 11.5≤$v_{\rm esc}$≤12 km/s. At $v_{\rm ej}$=16.4 km/s, the values of $p_{\rm ej}$ could reach 0.8, and at $v_{\rm ej}$=20 km/s and ejection from point W in the direction of the Earth's motion, all bodies could be ejected into hyperbolic orbits. When ejected from point F, the probability $p_{\rm ej}$ for $T$=10 million years, depending on the ejection velocity, took values from less than 0.01 at ($v_{\rm ej}$=11.5 km/s, 30°≤$i_{\rm ej}$≤60°) to 0.24-0.38 at ($v_{\rm ej}$=16.4 km/s, 30°≤$i_{\rm ej}$≤45°).

The values of the probability of collisions of bodies ejected from the Earth **with the Moon** in its present-day orbit were of the order of 0.01–0.02 at $v_{ej}$=11.3 km/s and 0.005–0.008 at $v_{ej}$=16.4 km/s. In [63], it was noted that for the Moon to contain the current amount of iron and to have acquired most of its mass from ejected Earth mantle, it must have acquired a significant part of its mass while moving close to the Earth.

The average collision velocities of ejected bodies with the Earth and the Moon are greater for greater ejection velocity. For ejection velocities of 11.3, 11.5, 12, 14, and 16.4 km/s, the impact velocities were approximately 13, 14–15, 14–16, 14–20, and 14–25 km/s for the Earth, respectively, and were within the range of 7–8, 10–12, 10–16, and 11–20 km/s for the Moon.

**Ejection of bodies from the Earth over the last tens of millions of years:** In [56] we found that 15-kilometer lunar craters were created by 1-kilometer impactors. The number of 15-kilometer lunar craters formed over the last billion years is about 50. The ratio of the probabilities of collisions of near-Earth objects with the Earth and the Moon was estimated in [56] to be 22. Thus, the number of 1-kilometer bodies that fell to the Earth over the last billion years can be estimated at about 1000, i.e., approximately 1 impact occurred every 1 million years. Impactors with a diameter of $d_i$=10-15 km formed the Chicxulub crater on Earth (with a diameter of $D_{cr}$=150-200 km and an age of $T_{cr}$=65.5 million years), the Vredefort crater (with $D_{cr}$=170-300 km and $T_{cr}$=2.02 billion years) and the Sudbury crater (with $D_{cr}$=130 km and $T_{cr}$=1.85 billion years). It is believed that $d_i$ may be 20-25 km for the Vredefort crater. The Popigai crater is the fourth largest confirmed impact crater on the Earth (with $D_{cr}$=90-100 km and $T_{cr}$=35.7 million years). It could have been formed by an impactor with a diameter of $d_i$=8 km. The impactors that created the above craters are assumed to have come from the asteroid belt. Some bodies ejected from the craters tens of millions of years ago may still be in the zone of the terrestrial planets and may fall onto these planets. When studying the natural transport of microbes between Mars and Earth, viable microbes are assumed to arrive after a flight time of less than 1 million years. In my calculations, out of 6,000 bodies ejected from Earth, only one impacted Mars within 1 million years (0.46 million years after ejection from point W). For 6,000 bodies, the probability of their impact on Mars within 1 million years is 0.00017 (it was about two parts in ten thousand).

**Ejection of bodies from the Moon and impacts of bodies ejected from planets on the Moon**: For ejections from the Moon **in its present orbit**, the probabilities of collisions of ejected bodies with planets were similar to the probabilities of collisions of bodies ejected from the Earth, if we consider lower ejection velocities from the Moon than from the Earth [124, 215, 514, 518, 519, 529, 532]. For example, the probability of a collision of bodies ejected from the Moon with the Earth was in the range of 0.27-0.35 for $v_{ej}$=2.5 km/s, 0.2-0.25 for $v_{ej}$=5 km/s, and 0.1-0.14 for $12 \le v_{ej} \le 16.4$ km/s. For bodies ejected from point F on the Moon, $T$=10 million years and $30^\circ \le i_{ej} \le 60^\circ$, the ratio $p_V/p_E$ of the probabilities of collisions of ejected bodies with Venus and Earth was about 0.6–0.9 for $v_{ej}$=2.5 km/s, 0.9–1.3 for $v_{ej}$=5 km/s, 1.3–1.5 for $11.5 \le v_{ej} \le 12$ km/s, and 0.6–0.8 for $v_{ej}$=16.4 km/s, respectively. The probabilities of collisions with the terrestrial planets for bodies ejected from the Moon in its present-day orbit at several ejection velocity values are presented in Table 2.

**Table 2.** Probabilities of collisions with planets for bodies ejected from the Moon in its current orbit (intervals for different ejection angles) for several ejection velocity values

| Ejection velocity, km/s | 2.5 | 5 | 12 | 16.4 | 2.5 -16.4 |
|---|---|---|---|---|---|
| Mercury | 0.044-0.072 | 0.052-0.108 | 0.048-0.072 | 0.048-0.076 | 0.04-0.1 |
| Venus | 0.28-0.324 | 0.276-0.312 | 0.204-0.232 | 0.172-0.18 | 0.2-0.3 |
| Earth | 0.272-0.324 | 0.2-0.252 | 0.108-0.164 | 0.1-0.132 | 0.1-0.2 |
| Mars | 0.008-0.016 | 0.004-0.016 | 0.004-0.008 | 0-0.012 | 0-0.02 |

In addition to studying the ejection of bodies from the Moon, located at its current distance from Earth on the Sun-Earth-Moon line, calculations were also conducted for a model when the Moon was closer to the Earth. It was assumed that at the initial moment of time, the bodies were located on the Sun-Earth-Moon line at a height $h$ above the Earth's surface equal to $\boldsymbol{h=3r_E}$, $\boldsymbol{h=5r_E}$, $\boldsymbol{h=7r_E}$, **or** $\boldsymbol{h=36r_E}$ (the distance $r_{ME}$ from the center of the Earth is greater than $h$ by $r_E$: $r_{ME}=h+r_E$), where $r_E$ is the radius of the Earth. These ejection calculations correspond to point F.

All bodies ejected from a lunar embryo moving close to the Earth fell onto the Earth and the Moon if their initial velocity was less than the corresponding parabolic velocity. At $v_{ej}$=2.5 km/s and the radius $r_{ME}$ of the lunar embryo's orbit equal to $4r_E$, the dynamic lifetime of the ejected bodies was less than 5 days.

At $r_{ME}$=6$r_E$ and $v_{ej}$=2.5 km/s, most of the ejected bodies quickly fell onto the Earth, but at $i_{ej}$≤30° some of the bodies left the Earth's Hill sphere. At $v_{ej}$=5 km/s (the Earth's parabolic velocity at $r_{ME}$=6$r_E$ is 4.57 km/s) and $T$=$T_{end}$ the $p_E$ values were in the range (for different $i_{ej}$) 0.26-0.37, 0.25-0.29 and 0.20-0.22 for $r_{ME}$=6$r_E$, $r_{ME}$=8$r_E$ and for the modern orbit, respectively, that is, the values of $p_E$ were slightly higher for the smaller orbit of the lunar embryo. At $v_{ej}$=12 km/s the $p_E$ values were practically independent of the initial orbit of the lunar embryo and were in the range 0.13-0.14, 0.11-0.14 and 0.11-0.16 for 4$r_E$≤$h$≤6$r_E$, $h$=8$r_E$ and the modern orbit of the Moon, respectively. At ejection velocities slightly greater than the parabolic velocity, the values of $p_E$ are approximately the same for ejections from the Earth and the Moon, but for different ejection velocities.

In several cases (with the radius of the orbit of the lunar embryo $r_{ME}$=4$r_E$, $v_{ej}$=5 km/s, $i_{ej}$=30°; $r_{ME}$=6$r_E$, $v_{ej}$=2.5 km/s, 15°≤$i_{ej}$≤30°; and $r_{ME}$=8$r_E$, $v_{ej}$=2.5 km/s, 15°≤$i_{ej}$≤45°) some (47-71%) of the ejected bodies continued to move around the Earth even after 100 million years. Collisions of bodies with the Moon were not taken into account in the calculations. Thus, in reality, such bodies should have collided with the Moon.

At $r_{ME}$=6$r_E$ and $v_{ej}$=2.5 km/s, no collisions of ejected bodies with Venus, Mercury, and Mars were obtained. At $r_{ME}$=6$r_E$ and $v_{ej}$=5 km/s, the probabilities of collisions of bodies with Mercury, Venus, and Mars were in the ranges of 0.036-0.072, 0.276-0.34, and 0-0.016, respectively. These ranges differ a little from the values for the present orbit of the Moon and $v_{ej}$=5 km/s, although the probabilities of collisions with the Earth at $r_{ME}$=6$r_E$ and $v_{ej}$=5 km/s are slightly higher (0.256-0.368) than the values (0.2-0.25) in Table 2 at $v_{ej}$=5 km/s and the present orbit of the Moon. At $v_{ej}$≥12 km/s, the collision probabilities of bodies for all terrestrial planets were similar for different $r_{ME}$≥6$r_E$. For $h$=36$r_E$, the collision probabilities $p_E$ of bodies ejected from the Moon with the Earth were in the range of 0.07-0.2 at 30°≤$i_{ej}$≤60° and 11.2≤$v_{ej}$≤16.4 km/s and were higher at lower ejection velocities.

At $v_{ej}$=5 km/s, 15°≤$i_{ej}$≤89°, $T$=$T_{end}$ and $h$=5$r_E$, the value of $p_E$ was about 0.3 (0.26-0.37). Such calculations at $h$>0 correspond to the motion of bodies ejected from the Moon, although the gravitational influence of the Moon and its motion around the Earth were not taken into account when integrating the equations of motion. In this case, the actual ejection velocities of bodies from the Moon differed slightly from the velocities used in the calculations. Calculations at 3$r_E$≤$h$≤36$r_E$ correspond to the case when the Moon has not yet reached its current orbit.

**The ratio $k_{em}$ of probabilities of collisions of bodies with the Earth and the Moon**, moving in its present orbit, **for bodies ejected from the terrestrial planets** were obtained mainly in the range from 15 to 30. This ratio varied for different bodies and was generally smaller at higher ejection velocities for bodies from Earth, Venus, and Mars. For $i_{ej}$=45° and ejection from point F on the planet's surface, this ratio was equal to 30, 29, and 18 for ejection from Venus at velocities of 10.5, 12, and 20 km/s, respectively. For ejection from Mars, this ratio was 23, 23, and 14 at $v_{ej}$ equal to 5.3, 6, and 20 km/s, while for ejection from Mercury, this ratio was 19, 16, and 21 with $v_{ej}$ equal to 5.3, 6, and 20 km/s, respectively.

The probabilities $p_E$ of collisions of bodies ejected from Mars and falling on the Earth for points C, D, F, U and W were mainly around 0.08-0.16 and 0-0.16 at 5.05≤$v_{ej}$≤10 and 15≤$v_{ej}$≤20 km/s, respectively. For point B, the value of $p_E$ could exceed 0.24. At $p_E$=0.1 and the ratio of the probabilities of collisions of bodies with the Earth and the Moon equal to $k_{em}$=20, the probability of a collision of a body ejected from Mars with the Moon can be estimated as $p_{moon}$=0.005. The probabilities $p_E$ of collisions of bodies ejected from Mercury and falling on the Earth were less than 0.03-0.05 (depending on the ejection point) at 4.23≤$v_{ej}$≤6 km/s. For high velocities, in some cases, $p_E$ reached 0.2. For $p_E$=0.04 and $k_{em}$=20, the probability of a collision of a body ejected from Mercury with the Moon can be estimated as $p_{moon}$=0.002. For the "average" points C, D, F, and U at 10.4≤$v_{ej}$≤16 km/s, the fraction $p_E$ of bodies ejected from Venus that collided with the Earth was about 0.04-0.1 at $i_{ej}$=45°, and could reach about 0.2 at $i_{ej}$=89°. For $p_E$=0.09 and $k_{em}$=30, the probability of a collision of a body ejected from Venus with the Moon can be estimated as $p_{moon}$=0.003.

The results of calculations of the ejection of bodies from Mars, Mercury, and Venus are presented briefly in abstracts and conference proceedings [121, 215, 519, 523, 525-529, 533, 536]. By the end of 2025, the calculations of the ejection of bodies from the terrestrial planets and the Moon were completed, while calculations of dust migration were still ongoing for large dust particles. Preparation of journal articles on the ejection of bodies from Mars, Mercury, Venus, and the Moon, as well as on the ejection of dust particles from the terrestrial planets, will begin in 2026.

**Ejection of bodies from Mars**: The ejection velocity $v_{ej}$ of bodies from Mars in the calculations performed was generally equal to 5.05, 5.1, 5.3, 5.5, 6, 8, 10, 15, and 20 km/s. The probability $p_{Ma}$ of a collision of a body ejected from Mars **with Mars** was significant only at an ejection velocity $v_{ej}$ close to parabolic velocity. For points C, D, F, and U (located at the intersection of the planet's surface and the plane passing through the Sun and the planet's center and perpendicular to the plane of the planet's orbit), the probabilities $p_{Ma}$ of a collision of a body ejected from Mars with Mars were generally around 0.04-0.25, 0.01-0.04, and 0-0.02 for $5.05 \le v_{ej} \le 5.3$, $5.5 \le v_{ej} \le 10$, and $15 \le v_{ej} \le 20$ km/s, respectively. For point B (on the planet's surface behind the planet's motion), the value of $p_{Ma}$ was usually smaller than for the above points. For point W (on the planet's surface ahead of the planet's motion), the value of $p_{Ma}$ could exceed 0.3 for $v_{ej}$=5.05 km/s and $p_{Ma}$=0 for $15 \le v_{ej} \le 20$ km/s.

The probabilities $p_E$ of collisions of bodies ejected from Mars **with the Earth** for points C, D, F, and U were generally around 0.08-0.16 and 0-0.16 for $5.05 \le v_{ej} \le 10$ and $15 \le v_{ej} \le 20$ km/s, respectively. For point B, the value of $p_E$ could exceed 0.24. For point W, the $p_E$ values ranged from 0 to 0.15. Some bodies ejected from Mars could impact the Earth after 0.1 million years, and some of the bodies could impact the Earth after hundreds of millions of years. For example, for ejection from point F with $v_{ej}$=6 km/s, $i_{ej}$=45°, and $N_o$=250, there were 32 collisions of bodies with the Earth between 0.22 and 270.9 million years. Among these collisions, there were 4, 7, 6, 7, 2, 4, and 2 collisions at times $t<1$, $1<t<5$, $5<t<20$, $20<t<50$, $50<t<200$, and $t>200$ million years, respectively. Moreover, about a half of the bodies ejected from Mars collided with the Earth after 20 million years, and some Martian meteorites could have traveled in space for tens of millions of years before their collisions with the Earth.

For points C, D, F, and U, the probabilities $p_V$ of collisions of bodies ejected from Mars **with Venus** were generally around 0.08-0.2 and 0.02-0.2 for $5.05 \le v_{ej} \le 10$ and $15 \le v_{ej} \le 20$ km/s, respectively. For $v_{ej} \ge 5.2$ km/s, the values of $p_V$ were often slightly larger than $p_E$. For ejection from point B, the $p_V$ value could exceed 0.3. For point W, $p_V$ values ranged from 0 to 0.18.

At $v_{ej} \ge 6$ km/s, the fraction of bodies ejected from Mars that subsequently collided **with Mercury** was usually less than 0.06 and exceeded the fraction of bodies that collided with Mars. For ejections from points C, D, F, and U, the fraction of bodies that collided with Mercury was mostly in the range 0.02-0.08 at $5.05 \le v_{ej} \le 20$ km/s. For point B, the value of $p_{me}$ was in a wider range (0.016-0.2) than for the four "average" points listed above. For ejection from point W, $p_{me}$=0 was obtained at $15 \le v_{ej} \le 20$ km/s, and $p_{me}$ could exceed 0.06 at $v_{ej}$=5.1 km/s.

For points C, D, F, U, and B, the fractions of bodies that impacted **the Sun** typically ranged from 0.2 to 0.9. Typically, for $5.1 \le v_{ej} \le 8$ km/s, more than a half of the ejected bodies impacted the Sun. The probability **of ejection of a body into a hyperbolic orbit** was less than 0.1 for $v_{ej} \le 6$ km/s, but could exceed 0.9 for $v_{ej}$=20 km/s. For point W and $15 \le v_{ej} \le 20$ km/s, almost all bodies were ejected into hyperbolic orbits. For point B, less than 10% of the bodies were ejected into hyperbolic orbits. For the remaining four points, the probability of such ejection ranged from 0.02 to 0.9, depending on the ejection velocity and ejection angle.

Ejection velocity $v_{ej}$ of bodies **from Mercury** was generally assumed to be 4.23, 4.25, 4.3, 4.5, 5, 6, 8, 10, 15, and 20 km/s in calculations. More than a half of the bodies ejected from Mercury typically fall back onto Mercury. The probabilities of collisions of ejected bodies with planets depend on the ejection velocity $v_{ej}$, the ejection angles $i_{ej}$, and the ejection points. For $4.23 \le v_{ej} \le 6$ km/s and point F, it was found that $p_{ej} \le 0.004$, $p_{Sun} \le 0.1$, $0.6 \le p_{Me} \le 0.97$, $0.02 \le p_V \le 0.3$, $p_E \le 0.04$, and $p_{Ma} \le 0.004$. The ejection probabilities from points C, D, and U did not differ significantly from those for point F. For $4.23 \le v_{ej} \le 6$ km/s and point B, it was found that $p_{ej} \le 0.01$, $p_{Sun} \le 0.1$, $0.65 \le p_{Me} \le 0.85$, $0.1 \le p_V \le 0.25$, $p_E \le 0.03$, and $p_{Ma} \le 0.004$. For $4.23 \le v_{ej} \le 6$ km/s and point W, it was found that $p_{ej} \le 0.004$, $p_{Sun} \le 0.2$, $0.3 \le p_{Me} \le 1$, $p_V \le 0.5$, $p_E \le 0.05$, and $p_{Ma}$=0. For point W, the probabilities $p_{Me}$ and $p_V$ can reach larger values than for other points. Large values of $p_{ej}$, $p_{Sun}$, $p_V$, $p_E$, and $p_{Ma}$ were often obtained at large values of $v_{ej}$, but this may not be true for $v_{ej}>8$ km/s. The values of $p_{Me}$ were usually smaller at larger $v_{ej}$. The values of $p_V$ were usually several times larger than $p_E$, with the difference being larger for lower ejection velocities. The probabilities of collisions of bodies ejected from Mercury with the Earth typically did not exceed 0.05 and 0.1 for $v_{ej}$ less than 8 km/s and 15 km/s, respectively. The probabilities $p_V$ of collisions of bodies ejected from Mercury with Venus typically ranged from 0.1 to 0.4 for ejection velocities from 4.3 to 10 km/s. For the "average" points C, D, F, and U, the values of probability $p_V$ were approximately 0.2-0.4 for $6 \le v_{ej} \le 10$ km/s. The fraction of bodies ejected into hyperbolic orbits did not exceed 0.02 for most calculation variants. No more than 20% of bodies typically collided with the Sun. When ejected from point W at $v_{ej} \ge 6$ km/s, for some values of $i_{ej}$ and $v_{ej}$, about a half of the bodies collided with Venus. Less frequently, most bodies ejected from point W collided with the Sun or left the Solar System.

For high ejection velocities, the collision probabilities of bodies with planets can vary greatly for points F, W, and B, as well as for ejection angles. For ejection from point F and $v_{ej}$=20 km/s, it was found that $0.004 \le p_{ej} \le 0.1$, $0.14 \le p_{Sun} \le 0.4$, $0.3 \le p_{Me} \le 0.44$, $0.12 \le p_V \le 0.4$, $0.02 \le p_E \le 0.1$, and $p_{Ma} \le 0.008$. For $v_{ej}$=20 km/s, $i_{ej} \ge 30°$ and ejection from point B, it was found that $0.9 \le p_{Me} \le 0.95$, and all other bodies collided with the Sun (there were no collisions of bodies with other planets, and ejection of bodies from the Solar System also did not occur). For $v_{ej}$=20 km/s and ejection from point W, the probabilities were different for different ejection angles: For $i_{ej}$=30° we obtained $p_{ej}$=0.044, $p_{Sun}$=0.268, $p_{Me}$=0.144, $p_V$=0.284, $p_E$=0.24 and $p_{Ma}$=0.02. For $i_{ej}$=89° most of the bodies were ejected from the Solar System ($p_{ej}$=0.744), and all the remaining bodies collided with the Sun.

**Ejection of bodies from Venus**: The ejection velocity $v_{ej}$ of bodies from Venus in calculations was generally taken to be 10.36, 10.4, 10.5, 10.6, 11, 12, 14, 16, and 20 km/s. Below in this paragraph, the probabilities of collisions with planets are discussed for bodies ejected from Venus from points C, D, F, and U at ejection velocities from 10.4 to 16 km/s. In this case, the fraction $p_{ej}$ of bodies ejected into hyperbolic orbits did not exceed 0.06, the fraction $p_{Sun}$ of bodies that collided with the Sun was about 0.1-0.4, the fraction $p_{Me}$ of bodies that collided with Mercury was about 0.01-0.2, the fraction $p_V$ of bodies that collided with Venus was about 0.3-0.8, and the fraction $p_E$ of bodies that collided with the Earth was about 0.04-0.2 (was several times less than $p_V$). The values of $p_{Sun}$ at $v_{ej}$=16 km/s were approximately 2 times greater than at $v_{ej}$=10.4 km/s. The values of $p_{Me}$ at $v_{ej}$=16 km/s were approximately several times (from 2 to 10 depending on the ejection point and $i_{ej}$) greater than at $v_{ej}$=10.4 km/s. The values of $p_V$ have some tendency (but not for all initial data) to be smaller for large $v_{ej}$. The probabilities of collisions of bodies with Mars often did not exceed 0.01 for all the initial data considered.

For the "average" ejection points C, D, F and U at $v_{ej}$=20 km/s, the $p_{ej}$ values were about 0.25-0.32 at $i_{ej}$=45° and could be about 0.02 at $i_{ej}$=89°; the $p_{Sun}$ values were about 0.15-0.3; $p_{Me}$ - about 0.1-0.15; $p_V$ - about 0.23-0.35; the $p_E$ values were about 0.04-0.1 at $i_{ej}$=45° and could be about 0.2 at $i_{ej}$=89°. For point B, the $p_{ej}$ values were generally less than 0.04, $p_{Sun}$ values were around 0.07-0.3, $p_{Me}$ values were around 0.02-0.35 (with a maximum at $v_{ej}$=20 km/s), $p_V$ values were around 0.4-0.8, $p_E$ values were around 0.04-0.2 (but equaled 0.01 at $v_{ej}$=20 km/s and $i_{ej}$=89°). For point W, at $v_{ej}$≤12 km/s, the $p_{ej}$ values were mostly less than 0.06, $p_{Sun}$ values were around 0.1-0.4, $p_{Me}$ values were around 0.01-0.1, $p_V$ values were around 0.3-0.8, and $p_E$ was around 0.1-0.2. More bodies were ejected from the Solar System at higher ejection velocity $v_{ej}$. For $v_{ej}$=20 km/s, the fraction of ejected bodies was about 0.8 at $i_{ej}$=45°, and all bodies were ejected at $i_{ej}$=89°.

**Ejection of dust particles from the terrestrial planets**. The migration of dust particles ejected at different velocities from points F, B, and W on the surfaces of the terrestrial planets was numerically studied. The ejection velocity values were the same as for the bodies. The ejection angle was 45°. The gravitational influence of the Sun and all planets, radiation pressure, the Poynting -Robertson effect, and the solar wind were taken into account when integrating the equations of motion. The relative error per integration step using the Bulirsch-Stoer method was less than $10^{-8}$. The evolution of the dust particle orbits was studied until the particles left the Solar System or collided with the Sun or planets. The calculations were carried out for β values (the ratio of the radiation pressure force to the gravitational force) equal to 0.00004, 0.0004, 0.004, 0.04 and 0.4 (that is, for silicate particle diameters of about 1 cm, 1 mm, 100, 10 and 1 micron). The results of these calculations for dust particles are so far presented only briefly in [527].

It was found that approximately a half of the micron-sized particles fell onto the Sun, and approximately a half of the micron-sized particles were ejected from the Solar System. When ejected from the Earth at low velocities, slightly more than a half of the micron-sized particles fell onto the Sun, while when ejected at velocities greater than 12 km/s, typically more than a half of the particles were ejected from the Solar System. When ejecting micron-sized particles from Mars and Mercury from point B (behind the planet's motion), typically more than a half of the particles fell onto the Sun, while when ejecting micron-sized particles from point W (in the direction of the planet's motion), typically more than a half of the particles were ejected from the Solar System. Moreover, at higher ejection velocities, all such particles could have left the Solar System.

Almost all 10-100 micron and millimeter particles ejected from the terrestrial planets typically fell onto the Sun unless their ejection velocities were high. At an ejection from point F on the Earth with a velocity of $v_{ej}$=16 km/s, approximately 10% of 10-micron particles and approximately 20% of millimeter particles left the Solar System. At an ejection from point W with a velocity of $v_{ej}$=20 km/s, all 100-micron particles ejected from the Earth left the Solar System. In most calculations, the dynamic lifetime of 1-100 micron particles did not exceed 1 million years, while some ejected bodies could move in the zone of the terrestrial planets for hundreds of millions of years. No more than 4% of millimeter particles ejected from the Earth, Venus, and Mars fell on the planets. The fraction of ejected bodies falling on the planets could be higher for planets closer to the Sun. When ejected at a velocity of $v_{ej}$=4.23 km/s from points F and W on Mercury, about 8-10% of the millimeter particles fell back onto Mercury.

## 16. Automatic removal of cosmic ray signatures from images taken by the Deep Impact spacecraft

In 2005–2006, S. I. Ipatov was a member of the Deep Impact team, led by Michael A'Hearn. NASA's Deep Impact spacecraft was the first in history to drop a probe on a comet, which collded with its surface after photographing it from close range. Ipatov worked on the automatic recognition and removal of cosmic ray signatures from the images taken by this spacecraft. The results of this research are presented in the article [34] and in the abstracts [356, 361, 368, 372]. Ipatov was also a co-author of the general papers [32, 35] of the Deep Impact team.

We analyzed the performance of several algorithms (imgclean, crfind, di_crrej and rmcr) written by several authors that recognized signatures of cosmic rays in a single image from the Deep Impact spacecraft. Images taken by different cameras (HRI, MRI, and ITS) during the Deep Impact spacecraft's flight to comet Tempel 1 were examined. The HRI camera (High Resolution Instrument) provided higher resolution images of the comet's nucleus, and the MRI camera (Medium Resolution Instrument) had a wider field of view. The ITS (Impactor camera Targeting Sensor) was installed on the impact module and worked only until the collision with the comet.

To automatically remove cosmic ray signatures from many images, we proposed using the imgclean algorithm. Other algorithms discussed above do not work well with a large number of images and do not complete their work for some images. However, other algorithms may be better used for analyzing specific images. Imgclean detects false cosmic ray trails near the edge of a comet's nucleus, and this algorithm often fails to recognize all pixels of long cosmic ray signatures. However, other algorithms discussed above sometimes fail to remove all cosmic ray signatures. Our rmcr algorithm is the only one among those mentioned above that works on raw images. Images with removed cosmic ray trails are useful for photometric and other studies.

For most visual HRI and MRI images acquired during low solar activity with exposure times $t$>4 s, the number of bright pixel clusters per image per second per $cm^2$ of CCD (charge-coupled device) was about 2–4, for both dark (i.e., with cosmic ray signatures only) and normal sky images. At high solar activity, this number of pixels sometimes exceeded 10. The ratio of the number of cosmic ray signatures consisting of n pixels obtained at high solar activity to the number at low solar activity was greater for a larger number of pixels. Due to higher background luminance variations in ITS images, all the considered algorithms detected too many false cosmic ray signatures with the default parameter settings. Clusters consisting of fewer than 4 pixels usually cannot be confidently identified as cosmic ray signatures on ITS CCDs at any parameter settings, since the luminance for such small cosmic ray signatures is low enough to be lost in the background noise. The number of clusters detected as cosmic ray signatures in a single infrared image using imgcleanf is at least several times higher than the actual number of cosmic ray signatures (other algorithms have detected even more false cosmic ray signatures), but the number of clusters detected by comparing two consecutive dark frames is consistent with the expected number of cosmic ray signatures. Some false cosmic ray signatures include bright pixels that are present repeatedly in different infrared images. We do not recommend using automatic cosmic ray signature removal on ITS and IR (infrared camera) images. Although we obtained many false cosmic ray signatures after removing cosmic ray signatures from ITS and IR images, such removal was useful for validating the pipeline. Such studies allowed us to see how improvements at different stages of the pipeline would reduce the number of false cosmic ray signatures and to understand what further changes are needed to the pipeline. Our interactive code, imr, allows the user to select regions of the image where glitches detected by imgclean as cosmic ray signatures are ignored. In other regions selected by the user, the brightness of some pixels is replaced by the local median brightness if the brightness of these pixels is several times greater than the median brightness. The interactive code allows to remove long cosmic ray signatures and prevents the removal of spurious cosmic ray signatures near the edge of the comet's nucleus. The interactive algorithm can be applied to editing any digital image. The results presented in [34] can be used for other cometary missions (including those with outbursts).

17. **Ejection of material from comet 9P/Tempel 1 after the collision of the Deep Impact impactor with the comet**

On July 4, 2005, the 370-kilogram impactor of the Deep Impact spacecraft collided with comet 9P/Tempel 1 at a velocity of 10.3 km/s at an angle to the comet's surface of approximately 20–35° [32]. The evolution of the ejecta cloud was observed by the Deep Impact (DI) mission's cameras, space telescopes (e.g., Rosetta, the Hubble Space Telescope, Chandra, and Spitzer), and more than 80 observatories on the Earth. Ejections similar to this type of ejecta can occur when a small celestial body collides with a comet at a high speed. Therefore, studying the ejecta after such an impact is important for understanding collisional processes in the Solar System. Studies of the velocities of ejecta from a comet provide insight into the size of the meteor shower generated by the comet.

We studied the velocities and relative amounts of material ejected from Comet Tempel 1 after the impactor's collision with the comet. These studies were based on the analysis of images taken by the Deep Impact spacecraft's cameras during the first 7 minutes after the collision. By analyzing these images, Ipatov concluded that comets may contain many cavities containing dust and gas under pressure at a depth of several meters below their surfaces. One of these cavities was opened during the impactor's collision with the comet, which caused additional ejection of material from the comet. The results obtained were published in papers [40, 41, 100] and abstracts [362, 369, 371, 380, 383, 387, 388, 390, 391, 394, 396, 398, 402, 406].

The velocities of the particles that made the main contribution to the brightness of the ejected material cloud, i.e., mainly particles with a diameter of less than 3 microns, were considered. After a time of $t_e$~0.2 sec, the velocities of most of the observed ejected material were of the order of 10 km/s. The velocities of some particles ejected during the first three seconds exceeded 1 km/s, but the fraction of material with such velocities among all ejected material was small. The velocities of most of the observed material ejected during $t_e$>4 sec, were an order of magnitude smaller than at $t_e$<1 sec (hundreds of m/s instead of several km/s).

When estimating the velocities of ejected particles, we analyzed particle motion over distances of several kilometers. Particle destruction, sublimation, and acceleration did not significantly affect our velocity estimates, as we analyzed particle motion over periods of no more than a few minutes. For particles with initial velocities $v_p$≥20 m/s, the increase in their velocity due to gas acceleration did not exceed a few meters per second.

The time-dependent variations in the velocities and amounts of material ejected after the Deep Impact impactor collided with the comet differed from those observed in experiments and theoretical models. For ground-based observations made several hours after impact, such differences could have been caused by the evaporation of ice in the ejecta and fast-moving gas. In our studies of particle motion over several minutes, the impact-induced extra ejecta (increasing the ejecta of small, bright particles) could have played a large role in this difference, and it might be possible to consider the ejecta as a superposition of the normal ejecta and the extra ejecta. The contribution of the extra ejecta to the cloud brightness could have been significant, but its contribution to the total ejected mass could have been relatively small, since the fraction of small particles observed among particles of all sizes was likely larger for the extra ejecta than for the normal ejecta. Our ejecta model considered only those particles that reached a distance $R$≥1 km from the ejecta location. Large areas of saturated pixels (saturated pixels are bright pixels that distort the image) in the DI images taken at time $t$ after impact more than 110 s did not allow us to draw unambiguous conclusions about the ejection velocities of all particles.

Our results showed that there was a local maximum in the ejection velocity at $t_e$~10 s with typical velocity projections $v_p$ onto the plane perpendicular to the line of sight of about 100-200 m/s (Fig. 1). At the same time, significant excess ejection began in several directions (ejection rays), a local increase in the brightness of the brightest pixel occurred, and the direction from the ejection site to the brightest pixel rapidly changed by about 50°. In images taken during the first 12 s and after the first 60 s, this direction was mostly close to the impact direction. For 1<$t_e$<3 and 8<$t_e$<60 s, the graph of the change in the estimated amount $r_{te}$ of ejected observed material versus time was greater than the exponential line connecting the $r_{te}$ values at 1 and 300 s. The above-mentioned features could have been caused by the impact triggering an additional ejection and the ejection of icier material. At $t_e$~55-60 s, the ejection velocity decreased sharply, and the direction from the ejection site to the brightest pixel quickly returned to the direction it had been before 10 s. This could have been caused by the sharp decrease in the additional ejection that began at $t_e$ ~10 s.

The additional ejecta could have originated from the entire surface of the crater, while the normal ejecta originated primarily from its edges. The "fast" ejecta could have been caused by the ejection of material from cavities containing material under gas pressure. The "slow" additional ejecta could have been similar to ejecta from the "fresh" surface of a comet and could have occurred long after the crater formed.

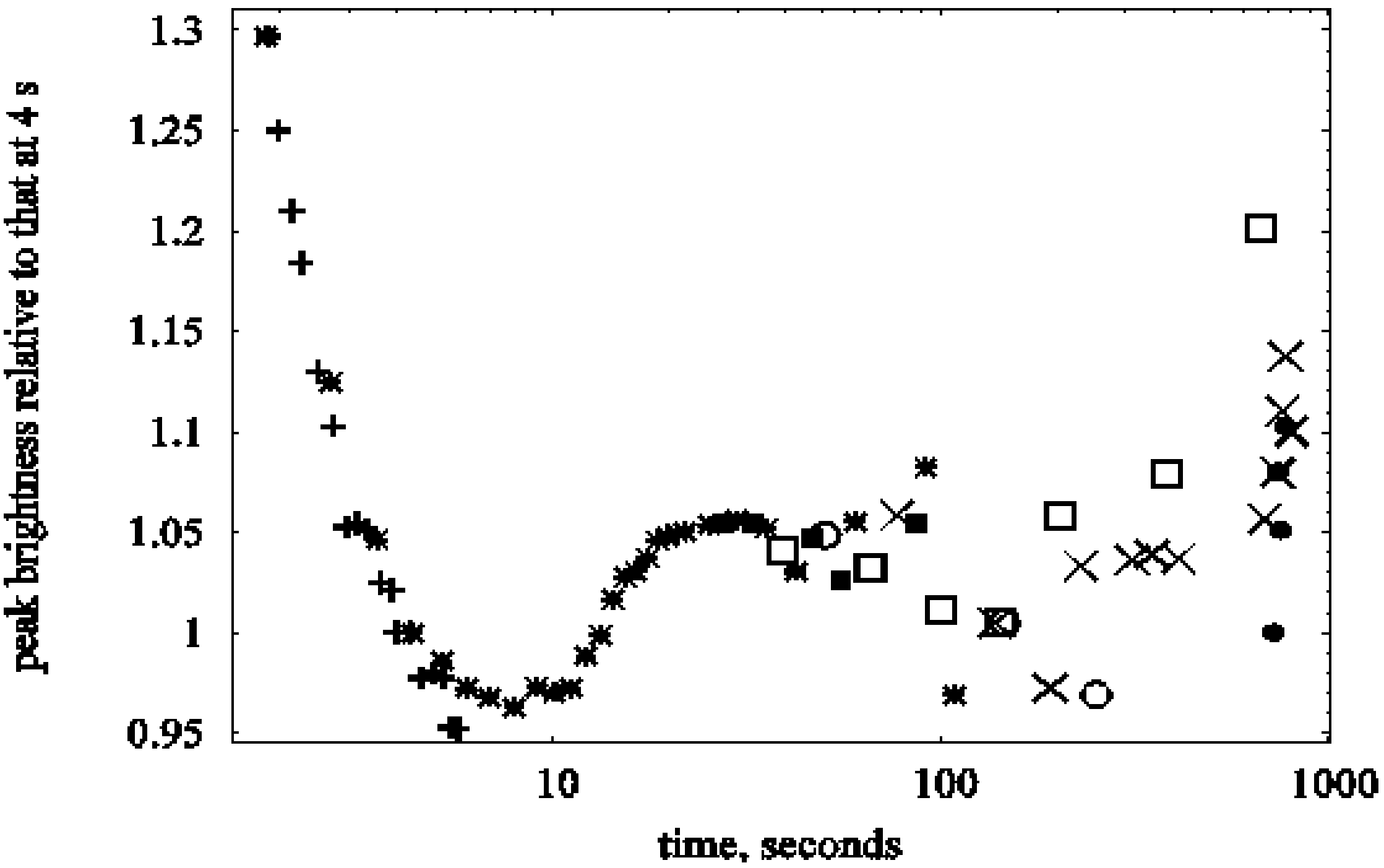

**Figure 1.** Change in the brightness of the brightest pixel as a function of time (in seconds) after the impact of the DI spacecraft's impactor with the comet. The maximum pixel brightness 4 seconds after the onset of the ejecta is taken as the unit of brightness.

Analysis of observations of the cloud that formed after the collision of the Deep Impact spacecraft's impactor (DI) with a comet, as well as of ejecta from various comets, supports the hypothesis that large cavities containing gas-pressured material may exist beneath a significant fraction of the comet's surface. The internal pressure of gas and material within these cavities can cause natural and induced ejections and may lead to comet breakup. The upper edge of the cavity, exposed within $t_e$ ~10 s, may be located several meters beneath the surface of Comet Tempel 1.

Our studies did not allow us to estimate precisely when the ejecta ended, but they are consistent with continuous ejecta for at least the first 10 minutes after the impact. The duration of the ejecta (up to 30-60 minutes) could have been longer than a typical ejecta, which could have lasted only a few minutes. Analysis of observations from the Deep Impact spacecraft's cameras supports the gravity-dominated cratering model (i.e., a larger amount of ejected material and a larger crater) and the idea that particles were ejected at different velocities and masses at the same time.

The velocity projections of most of the observed material ejected at $t_e$~0.2 s were found to be about 7 km/ s. Analysis of DI spacecraft observations using different approaches showed that for $1<t_e<115$ s, the time variations in the projections $v_p$ of the characteristic velocity of the observed particles onto the plane perpendicular to the line of sight can be considered proportional to $t_e^{-\alpha}$ with $\alpha$~0.7-0.75. For the *VExp* model with $v_p$ proportional to $t_e^{-\alpha}$ for any $t_e>1$ s, the fractions of observed (not total) material ejected (at $t_e\leq 6$ and $t_e\leq 15$ s) with $v_p\geq 200$ and $v_p\geq 100$ m/s were estimated to be approximately 0.1–0.15 and 0.2–0.25, respectively, when considering only the material observed during the first 13 minutes. The "fast" ejection with velocities of ~100 m/s could likely have lasted for at least several tens of seconds, and it could have significantly increased the fraction of particles ejected with velocities of ~100 m/s compared to the estimates for the *VExp* model and for the normal ejection. The above estimates are consistent with estimates (100-200 m/s) of the projected velocity of the leading edge of the Deep Impact comet dust cloud (DI cloud) made by other scientists and based on various ground-based and space telescope observations.

Excessive ejection of material in several directions (ejecta rays) was significant during the first 100 s and was still visible in images taken at $t$ ~500-770 s. These observations indicate that the ejection could have continued at $t_e$~10 min. The sharpest rays were caused by material ejected at $t_e$~20 s. In particular, excess ejecta were observed, especially in images taken at $t$~25-50 s after the impact, in directions perpendicular to the impact direction. The directions of the excess ejecta may have changed over time. The size of the DI dust cloud of significant opacity likely did not exceed 1 km.

On February 14, 2011, the Stardust spacecraft photographed a crater formed after the DI spacecraft impact with a comet. A number of authors have estimated the size of this crater. In [41], the location of the upper boundaries of cavities containing dust and gas under pressure in comets was discussed. The distance between the pre-impact surface of comet 9P/Tempel 1 and the upper boundary of the largest cavity opened during the ejection of material after the collision of the Deep Impact (DI) spacecraft with the comet is estimated to be approximately 5–6 m, if the diameter of the DI transit crater was approximately 150–200 m. The estimated depth was 4 m if the diameter was 100 m. These estimates used crater formation models and estimated the depth that the forming crater would reach 8 seconds after the onset of the cratering process. The largest cavity formed after the DI impact could have been relatively deep, since significant excess ejecta continued for about 50 s. These depth estimates are consistent with the depth (4–20 m) of the initial CO ice sublimation front in models of the 17P/Holmes comet explosion. Our research suggests that cavities containing dust and gas under pressure, located several meters below the comet surface, may be common.

## 18. **Observations of asteroids and comets**

In 1999, S. I. Ipatov worked for six months on a DWTC grant at the Royal Observatory of Belgium. Together with Eric Elst, who discovered about 4000 asteroids, and Thierry Pauwels he participated in observations of asteroids and comets using the 0.85-meter Schmidt telescope with a CCD camera (3072x2048 pixels) and was a co-discoverer of eight asteroids that were given numbers. In particular, Ipatov participated in the first observations of the comet named C/1999 T3 (LINEAR). The results of these observations of the comet are presented in Minor Planet Electronic Circ. [287] and IAU Circ. [288].

Information about the asteroids discovered by S.I. Ipatov is presented on the website: https://en.wikipedia.org/wiki/List_of_minor_planet_discoverers#S._I._Ipatov - a list of 8 minor planets discovered by S.I. Ipatov together with Eric Elst or with T. Pauwels. The website https://en.wikipedia.org/wiki/Category:Discoveries_by_Sergei_I._Ipatov provides the list of 5 minor planets discovered discovered by S. I. Ipatov that got names). These five asteroids have the following names:

41107 Ropakov - Ivan V. Ropakov (1892-1992), grandfather of the second discoverer, was a very brave man and the director of a small butter industry in the village of Pogorelka (Vologda region, Russia). https://ssd.jpl.nasa.gov/tools/sbdb_lookup.html#/?sstr=41107. https://minorplanetcenter.net/db_search/show_object?object_id=41107. [Ref: Minor Planet Circ. 64564]. The original version of the text about Ivan Ropakov included the words that he was a hero of the First World War. However, it turned out that the rules for naming asteroids prohibit mentioning military achievements. In the next application, the previous words had to be changed to "was a very brave man." Ivan Ropakov was awarded two St. George's Crosses.

43025 Valusha - Russian scientist Valentina I. Ipatova - Artioukhova (b. 1954) specializes in hydrobiology. She conducts biological monitoring of the environment in some regions of Russia, aiming with her work to protect the ecosystems from pollution and contamination. https://ssd.jpl.nasa.gov/tools/sbdb_lookup.html#/?sstr=43025. https://minorplanetcenter.net/db_search/show_object?object_id=43025 [Ref: Minor Planet Circ. 57951]. V. I. Ipatova – wife S. I. Ipatov.

49700 Mather - John C. Mather (born 1946), an American cosmologist and senior project scientist for the James Webb Space Telescope. He led the team that constructed the Cosmic Background Explorer (COBE). For his role in mapping microwave radiation and understanding the early universe, he received the 2006 Nobel Prize in Physics. https://ssd.jpl.nasa.gov/tools/sbdb_lookup.html#/?sstr=49700. https://minorplanetcenter.net/db_search/show_object?object_id=49700. [Ref: Minor Planet Circ. 57952]. S. I. Ipatov has several joint papers [27-30, 33, 37] with John Mather.

150316 Ivaniosifovich - Ivan Iosifovich Ipatov (1927–2015), father of the second discoverer, Sergei I. Ipatov, was one of the founders of the geodetic network of Russia. https://ssd.jpl.nasa.gov/tools/sbdb_lookup.html#/?sstr=150316. https://minorplanetcenter.net/db_search/show_object?object_id=150316. [Ref: WGSBN Bull. 1, #3, 12]. The original name application proposed Ivanipatov, but it turned out that Ipatov couldn't be used because an asteroid already bore that name. Therefore, in the second application, the surname was changed to a patronymic. Ivan Ipatov's father was named Osip, but his patronymic was listed as Iosifovich in Ivan's passport. Collective farmers like Osip typically didn't have passports until 1981.

337166 Ivanartioukhov - Ivan Semenovich Artioukhov (1908–1995), father of the second discoverer's wife, was the founder of the society for nature conservation in Russia. https://ssd.jpl.nasa.gov/tools/sbdb_lookup.html#/?sstr=337166. https://minorplanetcenter.net/db_search/show_object?object_id=337166.

Asteroid 14360 Ipatov was named in honor S. I. Ipatov. Sergej Ivanovich Ipatov (b. 1952) is a Russian scientist and specialist in the migration of minor planets. During his stay in 1999 at the Uccle Observatory, he was shown to be a very fine observer who made several discoveries with the Uccle Schmidt telescope. https://ssd.jpl.nasa.gov/tools/sbdb_lookup.html#/?sstr=14360. https://minorplanetcenter.net/db_search/show_object?object_id=14360. [Ref: Minor Planet Circ. 55721].

## 19. Models for determining the probability of detecting near-Earth objects in different regions of the sky

The problem of asteroid-comet hazard is actual now and has been addressed in numerous studies. The number of near-Earth objects (NEOs) with diameters greater than 1 km and 40 m is estimated at 1,000 and 1 million, respectively. A body with the size of the Chelyabinsk meteorite (20 m in diameter) having fallen onto the city at a steep angle could cause significant damage. To organize the monitoring of hazardous objects and counter the threat of their collision with Earth, it is essential to better understand which regions of the sky require greater attention during observations of NEOs. It is also important to compare the observational efficiency of planned space-based telescopes in various orbits, assess the most likely origins - and, consequently, the compositions - of NEOs moving in given orbits, and determine the typical velocities and impact angles at which such objects might collide with the Earth.

In the article [72] and abstracts [107, 457], together with L.V. Elenin, Ipatov discussed the possibilities of applying the results of computer modeling of the migration of small bodies from the Solar System to the construction of a model of the probability of the appearance and detection of NEOs in various regions of the sky, including potentially dangerous objects (for a number of telescopes). In constructing such a model, the following problems can also be solved: (1) Estimating the role of NEO replenishment sources (asteroids, comets, trans-Neptunian objects). (2) Calculating the probabilities of collisions and collision vectors of migrated bodies with the Earth. It is proposed to study the probabilities of the appearance and detection of NEOs, including those dangerous to inhabitants of the Earth, of various sizes in various parts of the starry sky not only for observations from the Earth, but also from several regions where it is planned to launch spacecraft with telescopes to observe NEOs.

When constructing a model for the probability of NEO appearance and detection in various regions of the sky, it is necessary to use the distribution of NEOs by their orbital elements and masses. Taking into account observational selection, the actual distribution of NEOs by their orbital elements can be estimated based on the population of known NEOs. Comparing observed NEOs with the results of migration calculations for various small bodies will provide a better understanding of the NEO distribution by orbital elements for small bodies, as well as the typical orbits of NEOs originating from different sources, and therefore with different compositions and albedos.

It would be useful to develop programs for constructing maps of the distribution of the probability of detecting NEOs in the celestial sphere for the current observing night. These programs and the models of the brightness of the starry sky used by them would be oriented towards Russian telescopes. It is proposed to calculate the probabilities of the appearance and detection of NEOs in the celestial sphere not only for all NEOs, but also separately for NEOs of different origins, as well as for NEOs passing at a short distance from the Earth and for objects that may collide with the Earth. In [72], calculations of the probability of detecting a celestial body and the brightness of the starry sky were discussed. When probabilistically choosing the orbital elements and the stellar magnitude of a celestial body, first, taking into account the distance from the observer to the orbital point, it is necessary to determine the positions of the body in its orbit at which it can be observed under ideal observing conditions. Then this interval, if non-zero, will be refined taking into account the brightness of the starry sky near the body in question and the ability of the telescope to observe a given region of the sky.

The proposed model for the probability of the appearance of potentially hazardous objects in various regions of the sky will allow observers to better understand which regions of the sky should be given more attention during observations at a given epoch. This model can be applied to more effectively operate optical instruments for the search and detection of NEOs (in particular, the ISON-NM and ISON-SSO telescopes). The proposed algorithms for calculating the brightness of the starry sky [433, 535] can be used by astronomers when planning observations of various objects. The proposed studies of the efficiency of NEO searches by various telescopes will allow us to compare the efficiency of observations by planned space telescopes in various orbits and provide recommendations for the creation and deployment of new telescopes, including space telescopes, designed for the search for NEOs. In 2015-2016, we submitted applications for a grant from the Russian Foundation for Basic Research on this topic. These applications were not supported, and we continued working on other tasks.

To organize observations of hazardous objects and prevent their collisions with the Earth, it is important to assess the likely origin (and therefore composition) of NEOs moving in given orbits, as well as the typical velocities and angles at which NEOs collide with Earth. The proposed studies will help answer these questions. The results of the proposed studies will help to clarify the mass and composition of matter delivered from various distances from the Sun to the NEO population, as well as to the terrestrial planets and the Moon. They will be important for specialists studying cratering and assessing the composition of matter in the upper layers and on the surfaces of the terrestrial planets and the Moon. These results will allow for a detailed study of the distribution of velocity vectors of bodies colliding with the terrestrial planets and the Moon.

# EXOPLANETS

## 20. Spectra of Earth-like exoplanets with various rotation periods

The first exoplanet was discovered in 1995. Currently, more than 5,000 exoplanets have been discovered. One can read about exoplanets, for example, in the monograph [Marov M.Ya., Shevchenko I.I. Exoplanets: Physics, Dynamics, Cosmogony. Moscow: Fizmatlit, 2022, in Russian].

The studies of the spectra of exoplanets similar to the Earth with different periods of axial rotation were carried out by S. I. Ipatov together with James Y-K. Cho in 2007-2008. The obtained results were initially briefly published in abstracts [384, 385, 435]. In 2014, Ipatov prepared and presented an article in the proceedings of the seminar "Exoplanet Studies" (June 3-4, 2014, Space Research Institute of the Russian Academy of Sciences, Moscow). However, these proceedings were not published, since the seminar participants presented a small number of articles. Below is the text of that unpublished article for the seminar proceedings. The results of research on this topic were published in more detail in the article [534] in 2025.

***Summary.*** Differences in the emission spectra of exoplanets that differ from the Earth only in their rotation period are comparable to differences associated with changes in the planet's observation angle. The largest differences in the spectra of Earth-like exoplanets were obtained for wavelengths of ~5-10 and ~13-16 microns. By analyzing the spectrum with a wavelength of approximately 9.4-10 microns, one can conclude whether the exoplanet's atmosphere contains ozone.

***Research models and methods***

Earth-like planets have now been discovered orbiting other stars. It is believed that about 1/3 of stars may contain planets similar to the Earth. To assess the possibility of detecting signs of life on such planets, we studied the spectra of planets similar to the Earth, but with different periods of axial rotation. Using the CCM3 atmospheric general circulation model and considering the atmospheric circulation over 2 years, we calculated [384, 385, 435] the spectra of the atmospheres of the Earth and exo-Earth rotating with periods $P$ equal to 1 or 100 days using the SBDART program. The CCM3 atmospheric general circulation model, in particular, takes into account radiation, water convection, cloud fraction, types of land surface, temperature and pressure profiles at each point of the planet (resolution - 128 longitude values $l_{on}$ and 64 latitude values $l_{at}$), and also calculates the spectrum at the planet's surface. The SBDART program calculates the radiative transfer in the atmosphere taking into account clouds [Ricchiazzi P., et al. BAMS, 1998, v. 79, 2101-2114]. This program calculates the spectrum for a single point on the Earth's surface (for a pair of longitude and latitude values), but it analyzes the atmosphere at different altitudes. Using SBDART as a subroutine, we calculated the average spectrum for some region on the planet. We analyzed the spectrum of the upward radiation at altitudes $h$ equal to 1 and 11 km (below, the spectrum at $h$=11 km is mainly discussed) at wavelengths from 0.3 to 1 micron and from 1 to 18 microns for the initial state, for the Earth, and the exo-Earth. The initial conditions for the Earth and the exo-Earth were the same. The following models were considered:

*Model 1.* In this model, the average radiative flux depends only on the region under consideration, not on the angle at which that region is viewed. For a single SBDART calculation, the weighting factor is proportional to the area of the region (2.8° by 2.8° at the equator) corresponding to the pair of longitude and latitude values $l_{on}$, $l_{at}$). This area is proportional to cos($l_{at}$), but is independent of $l_{on}$.

*Model 1a.* For model 1*a,* we considered fixed latitude and all longitude values (0-360°).

*Model 1b.* For model 1*b*, the average values of the radiation flux were calculated for all pairs ($l_{on}$, $l_{at}$), i.e. the entire surface of the planet was considered.

*Model 1c*. The southern hemisphere was considered (-90°≤$l_{at}$≤0, 0≤$l_{on}$≤360°).

*Model 2.* In this model, the average radiative flux depends on the angle at which different parts of the region under consideration, which constitutes half of the planet's surface, are visible. For example, when viewed from the equator at $l_{on}=l_{on0}$, then the weighting factor *k* for a single SBDART calculation is equal to abs (cos($l_{at}$)×cos($l_{at}$)×cos($l_{on}$-$l_{on0}$)). In this product, one cos($l_{at}$) is due to the fact that the region corresponding to the pair ($l_{on}$, $l_{at}$) is smaller for latitudes closer to the pole (as for model 1), and the other factor cos ($l_{at}$) is due to the fact that the observer sees regions at different latitudes at different angles. Looking from the pole, then *k*=abs(cos($l_{at}$)×sin($l_{at}$)).

***Cloud and temperature maps***

Cloud maps differed for different rotation periods of the planet (Fig. 1). For the Earth, there were three large cloud regions (blue) in the meridian direction, while for the exo-Earth (with *P*=100 days), such a region was only near the equator. For this exoplanet, there were almost no clouds near the south pole, while for the Earth and the initial state, there were many clouds near the poles. Colored figures can be viewed online in the file for this book, as well as in the article [534].

In the equatorial region, the temperature maps (Fig. 2) differed a little at *P*=1 day and *P*=100 days. For the exo-Earth, the temperature values *T* were more uniform than for the Earth. For example, the region *T*>290° K for the exo-Earth was smaller than for the Earth, and the region *T*<230° K was absent for the exo-Earth. The main difference in the *T*-maps was near the South Pole. At the South Pole, the temperature was higher for the exo-Earth than for the Earth and the initial state.

***Radiation spectra of the Earth and exo-Earth***

Examples of the radiation spectra of the Earth and exo-Earth (at *P*=100 days) are shown in Figs. 3-4. For the considered periods of axial rotation in model 1a, the radiation flux was maximum at the equator and decreased towards the poles. The maximum value $F_{max}$ of the flux was typically achieved for a wavelength of λ≈10 microns. For the North Pole, the maximum flux values were similar for the Earth, exo-Earth, and the initial state. For the South Pole, the $F_{max}$ values were greater for the Earth and exo-Earth than for the initial state, by a factor of 1.3 and 1.7, respectively. The radiations of both planets differed significantly for wavelengths of ~5-10 and ~13-16 microns, for example, near the poles and at $l_{at}$≤-43° and $l_{at}$≥77°. For all graphs at *h*=11 km, a local minimum was observed at λ~14-16 microns and a smaller local minimum at λ~9.5 microns.

For the entire surface of the planet (model 1b), the graphs of the dependence of the radiative flux on λ were very close for the Earth, the exo-Earth, and the initial state. The value of $F_{max}$ for the entire surface of the southern hemisphere (model 1c) was 15 watts m$^{-2}$microns$^{-1}$ and was 7% less than for the entire surface (16 watts m$^{-2}$ microns$^{-1}$) (model 1b) (these values are 2 times less than at the equator). For model 2, when looking at the planet from the North Pole, $F_{max}$ was 1.2 times greater than from the South Pole (see Table 1). The difference between the radiation from the northern and southern hemispheres could be caused by Antarctica. The results obtained for models 1b,c and 2 indicate that the influence of the line of sight on the observed radiation flux from the entire planet may be greater than the influence of the rotation period. For the southern hemisphere, at a wavelength of 14-16 microns and *h*=11 km, the flux for the exo-Earth was 1.2 times greater than for the Earth and the initial state. Differences in flux values may be related to cloud distribution patterns and the fact that $F_{max}$ is higher in hotter zones. For the exo-Earth, cloud coverage was greatest over a broad zone near the equator and lowest near the South Pole.

The following common features were obtained for the Earth and exo-Earth: (1) the planets have a wide range of $CO_2$ absorption of about 14 microns; (2) there are no significant differences for the spectra near the equator (however, for some regions, for example, near the poles, there may be significant differences in the spectra); (3) if the spectrum is integrated over the entire disk of the planet, the difference in the spectrum obtained by observing the Earth/exo-Earth from different directions is significantly reduced compared to observations of individual regions of the planets, but there is still a noticeable difference in the integrated spectrum signal for the Earth and exo-Earth (for example, this difference is noticeable in the case of the spectrum at an altitude of 11 km observed from the South Pole, but the difference is small if the entire disk is observed from different directions of the equator). Our results show that the spectral signal cannot be used to determine the rotation velocity of the exoplanet, since the observation angle of the exoplanet is generally not known in advance. In the case of Earth-like exoplanets, the greatest differences in spectra are observed at wavelengths of ~5-10 and ~13-16 microns.

By analyzing the spectrum with a wavelength of approximately 9.4-10 microns, we can determine whether the exoplanet's atmosphere contains ozone. A local minimum at a wavelength of 9.4-10 microns is absent (Figure 4) in the upward radiation plots at an altitude of 11 km in the absence of ozone, but this minimum is present in the plots of the models with ozone. At wavelengths shorter than 0.35 microns (SBDART does not consider wavelengths shorter than 0.25 microns), some differences were also observed between the models with and without ozone. Since ozone is important for life, the 9.4-10 microns band may be important for future observations of Earth-like exoplanets.

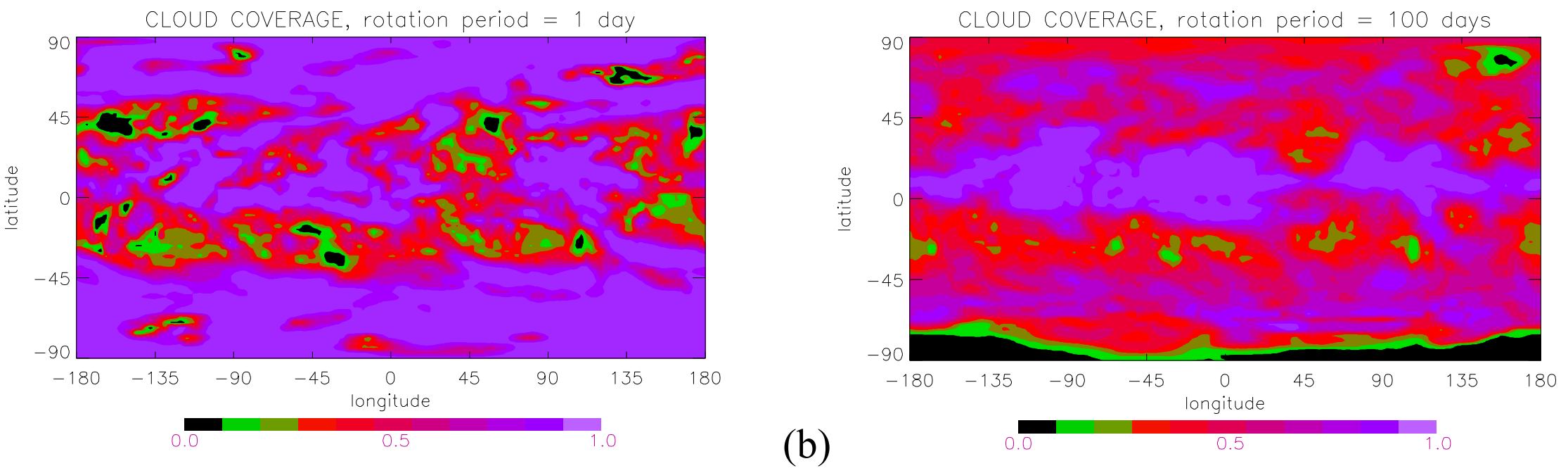

**Figure 1.** Cloud maps (a) for the Earth model (*P*=1 day), (b) for the exo-Earth model (*P*=100 days).

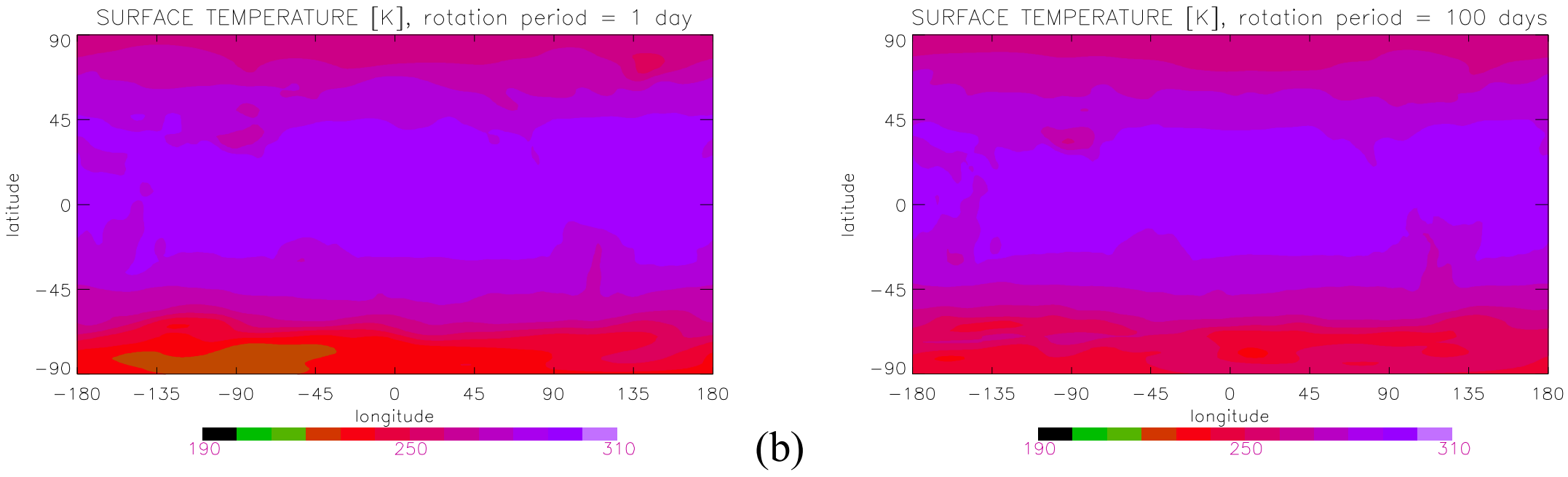

**Figure 2.** Surface temperature (K) maps (a) for the Earth model (*P*=1 day), (b) for the exo-Earth model (*P* =100 days).

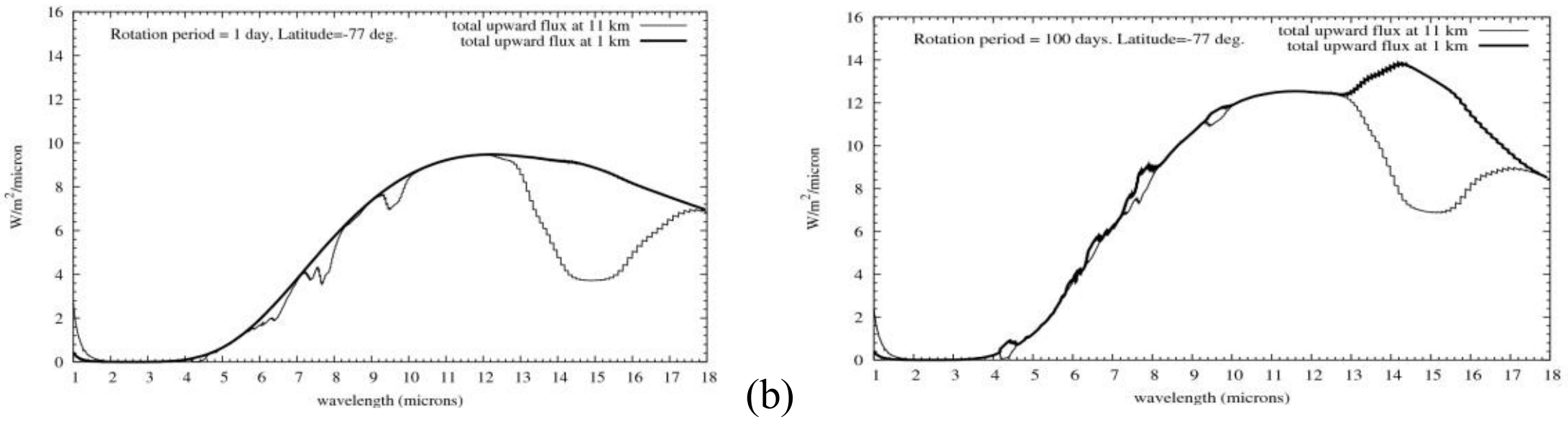

**Figure 3.** Longitude-averaged upwelling spectrum at -77° latitude (model 1a) (a) for the Earth model (*P*=1 day) and (b) for the exo-Earth model (*P*=100 days). The upper line corresponds to an altitude of 1 km, and the lower line to an altitude of 11 km.

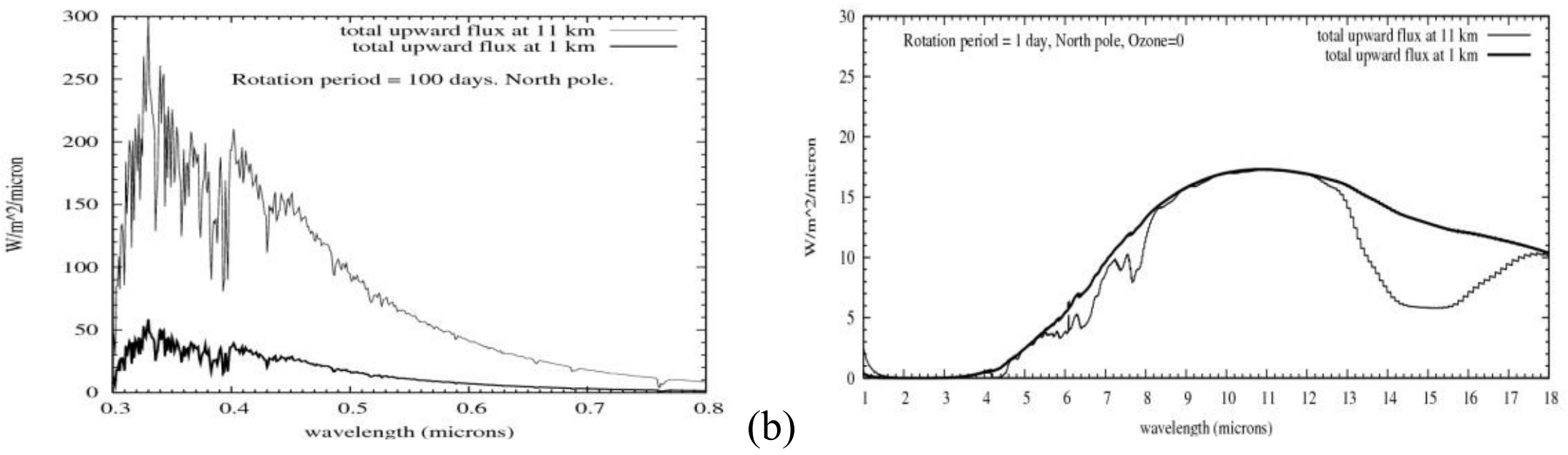

**Figure 4.** Spectra seen from the North Pole, (a) for the exo-Earth (*P*=100 days, model 1a, the spectra are almost the same for the Earth (*P*=1 day) and for the initial state), (b) for the Earth (*P*=1 day) in the absence of ozone. The upper line in the graphs corresponds to an altitude of 1 km, and the lower line to an altitude of 11 km.

**Table 1.** The maximum radiation flux (in watts m$^{-2}$ microns$^{-1}$) from the hemisphere of the planet at a wavelength of 10.0-10.4 microns at an altitude of 1 km (at an altitude of 11 km the flux is usually 1.01-1.02 times less) for several directions of viewing the planet (model 2).

| | South Pole | North Pole | Equator. $l_{on}$=90° | Equator. $l_{on}$=270° |
|---|---|---|---|---|
| Initial state | 6.61 | 8.09 | 8.27 | 8.49 |
| Earth | 6.63 | 7.97 | 8.05 | 8.31 |
| Exo-Earth | 6.61 | 7.77 | 8.25 | 8.09 |

## 21. Efficiency of exoplanet searches using microlensing observations with different telescopes

Studies of the efficiency of searching for exoplanets using the microlensing method using various telescopes was carried out by S.I. Ipatov together with Keith Horne in 2011-2013. The obtained results were published briefly in papers [103] and in abstracts [417, 423, 430, 436]. In 2014, Ipatov prepared and presented an article in the proceedings of the seminar "Exoplanet Studies" (June 3-4, 2014, Space Research Institute of the Russian Academy of Sciences, Moscow). However, these proceedings were not published, since the seminar participants presented a small number of articles. Below is the text of that unpublished article for the seminar proceedings. The results of research on this topic are presented in more detail in the article [535] in 2025. Ipatov was also a co-author of articles [42-50] devoted to the search for planets using microlensing method.

***Resume***

Microlensing event observations using various telescopes and several approaches to selecting observed events. In developing an algorithm for comparing these efficiencies, we considered sky brightness models consistent with OGLE and RoboNet observations from 2011.

***1. Introduction***

If a ray of light from a source star passes a lens star on its way to the Earth, the observed brightness of the source star may increase (due to the bending of the light's trajectory). This phenomenon is called microlensing. If the lens star has a planet, it may cause a change in the brightness of the source star. The microlensing method makes it possible to find planets with masses close to the mass of the Earth, located at a distance from the star at which life is possible on these planets. We have compared [103] the relative efficiency of detecting exoplanets in microlensing observations carried out on different telescopes for several approaches to selecting the observed microlensing events. The time intervals at which microlensing events can be observed are determined based on the positions of the Sun and Moon and other constraints on the direction of the telescope. Based on the approach presented in [Horne K., Snodgrass C., and Tsapras Y. MNRAS, 2009, v. 396, 2087-2102], at each time step we calculated the comparative efficiency of exoplanet detection. At each time step, the event with the highest detection probability was selected. We considered the following 13 telescopes, numbered from $N_t$=1 to $N_t$=13:

1. 2 m FTS - Faulkes Telescope South - Siding Spring Observatory, Australia.
2. 2 m FTN - Faulkes Telescope North - Haleakela, Hawaii, USA.
3. 2 m LT - Liverpool Telescope - La Palma, Canary Islands, Spain.
4. 1.3 m OGLE - The Optical Gravitational Lensing Experiment - Las Campanas, Chile.

5-7. Three 1 m CTIO - Cerro Tololo Inter-American Observatory, Chile.

8. 1 m MDO - McDonald observatory, Texas, USA.

9-11. Three 1 m SAAO - South African Astronomical Observatory, South Africa.

12-13. Two 1 m SSO - Siding Spring Observatory, Australia.

***2. Sky brightness models***

As part of our studies to create an algorithm for optimal selection of observed microlensing events to maximize exoplanet detection efficiency, we constructed a model of the sky brightness [433] (based on [Krisciunas K., Schaefer B. PASP, 1991, v. 103, 1033–1039]) and seeing (a characteristic of image distortion due to atmospheric turbulence) using infrared observations from the OGLE, FTS, FTN, and LT telescopes in 2011. $\chi^2$ optimization was used to analyze the observations. The seeing values for FTN were almost a half of those for the other three telescopes considered (FTS, LT, and OGLE). The resulting sky brightness at the zenith during infrared observations were 19.0, 18.7, 19.6, and 18.1 mag arcsec$^{-2}$ for FTS, FTN, LT, and OGLE, respectively. For the same airmass values (airmass=1/cos(the angle between the directions to the point in the sky under consideration and to the zenith)), the typical sky brightness in different regions of the sky usually varied by less than 1 mag if images were considered in which the Moon was below the horizon. Most deviations of the sky brightness from the best model for each event (observations minus $\chi^2$ optimization, which is different for each event) were in a small interval (from -0.4 to 0.4 mag) even for all positions of the Moon and the Sun; for the Moon below the horizon, there were many values in the interval [-0.2, 0.2], but the absolute values of some deviations could reach 4 mag if all observations were considered. The influence of the Sun's altitude on the brightness of the starry sky began to play a role for the Sun's altitude $\theta_{Sun}$>-14° and was significant at $\theta_{Sun}$>-7°. When considering FTS observations with the Moon below the horizon, the deviations $s_{br}$ of the brightness of the starry sky could reach -3 mag. at -8°<$\theta_{Sun}$<-7°, $s_{br}$>-1 mag. at $\theta_{Sun}$<-8°, and $s_{br}$>-0.4 mag. at $\theta_{Sun}$<-14°.

***3. Observations that increase the efficiency of exoplanet detection***

The algorithm we developed shows which microlensing events are best observed at given times using a given telescope in order to increase the efficiency of exoplanet detection during observations. To estimate the comparative efficiency of exoplanet detection, for the "best" events (i.e., for the events selected for observation at the current time), we considered the value $w_{sum}=\sum g_i[(\Delta t+t_{done})^{1/2}-t_{done}^{1/2}]$ (where $\Delta t=2t_{slew}$ for the event observed at the current moment in time, and $\Delta t=t_{slew}$ and $t_{done}$=0 for other events, $t_{slew}$ is the time of rotation of the telescope from one target to another, $t_{done}$ is the elapsed exposure time, the coefficients $g_i$ are discussed in [535] when describing the algorithm and the detection region of the $i$ -th event). An example of the algorithm's operation is shown in Fig. 1. Color figures 1-2 can be viewed on the website with this book or in the article [535].

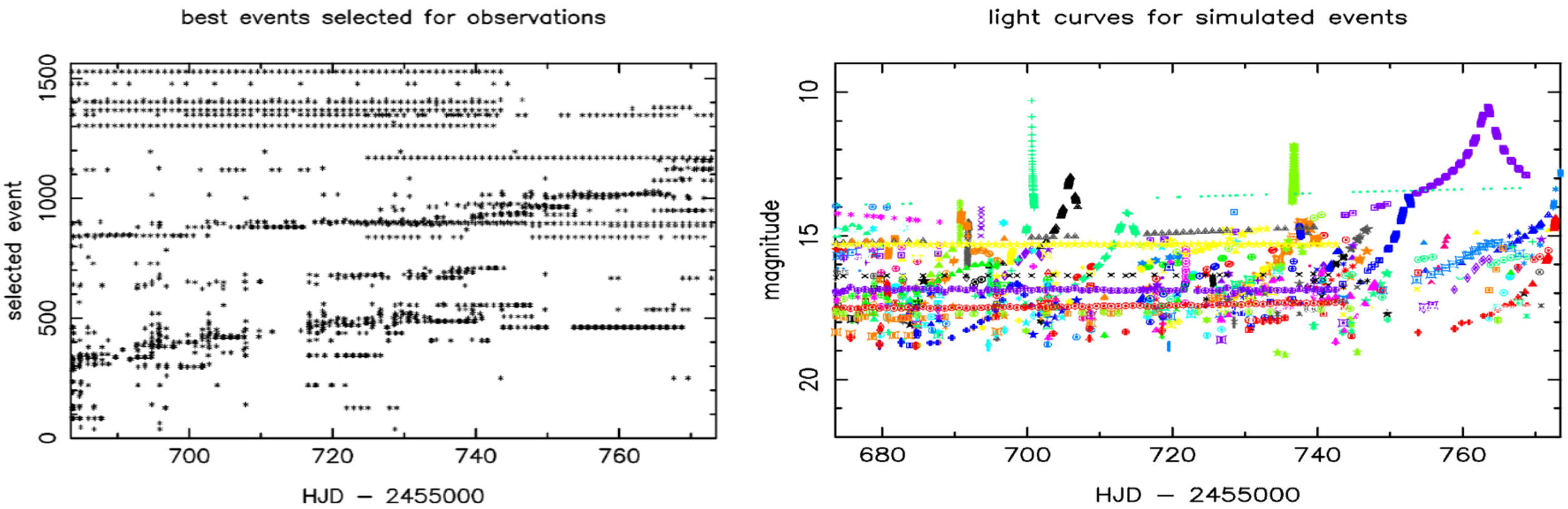

(a) (b)

**Figure 1. (a) Time intervals for events selected by our algorithm for** OGLE observations. Considered microlensing events: 1110001-111562. (b) Light curves for microlensing events selected for OGLE observations. Considered events: 110001-111562.

For 13 telescopes, the values of $r_{wsumt}=(w_{sum}/w_{sumOGLE})/(t_{sum}/t_{sumOGLE})$ were calculated and compared, where $t_{sum}$ is the total time over the considered time interval during which at least one event can be observed ($t_{sumOGLE}$ is the value for OGLE). The obtained results show (Fig. 2) that for the search for exoplanets by the microlensing method based on already known microlensing events, the comparative efficiency $r_{wsumt}$ of detecting exoplanets per unit time for 2-meter telescopes of the RoboNet network (FTS, FTN or LT telescopes) is greater (usually by a factor of 1.4-2.2) than for OGLE. For SBIG CCD camera, the $w_{sum}$ values were ~1.2 times smaller than for Sinistro. For the 1-m telescope with the Sinistro CCD camera, this efficiency/probability was often approximately 0.8 of the OGLE value, but in some calculations it was greater than for OGLE. The above estimates of the telescope efficiency comparison for exoplanet searches are given for the case in which each telescope is pointed at the microlensing event for which the calculated detection efficiency is the maximum for the events under consideration. Telescopes with a wider field of view, such as OGLE, are more effective at searching for new microlensing events. The probability of detecting an exoplanet is usually proportional to the telescope mirror diameter.

Our analysis shows that when observing with multiple telescopes located at the same observatory, it is better to observe different microlensing events with different telescopes for most of the observing time, but at times corresponding to the peak of the event's light curve, it is often better to observe the same event using all telescopes. For the selection of events among 1562 OGLE events (numbered 110001-111562) and a 90-day (or longer) observing period, a significant part (often more than 50%) of the contribution to the exoplanet detection efficiency (in $w_{sum}$) came from short time intervals corresponding to the peaks of the light curves. The relative efficiency of a telescope compared to OGLE can differ by a factor of up to 3, depending on the presence of peaks in the light curves in the considered time interval. Our algorithm can be used to schedule various observations (and not just microlensing event observations).

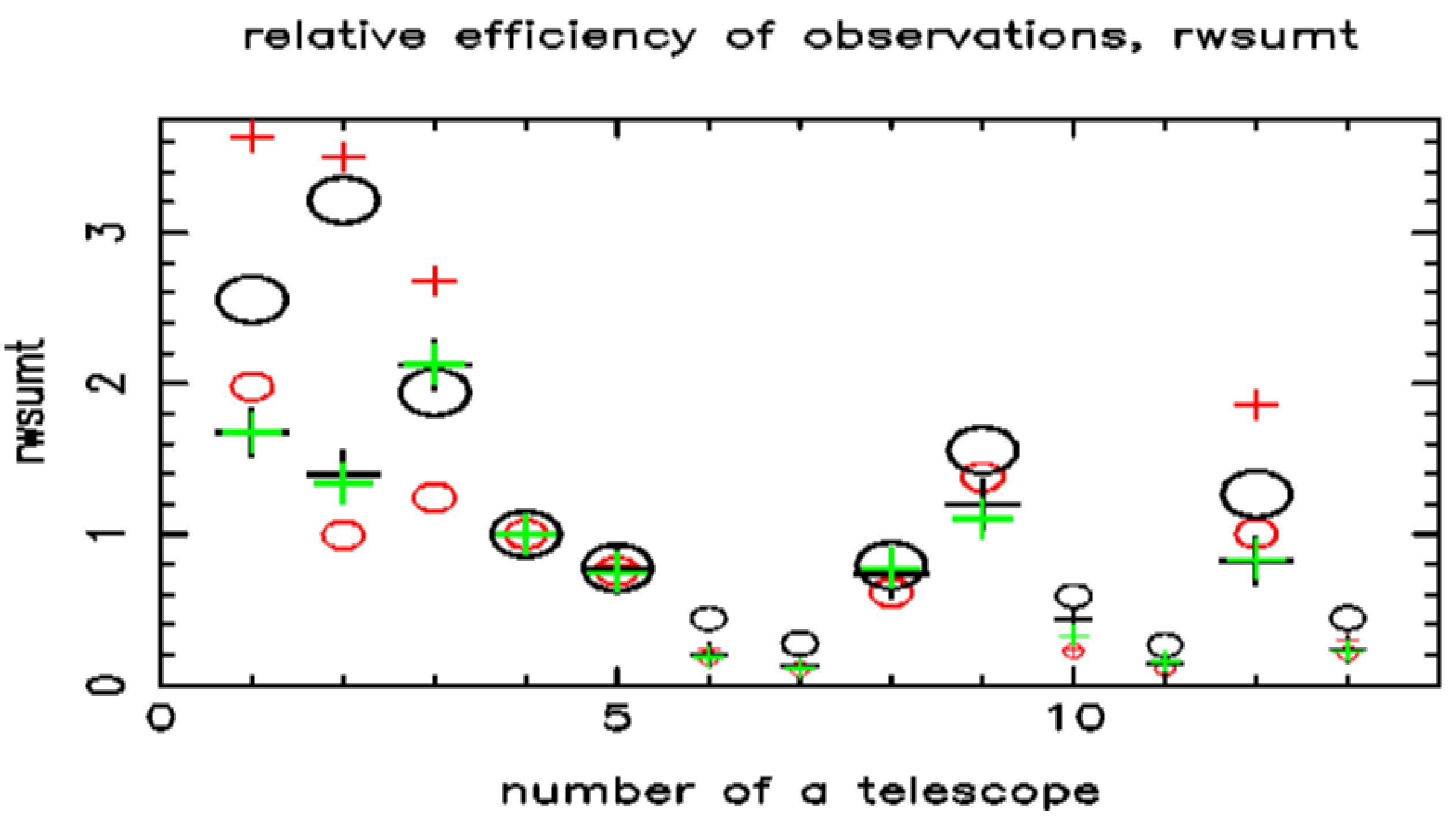

**Figure 2. Relative efficiency** $r_{wsumt}=(w_{sum}/w_{sumOGLE})/(t_{sum}/t_{sumOGLE})$ **of the exoplanet detection probability (in the case of 1562 microlensing events available for observation) as a function of the telescope number** in the case when 1 m telescopes (with the Sinistro CCD camera) located at the same location observe different events at the same time. Black or red crosses and ellipses are given for a 100-day interval starting on April 22 and August 1, respectively. The symbols for calculations with real values of $t_0$ ($t_0$ is the moment of time corresponding to the peak of the light curve) and with random values of $t_0$ ($t_0=R_{NDM}\cdot(t_{mx}+2t_E)-t_E$, where $t_E$ is the time equal to the ratio of the Einstein angular radius to the corresponding relative velocity, $R_{NDM}$ is a random value from 0 to 1, $t_{mx}$ is the duration of the time interval under consideration, $t_0$ is the beginning of the interval) are large black and small red, respectively. Green crosses are given for real values of $t_0$ and the 90-day interval starting on April 22, 2011. Small symbols are given for non-priority telescopes (not entitled to observe the same events at a particular time as higher-priority telescopes from the same observatory). With the random choice of $t_0$ considered, the number of peaks in the light curves was greater.

The ratio $r_{50}$ of the value of $w_{sum}$ (characterizing the efficiency of exoplanet detection) in the case of observations of only events with the maximum increase in event brightness $A_{max}>50$ (4% of all events) to the value of $w_{sum}$ in the case of the possibility of observing all (at any values of $A_{max}$) 1562 events may differ by an order of magnitude for the different time intervals under consideration. For example, for observations with the OGLE telescope $r_{50}$ is 0.07 and 0.99 for 5-day intervals starting on April 22 and August 1, 2011, respectively. For 100-days intervals, $r_{50}$ is 0.63 and 0.83, respectively. For $50<A_{max}<200$ (2% of all events), the corresponding $r_{50\text{-}200}$ values were 0.056, 0.78, 0.09, and 0.50, respectively. For the 8 telescopes considered and the 100-days time interval, the values of $r_{50}$ and $r_{50\text{-}200}$ were in the range of 0.58–0.74 and 0.06–0.21 for the time interval starting on April 22, and in the range of 0.73–0.88 and 0.29–0.52 for the interval starting on August 1, 2011. The values of $w_{sum}$ obtained in our algorithm are close to the values obtained when selecting events for observations according to [Dominik et al. Astron. Nachr, 2010, v. 331, No. 7, 671-691].

## 22. Migration of bodies and dust in the Proxima Centauri exoplanetary system

Star Proxima Centauri is the third stellar component (c) in the Alpha Centauri system. This red dwarf has a mass of 0.12 solar masses. The planet Proxima Centauri *b* has been discovered orbiting this star. This planet ($a_b$=0.04857 AU, $e_b$=0.11, $m_b$=1.17$m_E$) has a mass close to that of the Earth. With a much lower luminosity of the star (than the Sun), this planet is in the habitable zone, which has a radial extension from ~0.042 to ~0.082 AU. Other planets, Proxima Centauri *c* and *d* are outside this zone. Planet *c* ($a_c$=1.489 AU, $e_c$=0.04, $m_c$=7$m_E$) is beyond the ice line.

In the papers [60-62], as well as in [119-120, 123, 195, 204, 207, 213, 489, 504, 507, 508, 510-511, 517, 522, 530, 531], the results of calculations of the migration of planetesimals from the feeding zone of

an exoplanet *c,* including to the inner exoplanet *b* ($a_b$=0.04857 AU, $e_b$=0.11, $m_b$=1.17$m_E$), which may be located in the habitable zone, were presented. Calculations were also carried out for a smaller mass of the planet *c*. The symplectic integrator SWIFT was used for the integration. Two series of calculations were carried out **for the *MP model***. In the **first series** of calculations, the following initial values of the semi-major axes of the orbits and masses of the two exoplanets were adopted: $a_b$=0.0485 AU, $a_c$=1.489 AU, $m_b$=1.27$m_E$ and $m_c$=12$m_E$. The initial eccentricity $e_b$ and inclination $i_b$ of the orbit of exoplanet *b* were considered to be equal to 0, and for exoplanet *c* eccentricity $e_c$ was equal to 0 or 0.1 and $i_c$ =$e_c$/2 rad. In the **second series** of calculations, taking into account more recent observational data, the values $a_b$=0.04857 AU, $e_b$=0.11, $m_b$=1.17$m_E$, $a_c$=1.489 AU, $e_c$=0.04, $m_c$=7$m_E$ and $i_b$=$i_c$=0 were adopted. In both series of calculations, the densities of exoplanets *b* and *c* were considered to be equal to the densities of Earth and Uranus, respectively. In each variant of the calculations, the initial values of the semi-major axes of the orbits of 250 planetesimals were in the range from $a_{min}$ to $a_{min}$ +0.1 AU, where $a_{min}$ varied from 0.9 to 2.2 AU with a step of 0.1 AU. The initial eccentricities $e_o$ of the planetesimals' orbits were equal to 0 or 0.15 for the first series of calculations and $e_o$=0.02 or $e_o$ =0.15 for the second series of calculations, and the initial inclinations of their orbits were equal to $e_o$/2 rad.

For the *MP* model planetesimals and exoplanets were considered as point masses and their collisions were not modeled. The resulting arrays of planetesimal orbital elements with a step of 100 years were used to calculate the probabilities of their collisions with exoplanets. The collision probabilities were calculated using a method similar to that used in [55], but taking into account the corresponding masses of the exoplanets and the star. In the case where the collision probability reached 1 (such cases were not obtained in the calculations for the Solar System), this planetesimal was no longer taken into account when calculating the total probability of collisions with an exoplanet. The probability $p_d$ of a collision of a planetesimal migrating from the feeding zone of exoplanet *c* with exoplanet *d* ($a_d$=0.02895 AU, $m_d$=0.29$m_E$, $e_d$=$i_d$=0) was also calculated, although this exoplanet was not considered when integrating the equations of motion. Only one of several hundred planetesimals has ever had its orbit crossed that of an exoplanet *b*, however, such a planetesimal has collided with this exoplanet quite frequently. When considering thousands of planetesimals, the value of $p_b$ was geater than the probability of a collision with the Earth of planetesimals from the feeding zone of giant planets in the Solar System.

In the second series of *MP* calculations (with 250 initial planetesimals in each variant) at $e_o$=0.02 the total number of planetesimals was 4500 and only in 5 of 16 variants non-zero probabilities of planetesimal collisions with exoplanet *b* were obtained. In two variants $p_b$=0.004 was obtained, while the average value of $p_b$ for one of the 4500 planetesimals was 4.7×10$^{-4}$, but among them there were two planetesimals with $p_b$=1. Out of 16 variants of the second series of *MP* calculations there were 4 cases with non-zero value of $p_d$. On average, for one of the 4500 planetesimals the probability of collision with exoplanet *d* was $p_d$=2.7×10$^{-4}$, but among them there was one planetesimal with $p_d$=1. At $e_o$=0.15, only in 3 of the 6 variants of the second series of *MP* calculations with 250 initial planetesimals (with a total number of 1500 planetesimals) non-zero values of probability $p_b$ were obtained. For planet *b,* the average value of probability for 1500 planetesimals turned out to be $p_b$=2.0×10$^{-3}$ and $p_b$=1 for three of them, and for planet *d,* the average value $p_d$=2.0×10$^{-3}$ and also for three planetesimals $p_d$=1. In the first series of *MP* calculations at $i_c$=$e_c$=0 and $e_o$=0.15, the values of the probability $p_c$ of a collision of one planetesimal, initially located in the vicinity of an exoplanet *c*, with this exoplanet were of the order of $p_c$=0.06-0.1. At $i_c$=$e_c$/2=0.05 and $e_o$=0.15, it was obtained that $p_c$=0.02-0.04. In the second series of calculations *MP,* the value of $p_c$ was mostly in the range from 0.1 to 0.3, except for the variants with $a_{min}$=1.4 AU and $e_o$=0.02, in which $p_c$=0.4-0.8. After 20 million years, the increase in $p_c$ was usually small, since by this time few planetesimals remained in elliptical orbits.

Calculations for the main **model *C***, in which planetesimals that collided with the planet were excluded from further integration, were performed for initial data similar to those for the second series of *MP* calculations. In these calculations, $a_{min}$ varied from 0.9 to 2.2 AU, and the remaining initial data were the same as for the second series of *MP* calculations. Calculations were also performed for a model in which the mass of the planet *c* was 10 times smaller than the present-day mass of this planet. For both models (*C* and *MP*), bodies that collided with the star or reached the boundary of the star's Hill sphere (1200 AU) were also excluded from integration. The time interval considered was typically equal to several hundred million years for model *C* and was at least 50 million years for model *MP*.

For model *C* the probability $p_b$ of a collision of a planetesimal from the planet's feeding zone with with planet *b* was approximately two times smaller, and the values of $p_c$ were, on average, almost 2 times greater than for the *MP* model. The probability of $p_b$ for model *C* was estimated to be 2.0×10$^{-4}$ at $e_o$=0.02 and 10$^{-3}$ at $e_o$=0.15. The total mass of planetesimals delivered from the feeding zone of planet *c* to planet *b,* is equal to $m_{c\text{-}b}=p_b \cdot m_{fzc}$, and the mass of water in these planetesimals equalrd to $m_{ice}=p_b \cdot k_{ice} \cdot m_{fzc}$, where $m_{fzc}$ is the total mass of planetesimals beyond the ice line that fell into the feeding zone of the exoplanet Proxima Centauri *c*, and $k_{ice}$ is the fraction of water in the planetesimals. An example of the evolution of the orbit of a planetesimal that collided with planet *b* is shown in Fig. 1.

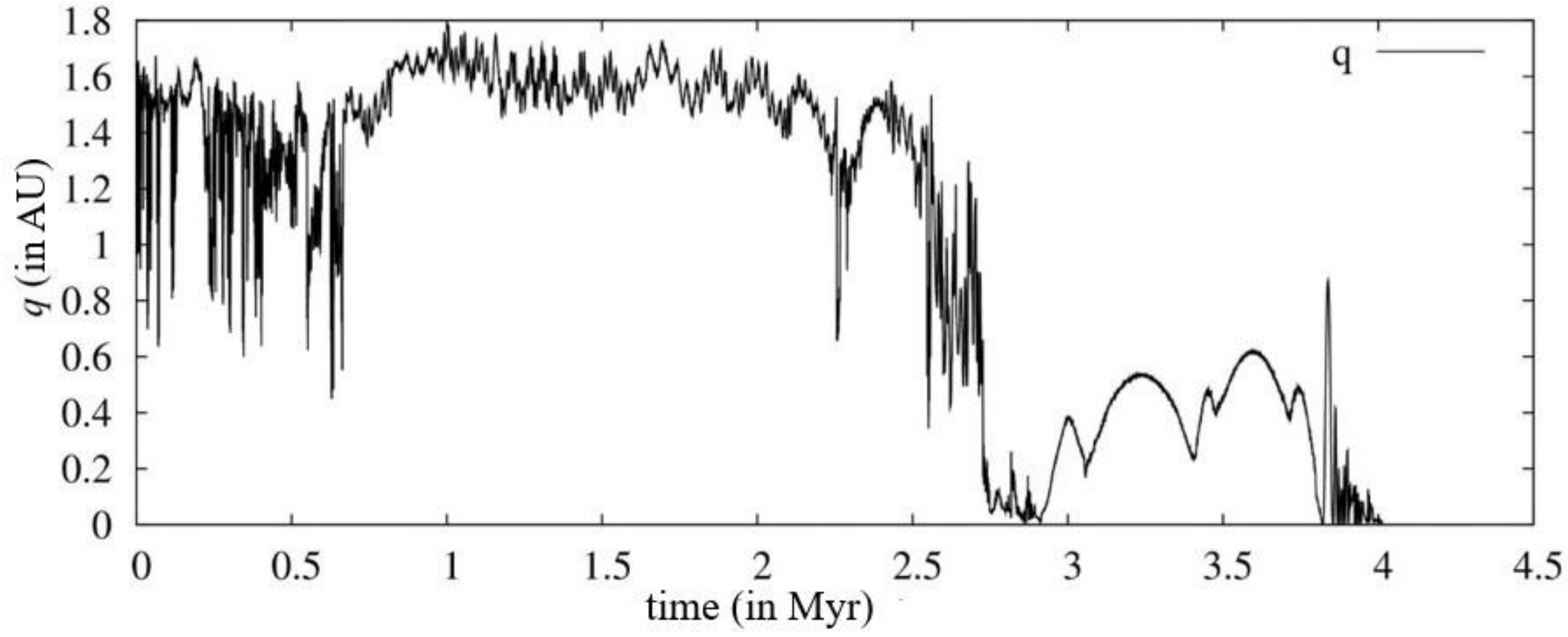

**Figure 1.** Plot of the variation with time (in million years) of the perihelion distance ($q$ in AU) of the orbit of planetesimals initially located near the orbit of the planet Proxima Centauri *c*, before its collision with the planet Proxima Centauri *b*.

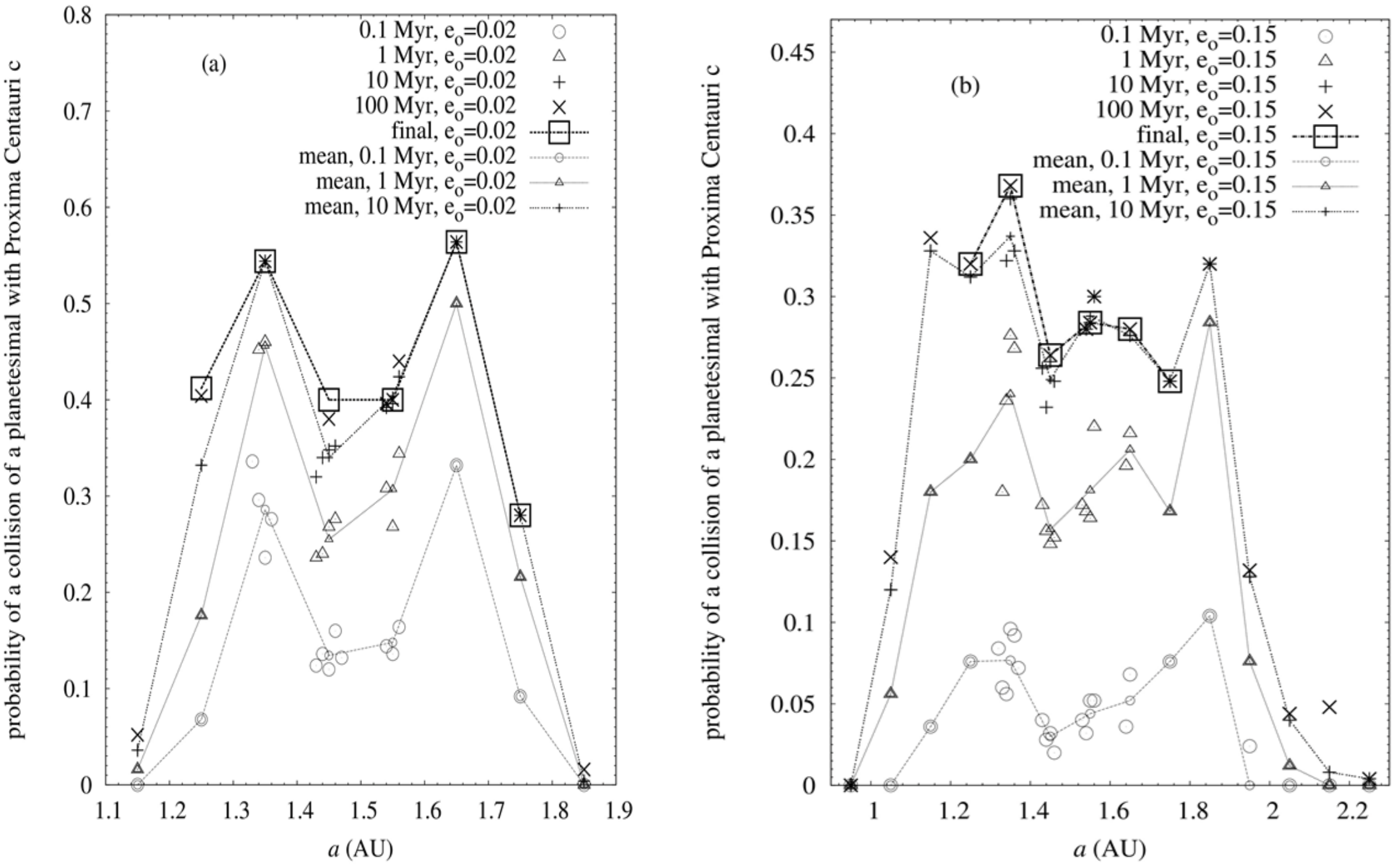

**Figure 2.** The probability $p_c$ of a collision of a planetesimal with the planet Proxima Centauri *c* for $a=a_{min}+0.05+a_{ts}$, where $a_{ts}$ is equal to 0, -0.03, -0.02, -0.01, and 0.01 AU with an integration step $t_s$ equal to 1, 0.1, 0.2, 0.5, and 2 Earth days, respectively (the values of $a_{ts}$ are used only to show results for variants which differ only in an integration step). The probability values are given for the entire time interval considered, as well as for 0.1, 1, 10, and 100 million years. The initial eccentricities $e_o$ of the planetesimal orbits were equal to 0.02 in Fig. (a) and 0.15 in Fig. (b).

At $e_o$=0.02, the final values of $p_c$ of the probability of planetesimal collisions with a planet *c* were maximum (about 0.55) at $a_{min}$=1.3 AU and $a_{min}$=1.6 AU and exceeded 0.4 at $1.2 \le a_{min} \le 1.6$ AU (Fig. 2). At $e_o$=0.15, the final values of $p_c$ were usually smaller than at $e_o$=0.02 and reached a maximum (about 0.37) at $a_{min}$=1.3 AU and exceeded 0.25 at $1.2 \le a_{min} \le 1.8$ AU. The ratio $p_{cej}=p_c/p_{ej}$ of the probability of a collision of a planetesimal with a planet $c$ to the probability $p_{ej}$ of ejection of a planetesimal into a hyperbolic orbit at $e_o$=0.02 and $e_o$=0.15 was in the ranges of 0.8-1.3 and 0.4-0.6, respectively, in calculations for 1.2<$a_o$<1.8 AU and the current mass of the planet $c$ (Fig. 3). This ratio was in the intervals of 1.3-1.5 and 0.5-0.6 for the mass of the planet $c$, which was a half of its present mass.

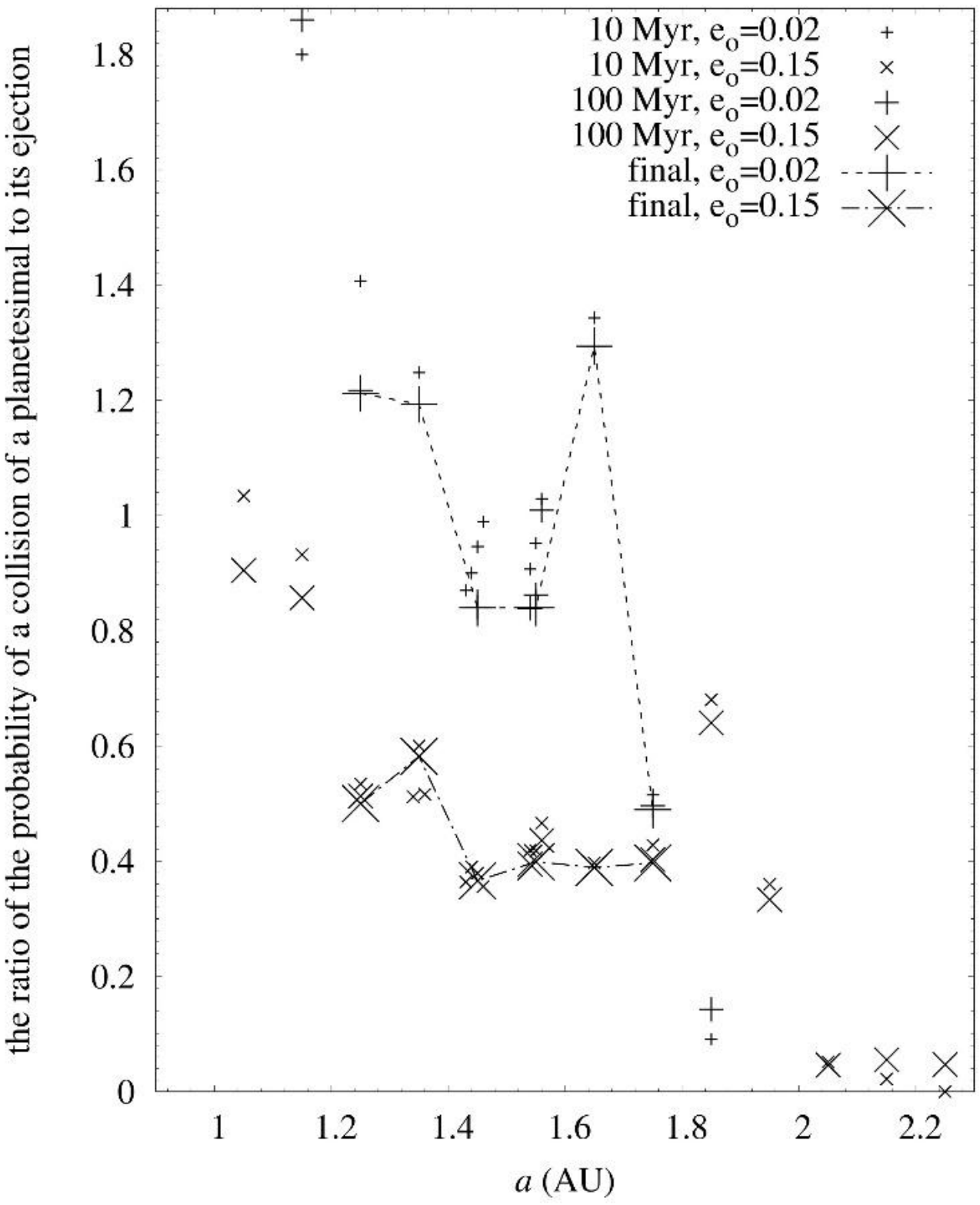

**Figure 3.** Graph of the ratio of the probability of a planetesimal collision with a planet to the probability of its ejection into a hyperbolic orbit depending on the initial value of $a_o$ of the semi-major axis of its orbit in the vicinity of the orbit of the planet Proxima Centauri $c$ for the entire calculation period (usually after 1 billion years), as well as for 10 and 100 million years.

The total mass of planetesimals ejected into hyperbolic orbits exceeded the mass of planetesimals that entered into the planets. Based on the obtained values of $p_{ej}$ from the law of conservation of energy, it can be estimated that the semimajor axis of the orbit of planet $c$ during its accumulation could have decreased by at least one and a half times. The total mass of planetesimals ejected into hyperbolic orbits by planet $c$ could have been about (3.5-7)$m_E$. Taking into account the mass of planet $c$ (equal to 7$m_E$), we obtain that the mass $m_{fzc}$ planetesimals in the feeding zone of planet $c$ could be no less than 10$m_E$ and 15$m_E$ at $e_o$=0.02 and $e_o$=0.15 respectively.

Calculations showed that planetesimals could have collided with planet $b$ even when the mass of the embryo of planet $c$ was ten times smaller than the current mass of planet $c$. With $m_{fzc}$=10$m_E$ and $p_b$=2 $10^{-4}$ we obtain the minimum estimate $m_{c\text{-}b}$=$2\cdot10^{-3}m_E$ (corresponding to $e_o$=0.02) of the total mass of planetesimals delivered to planet $b$ from the feeding zone of planet $c$. For $e_o$=0.15, $m_{fzc}$=15$m_E$ and $p_b$=$10^{-3}$ this estimate is equal to $m_{c\text{-}b}$=$1.5\cdot10^{-2}m_E$. Large values of $e_o$ correspond to the growth of the orbital eccentricities of the planetesimals due to their mutual gravitational influence. With greater planetesimal masses, the growth of the orbital eccentricities of the planetesimals could be greater. The estimates of $m_{c\text{-}b}$ made above for the current mass of planet $b$ could be smaller if, during the bombardment under consideration, the mass of planet $b$ was less than its current mass. However, it can be assumed that planet $b$ formed faster than planet $c$, since planet $b$ is much closer to the star than planet $c$. The amount of matter delivered to planet $d$ could be slightly less than to planet $b$. The probability of a planetesimal collision with planet $b$, averaged over all planetesimals considered, was ~$10^{-4}$–$10^{-3}$. The total mass of planetesimals from the feeding zone of planet $c$ delivered to planet $b$ was estimated to be 0.002$m_E$ and 0.015$m_E$ for initial planetesimal orbital eccentricities of $e_o$=0.15 and $e_o$=0.02, respectively. We found that the collision probabilities of planetesimals from the feeding zone of the giant planets with the Earth are of the order of $10^{-6}$–$10^{-5}$, i.e., significantly less than the values of $p_b$ and $p_d$. Therefore, the influx of icy planetesimals to exoplanets Proxima Centauri *b* and *d* could be greater than a similar influx to the Earth. Some of the material from a planetesimal that collided with an exoplanet was ejected from the exoplanet. Summarizing data from a number of publications, it can be assumed that the fraction of water in planetesimals in the feeding zone of the planet Proxima Centauri *c* could have been between 10 and 50%. The mass of water delivered to the exoplanet Proxima Centauri *b* could have exceeded the mass of Earth's oceans.

In [60], estimates are given of the size **of the feeding zone of the planet** Proxima Centauri *c* for initial orbital eccentricities of planetesimals equal to 0.02 or 0.15. The studies are based on the results of modeling the evolution of the orbits of planetesimals under the influence of the star and the planets Proxima Centauri *c* and *b*. The time interval under consideration reached a billion years. If we consider the initial values of the product $a\ e$, which characterizes the changes in the distance from the star during the motion of the planet Proxima Centauri *c* and the planetesimal (with an initial eccentricity $e_o$), (for example, $e_c \cdot a_c$ and $e_o \cdot a_{min002}$), then we have $a_c - a_{min002} = 0.04a_c + 0.02a_{min002} + 2.54a_c \cdot \mu^{1/3}$, $a_{max002} - a_c = 0.04a_c + 0.02a_{max002} + 2.40a_c \cdot \mu^{1/3}$, $a_c - a_{min015} = 0.04a_c + 0.15a_{min015} + 2.23a_c \cdot \mu^{1/3}$, and $a_{max015} - a_c = 0.04a_c + 0.15a_{max015} + 4.3a_c \cdot \mu^{1/3}$, where $a_c$ and $e_c=0.04$ are the semi-major axis and the eccentricity of the orbit of planet *c*, $\mu$ is the ratio of the mass of planet *c* to the mass of the star, $a_{min002}$, $a_{max002}$, $a_{min015}$ and $a_{max015}$ are the minimum and maximum initial values of the semi-major axes of the considered orbits of planetesimals for the feeding zone of planet *c* at initial orbital eccentricities of planetesimals equal to 0.02 and 0.15, respectively. The coefficients in front of $a_c\mu^{1/3}$ in the three formulas above are approximately 2.2–2.5, i.e., close to the coefficients in $\gamma=(2.1–2.45)\mu^{1/3}$ for circular initial orbits.

After the accumulation of Proxima Centauri *c,* some planetesimals could have continued to move in stable elliptical orbits within its feeding zone, which was mainly cleared of planetesimals (Fig. 4). Such planetesimals typically move in certain **resonances** with planet *c*, such as 1:1 (like Jupiter's Trojans), 5:4, and 3:4, and typically have small eccentricities (Fig. 5). Some planetesimals, which had been moving in chaotic orbits for a long time (1-2 million years), fell into 5:2 and 3:10 resonances with planet *c* and moved in these resonances for at least tens of millions of years.

The motion of planetesimals initially located in the feeding zone of the planet Proxima Centauri *c* was considered, **at distances from the star from 500 AU up to the radius of the star's Hill sphere**, equal to 1200 AU [62]. In the gasless model considered, the main ejection of planetesimals from most of the feeding zone of a nearly formed planet *c* to orbits with apocentric distances $a(1+e)$ greater than 500 AU occurred during the first 10 million years. Only for planetesimals initially located at the edges of the planet's feeding zone, the fraction of planetesimals that reached 500 AU over a time greater than 10 million years was more than a half. Some planetesimals could reach the outer part of the star's Hill sphere even after hundreds of millions of years. About 90% of the planetesimals that reached 500 AU from Proxima Centauri reached 1200 AU from the star in less than 1 million years, with the current mass of the planet *c*. Moreover, no more than 2% of planetesimals with apocentric distances of their orbits between 500 and 1200 AU moved in such orbits for more than 10 million years (but less than several tens of millions of years). For a planet's mass equal to a half of the mass of planet *c*, the fraction of planetesimals that increased the apocentric distances of their orbits from 500 to 1200 AU in less than 1 million years was about 70-80%. For the current mass of planet *c*, among planetesimals that reached 500 AU from the star, the fraction of planetesimals with orbital eccentricities greater than 1 was 0.05 and 0.1, at initial orbital eccentricities of $e_o=0.02$ and $e_o=0.15$, respectively. Among planetesimals that reached 1200 AU, from the star, this fraction was about 0.3 for both values of $e_o$. The minimum values of the orbital eccentricity of planetesimals that reached 500 and 1200 AU from the star were 0.992 and 0.995, respectively.

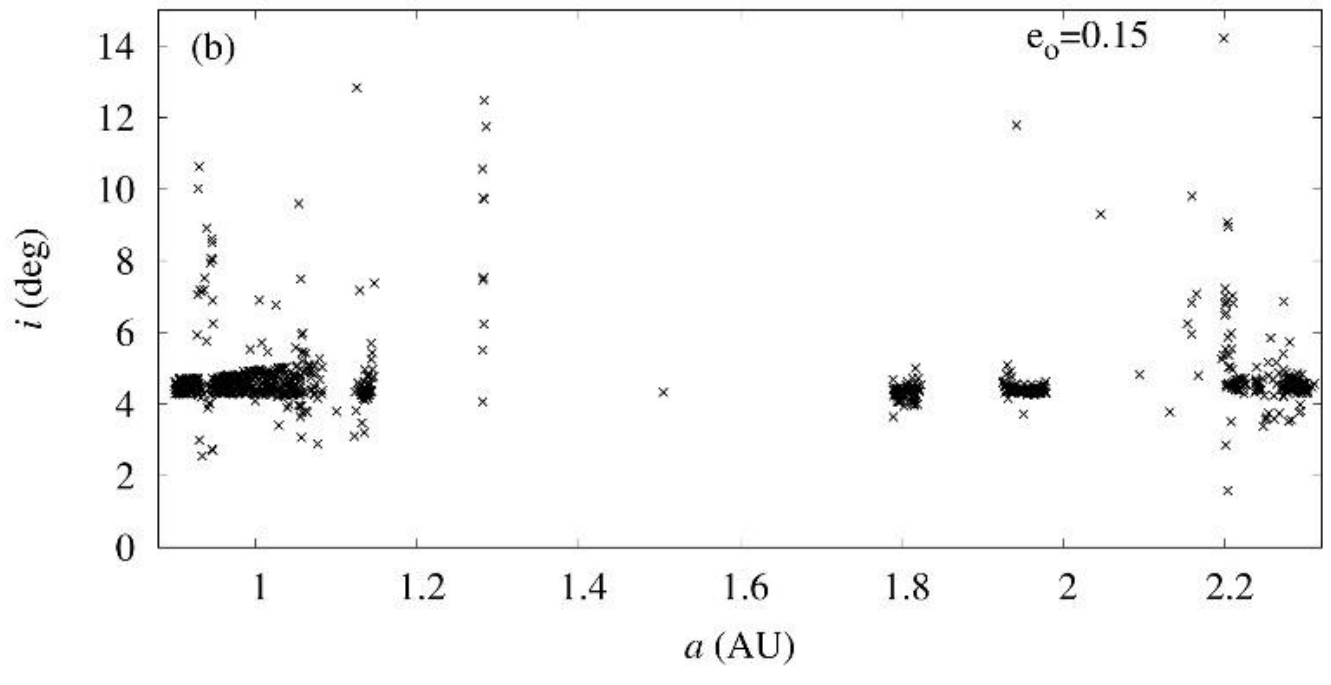

(d)

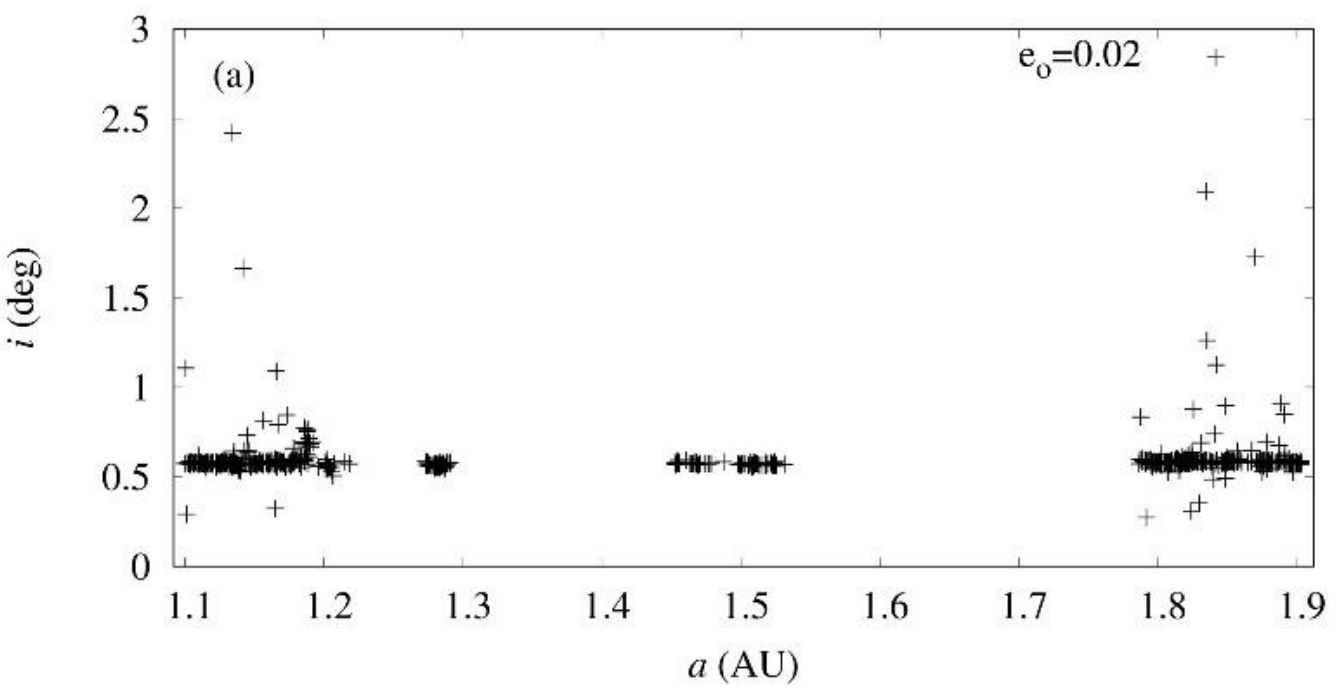

(c)

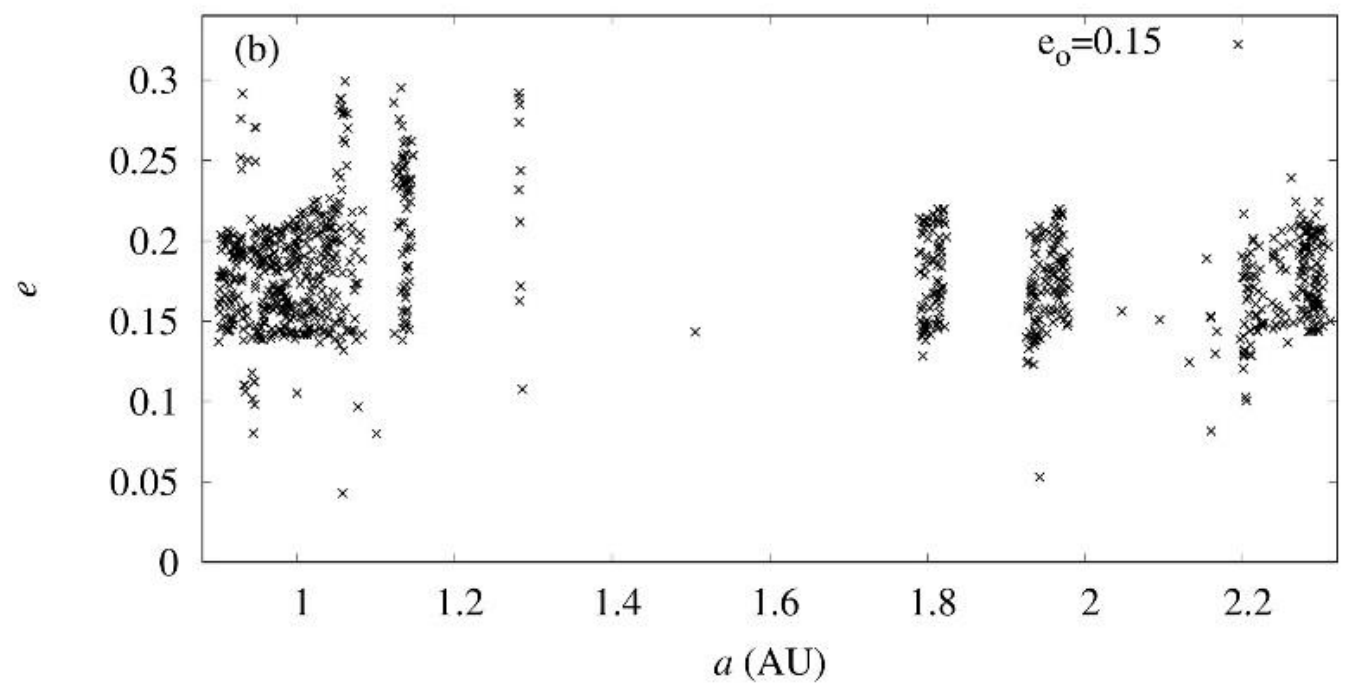

(b)

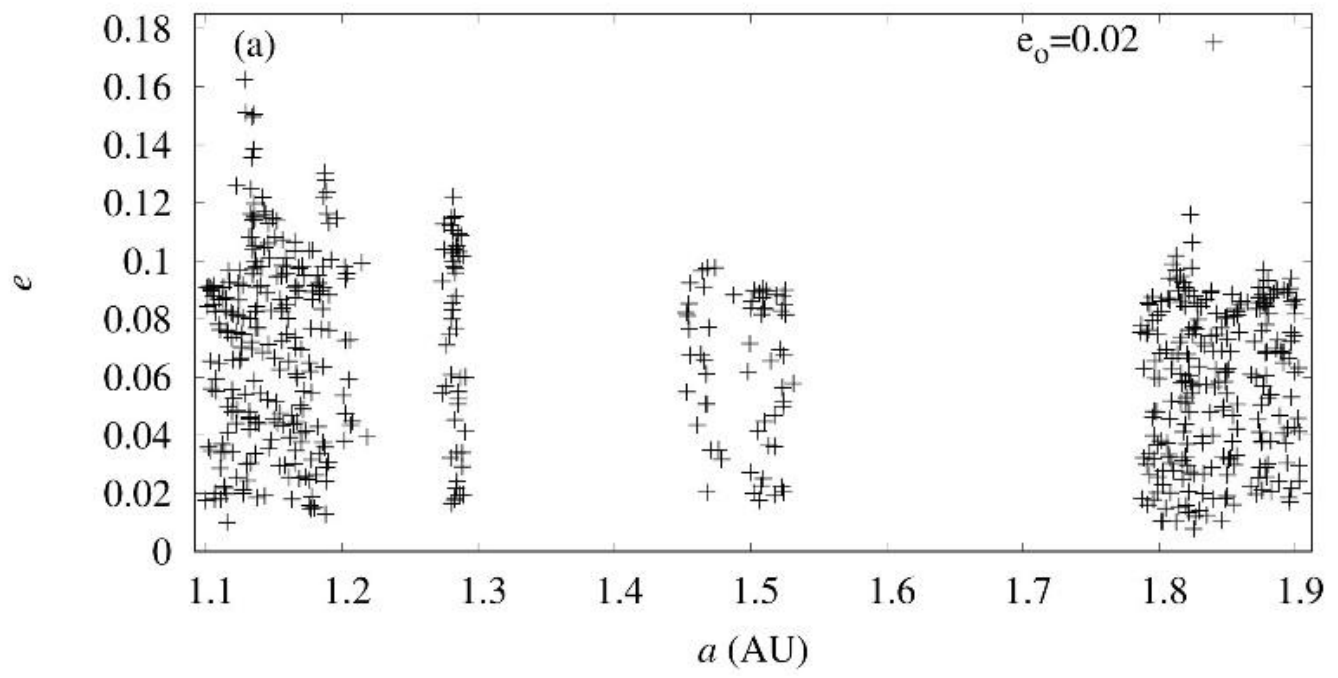

(a)

**Figure 4.** Plots of the distribution of planetesimals by semi-major axes (in AU) and their orbital eccentricities after 1 billion years for the initial orbital eccentricities of planetesimals equal to 0.02 (a) and 0.15 (b), as well as the distribution of planetesimals by semi-major axes and inclinations (in degrees) of their orbits for $e_o$=0.02 (c) and $e_o$=0.15 (d). The crosses correspond to the different remaining planetesimals. The initial values of the semi-major axes of the planetesimals' orbits varied from 0.9 to 2.3 AU and were located in the feeding zone of the planet Proxima Centauri *c*. Although the feeding zone of this planet was mostly cleared of planetesimals, individual bodies continued to move in it along resonant orbits even after 1 billion years.

For the considered model, disk of planetesimals in the outer part of the star's Hill sphere was quite flat. The orbital inclinations *i* of more than 80% of planetesimals that reached 500 or 1200 AU from the star did not exceed 10°. For the current mass of planet *c,* on average, across all calculation variants, the fraction of such planetesimals with $i$>20° did not exceed 1%. The obtained results may be of interest for understanding the motion of bodies in some other exoplanetary systems, especially in systems with a single dominant planet. They can be used to specify the initial data for models of the evolution of the disk of bodies in the outer part of the Hill sphere of Proxima Centauri, which take into account gravitational interactions and collisions of bodies with each other, as well as the influence of other stars. Highly inclined orbits of bodies in the outer part of the Hill sphere of Proxima Centauri can only be obtained mainly due to bodies entering the Hill sphere from outside. The radius of Proxima Centauri's Hill sphere is an order of magnitude smaller than the radius of the outer boundary of the Solar System's Hills cloud and two orders of magnitude smaller than the radius of the Sun's Hill sphere. The Hills cloud is the inner (flattened) portion of the Oort cloud (from 1,000–5,000 to 20,000 AU). Therefore, it is unlikely that Proxima Centauri would have an Oort cloud analog as massive as the Sun's.

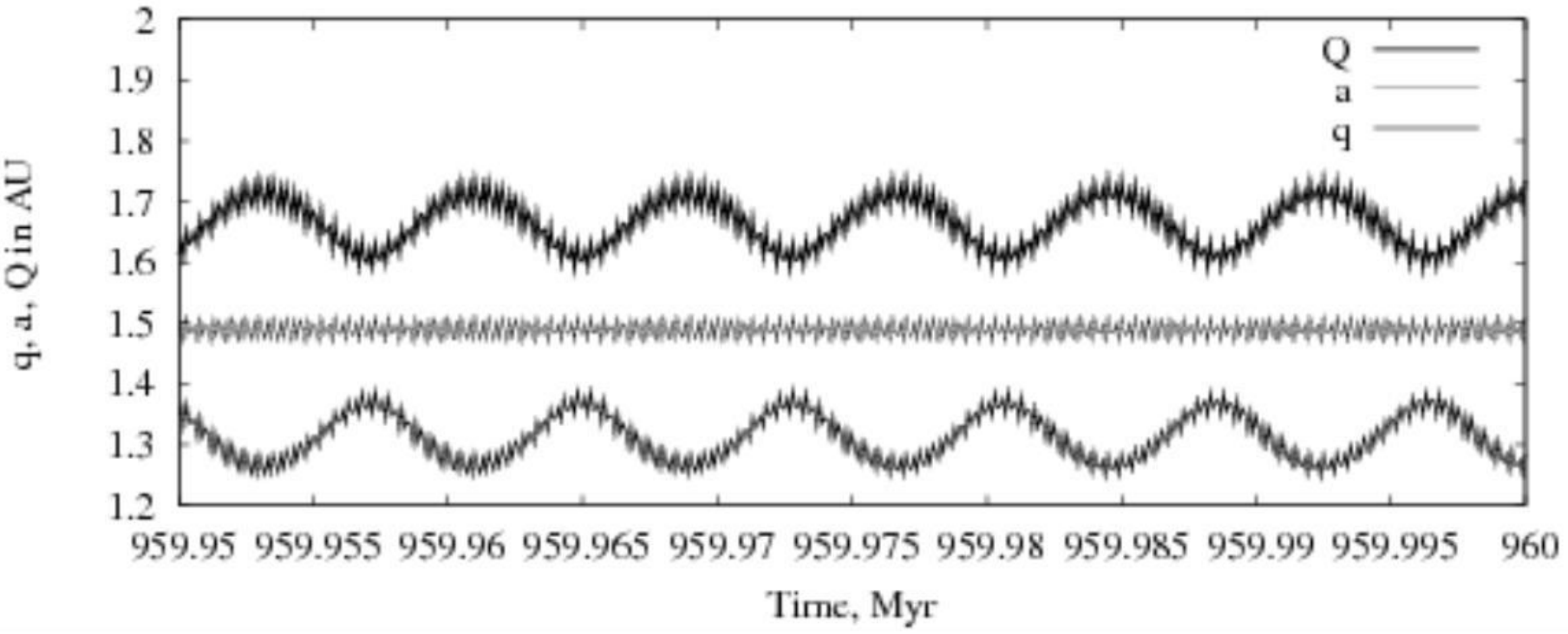

**Figure 5.** Changes in the semi-major axis *a*, perihelion and aphelion distances *q* and *Q* of the planetesimal's orbit with time (in million years) at $a_o$=1.50041 AU and $e_o$=0.15. The planetesimal moved in a 1:1 resonance with the planet Proxima Centauri *c*.

In [119, 520] **the migration of dust particles** whose initial orbits were close to the orbit of the planet Proxima Centauri *c* was considered. The initial eccentricities of the particle orbits $e_o$ were equal to 0.02 or 0.15. The Bulirsh-Stoer method from the SWIFT package was used for integration. The relative error at the integration step did not exceed $10^{-8}$. Dust migration was studied similarly to calculations [30] for the Solar system. The gravitational influence of the star and planets *b* and *c,* the Poynting -Robertson effect, radiation pressure and the pressure of the stellar wind. The ratio of the stellar wind pressure to the pressure due to the Poynting-Robertson effect, as for the Solar System, was considered to be 0.35. In different variants, the ratio β of the radiation pressure force to the gravitational force varied from 0.0002 to 1. For silicate particles in the Solar System, such β values correspond to particle diameters *d* from 2000 to 0.4 microns; and *d* is proportional to 1/β. In the Solar System, silicate particles with 0.001≤β≤0.1 correspond to diameters from 4 to 400 microns. With such β values, dust particles effectively deliver matter (including volatiles) to the planet *b*. If the radiation pressure in the Proxima Centauri system is less than in the Solar system (the mass of this star is an order of magnitude less than the mass of the Sun), then for the same particle sizes the value of β will be smaller, meaning that the same β will correspond to larger particles in the Proxima Centauri system. In the Solar System, for silicate particles with a density of 2.5 g/cm$^3$, β values of 0.001, 0.01, and 0.1 correspond to particle diameters of 500, 50, and 5 microns, respectively. At a density of 1 g/cm$^3$, for the same β values, the diameters are 2.5 times larger.

The probabilities of collisions of particles with planet *b* at $e_o$=0.02 were about 0.15-0.2, 0.1, 0.06-0.08, and 0.016-0.03 for 0.001≤β≤0.004, β=0.01, β=0.02, and 0.04≤β≤0.1, respectively. For $e_o$=0.15, the particle collision probabilities with planet *b* were about 0.07-0.15, 0.04, and 0.01-0.03 for 0.001≤β≤0.01, β=0.02, and 0.04≤β≤0.1, respectively. For $e_o$=0.02, the particle collision probabilities with planet *c* were about 0.016-0.05, 0.02, and 0.01-0.02 for 0.001≤β≤0.004, β=0.01, and 0.02≤β≤0.1, respectively. For $e_o$=0.15, the particle collision probabilities with planet *c* did not exceed 0.03 for all considered variants. Although the initial orbits of the dust particles were close in the orbit of planet *c* and the mass of planet *c* was greater than the mass of planet *b*, for 0.001≤β≤0.1 (in the Solar System with particle diameters of the order of 10-100 microns), more particles collided with the inner planet *b* than with planet *c*.

At β≥0.4, most particles were ejected into hyperbolic orbits. At 0.004≤β≤0.2, most particles collided with the star, with the collision probability being highest at β=0.04. The evolution times of the dust disks under consideration were generally shorter at higher β (with smaller particle diameters). They were 300 years at β=1 and several million years at 0.004≤β≤0.04. The probability of a dust grain from the vicinity of the orbit of the planet Proxima Centauri *c* to collide with planet *b* at 0.001≤β≤0.01 was generally around 0.1-0.2, which is significantly higher than the similar probability for planetesimals (~$10^{-4}$-$10^{-3}$). A lot of icy material and volatiles could have been delivered to the inner exoplanets Proxima Centauri *b* and *d* (with semi-major axes of orbits of 0.05 and 0.03 AU).

## 23. Migration of bodies in the TRAPPIST-1 exoplanetary system

Mixing of planetesimals-bodies in the TRAPPIST-1 exoplanetary system was studied in [123, 205, 212, 214, 498, 508, 511, 517, 521, 530, 531] during the late gasless stage of formation of nearly formed planets. A more detailed journal article is planned for publication on this topic. The TRAPPIST-1 exoplanetary system consists of a star with a mass equal to 0.0898 solar masses and 7 planets (from *b* to *h*) with masses from 0.33 to 1.37 Earth masses. The semi-major axes of the planets' orbits range from 0.012 to 0.062 AU. The semi-major axes and eccentricities of the orbits and the masses of the planets are presented in Table. 1. It is believed that planets *e*, *f* and *g* may have liquid water.

The late, gasless stage of formation of the nearly formed planets of the TRAPPIST-1 system was studied. The considered model of mixing of bodies in the TRAPPIST-1 planetary zone can also characterize the migration of bodies ejected from some planets after collisions of these planets with some planetesimals or other bodies. The motion of planetesimals under the gravitational influence of the star and seven TRAPPIST-1 planets (from *b* to *h*) was calculated using a symplectic integrator RMVS3 from the SWIFT package is similar to Model *C* for the Proxima-Centauri system. Planetesimals that collided with planets or the star or were ejected into hyperbolic orbits (reaching 50 AU from the star) were excluded from the integration. In each calculation variant, the initial values of the semi-major axes of the orbits $N_{p0}$=250 or $N_{p0}$=1000 planetesimals varied from $a_{min}$ to $a_{max}$ near the semi-major axis of the orbit of one of the planets, the initial eccentricities of their orbits were equal to $e_o$=0.02 or $e_o$=0.15, and their initial inclinations were equal to $e_o$/2 rad. The values of $a_{min}$ and $a_{max}$ are presented in Table 1. The considered disk of planetesimals was located near the orbit of one of the considered planets. Calculations were also carried out for $a_{min}$=0.07 AU. In these calculations for the "outer" disk, $a_{max}$=0.1 AU for $N_{p0}$=250 and $a_{max}$=0.09 AU for $N_{p0}$=1000. Calculations with a step of 0.01, 0.02 and 0.1 Earth days gave similar results.

**Table 1.** Orbital elements, masses *m* (in Earth mass $m_E$) of exoplanets in the TRAPPIST-1 system and the values of $a_{min}$ and $a_{max}$ for the disks under consideration near planets *b, c, d, e, f, g,* and *h*.

| planets | $m/m_E$ | *a,* AU | *e* | $a_{min}$, AU | $a_{max}$, AU |
|---|---|---|---|---|---|
| *b* | 1.37 | 0.0115 | 0.0062 | 0.0094 | 0.0137 |
| *c* | 1.31 | 0.0158 | 0.0065 | 0.0137 | 0.0190 |
| *d* | 0.39 | 0.0223 | 0.0084 | 0.0190 | 0.0258 |
| *e* | 0.69 | 0.0292 | 0.0051 | 0.0258 | 0.0339 |
| *f* | 1.04 | 0.0385 | 0.0101 | 0.0339 | 0.0427 |
| *g* | 1.32 | 0.0468 | 0.0021 | 0.0427 | 0.0544 |
| *h* | 0.33 | 0.0619 | 0.0057 | 0.0544 | 0.0694 |

In the calculations performed (see Tables 2 and 3), no more than 3.2% of planetesimals were ejected into hyperbolic orbits. There was no ejection of planetesimals into hyperbolic orbits for *b* disks, and for *c* - *d* disks, the fraction of ejected planetesimals did not exceed 0.01. There were no collisions of planetesimals with the parent star in the considered calculation variants. More than a half of the planetesimals from disks initially located near the orbits of planets from *b* to *g* collided with planets in less than 1000 years, and from disks *b* - *d* even in 250 years. The number of collisions of planetesimals with planets *c* - *h* over the entire considered time interval was greater than for 100 thousand years by no more than 4%. Planetesimals with a dynamic lifetime of more than 100 thousand years were usually ejected into hyperbolic orbits or collided with planet *h*. The evolution times of planetesimal disks around planets *b* - *h* varied from 11 thousand years to more than 60 million years at $N_{p0}$=250 and from 33 thousand years to more than 48 million years (in three *h* disks one body remained) at $N_{p0}$=1000.

The fraction of planetesimals that collided with the host planet was 0.36–0.8 at $e_o$=0.02 and 0.22–0.75 at $e_o$=0.15. This fraction (compared to collisions with all planets) generally decreased with increasing time interval under consideration. In each calculation variant, there was at least one planet for which the number of planetesimal collisions exceeded 25% of the number of collisions of planetesimals with the host planet. The fraction of planetesimal collisions with the host planet was usually smaller for disks located further from the star. For the initial disk near the orbit of planet *h* (with mass $m_h$=0.33$m_E$, where $m_E$ is the mass of the Earth) the number of collisions of planetesimals with planet *g* (with mass $m_g$=1.32$m_E$) was approximately the same as with planet *h*.

**Table 2.** The number of collisions of planetesimals with planets (from *b* to *h*) in the TRAPPIST-1 system for different disks (from *b* to *h*). The initial orbital eccentricities of the planetesimals were equal to $e_o$. Each initial disk contained 250 planetesimals. The variants with the integration step $t_s$=0.01 Earth days are marked with bold letters in the left column (the bottom row of two adjacent rows); for other variants $t_s$=0.1 Earth days.

| | $e_o$ | *b* | *c* | *d* | *e* | *f* | *g* | *h* | *thrown out* |
|---|---|---|---|---|---|---|---|---|---|
| *b* | 0.02 | **199** | 4 7 | 2 | 2 | 0 | 0 | 0 | 0 |
| ***b*** | 0.02 | **187** | 54 | 2 | 3 | 2 | 1 | 0 | 0 |
| *b* | 0.15 | **188** | 55 | 6 | 1 | 0 | 0 | 0 | 0 |
| ***b*** | 0.15 | **180** | 63 | 5 | 2 | 0 | 0 | 0 | 0 |
| *c* | 0.02 | 49 | **153** | 32 | 9 | 3 | 4 | 0 | 0 |
| ***c*** | 0.02 | 56 | **148** | 26 | 11 | 3 | 5 | 0 | 0 |
| *c* | 0.15 | 77 | **140** | 13 | 10 | 6 | 2 | 2 | 0 |
| ***c*** | 0.15 | 73 | **141** | 23 | 7 | 0 | 5 | 1 | 0 |
| *d* | 0.02 | 13 | 42 | **120** | 43 | 18 | 11 | 3 | 0 |
| ***d*** | 0.02 | 13 | 47 | **128** | 29 | 16 | 14 | 3 | 0 |
| *d* | 0.15 | 2 8 | 6 9 | **68** | 45 | 16 | 19 | 3 | 2 |
| ***d*** | 0.15 | 30 | 71 | **59** | 43 | 23 | 19 | 5 | 0 |
| *e* | 0.02 | 5 | 20 | 31 | **110** | 38 | 39 | 3 | 3 |
| ***e*** | 0.02 | 7 | 27 | 24 | **105** | 46 | 34 | 4 | 3 |
| *e* | 0.15 | 11 | 38 | 32 | **69** | 53 | 32 | 12 | 3 |
| ***e*** | 0.15 | 7 | 29 | 22 | **72** | 57 | 53 | 3 | 7 |
| *f* | 0.02 | 6 | 11 | 10 | 37 | **98** | 75 | 11 | 2 |
| ***f*** | 0.02 | 4 | 7 | 14 | 43 | **108** | 58 | 14 | 2 |
| *f* | 0.15 | 6 | 2 1 | 26 | 42 | **61** | 77 | 12 | 5 |
| ***f*** | 0.15 | 7 | 20 | 23 | 45 | **57** | 80 | 10 | 8 |
| *g* | 0.02 | 3 | 12 | 12 | 33 | 45 | **110** | 31 | 4 |
| ***g*** | 0.02 | 1 | 11 | 16 | 20 | 48 | **114** | 33 | 7 |
| *g* | 0.15 | 2 | 1 2 | 20 | 3 6 | 62 | **94** | 18 | 6 |
| ***g*** | 0.15 | 5 | 16 | 16 | 26 | 68 | **95** | 22 | 2 |
| *h* | 0.02 | 1 | 10 | 7 | 15 | 43 | 79 | **90** | 5 |
| ***h*** | 0.02 | 6 | 3 | 6 | 24 | 31 | 71 | **105** | 4 |
| *h* | 0.15 | 6 | 9 | 8 | 22 | 48 | 83 | **66** | 8 |
| ***h*** | 0.15 | 1 | 6 | 14 | 31 | 51 | 86 | **54** | **7** |

Planetesimals migrating from distances greater than 0.7 AU could have collided with all planets, but most of their collisions (>70%) were with *g* and *h* planets, and the fraction of collisions was higher for planets further from the star. Planetesimals could have collided with all planets for *d* - *h* disks and at least with *b* - *e* planets for *b* - *c* disks. Therefore, the outer layers of neighboring planets in the TRAPPIST-1 system may include similar material if there were many planetesimals near their orbits during the late stages of planetary accumulation. A similar mixing of planetesimals was obtained by me in modeling the accumulation of terrestrial planets (see [15] and Section 5 of this book).

**Table 3.** The number of collisions of planetesimals with planets (from *b* to *h*) in the TRAPPIST -1 system for different disks (from *b* to *h*). The initial eccentricities of the orbits of planetesimals were equal to $e_o$. Each initial disk contained 1000 planetesimals. The variants with the integration step $t_s$=0.01 Earth days are marked with bold letters in the left column (the bottom row of two adjacent rows); for other variants $t_s$=0.02 Earth days. $N_l$ is the number of bodies remaining in elliptical orbits after 1000 years. The evolution of the disk ended after *T* million years.

| | $e_o$ | *b* | *c* | *d* | *e* | *f* | *g* | *h* | *thrown out* | $N_l$ | *T,* million years |
|---|---|---|---|---|---|---|---|---|---|---|---|
| *b* | 0.02 | **780** | 199 | 12 | 3 | 3 | 1 | 2 | 0 | 293 | 8.45 |
| ***b*** | 0.02 | **802** | 178 | 10 | 6 | 4 | 0 | 0 | 0 | 318 | 5.51 |
| *b* | 0.15 | **743** | 219 | 21 | 8 | 4 | 2 | 2 | 0 | 47 | 0.142 |
| ***b*** | 0.15 | **741** | 228 | 15 | 10 | 4 | 1 | 1 | 0 | 53 | 0.053 |
| *c* | 0.02 | 230 | **587** | 111 | 46 | 12 | 14 | 0 | 0 | 227 | 0.033 |
| ***c*** | 0.02 | 175 | **633** | 121 | 42 | 12 | 14 | 2 | 1 | 245 | 0 .065 |
| *c* | 0.15 | 321 | **518** | 77 | 32 | 15 | 5 | 0 | 0 | 32 | 0.060 |
| ***c*** | 0.15 | 313 | **529** | 78 | 44 | 15 | 19 | 0 | 2 | 3 5 | 0.103 |
| *d* | 0.02 | 58 | 205 | **445** | 155 | 62 | 64 | 9 | 2 | 104 | 0.888 |
| ***d*** | 0.02 | 52 | 197 | **459** | 167 | 69 | 48 | 8 | 0 | 98 | 2.60 |
| *d* | 0.15 | 104 | 314 | **241** | 168 | 83 | 67 | 21 | 2 | 132 | 0.286 |
| ***d*** | 0.15 | 109 | 304 | **282** | 177 | 63 | 55 | 6 | 4 | 108 | 0.332 |
| *e* | 0.02 | 14 | 73 | 133 | **432** | 171 | 143 | 30 | 4 | 196 | 0.620 |
| ***e*** | 0.02 | 29 | 88 | 114 | **420** | 187 | 132 | 19 | 11 | 204 | 0.245 |
| *e* | 0.15 | 27 | 132 | 160 | **267** | 200 | 179 | 27 | 0 | 240 | 0.555 |
| ***e*** | 0.15 | 32 | 109 | 163 | **291** | 203 | 170 | 21 | 11 | 260 | 0.561 |
| *f* | 0.02 | 10 | 47 | 59 | 168 | **371** | 290 | 44 | 11 | 291 | 0.818 |
| ***f*** | 0.02 | 27 | 50 | 56 | 159 | **371** | 267 | 55 | 15 | 330 | 33.23 |
| *f* | 0.15 | 21 | 74 | 76 | 173 | **261** | 310 | 57 | 28 | 359 | 1.30 |
| ***f*** | 0.15 | 24 | 75 | 80 | 186 | **261** | 305 | 53 | 16 | 391 | 0.436 |
| *g* | 0.02 | 15 | 32 | 50 | 116 | 188 | **457** | 124 | 18 | 398 | 1.78 |
| ***g*** | 0.02 | 9 | 37 | 45 | 98 | 202 | **475** | 123 | 11 | 424 | 4.81 |
| *g* | 0.15 | 10 | 50 | 67 | 120 | 227 | **402** | 101 | 23 | 484 | 0.975 |
| ***g*** | 0.15 | 17 | 45 | 59 | 118 | 252 | **405** | 81 | 23 | 456 | 22.60 |
| *h* | 0.02 | 9 | 31 | 28 | 77 | 143 | 282 | **407** | 22 | 590 | >48 |
| ***h*** | 0.02 | 5 | 31 | 31 | 80 | 153 | 279 | **406** | 14 | 598 | >44 |
| *h* | 0.15 | 13 | 45 | 49 | 90 | 191 | 338 | **252** | 21 | 726 | >25 |
| ***h*** | 0.15 | 9 | 37 | 47 | 89 | 165 | 376 | **246** | 31 | 726 | 11.28 |

## 24. Migration of bodies in the Gliese 581 exoplanetary system

Gliese 581 was originally thought to contain a star with a mass equal to 0.307 solar masses and five planets (*e*, *b*, *c*, *g*, *d*, in order of distance from the star). It is now believed that only three planets (*e*, *b*, *c*) exist close to the star. The unconfirmed planet Gliese 581g (with a mass of about two Earth masses) would be in the habitable zone of its star. Calculations for this planetary system were performed by Ipatov [209, 212, 511, 530, 531] (a more detailed journal article is planned for publication on this topic) for two models: (1) for five planets, including the unconfirmed planets (with circular initial orbits of the planets) and (2) with three confirmed planets (with known orbital eccentricities). In each **calculation variant,** the initial values of the semi-major axes of the orbits of $N_o$=250 or $N_o$=1000 planetesimal-bodies varied from $a_{min}$ to $a_{max}$ near the semi-major axis of the orbit of one of the planets, the initial eccentricities of their orbits were equal to $e_o$=0.02 or $e_o$=0.15, and their initial inclinations were equal to $e_o$/2 rad. The values of $a_{min}$ and $a_{max}$ and the semi-major axes and eccentricities of the orbits and masses of the planets are presented in Table 1. In my calculations with five planets, the eccentricities and inclinations of the orbits of the planets were considered to be equal to 0. The considered disk of planetesimals was located near the orbit of one of the considered planets. An outer disk *h* was also considered, for which $a_{min}$=0.25 AU and $a_{max}$=0.3 AU. The motion of planetesimals under the gravitational influence of the star and planets was modeled using the RMVS3 symplectic integrator from the SWIFT package. Planetesimals that collided with planets or the star or were ejected into hyperbolic orbits (reaching within 50 AU of the star) were excluded from the integration.

**Table 1.** Orbital elements and masses $m$ (in Earth mass $m_E$) of exoplanets in the Gliese 581 system and the values of $a_{min}$ and $a_{max}$ for the disks under consideration near planets *e, b, c, g,* and *d*. On the left are the initial data for calculation options with five planets, on the right – with three planets.

| planets | $m/m_E$ | *a,* AU | *e* | $a_{min}$, AU | $a_{max}$, AU |
|---|---|---|---|---|---|
| *e* | 1.7, 1.657 | 0.02815, 0.029 | 0.0, 0.125 | 0.022 | 0.0344 |
| *b* | 15.8, 15.2 | 0.0406, 0.041 | 0.0, 0.022 | 0.0344 | 0.0563 |
| *c* | 5.5, 5.65 | 0.0721, 0.074 | 0.0, 0.087 | 0.0563 | 0.1 |
| *g* | 2.2 | 0.13 | 0.0 | 0.1 | 0.174 |
| *d* | 6.98 | 0.218 | 0.0 | 0.174 | 0.25 |

First, below I discuss **the results of calculations that are similar for calculations with 3 and 5 planets**. In these calculations, there were no planetesimal collisions with the star. The fraction $p_{ej}$ of planetesimals ejected into hyperbolic orbits was higher for disks located farther from the star. For planetesimals from the feeding zones of planets located close to the star, most of collisions of planetesimals with planets occurs within the first thousand years. However, a few planetesimals could remain in elliptical orbits for millions or tens of millions of years. Some planetesimals, initially located near the orbit of one planet, could also impact most of the other planets.

Some calculations for the Glisse 581 system with $N_o$=250 for 5 planets and 10 million years were discussed in [209]. Later, such calculations were made for larger time intervals. **For the 5-planet calculations,** about 20-60% and less than 10% of the planetesimals were still moving in elliptical orbits after several hundred million years at $e_o$=0.02 and $e_o$=0.15, respectively.

The following three paragraphs discuss the results of disk evolution over **the first 10 million years**. Table 2 lists the numbers $N_e$, $N_b$, $N_c$, $N_g$ and $N_d$ of collisions of planetesimals with planets *e, b, c, g* and *d* over 10 million years. The fraction $p_{ej}$ of ejected planetesimals was higher for disks located further from the star. For disks corresponding to planets, the fraction $p_{ej}$ did not exceed 0.05 and 0.15 at $e_o$ equal to 0.02 and 0.15, respectively. For disk *h,* the fraction $p_{ej}$ was about 0.01 and 0.27-0.4 at $e_o$=0.02 and $e_o$=0.15, respectively. At $e_o$=0.15, the fraction of planetesimals remaining after 10 million years did not exceed 0.03 for disks *e, b, c,* and *g* and 0.1 for disk *d.* At $e_o$=0.02 this fraction did not exceed 0.24 for disk *e*, 0.46 for disks *b* and *d*, and 0.64 for disks *c* and *g*.

At $e_o$=0.02, the fraction of initial planetesimals that collided with the host planet was about 0.4-0.46 for disks *e*, *b*, and *d*, about 0.3 for disk *c*, and 0.18 for disk *g* (Table 2). At $e_o$=0.15, this fraction was 0.5, 0.8, 0.4, 0.2, and 0.5 for disks *e, b, c, g* and *d,* respectively. The fraction of planetesimals that collided with one of the neighboring planets (close to the host planet) for disks *e, b, c, g* and *d* was 0.26, 0.06, 0.04, 0.09, and 0.04 at $e_o$=0.02. It was 0.5, 0.12, 0.4, 0.35 and 0.15 at $e_o$=0.15.

Some planetesimals, initially located near one of the planets, could also fall onto other planets. The highest ratio $K_a$ of the number of planetesimals falling onto another planet to the number of planetesimals falling onto the host planet and coming from the initial disk at $e_o$=0.02 was obtained for planet *e*, which is closest to the star. For disk *e* the ratio $N_b/N_e$ was about 0.55-0.75 and 0.98-1.2 at $e_o$=0.02 and $e_o$=0.15, respectively. At $e_o$=0.02, the ratio $K_a$ for disk *g* was equal to 0.5, and it was less than 0.15 for disks *b*, *c*, and *d*. At $e_o$=0.15 for disk *g,* the value of $N_g$ was even approximately two times smaller than $N_d$. At $e_o$=0.15, the ratio $K_a$ for disk *b* was less than 0.21, but for disk *c* it was about 0.8-0.9. At $e_o$=0.15 for disk *d,* the ratio $N_c/N_d$ was approximately 0.3, although planets *c* and *d* were not close to each other. Planetesimals from the *h* disk fell onto four planets at $e_o$=0.15, but did not collide with any planet at $e_o$=0.02.

For disks close to the star, most of collisions of planetesimals with planets occurred within **the first thousand years**. At $e_o$=0.02, the ratio $k_1$ of the total number of planetesimal–planet collisions in the first thousand years to the number of collisions over 10 million years was between 0.6 and 0.67 for disks *e, b,* and *c*; it was equal to 0.3 for disk *g* and was between 0.23 and 0.49 for disk *d*. The latter range was obtained for different values of the integration step $t_s$. For all other disks, this range of $k_1$ values was narrow for the different considered $t_s$. At $e_o$=0.15, the ratio $k_1$ was 0.85, 0.8, 0.45, 0.13–0.15, and 0.06–0.08 for disks *e, b, c, g, d,* and *h,* respectively.

Note that for disk *g* at $e_o$=0.15, the fractions of planetesimals that impacted planets *c* and *d* (each) were greater than the fraction of planetesimals that impacted the host planet *g*. In calculations with five planets, about 5% of the planetesimals from the orbital vicinity of the outer planet *d* (with a mass of 7 Earth masses) fell onto planet *g*. Therefore, such a planet *g* could have experienced an influx of water-containing bodies.

For the variants presented in Table 2 for $T$=10 million years, calculations were also carried out for time intervals **greater than 10 million years**. With an integration step $t_s$ equal to 0.01 Earth days, they were made during several tens of millions of years (usually slightly less than 100 million years), and with $t_s$ equal to 0.04 Earth days, they were made during several hundred million years. The calculations were suspended since the number of remaining planetesimals in the variants did not change over a long period of time. Compared with the data in Table 2, for large time intervals the values of the number of collisions of planetesimals and planets differed from the values in Table 2 by a maximum of a few units, and values less than 40 did not change. Changes in the number of remaining planetesimals did not exceed 10% at $e_o$=0.02, and at $e_o$=0.15 the number of remaining planetesimals usually differed by only one compared to the data in Table 2. Only for the *h* disk and $e_o$=0.15 the number of remaining planetesimals over a large time interval was less than the values in Table 2 by approximately 30 due to the ejection of planetesimals into hyperbolic orbits.

**At $e_o$=0.15** and considering the disks of planetesimals around the planets, the remaining planetesimals moved along stable orbits in **resonances with the planets**. During the evolution **of the disk *e*** at calculations with a step $t_s$ of 0.01 and 0.04 Earth days, after a time of about 80 and 300 million years, respectively, in both variants, 7 identical (for both variants) initial planetesimals remained in identical orbits. Three planetesimals moved in a 4:3 resonance with the motion of planet *e* inside its orbit. The other three planetesimals moved in a 5:4 resonance, and one planetesimal - in a 1:1 resonance with the motion of planet *e*. After an evolution time **of disk *b*** of about 100 and 350 million years, in calculations with $t_s$ equal to 0.01 and 0.04 Earth days, respectively, in both variants, 2 identical initial planetesimals remained, moving in identical orbits behind the orbit of planet *b* in a 4:5 resonance with this planet. After about 100 and 360 million years of evolution **of disk *c***, calculated with $t_s$ equal to 0.01 and 0.04 Earth days, respectively, in both variants there remained three identical initial planetesimals moving in identical orbits in 1:1 resonance with planet *c*. After about 100 and 330 million years of evolution of **disk *g***, at calculations with $t_s$ equal to 0.01 and 0.04 Earth days, respectively, in both variants there remained three identical initial planetesimals moving in identical orbits in 1:1 resonance with planet *g*, as well as one planetesimal moving beyond the orbit of planet *g* in 4:5 resonance with this planet. During the evolution **of disk *d*** in calculations with $t_s$ equal to 0.01 and 0.04 Earth days, after a time of approximately 50 and 200 million years, respectively, in both variants, 16 identical (for both variants) initial planetesimals remained, moving in a 1:1 resonance with the motion of planet *d*. In the external 5:6 resonance with planet *d,* 7 planetesimals moved at $t_s$=0.01 Earth days and 4 planetesimals at $t_s$=0.04 Earth days.

During the evolution **of the disk *h*** after 20 and after 150 million years, 63 and 32 planetesimals remained in calculations with $t_s$ equal to 0.01 and 0.04 Earth days, respectively. Not all of them moved in resonant orbits. In the outer 4:5 resonance with planet *h,* 13 planetesimals moved in calculations with both values of $t_s$, and in the outer 3:4 resonance with planet *h,* 12 planetesimals moved in calculations with $t_s$ equal to 0.01 Earth days, and 11 planetesimals at $t_s$=0.04 Earth days. The number of planetesimals moving in 5:8 resonance was greater (6 instead of 2) in the calculation with a smaller value of $t_s$. The difference in these calculations is due to the different time intervals considered. Over a longer time interval, 104 planetesimals were ejected instead of 89.

**At $e_o$=0.02,** the orbits of the remaining planetesimals have not yet been analyzed. These calculations also include cases of 1:1 resonances with planets. The remaining planetesimals moved primarily in orbits quite distant from the orbits of the planets. In this case, it is of interest to study the initial zones in the disk from which the planetesimals either collided with the planets or left the planetary system, and the resonances with the planets that lead to this.

**For the 3-planet simulations** with $N_o$=1000, the fraction of planetesimals remaining in elliptical orbits after 1 million years was much smaller than for the 5-planet simulations, mainly due to the eccentric orbits of the planets for the 3-planet simulations. The results of the evolution of such disks are presented in Table 3. For calculations with three planets, the evolution of all three disks (*e*, *b*, *c*) was completed in less than 1.3 million years at $e_o$=0.15 and in no more than 1.1 million years for disks *e* and *b* at $e_o$=0.02. At the end of the evolution of disk *c* and $e_o$=0.02, the six remaining planetesimals were moving in 1:1 resonance with planet *c*, and one planetesimal was in an external 4:5 resonance with this planet.

At the end of the evolution, at $e_o$=0.02, the fraction of the initial planetesimals that collided with the host planet was about 0.44, 0.74, and 0.57 for the *e*, *b*, and *c* disks, respectively. At $e_o$=0.15, it was 0.4, 0.79, and 0.52, respectively. At $e_o$=0.02, the fraction of planetesimals that collided with one of the neighboring planets (close to the host planet) was 0.50, 0.20, and 0.36 for the *e*, *b,* and *c* disks*,* respectively.

At $e_o$=0.15, it was 0.52, 0.15, and 0.40, respectively. When considering planetesimal migration from distances greater than $a_{max}$ for disk *c,* the ratio of planetesimal impacts with planets *c* and *b* was 2 and 2.9 at $e_o$=0.02 and $e_o$=0.15, respectively. These results indicate that the outer layers of neighboring exoplanets in the Gliese 581 exoplanet system may contain similar material. Disk *g* in the three-planet calculations was external to these planets, and more than half of the planetesimals in it remained in elliptical orbits.

**Table 2.** Number $N_e$, $N_b$, $N_c$, $N_g$ and $N_d$ of collisions of planetesimals with planets *e, b, c, g* and *d* over 10 million years with an integration step $t_s$ equal to 0.01, 0.04, or 0.1 Earth days. The initial orbital eccentricities of the planetesimals were equal to $e_o$ (0.02 or 0.15). The number of initial planetesimals in each variant was 250. $N_{ej}$ is the number of planetesimals that reached 50 AU from the star. $N_{el}$ is the number of planetesimals that remained in elliptical orbits.

| disk | $e_o$ | $t_s$ | $N_e$ | $N_b$ | $N_c$ | $N_g$ | $N_d$ | $N_{ej}$ | $N_{el}$ |
|---|---|---|---|---|---|---|---|---|---|
| *e* | 0.02 | 0.01 | **116** | 66 | 5 | 0 | 2 | 3 | 58 |
| *e* | 0.02 | 0.04 | **106** | 80 | 5 | 0 | 0 | 0 | 59 |
| *e* | 0.02 | 0.1 | **120** | 66 | 3 | 1 | 1 | 0 | 59 |
| *b* | 0.02 | 0.01 | 10 | **117** | 12 | 0 | 0 | 2 | 109 |
| *b* | 0.02 | 0.04 | 7 | **111** | 16 | 1 | 1 | 2 | 112 |
| *c* | 0.02 | 0.01 | 1 | 13 | **79** | 1 | 3 | 4 | 149 |
| *c* | 0.02 | 0.04 | 1 | 9 | **73** | 7 | 6 | 4 | 150 |
| *g* | 0.02 | 0.01 | 0 | 3 | 17 | **51** | 14 | 5 | 160 |
| *g* | 0.02 | 0.04 | 0 | 6 | 12 | **44** | 22 | 6 | 160 |
| *d* | 0.02 | 0.01 | 0 | 4 | 11 | 8 | **104** | 9 | 114 |
| *d* | 0.02 | 0.04 | 0 | 6 | 4 | 5 | **118** | 12 | 105 |
| *d* | 0.02 | 0.1 | 0 | 3 | 6 | 8 | **114** | 13 | 106 |
| *h* | 0.02 | 0.04 | 0 | 0 | 0 | 0 | 0 | 2 | 248 |
| *e* | 0.15 | 0.01 | **120** | 117 | 4 | 1 | 0 | 1 | 7 |
| *e* | 0.15 | 0.04 | **105** | 125 | 8 | 2 | 1 | 2 | 7 |
| *b* | 0.15 | 0.01 | 8 | **200** | 29 | 3 | 3 | 4 | 3 |
| *b* | 0.15 | 0.04 | 17 | **183** | 38 | 1 | 1 | 7 | 3 |
| *c* | 0.15 | 0.01 | 1 | 85 | **111** | 16 | 16 | 17 | 4 |
| *c* | 0.15 | 0.04 | 4 | 95 | **105** | 13 | 16 | 13 | 4 |
| *g* | 0.15 | 0.01 | 0 | 35 | 54 | **45** | 88 | 21 | 7 |
| *g* | 0.15 | 0.04 | 5 | 41 | 52 | **34** | 83 | 30 | 5 |
| *d* | 0.15 | 0.01 | 1 | 14 | 35 | 16 | **123** | 37 | 24 |
| *d* | 0.15 | 0.04 | 1 | 20 | 36 | 21 | **118** | 30 | 24 |
| *h* | 0.15 | 0.01 | 0 | 8 | 10 | 13 | 53 | 68 | 98 |
| *h* | 0.15 | 0.04 | 0 | 7 | 8 | 12 | 84 | 82 | 63 |

**Table 3.** The number of collisions ($N_e$, $N_b$ and $N_c$) of planetesimals with planets (*e*, *b* and *c*) over *T* million years with an integration step $t_s$, equal to 0.04 Earth days, for calculations with 3 planets. The initial orbital eccentricities of the planetesimals were equal to $e_o$ (0.02 or 0.15). The number of initial planetesimals in each variant was equal to 1000. $N_{ej}$ is the number of planetesimals that reached 50 AU from the star. $N_{el}$ is the number of planetesimals that remained in elliptical orbits.

| disk | $e_o$ | $t_s$ | $N_e$ | $N_b$ | $N_c$ | $N_{ej}$ | $N_{el}$ | *T,* million years |
|---|---|---|---|---|---|---|---|---|
| *e* | 0.02 | 0.04 | **445** | 501 | 47 | 9 | 0 | 0.11 |
| *b* | 0.02 | 0.04 | 28 | **742** | 205 | 25 | 0 | 0.91 |
| *c* | 0.02 | 0.04 | 22 | 363 | **568** | 40 | 7 | 146. |
| *g* | 0.02 | 0.04 | 6 | 102 | 211 | 2 9 | 652 | 15. |
| *e* | 0.15 | 0.04 | **401** | 524 | 65 | 10 | 0 | 0.13 |
| *b* | 0.15 | 0.04 | 45 | **787** | 150 | 18 | 0 | 0.25 |
| *c* | 0.15 | 0.04 | 26 | 403 | **524** | 47 | 0 | 1.28 |
| *g* | 0.15 | 0.04 | 7 | 8 7 | 253 | 31 | 622 | 13. |

## NON-ASTRONOMICAL STUDIES

### 25. Minimizing the number of transitions when routing two-layer microcircuits

Together with N.N. Kozlov and V.N. Toroptseva, S.I. Ipatov participated in the development of algorithms and computer programs for minimizing the number of transitions during the routing of two-layer microcircuits. A preprint on this topic was published in 1990 [231]. In the 1990s, such topics in Russia became irrelevant due to the collapse of the Russian microelectronics industry.

### 26. Studies of the generation of acoustic waves under the influence of fluids on pore walls and their propagation in a porous medium with fluids and gas

S.I. Ipatov was responsible for mathematical modeling in the grant from the oil company Schlumberger. The grant "Study of the generation of acoustic waves under the influence of fluids on the walls of pores and their propagation in a porous medium with fluids and gas" was in the Gubkin Russian State University of Oil and Gas (1999-2002). Based on the results of this research, a joint article [69] "Study of the amplitude-frequency spectra of acoustic and electromagnetic noise signals during fluid filtration in rocks" was published in the journal Geophysics. The results of this research were also included in sections [79, 80, 81, 83] of monographs by M.I. Kremenetsky and A.I. Ipatov (A.I. Ipatov is S.I. Ipatov's brother).

*Abstract*. The phenomena causing both acoustic and electromagnetic emissions during core filtration have been studied in detail experimentally. The shape and magnitude of the amplitude-frequency characteristic (AFC) anomalies recorded in the spectra depend primarily on the reservoir type and the structure of its pore space. AFC spectra of acoustic emission signals obtained for dry samples with clearly defined large filtration channels and high permeabilities $k$ are easily distinguished from similar results in rocks with lower $k$ and a more complex pore structure (containing medium and small pores).

Borehole surveys based on the analysis of the spectral characteristics of emission noise in certain physical fields allow us to assess the type of fluid being drained and the nature of the pore space in specific formation thicknesses during rock filtration. The observed phenomena were substantiated through physical and mathematical modeling of liquid and gas filtration in rocks over a wide range of petrophysical properties and filtration rates, as well as for various pressure-temperature conditions.

## LIST OF PUBLICATIONS BY S.I. IPATOV

The current list of publications in Russian is updated over time and can be found on the websites https://sites.google.com/view/siipatov/publications-in-russian and https://1drv.ms/w/c/c67d93a65f0a2a17/ERcqCl-mk30ggMYFBAAAAABye6tCmVUGVWRphNZ_k-WIg?e=DjHucX. Most of the articles in English can be downloaded free of charge using the links below. In particular, many articles are available on the websites https://www.researchgate.net/profile/S_Ipatov/publications, https://independent.academia.edu/SergeiIpatov and http://arxiv.org .

List of my publications in English can be found on http://siipatov.webnode.ru/publications/ and in the file https://1drv.ms/w/c/c67d93a65f0a2a17/ERcqCl-mk30ggMahBwAAAAABRRn0ODVZPKj5m4sW9wQodg?e=zLQQFp. Most of my papers in English can be downloaded free from https://www.researchgate.net/profile/S_Ipatov/publications, https://independent.academia.edu/SergeiIpatov and http://arxiv.org. Most of the journal papers can be downloaded for personal (non-commercial) use only.

**Websites with a list of publications by S.I. Ipatov**:

https://www.scopus.com/authid/detail.uri?authorId=6603765191

https://www.webofscience.com/wos/author/record/O-2302-2014. WoS Researcher ID: O-2302-2014.

https://publons.com/researcher/1711006/sergei-ipatov/publications/

https://scholar.google.com/citations?user=496SwBAAAAAJ (on this website the number of citations is greater than 3000 and h-index=28) and http://scholar.google.com/scholar?hl=en&lr=&q=si+ipatov&btnG=Search – lists most of my English publications (with citations and links to abstracts of my papers). The links to papers are for journals, which can not be free to download papers.

https://elibrary.ru/author_profile.asp?authorid=18595. e-library SPIN- code : 7182-9350. The list of articles in English is not complete.

https :// orcid . org /0000-0002-1413-9180

https://ui.adsabs.harvard.edu/search/q=author%3A("ipatov%2C%20s.")&sort=date%20desc%2C%20bibcode%20desc&p_=0 or http://adsabs.harvard.edu/abstract_service.html (and print: ipatov, s.) – the list of publications that includes abstracts of publications (including abstracts of all papers in English and preprints; may include some abstracts for which there are no links on my website and in the text below).

https://openalex.org/authors?page=1&filter=ids.openalex%3Aa5023189280

https://www.researchgate.net/profile/S_Ipatov/publications (you can see the links to files with papers on this website). Many of my publications can be downloaded for free from this site and from https://independent.academia.edu/SergeiIpatov.

https://independent.academia.edu/SergeiIpatov (below in references you can use web addresses without text with the name of publication after the number; if the reference will not work you can find the correct reference in the list of all publications on the website). May be there are some files in academia and researchgate (for nonjournal publications) to which the links are not listed below.

https://scinapse.io/authors/2515993428 may be now it is needed to pay to search on scinapse.io?

https://arxiv.org/search/astro-ph?searchtype=author&query=Ipatov,+S – with files of my papers

Information about publications published both in English and Russian below is presented only in English (in the Russian book such information is presented in both languages)

### Articles in peer-reviewed journals indexed in WoS and/or Scopus

1. Ipatov S.I. On the gravitational interaction of two planetesimals. *Soviet Astronomy* (translated from Astronomicheskii Zhurnal), 1981, v. 25 (58), N 3, pp. 352-357. https://www.researchgate.net/publication/234297279. https://www.academia.edu/44447765. https://articles.adsabs.harvard.edu/pdf/1981SvA....25..352I.

2. Ipatov S.I. Computer modeling of the evolution of the plane rings of gravitating particles moving around the Sun, *Soviet Astronomy* (translated from Astronomicheskii Zhurnal), 1981, v. 25 (58), N 5, pp. 617-623. https://www.researchgate.net/publication/234297279, https://www.academia.edu/44448038https://articles.adsabs.harvard.edu/pdf/1981SvA....25..617I

3. Ipatov S.I. *Solar System Research* (translated from Astronomicheskii Vestnik), 1987, v. 21, N 3, pp. 129-135. https://www.researchgate.net/publication/234436814 . https://www.academia.edu/59079931.

4. Ipatov S.I., Accumulation and migration of the bodies from the zones of giant planets, *Earth, Moon, and Planets*, 1987, v. 39, N 2, pp. 101-128. https://www.researchgate.net/publication/227047521. https://www.academia.edu/44447538. https://doi.org/10.1007/BF00054059. http://adsabs.harvard.edu/abs/1987EM%26P...39..101I.

5. Ipatov S.I. Computer simulation of the possible evolution of the orbits of Pluto and bodies of the trans-Neptune belt, *Kinematics and Physics of Celestial Bodies* (translated from Kinematika i Physics Heavenly Tel), 1988, v. 4, *N* 6, pp. 76-82. https://www.researchgate.net/publication/253284007. https://www.academia.edu/126485666.

6. Ipatov S.I. The gap problem and asteroid-type resonant orbit evolution, *Kinematics and Physics of Celestial Bodies* (translated from Kinematika i Physics Heavenly Tel), 1988, v. 4, *N* 4, pp. 49-57. https://www.researchgate.net/publication/238555999. https://www.academia.edu/126485853.

7. Ipatov S.I. Evolution times for disks of planetesimals, *Soviet Astronomy*, 1988, v. 32 (65), N 5, pp. 560-566. https://www.researchgate.net/publication/241265869. https://www.academia.edu/44452977. https://articles.adsabs.harvard.edu/pdf/1988SvA....32..560I.

8. Ipatov S.I. *Soviet Astron. Letters* (translated from Pis'ma v Astronomicheskii Zhurnal), 1989, v. 15, N 4 (8), pp. 324-328. https://www.researchgate.net/publication/234455005. https://www.academia.edu/44452706. http://adsabs.harvard.edu/abs/1989SvAL...15..324I.

9. Ipatov S.I. Evolution of the orbital eccentricities of planetesimals during formation of the giant planets, *Solar System Research* (translated from Astronomicheskii Vestnik), 1989, v. 23, No. 3, pp. 119-125. https://www.researchgate.net/publication/241498872. https://www.academia.edu/works/59079941.

10. Ipatov S.I. Planetesimal migration during the last stages of accumulation of the giant planets, *Solar System Research* (translated from Astronomicheskii Vestnik), 1989, v. 23, No. 1, pp. 16-23. https://www.researchgate.net/publication/258976468. https://www.academia.edu/59079957.

11. Ipatov S.I. Evolution of initially highly eccentric orbits of the growing nuclei of the giant planets, *Soviet Astron. Letters* (translated from Pis'ma v Astronomicheskii Zhurnal), 1991, v. 17, No. 2, pp. 113-119. https://www.researchgate.net/publication/253317841. https://www.academia.edu/44452706. http://adsabs.harvard.edu/abs/1989SvAL...15..324I.

12. Ipatov S.I. Numerical model of the evolution of asteroid orbits at the 2:5 resonance, *Solar System Research* (translated from Astronomicheskii Vestnik), 1992, v. 26, N 6. pp. 520-541. https://www.researchgate.net/publication/259022526. https://www.academia.edu/59079955.

13. Ipatov S.I. Migration and accumulation of bodies in the forming solar system, *International Applied Mechanics* (translated from Prikladnaya Mekhanika), 1992, v. 28, N 11, pp. 771-774. https://www.researchgate.net/publication/256171667. https://www.academia.edu/49325518. https://link.springer.com/article/10.1007/BF00847313, DOI 10.1007/BF00847313.

14. Ipatov S.I. Evolution of asteroidal orbits at the 5:2 resonance. *Icarus*, v. 95, N 1. pp. 100-114 (1992), https://www.sciencedirect.com/science/article/abs/pii/001910359290194C?via%3Dihub, https://doi.org/10.1016/0019-1035(92)90194-C, https://www.researchgate.net/publication/253663007, https://www.academia.edu/44448338.

15. Ipatov S.I. Migration of bodies in the accretion of planets, *Solar System Research* (translated from Astronomicheskii Vestnik), 1993, v. 27, N 1. pp. 65-79. https://www.researchgate.net/publication/234281490. https://www.academia.edu/44448077.

16. Ipatov S.I. Gravitational interaction of two planetesimals moving in close orbits, *Solar System Research* (translated from Astronomicheskii Vestnik), 1994, v. 28, N 6, pp. 494-512. https://www.researchgate.net/publication/234445433. https://www.academia.edu/59080069 .

17. Ipatov S.I. Migration of bodies in the accumulation of planets, *Earth, Moon, and Planets*, 1995, v. 67, N 1-3, pp. 217-219. DOI: 10.1007/BF00613304. https://www.researchgate.net/publication/252149292. https://www.academia.edu/44452347. http://adsabs.harvard.edu/full/1995EM%26P...67..217I.

18. Ipatov S.I., Gravitational interaction of objects moving in crossing orbits, *Solar System Research* (translated from Astronomicheskii Vestnik), 1995, v. 29, N 1. pp. 9-20. https://www.researchgate.net/publication/253123243. https://www.academia.edu/126464286.

19. Ipatov S.I., Migration of small bodies to the Earth, *Solar System Research* (translated from Astronomicheskii Vestnik), 1995, v. 29. N 4. pp. 261-286. https://www.researchgate.net/publication/225900805. https://www.academia.edu/49336026.

20. Ipatov S.I. Migration of small bodies in the Solar system, *Earth, Moon, and Planets*, 1996, v. 72. N 1-3. pp. 211-214. DOI: 10.1007/BF00117519. https://www.researchgate.net/publication/225900805. https://www.academia.edu/44452474. https://articles.adsabs.harvard.edu/pdf/1995EM%26P...67..217I.

21. Ipatov S.I. and Hahn, G.J. Orbital evolution of the P/1996 R2 and P/1996 N2 objects, *Solar System Research* (translated from Astronomicheskii Vestnik), v. 33, N 6, pp. 487-500 (1999). https://arxiv.org/abs/2412.12939. https://doi.org/10.48550/arXiv.2412.12939. https://www.researchgate.net/publication/224789085. https://www.academia.edu/59079985.

22. Ipatov S.I. Migration of trans-Neptunian objects to the Earth, *Celestial Mechanics and Dynamical Astronomy*, 1999, v. 73, N 1-4, pp. 107-116. 3, N 1-4, pp. 107-116 (1999) https://doi.org/10.1023/A:1008386727899. https://link.springer.com/article/10.1023%2FA%3A1008386727899. http://adsabs.harvard.edu/full/1999imda.coll..107I. https://www.researchgate.net/publication/225246530. https://www.academia.edu/44448423. https://1drv.ms/b/c/c67d93a65f0a2a17/ERcqCl-mk30ggMbBBwAAAAABFh-oLcdO2GOETRrOiyL_Eg?e=eTgrat.

23. Ipatov S.I. Accessing astronomical data over the Internet, *Solar System Research* (translated from Astronomicheskii Vestnik), 1999, v. 33, N 3, P. 247-251. https://www.researchgate.net/publication/253591120. https://www.academia.edu/59079984.

24. Ipatov S.I. Variations in orbital elements of planets, *Solar System Research* (translated from Astronomicheskii Vestnik), 2000, v. 34, N 3, pp. 179-185. https://arxiv.org/abs/2412.12925. https://doi.org/10.48550/arXiv.2412.12925, https://www.researchgate.net/publication/253300880. https://www.academia.edu/59080053.

25. Ipatov S.I., Henrard J. Evolution of orbits of trans-Neptunian bodies at the 2:3 resonance with Neptune, *Solar System Research* (translated from Astronomicheskii Vestnik), 2000, v. 34, N 1, pp. 61-74. http://arxiv.org/abs/2412.13297. https://doi.org/10.48550/arXiv.2412.13297. https://www.researchgate.net/publication/252901582. https://www.academia.edu/59080057. https://1drv.ms/b/c/c67d93a65f0a2a17/ERcqCl-mk30ggMa6BwAAAAABU04mw-I5GBoTsgLHM9Zecg?e=xdnA1e - pdf in Russian.

26. Ipatov S.I. Comet hazard to the Earth, *Advances in Space Research*, Elsevier, 2001, v. 28, N 8, pp. 1107-1116. DOI: 10.1016/S0273-1177(01)00494-X. https://www.researchgate.net/publication/2231935. https://www.academia.edu/44452766. initial version on http://arXiv.org/format/astro-ph/0108187. https://www.sciencedirect.com/science/article/abs/pii/S027311770100494X?via%3Dihub.

27. Ipatov S.I., Mather J.C. Migration of trans-Neptunian objects to the terrestrial planets, *Earth, Moon, and Planets*, 2003, v. 92, p. 89-98. https://doi.org/10.1023/B:MOON.0000031928.45965.7b. http://www.roe.ac.uk/~jkd/kbo_proc/ipatov.pdf; initial version on http://arXiv.org/format/astro-ph/0305519. https://www.researchgate.net/publication/1931873. https://www.academia.edu/44452493.

28. Ipatov S.I., Mather J.C. Comet and asteroid hazard to the terrestrial planets, *Advances in Space Research*, Elsevier, 2004, v. 33, N 9, p. 1524-1533. https://doi.org/10.1016/S0273-1177(03)00451-4. https://www.researchgate.net/publication/1928889. https://www.academia.edu/44452546. Initial version on http://arXiv.org/format/astro-ph/0212177.

29. Ipatov S.I., Mather J.C. Migration of Jupiter-family comets and resonant asteroids to near-Earth space, *Annals of the New York Academy of Sciences*, 2004, v. 1017, pp. 46-65. https://nyaspubs.onlinelibrary.wiley.com/doi/abs/10.1196/annals.1311.004. https://elibrary.ru/item.asp?id=13462803. https://doi.org/10.1196/annals.1311.004. initial version on http://arXiv.org/format/astro-ph/0308448, https://www.researchgate.net/publication/8488458. https://www.academia.edu/44447592.

30. Ipatov S.I., Mather J.C., Taylor P.A. Migration of interplanetary dust, *Annals of the New York Academy of Sciences*, 2004, v. 1017, pp. 66-80. https://nyaspubs.onlinelibrary.wiley.com/doi/abs/10.1196/annals.1311.005. https://elibrary.ru/item.asp?id=13469889. https://doi.org/10.1196/annals.1311.005. http://www.findthatfile.com/search-66853853-hPDF/download-documents-ipatov-nyas-dust.pdf.htm. initial version on http://arXiv.org/format/astro-ph/0308450; DOI: 10.1196/annals.1311.005, https://www.researchgate.net/publication/8488459. https://www.academia.edu/44447510.

31. Marov M.Ya., Ipatov S.I. Migration of dust particles and volatiles delivery to the terrestrial planets, *Solar System Research* (translated from Astronomicheskii Vestnik), 2005, v. 39, N 5, 374-380. https://link.springer.com/article/10.1007/s11208-005-0050-1. https://www.researchgate.net/publication/225445817. https://www.academia.edu/44453038. Russian text: https://1drv.ms/b/c/c67d93a65f0a2a17/ERcqCl-mk30ggMYcCAAAAAAByoUVYwudc4X0JMoLf6oapg?e=5Yku1j.

32. A'Hearn M.F., M. J. S. Belton, W. A. Delamere, J. Kissel, K. P. Klaasen, L. A. McFadden, K. J. Meech, H. J. Melosh, P. H. Schultz, J. M. Sunshine, P. C. Thomas, J. Veverka, D. K. Yeomans, M. W. Baca, I. Busko, C. J. Crockett, S. M. Collins, M. Desnoyer, C. A. Eberhardy, C. M. Ernst, T. L. Farnham, L. Feaga, O. Groussin, D. Hampton, **S. I. Ipatov,** J.-Y. Li, D. Lindler, C. M. Lisse, N. Mastrodemos, W. M. Owen Jr., J. E. Richardson, D. D. Wellnitz, R. L. White, Deep Impact: Excavating Comet Tempel 1, *Science*, 14 October 2005, v. 310, pp. 258-264. https://science.sciencemag.org/content/310/5746/258.full.pdf+html. https://doi.org/10.1126/science.1118923. https://www.academia.edu/99306299. https://www.researchgate.net/publication/7611098.

33. Ipatov S.I., Mather J.C. Migration of small bodies and dust to near-Earth space. *Advances in Space Research*, 2006, v. 37, *N* 1, 126-137. https://doi.org/10.1016/j.asr.2005.05.076, https://www.researchgate.net/publication/1802821. Initial version on http://arXiv.org/format/astro-ph/0411004. https://www.academia.edu/49324706.

34. Ipatov S.I., A'Hearn M.F., Klaasen K.P. Automatic removal of cosmic ray signatures on Deep Impact images, *Advances in Space Research*, 2007, v. 40, 160-172. https://www.sciencedirect.com/science/article/abs/pii/S0273117707003493?via%3Dihub. https://doi.org/10.1016/j.asr.2007.04.012. https://www.researchgate.net/publication/228922305, Initial version on http://arXiv.org/format/astro-ph/0610931. https://www.academia.edu/44448364.

35. Klaasen K.P., A'Hearn M.F., Baca M.W., Delamere W.A., Desnoyer M., Farnham T.L., Groussin O., Hampton D., **Ipatov S.I.,** Li J.-Y., Lisse C.M., Mastrodemos N., McLaughlin S., Sunshine J.M., Thomas P.C., Wellnitz D.D. Invited article: Deep Impact instrument calibration. *Review of Scientific Instruments Journal*, 2008, v. 79, issue 9, pp. 091301-091301-77 (77 pages). https://doi.org/10.1126/10.1063/1.2972112. https://www.researchgate.net/publication/23557358. https://www.academia.edu/59080018. http://link.aip.org/link/?RSI/79/091301.

36. Boss A.P., **Ipatov S.I.,** Keiser S.A., Myhill E.A., Vanhala H.A.T., Simultaneous triggered collapse of the presolar dense cloud core and injection of short-lived radioisotopes by a supernova shock wave, *Astrophysical Journal Letters*, 2008, v. 686, pp. L119-L123. DOI: 10.1086/593057. http://arxiv.org/abs/0809.3045. http://iopscience.iop.org/1538-4357/686/2/L119/pdf. https://www.researchgate.net/publication/1775247, https://www.academia.edu/70231160.

37. Ipatov S.I., Kutyrev A., Madsen G.J., Mather J.C., Moseley S.H., Reynolds R.J., Dynamical zodiacal cloud models constrained by high resolution spectroscopy of the zodiacal light, *Icarus*, 2008, v. 194, N. 2, pp. 769-788. https://doi.org/10.1016/j.icarus.2007.11.009; https://www.researchgate.net/publication/222219285, https://www.sciencedirect.com/science/article/abs/pii/S0019103507005623?via%3Dihub, http://arxiv.org/abs/0711.3494 ; an old version of the paper with a greater number of figures, but with less analysis is on http://arXiv.org/format/astro-ph/0608141. https://www.academia.edu/44448220.

38. Boss A.P., Keiser S.A., **Ipatov S.I**., Myhill E.A., Vanhala H.A.T. Triggering collapse of the presolar dense cloud core and injecting short-lived radioisotopes with a shock wave. I. Varied shock speeds, *Astrophysical Journal*, 2010, v. 708, pp. 1268-1280, doi 10.1088/0004-637X/708/2/1268, http://arxiv.org/abs/0911.3417. http://iopscience.iop.org/0004-637X/708/2/1268/pdf/0004-637X_708_2_1268.pdf. https://www.researchgate.net/publication/45885018, https://www.academia.edu/126486109.

39. Ipatov S.I. The angular momentum of two collided rarefied preplanetesimals and the formation of binaries, *Monthly Notices of the Royal Astronomical Society*, 2010, v. 403, 405-414. http://arxiv.org/abs/0904.3529. https://doi.org/10.1111/j.1365-2966.2009.16124.x, https://academic.oup.com/mnras/article/403/1/405/1019032, https://www.researchgate.net/publication/227698427, http://mnras.oxfordjournals.org/content/403/1/405.full.pdf+html. https://www.academia.edu/44448131.

40. Ipatov S.I., A'Hearn M.F. The outburst triggered by the Deep Impact collision with Comet Tempel 1, *Monthly Notices of the Royal Astronomical Society*, 2011, v. 414, N 1, 76-107. DOI: 10.1111/j.1365-2966.2010.18079.x. http://arxiv.org/abs/0810.1294. http://mnras.oxfordjournals.org/content/414/1/76.full.pdf+html. https://www.researchgate.net/publication/47822083, https://academic.oup.com/mnras/article/414/1/76/1085918, https://www.academia.edu/44447456.

41. Ipatov S.I. Location of upper borders of cavities containing dust and gas under pressure in comets, *Monthly Notices of the Royal Astronomical Society*, 2012, v. 423, 3474-3477. DOI: 10.1111/j.1365-2966.2012.21144.x. http://mnras.oxfordjournals.org/content/423/4/3474.full.pdf,

http://arxiv.org/abs/1205.6000. https://academic.oup.com/mnras/article/423/4/3474/1747981. https://www.researchgate.net/publication/225046674, https://www.academia.edu/44448206.

42. Shin I.-G. et al., including **Ipatov S.** (157 co-authors). Microlensing binaries with candidate brown dwarf companions, *The Astrophysical Journal*, 2012, vol. 760, Issue 2, article id. 116, 10 pp. DOI: 10.1088/0004-637X/760/2/116. http://arxiv.org/abs/1208.2323. http://iopscience.iop.org/article/10.1088/0004-637X/760/2/116/pdf.

43. Street R.A. et al., including **Ipatov S.** (137 co-authors), MOA-2010-BLG-073L: An M-dwarf with a substellar companion at the planet/brown dwarf boundary, *The Astrophysical Journal*, 2013, vol. 763, Issue 1, article id. 67, 13 pp. DOI: 10.1088/0004-637X/763/1/67. http://arxiv.org/abs/1211.3782. http://iopscience.iop.org/article/10.1088/0004-637X/763/1/67/pdf.

44. Hwang K.-H. et al., including **Ipatov S**. (84 co-authors). Interpretation of a short-term anomaly in the gravitational microlensing event MOA-2012-BLG-486, *The Astrophysical Journal,* 2013, v. 778, article id. 55, 6 p, DOI: 10.1088/0004-637X/778/1/55. http://arxiv.org/abs/1308.5762, http://iopscience.iop.org/article/10.1088/0004-637X/778/1/55/pdf.

45. Han C. et al., including **Ipatov S.** (73 co-authors). Microlensing discovery of a tight, low mass-ratio planetary-mass object around an old, field brown dwarf. *The Astrophysical Journal,* 2013, vol. 778, issue 1, article id. 38, 6 pp. DOI: 10.1088/0004-637X/778/1/38. http://arxiv.org/abs/1307.6335, http://iopscience.iop.org/article/10.1088/0004-637X/778/1/38/pdf.

46. Kains N. et al., including **Ipatov S.** (130 co-authors). A giant planet beyond the snow line in microlensing event OGLE-2011-BLG-0251, *Astronomy & Astrophysics*, 2013, Volume 552, id. A70, 10 pp. DOI: 10.1051/0004-6361/201220626. http://arxiv.org/abs/1303.1184, http://www.aanda.org/articles/aa/pdf/2013/04/aa20626-12.pdf.

47. Yee J.C. et al., including **Ipatov S.** (76 co-authors). MOA-2013-BLG-220Lb: Massive planetary companion to galactic-disk host, *The Astrophysical Journal*, 2014, v. 790, no 1, article id 14, 7 pp. DOI: 10.1088/0004-637X/790/1/14. http://arxiv.org/pdf/1403.2134v1.pdf, http://iopscience.iop.org/article/10.1088/0004-637X/790/1/14/pdf.

48. Tsapras Y. et al., including **Ipatov S.** (131 co-authors). Super-Jupiter orbiting a late-type star: A refined analysis of microlensing event OGLE-2012-BLG-0406*, Astrophysical Journal,* 2014, v. 782, issue 1, article id. 48, 9 pp., DOI: 10.1088/0004-637X/782/1/48. http://arxiv.org/abs/1310.2428, http://iopscience.iop.org/article/10.1088/0004-637X/782/1/48/pdf.

49. Fukui A. et al., including **Ipatov S.** (65 co-authors). A Saturn-mass planet around an M dwarf with the mass constrained by Subaru AO Imaging, *The Astrophysical Journal*, 2015. Volume 809, Issue 1, article id. 74, 16 pp. DOI: 10.1088/0004-637X/809/1/74. http://arxiv.org/abs/1506.08850, http://iopscience.iop.org/article/10.1088/0004-637X/809/1/74/pdf .

50. Kains N. et al., including **Ipatov S**. (36 co-authors). A Census of variability in globular cluster M68 (NGC 4590), *Astronomy & Astrophysics*, 2015. Volume 578, id. A128, 23 pp. DOI: 10.1051/0004-6361/201424600. http://arxiv.org/abs/1502.07345, http://www.aanda.org/articles/aa/pdf/2015/06/aa24600-14.pdf. Also published in VizieR On-line Data Catalog: J/A+A/578/A128. 2015. V. 357.

51. Ipatov S.I. Formation of trans-Neptunian satellite systems at the stage of condensations, *Solar System Research* (translated from Astronomicheskii Vestnik), 2017, v. 51, N 4, P. 294-314. DOI: 10.1134/S0038094617040013. https://arxiv.org/abs/1801.05217. https://link.springer.com/article/10.1134/S003809461704001. https://www.researchgate.net/publication/318595551. https://www.academia.edu/39245929.

52. Ipatov S.I. Origin of orbits of secondaries in the discovered trans-Neptunian binaries, *Solar System Research* (translated from Astronomicheskii Vestnik), 2017, v. 51, N 5, 409-416. doi 10.1134/S0038094617050045. https://arxiv.org/abs/1801.05254. https://link.springer.com/article/10.1134/S0038094617050045. https://www.researchgate.net/publication/319885905. https://www.academia.edu/39246033.

53. Marov M.Ya., Ipatov S.I. Delivery of water and volatiles to the terrestrial planets and the Moon, *Solar System Research*, 2018, v. 52, N 5, p. 392-400). DOI: 10.1134/S0038094618050052. https://arxiv.org/abs/2003.09982. https://link.springer.com/article/10.1134/S0038094618050052. https://www.academia.edu/39245748. https://www.researchgate.net/publication/340115418.

54. Ipatov S.I. Formation of embryos of the Earth and the Moon from the common rarefied condensation and their subsequent growth. *Solar System Research*, 2018, v. 52, N 5, p. 401-416. DOI 10.1134/S0038094618050040. https://doi.org/10.1134/S0038094618050040. https://link.springer.com/article/10.1134%2FS0038094618050040, WOS: 000451063100004. ISSN:

0038-0946. eISSN: 1608-3423; http://arxiv.org/abs/2003.09925. https://www.researchgate.net/publication/340115212.

55. Ipatov S.I. Probabilities of collisions of planetesimals from different regions of the feeding zone of the terrestrial planets with the forming planets and the Moon. *Solar System Research*, 2019, v. 53, N 5, p. 332-361. DOI: 10.1134/S0038094619050046. https://rdcu.be/bRVA8. http://arxiv.org/abs/2003.11301. https://link.springer.com/article/10.1134/S0038094619050046. https://www.researchgate.net/publication/340173252, https://www.academia.edu/42216860. https://www.webofscience.com/wos/woscc/full-record/WOS:000511210600003.

56. Ipatov S.I., Feoktistova, E.A., Svetsov V.V. Number of near-Earth objects and formation of lunar craters over the last billion years. *Solar System Research*, 2020, v. 54, N 5, P. 384-404. DOI: 10.1134/S0038094620050019. https://rdcu.be/b7ZdT. https://arxiv.org/abs/2011.00361. https://link.springer.com/article/10.1134/S0038094620050019. https://www.researchgate.net/publication/345500846, https://www.academia.edu/44441315. https://www.webofscience.com/wos/woscc/full-record/WOS:000574226000003.

57. Feoktistova E.A., Ipatov S.I. Depths of Copernican craters on lunar maria and highlands. *Earth, Moon, and Planets*, 2021. v. 125, Article id. 1, 23 p. https://doi.org/10.1007/s11038-021-09538-y. http://arxiv.org/abs/2103.00291. https://link.springer.com/article/10.1007/s11038-021-09538-y. https://www.researchgate.net/publication/349703674, https://www.academia.edu/49324295, https://www.webofscience.com/wos/woscc/full-record/WOS:000612895400001. https://elibrary.ru/item.asp?id=44979320.

58. Marov M.Ya, Ipatov S.I. Formation of the Earth and Moon: Influence of Small Bodies. *Geochemistry International*, 2021, Vol. 59, No. 11, pp. 1010–1017. ISSN 0016-7029, DOI: 10.1134/S0016702921110070, http://arxiv.org/abs/2112.06047. https://link.springer.com/article/10.1134/S0016702921110070, https://trebuchet.public.springernature.app/get_content/29ea1f0f-1e09-4ad2-806f-3589b28788c3. https://www.researchgate.net/publication/357014173, https://www.academia.edu/99306270, https://www.webofscience.com/wos/woscc/full-record/WOS:000728564800002 .

59. Marov M.Ya, Ipatov S.I. Migration processes in the Solar System and their role in the evolution of the Earth and planets. *Physics – Uspekhi*. 2023. V. 66, N 1, p. 2-31. http://www.mathnet.ru/php/archive.phtml?wshow=paper&jrnid=ufn&paperid=7321&option_lang=eng. https://doi.org/10.3367/UFNe.2021.08.039044. https://arxiv.org/abs/2309.00716, https://www.academia.edu/99306270/Migration_processes_in_the_Solar_System_and_their_role_in_the_evolution_of_the_Earth_and_planets. https://www.researchgate.net/publication/373685580, https://www.webofscience.com/wos/woscc/full-record/WOS:001112589600001.

60. Ipatov S.I. Stable orbits in the feeding zone of the planet Proxima Centauri c. *Solar System Research*, 2023, v. 57, N 3, P. 236-248. DOI: 10.1134/S0038094623030036. https://doi.org/10.1134/S0038094623030036. http://arxiv.org/abs/2309.00492, https://link.springer.com/article/10.1134/S0038094623030036. https://rdcu.be/df4dl. https://www.academia.edu/104388702/. https://www.researchgate.net/publication/373641778, https://www.webofscience.com/wos/woscc/full-record/WOS:001024449300005

61. Ipatov S.I. Delivery of icy planetesimals to inner planets in the Proxima Centauri planetary system. *Meteoritics and Planetary Science*. 2023. V. 58. P. 752-774. https://doi.org/10.1111/maps.13985, https://onlinelibrary.wiley.com/doi/10.1111/maps.13985. https://arxiv.org/abs/2309.00695, https://www.academia.edu/104387946/. https://www.researchgate.net/publication/373685582. https://www.webofscience.com/wos/woscc/full-record/WOS:000994800100001.

62. Ipatov S.I. Motion of planetesimals in the Hill sphere of the star Proxima Centauri // *Solar System Research*, 2023, v. 57, N 6. P. 612-628. https://link.springer.com/article/10.1134/S0038094623060047, https://doi.org/10.1134/S0038094623060047. http://arxiv.org/abs/2401.09086. https://www.researchgate.net/publication/377419089. https://www.academia.edu/113590871.

63. Ipatov S.I. Growth of the Moon due to bodies ejected from the Earth. *Solar System Research*, 2024, v. 58, N 1. P. 94-111. DOI: 10.1134/S0038094624010040. https://doi.org/10.1134/S0038094624010040, http://arxiv.org/abs/2405.19797. https://www.researchgate.net/publication/381005974, https://www.academia.edu/122432299. https://rdcu.be/dGWze.

64. Ipatov S.I. Migration of celestial bodies in the Solar system and in several exoplanetary systems. *Solar System Research*, 2024, v. 58, Suppl. 1, pp. S50-S63. https://doi.org/10.1134/S0038094623600105. https://rdcu.be/d04Q0. http://arxiv.org/abs/2411.05436.

https://www.researchgate.net/publication/385627526. https://www.academia.edu/125363162. https://www.webofscience.com/wos/woscc/full-record/WOS:001361601400008 .

65. Ipatov S.I. Probabilities of collisions of bodies ejected from forming Earth with the terrestrial planets. *Icarus*, 2025., v. 425, id. 116341. (24 p.) https://doi.org/10.1016/j.icarus.2024.116341. http://arxiv.org/abs/2411.04218. https://www.researchgate.net/publication/384776785. https://www.academia.edu/125337439

**Articles in peer-reviewed journals indexed only in the Russian Science Citation Index.**

66. Ipatov S.I. The origin of one of Kirkwood gaps, *Zemlya i Vselennaya* (Earth and Universe), N 1, pp. 86-88 (1991). https://publ.lib.ru/ARCHIVES/Z/"Zemlya_i_Vselennaya"_(jurnal)/. In Russian.

67. Ipatov S.I. Methods of choosing the pairs of contacting bodies for investigations of the evolution of discrete systems with binary interactions, *Matematicheskoe Modelirovanie* (Mathematical Modeling), v. 5, N 1, pp. 35-59 (1993). https://www.researchgate.net/publication/265479283. http://www.mathnet.ru/php/getFT.phtml?jrnid=mm&paperid=1948&what=fullt&option_lang=rus or http://www.mathnet.ru/php/getFT.phtml?jrnid=mm&paperid=1948&volume=5&year=1993&issue=1&fpage=35&what=fullt&option_lang=eng. In Russian.

68. Ipatov S.I. Migration of celestial bodies in the Solar System, *Astronomical and Astrophysical Transactions*, 1998, v. 15. pp. 241-247. http://images.astronet.ru/pubd/2008/09/28/0001230313/241-247.pdf, https://www.elibrary.ru/download/elibrary_29914114_10473458.pdf. https://elibrary.ru/download/elibrary_29914114_29135110.pdf. http://adsabs.harvard.edu/full/1998A%26AT...15..241I. https://www.researchgate.net/publication/233045896, https://www.academia.edu/44453108.

69. Ipatov A.I., Gorodnov A.V., **Ipatov S.I.**, Maryenko N.N., Petrov L.P., Skopintsev S.P. Study of amplitude-frequency spectra of acoustic and electromagnetic noise signals during fluid filtration in rocks. *Geofizika* (Geophysics, the journal of Euro-Asian geophysical society). 2004, N 2, 25-30. https://1drv.ms/w/c/c67d93a65f0a2a17/ERcqCl-mk30ggMbYBwAAAAABKBFoeNyKGn9pUonfilMyWg?e=Gncgdt - summary. https://elibrary.ru/item.asp?id=22294745 . In Russian.

70. Ipatov S.I. Migration of trans-Neptunian objects to the Earth, *Astronomical and Astrophysical Transactions,* 2005, v. 24, pp. 35-38. http://images.astronet.ru/pubd/2008/09/28/0001230841/35-38.pdf, https://www.researchgate.net/publication/228677695, https://elibrary.ru/download/elibrary_29768969_24725472.pdf. https://www.elibrary.ru/download/elibrary_29768969_98652474.pdf. https://www.academia.edu/49324809.

71. Ipatov S.I. Formation of celestial bodies with satellites at the stage of rarefied condensations. *Mechanics, Control and Informatics*. 2015. Vol. 7. No. 3 (56). pp. 386-399. http://www.iki.rssi.ru/books/2015marov.pdf. https://elibrary.ru/download/elibrary_23244550_77400796.pdf, https://1drv.ms/b/c/c67d93a65f0a2a17/ERcqCl-mk30ggMabBAAAAAABtnwd6SpCO6VP-kRL0SeBBw?e=ipTzwx, https://1drv.ms/b/c/c67d93a65f0a2a17/ERcqCl-mk30ggMYYDwAAAAABjMkb-hdCI4JuwHXCrzU5LA?e=dBZ0RU. In Russian

72. Ipatov S.I., Elenin L.V. Models for determining the probability of detection of near-Earth objects in different regions of the sky. *Ecological Bulletin of the Research Centers of the Black Sea Economic Cooperation.* 2017, No. 4, issue 3. pp. 69-75. http://vestnik.kubsu.ru/archive/2017-issue-4v3/. https://elibrary.ru/item.asp?id=32331347. http://vestnik.kubsu.ru/article/view/810. https://1drv.ms/b/c/c67d93a65f0a2a17/ERcqCl-mk30ggMYXDwAAAAABCSgKH3zrAtSPBAZjMiRkDA?e=L4dequ. In Russian.

73. Ipatov S.I. Collisions of planetesimals from different parts of the feeding zone the terrestrial planets with the forming planets and the Moon embryo. *INASAN Science Reports*, 2020, v. 5, N 3, p. 97-99. DOI: 10.26087/INASAN.2020.5.3.004. https://elibrary.ru/download/elibrary_44093126_92967049.pdf. http://www.inasan.ru/wp-content/uploads/2020/12/ntr53.pdf#page=15. In Russian.

74. Marov M.Ya., Ipatov S.I. Migration of planetesimals from the feeding zone of the giant planets to the terrestrial planets and the Moon. *INASAN Science Reports*, 2020, v. 5, N 3, p. 94-96. DOI: 10.26087/INASAN.2020.5.3.003. https://elibrary.ru/download/elibrary_44093125_77520272.pdf. http://www.inasan.ru/wp-content/uploads/2020/12/ntr53.pdf#page=12. In Russian.

75. Ipatov S.I., Feoktistova E.A., Svettsov V.V. Estimates of variations in the number of near-Earth objects based on the ages of lunar craters over the past billion years. *INASAN Science Reports*, 2020, v. 5,

N 3, p. 100-102. DOI: 10.26087/INASAN.2020.5.3.005. https://elibrary.ru/download/elibrary_44093127_37970864.pdf. http://www.inasan.ru/wp-content/uploads/2020/12/ntr53.pdf#page=18. In Russian.

76. Ipatov S.I. Contribution of T.M. Eneev to planetary cosmogony. *Zemlya i Vselennaya* (the Earth and Universe). 2024, V. 5, 60-66. DOI: 10.7868/S0044394824050050. https://www.researchgate.net/publication/389497841. https://www.academia.edu/127962691/. https://doi.org/10.7868/S0044394824050050 . https://elibrary.ru/download/elibrary_80373069_10087563.pdf. In Russian.

77. Ipatov S.I. Processes of migration in the Solar system and in several exoplanetary systems. *Zemlya i Vselennaya* (the Earth and Universe). 2025, № 4, 2025. № 4, p. 18-43. DOI: 10.7868/S0044394825040024. https://www.academia.edu/145078900. https://www.researchgate.net/publication/397870089 . https://www.elibrary.ru/item.asp?id=88754279. In Russian.

**Monograph in Russian**

78. Ipatov S.I. *Migration of celestial bodies in the solar system*. URSS Publishing House. 2000. 320 p. ISBN 5-8360-0137-5. Stereotype Publ.: LENAND. 2021. https://www.rfbr.ru/view_book/1277/. https://elibrary.ru/download/elibrary_46237738_86402718.pdf. https://elibrary.ru/item.asp?id=46237738 , http://booksee.org/book/1472075. https://www.researchgate.net/publication/369586110. https://www.academia.edu/49357889. https://1drv.ms/b/c/c67d93a65f0a2a17/ERcqCl-mk30ggMYaDwAAAAAB40JEdQ5xs141snrwH27c4w?e=mGeHv8.

**Sections in monographs** (79-83 in Russian)

79. Kremenetsky M.I., Ipatov A.I., **Ipatov S.I.** Co-author of sections 7.5.1.2 and 7.5.1.3 (pp. 312-315) in the monograph by A.I. Ipatov and M.I. Kremenetsky “*Geophysical and hydrodynamic control of hydrocarbon deposit development*”, Moscow, Academy Publishing House, 2005 (second edition in 2006, third edition in 2010).

80. Kremenetsky M.I., Ipatov A.I., **Ipatov S.I.** Mechanism of acoustic emission noise generation in a pore matrix due to compression in natural constrictions. Section 5.3.1.2 in the monograph by M.I. Kremenetsky, A.I. Ipatov “*Stationary hydrodynamic and geophysical monitoring of oil and gas field development*”. Golden Fund of Oil and Gas Science. Gazprom. 2018. pp. 474-476.

81. Kremenetsky M.I., Ipatov A.I., **Ipatov S.I.** Turbulent mechanism of occurrence of standing waves of acoustic noise in calibrated channels (resonance effect). Section. 5.3.1.3 in the monograph by M.I. Kremenetsky, A.I. Ipatov "*Stationary hydrodynamic and geophysical monitoring of oil and gas field development*". Golden Fund of Oil and Gas Science. Gazprom. 2018. pp. 476-478.

82. Ipatov S.I. Chapter 2. Formation of the Earth-Moon system. P. 34-84. In the monograph: Marov M.Ya., Voropaev S.A., **Ipatov S.I.**, Badyukov D.D., Slyuta E.N., Stennikov A.V., Fedolov V.S., Dushenko N.V., Sorokin E.M., Kronrod E.V. “*Formation of the Moon and early evolution of the Earth*”. URSS. 2019. 320 p. ISBN 978-5-9710-7283-6. https://urss.ru/cgi-bin/db.pl?lang=Ru&blang=ru&page=Book&id=257942#FF1– Contents and preface. https://elibrary.ru/item.asp?id=41589572

83. Kremenetsky M.I., Ipatov A.I., **Ipatov S.I.** The mechanism of occurrence of acoustic emission noise in the pore matrix due to compression in natural constrictions. Section. 3.3.2. pp. 331-333. In the book: Kremenetsky M.I., Ipatov A.I. *Application of production geophysical control for optimization of oil and gas fields*. Vol. 2. The role of hydrodynamic -geophysical monitoring in development management. M. Izhevsk. Institute computer research. 2020. 756 p. ISBN 978-5-4344-0887-5.

84. Dorofeeva V.A., Dunaeva A.N., **Ipatov S.I**., Kronrod V.A., Kronrod E.V., Kuskov O.L., Marov M.Ya., Rusol A.V. Studies of the problems of planetary cosmogony, geochemistry and cosmochemistry by methods of mathematical modeling // *Advances in Geochemistry, Analytical Chemistry, and Planetary Sciences*: 75th Anniversary of the Vernadsky Institute of the Russian Academy of Sciences. Ed. by V.P. Kolotov and N.S. Bezaeva. Springer. Cham. 2023. P. 263-295. DOI: 10.1007/978-3-031-09883-3_14. https://link.springer.com/chapter/10.1007/978-3-031-09883-3_14 . https://link.springer.com/content/pdf/10.1007/978-3-031-09883-3_14?pdf=chapter%20toc. https://www.academia.edu/102439508/. https://www.researchgate.net/publication/368919287 – pdf file with the paper. https://link.springer.com/book/10.1007/978-3-031-09883-3 - a whole book. (indexed in Scopus)

**Non-journal publications in scientific collections and periodicals indexed in WoS and/or Scopus**

85. Ipatov S.I. Computer modelling of the process of solar system formation, *Proc. Intern. IMACS*

*Conference on Mathematical modelling and applied mathematics* (June 18-23, 1990, Moscow). Ed. by A.A. Samarskii and M.P. Sapagovas. Elsevier. Amsterdam, 1992, pp. 245-252. https://www.academia.edu/44453062. https://www.amazon.com/Mathematical-Modelling-Applied-Mathematics-International/dp/0444891803. https://www.ebay.com/itm/296620800951. (indexed in WoS)

86. Ipatov S.I. Migration of matter from the Edgeworth-Kuiper and main asteroid belts to the Earth. Proceedings of the IAU Colloquium No 181 and COSPAR Colloquium No. 11 "*Dust in the solar system and other planetary systems*" (April 10-14, 2000, Canterbury, UK), ed. by S.F. Green, I.P. Williams, J.A.M. McDonnell and N. McBride, *COSPAR Colloquia Series*, Pergamon, 2002, V. 15, pp. 233-236, http://arXiv.org/format/astro-ph/0205250. https://elibrary.ru/item.asp?id=17810354, https://ntrs.nasa.gov/citations/20020067409. https://www.sciencedirect.com/science/article/abs/pii/S0964274902803481?via%3Dihub, DOI: 10.1016/S0964-2749(02)80348-1 (indexed in WoS и Scopus).

87. Ipatov S.I. Formation and migration of trans-Neptunian objects and asteroids. Proceedings of the conference "*Asteroids, comets, meteors, 2002*" (July 29 - August 2, 2002, Berlin), ed. by B. Warmbein. SP-500, European Space Agency, 2002, pp. 371-374, http://arXiv.org/format/astro-ph/0211618. https://www.academia.edu/44453287. https://ui.adsabs.harvard.edu/abs/2002ESASP.500..371I/abstract, https://articles.adsabs.harvard.edu/pdf/2002ESASP.500..371I. https://1drv.ms/b/c/c67d93a65f0a2a17/ERcqCl-mk30ggMaqDAAAAAABSyXFX04dUGtO2g1ER2_ZUw?e=OAfF8m (indexed in Scopus)

88. Ipatov S.I. Evolution of the Edgeworth-Kuiper belt. Highlights of astronomy JD4 ASP (Proceedings of JD-4 of *IAU General Assembly*, August 7-18, 2000, Manchester, UK), ed. H. Rickman, *IAU Symposia. Highlights of astronomy*, 2002. V. 12, pp. 247-248. https://www.cambridge.org/core/services/aop-cambridge-core/content/view/1588AB52442DCC86C4084C1C9EB5F1EC/S153929960001340Xa.pdf (indexed in WoS).

89. Ipatov S.I. Formation and migration of trans-Neptunian objects. *Proc. of international conference* «*Scientific Frontiers in Research of Extrasolar Planets*» (18-21 June 2002, Washington D.C., USA), D. Deming and S. Seager, eds., *ASP Conference Series*, 2003, v. 294, pp. 349-353. http://arXiv.org/format/astro-ph/0210131. http://adsabs.harvard.edu/full/2003ASPC..294..349I, https://articles.adsabs.harvard.edu/pdf/2003ASPC..294..349I (indexed in WoS).

90. Marov M.Ya., Ipatov S.I. Migration processes and volatiles delivery, *Proc. of the 213th IAU Symposium* "*Bioastronomy 2002: Life among the stars*" (8-12 July 2002, Great Barrier Reef, Australia). Edited by R. Norris, and F. Stootman. San Francisco: *Astronomical Society of the Pacific*, 2004, pp. 295-298. https://1drv.ms/b/c/c67d93a65f0a2a17/ERcqCl-mk30ggMb2BwAAAAABTMShPO8XhP5WWyFwxGP1MQ?e=pqGWye. https://www.cambridge.org/core/services/aop-cambridge-core/content/view/163E12E5568952F21757F517B6CE5EA5/S007418090019343Xa.pdf. https://articles.adsabs.harvard.edu/full/2004IAUS..213..295M. (indexed in WoS).

91. Ipatov S.I. Migration of celestial bodies in the Solar system. "*Planetary Systems in the Universe: Observation, Formation and Evolution*" *Proc. of IAU Symposium N 202*, August 7-10, 2000, Manchester, UK), Astronomical Society of the Pacific Conference Series, *IAU Symposium Volumes*, ed. by A. Penny, P. Artymowicz, A.-M. Lagrange, and S. Russell, 2004, pp. 190-192. http://articles.adsabs.harvard.edu/pdf/2004IAUS..202..190I. https://1drv.ms/b/c/c67d93a65f0a2a17/ERcqCl-mk30ggMbzBwAAAAABtxSmb15m7hpzVwylhGi3LQ?e=CA9dLF. https://www.cambridge.org/core/services/aop-cambridge-core/content/view/D1EE6B25B9435059539599D5FCE69125/S0074180900217816a.pdf (indexed in WoS).

92. Ipatov S.I. Formation and migration of trans-Neptunian objects, Proc. of the 14th Annual Astrophysics Conference in Maryland "*The Search for Other Worlds*" (October 13-14, 2003, University of Maryland, College Park, MD, USA), ed. by S.S. Holt and D. Deming, American Institute of Physics, *AIP Conference Proceedings*, 2004, Volume 713, pp. 277-280. http://arXiv.org/format/astro-ph/0401279. DOI 10.1063/1.1774538. http://dx.doi.org/10.1063/1.1774538. https://pubs.aip.org/aip/acp/article-abstract/713/1/277/720350/Formation-and-Migration-of-Trans-Neptunian-Objects?redirectedFrom=fulltext. (indexed in WoS).

93. Ipatov S.I., Mather J.C. Migration of small bodies and dust to the terrestrial planets, *Proc. of the IAU Colloq. N 197* "*Dynamics of populations of planetary systems*" (Belgrade, Serbia and Montenegro, 31

August – 4 September, 2004), ed. by Z. Knezevic and A. Milani, 2005, Proceedings of the International Astronomical Union, pp. 399-404. http://arXiv.org/format/astro-ph/0411005. https://1drv.ms/b/c/c67d93a65f0a2a17/ERcqCl-mk30ggMbwBwAAAAABhpLEF5J2NRwhrdvO6Lwxjg?e=zib1eD. or https://www.cambridge.org/core/services/aop-cambridge-core/content/view/C1924E278F759197F592DBAF63312A29/S1743921304008907a.pdf. https://doi.org/10.1017/s1743921304008907. (indexed in WoS и Scopus).

94. Ipatov S.I., Mather J.C. Migration of comets to the terrestrial planets, *Proceedings of the IAU Symposium No. 236* "*Near-Earth Objects, Our Celestial Neighbors: Opportunity and Risk*" (14-18 August 2006, Prague, Czech Republic), ed. by A. Milani, G.B. Valsecchi, & D. Vokrouhlický, Cambridge Univ. Press, Cambridge, Proceedings of the International Astronomical Union, 2007, pp. 55-64. DOI: 10.1017/S1743921307003067. http://arXiv.org/format/astro-ph/0609721 – initial version. http://articles.adsabs.harvard.edu/pdf/2007IAUS..236...55I. https://www.academia.edu/44452956. https://1drv.ms/b/c/c67d93a65f0a2a17/ERcqCl-mk30ggMbuBwAAAAABzP-q7brxfEWdsdrA-uGBzA?e=ZHpj5N, or http://www.journals.cambridge.org/action/displayFulltext?type=1&fid=997604&jid=&volumeId=&issueId=S236&aid=997596 (indexed in Scopus).

95. Madsen G.J., Reynolds R.J., **Ipatov S.I.**, Kutyrev A., Mather J.C., Moseley S.H. New observations and models of the kinematics of the zodiacal dust cloud, *Proceedings of the workshop "Dust in Planetary Systems*" (September 26-30, 2005, Kaua'i, Hawaii, USA), ed. by H. Krüger and A. Graps, *ESA Publications,* SP-643, 2007, v. 643. p. 61-64. http://articles.adsabs.harvard.edu/pdf/2007ESASP.643...61M. Initial version on http://arXiv.org/format/astro-ph/0604229. (indexed in WoS и Scopus).

96. Ipatov S.I., Mather J.C. Migration of dust particles to the terrestrial planets, *Proceedings of the workshop "Dust in Planetary Systems*" (September 26-30, 2005, Kaua'i, Hawaii, USA), ed. by H. Krüger and A. Graps, *ESA Publications*, SP-643, 2007, v. 643. pp. 91-94. ISBN: 92-9092-207-9. https://www.lpi.usra.edu/meetings/dust2005/pdf/4049.pdf - abstracts. http://www.mpi-hd.mpg.de/dustgroup/~graps/dips2005/370_Ipatov_FINAL.pdf, or https://1drv.ms/b/c/c67d93a65f0a2a17/ERcqCl-mk30ggMbvBwAAAAABuMA_vLJ3VItX7iMUc4P1Og?e=a74D5R, initial version on http://arXiv.org/format/astro-ph/0606434. https://articles.adsabs.harvard.edu/pdf/2007ESASP.643...91I (indexed in WoS и Scopus).

97. Boss A.P., **Ipatov S.I.,** Myhill E.A. Triggering presolar cloud collapse and injection of short-lived radioisotopes by a supernova shock wave, *70th Annual Meteoritical Society Meeting* (Tucson, Arizona, August 13-17, 2007). *Meteoritics and Planetary Science,* 2007, Vol. 42, Supplement, p. A23, #5011. https://www.lpi.usra.edu/meetings/metsoc2007/pdf/5011.pdf. WOS:000248755800029. (indexed in WoS).

98. Ipatov S.I. Angular momentum of two collided rarefied preplanetesimals and formation of binaries, *Proceedings of the International Astronomical Union*, IAU Symposium, IAU vol. 5, *Symposium S263, "Icy bodies in the Solar System*" (Rio de Janeiro, Brazil, 3-7 August, 2009), ed. by J.A. Fernandez, D. Lazzaro, D. Prialnik, R. Schulz, Cambridge University Press, 2010, pp. 37-40. https://doi.org/10.1017/S1743921310001468. http://arxiv.org/abs/1011.5544. https://1drv.ms/b/c/c67d93a65f0a2a17/ERcqCl-mk30ggMalDAAAAAABTkEoY0RWGjchGGeZPZGl4g?e=8kZol1. https://www.cambridge.org/core/services/aop-cambridge-core/content/view/FB1F3086D4A9C17A484D831192D83B4C/S1743921310001468a.pdf. https://articles.adsabs.harvard.edu/pdf/2010IAUS..263...37I (indexed in WoS).

99. Ipatov S.I. Collision probabilities of migrating small bodies and dust particles with planets, *Proceedings of the International Astronomical Union,* IAU Symposium, IAU vol. 5, *Symposium S263, "Icy bodies in the Solar System*" (Rio de Janeiro, Brazil, 3-7 August, 2009), ed. by J.A. Fernandez, D. Lazzaro, D. Prialnik, R. Schulz, Cambridge University Press, 2010, pp. 41-44. DOI: 10.1017/S174392131000147X. http://arxiv.org/abs/0910.3017. https://articles.adsabs.harvard.edu/pdf/2010IAUS..263...41I, https://www.cambridge.org/core/services/aop-cambridge-core/content/view/D26F868B87FD73D83AC561A83906A3FD/S174392131000147Xa.pdf. https://www.academia.edu/44453257. https://arxiv.org/abs/0910.3017 (indexed in WoS).

100. Ipatov S.I., A'Hearn M.F. Deep Impact ejection from Comet 9P/Tempel 1 as a triggered outburst, *Proceedings of the International Astronomical Union*, IAU Symposium, IAU vol. 5, *Symposium S263,* "*Icy bodies in the Solar System*" (Rio de Janeiro, Brazil, 3-7 August, 2009), ed. by J.A. Fernandez,

D. Lazzaro, D. Prialnik, R. Schulz, Cambridge University Press, 2010, pp. 317-321. doi:10.1017/S1743921310002000. https://articles.adsabs.harvard.edu/pdf/2010IAUS..263..317I, http://arxiv.org/abs/1011.5541. https://1drv.ms/b/c/c67d93a65f0a2a17/ERcqCl-mk30ggMakDAAAAAABEYNUCouSH_ETUtOLvP8jfg?e=9IQ7Hf. https://www.cambridge.org/core/services/aop-cambridge-core/content/view/8EFB2ACEAE31DAAE10057631239CB599/S1743921310002000a.pdf. (indexed in WoS).

101. Ipatov S.I. Cavities as a source of outbursts from comets (chapter 3), In "*Comets: Characteristics, Composition and Orbits*", ed. by Peter G. Melark, Nova Science Publishers, ISBN 978-1-61324-658-0, 2012, pp. 101-112, http://arxiv.org/abs/1103.0330. https://www.academia.edu/44453155. https://1drv.ms/b/c/c67d93a65f0a2a17/ERcqCl-mk30ggMahDAAAAAABy9WWhxYYpXcyCFOUoqOA-w?e=U9ytQM. (indexed in Scopus).

102. Ipatov S.I. Angular momenta of collided rarefied preplanetesimals, *Proc. IAU Symp. No. 293* "*Formation, detection, and characterization of extrasolar habitable planets*" (20-31 August 2012, Beijing, China), ed. by Nader Haghighipour, *Proceedings of the International Astronomical Union*, vol. 8, Symposium S293, Cambridge University Press. DOI: 10.1017/S1743921313013008, 2014, pp. 285-288, http://arxiv.org/abs/1412.8445. https://www.cambridge.org/core/services/aop-cambridge-core/content/view/491545577B112C7A6B7739EC73B61681/S1743921313013008a.pdf/angular_momenta_of_collided_rarefied_preplanetesimals.pdf. (indexed in Scopus).

103. Ipatov S.I., Horne K., Alsubai K., Bramich D., Dominik M., Hundertmark M., Liebig C., Snodgrass C., Street R., Tsapras Y. Simulator for Microlens Planet Surveys, *Proc. IAU Symp. No. 293 "Formation, detection, and characterization of extrasolar habitable planets*" ((20-31 August 2012, Beijing, China), ed. by Nader Haghighipour, Cambridge University Press. *Proceedings of the International Astronomical Union*, Vol. 8, Issue S293, 2014, pp. 416-419, DOI: 10.1017/S1743921313013306, http://dx.doi.org/10.1017/S1743921313013306. http://arxiv.org/abs/1308.6159. https://www.cambridge.org/core/services/aop-cambridge-core/content/view/41D9134736C10CB1172B91F8A4C431CA/S1743921313013306a.pdf/simulator_for_microlens_planet_surveys.pdf. https://www.academia.edu/44452515. (indexed in Scopus).

104. Ipatov S.I. Location of the upper border of the cavity excavated after the Deep Impact collision, *XXVIIIth IAU General Assembly*, August 2012, ed. Editor-in-Chief: Thierry Montmerle, Publisher: Cambridge University Press, *Highlights of Astronomy*, 2015, Vol. 16, p. 157. https://doi.org/10.1017/S1743921314005122. http://www.cambridge.org/it/academic/subjects/astronomy/astronomy-general/highlights-astronomy-volume-16. https://www.cambridge.org/core/services/aop-cambridge-core/content/view/1CC7856F6308098EC47343C2C760C0B9/S1743921314005122a.pdf. (indexed in Scopus).

105. Ipatov S.I. Formation and growth of the embryos of the Earth and the Moon. *81st Annual Meeting of the Meteoritical Society* (July 22-27, 2018, Moscow, Russia). LPI Contrib. No. 2067. *Meteoritics and Planetary Science*. 2018. V. 53. Issue S1, #6024, p. 114. https://www.hou.usra.edu/meetings/metsoc2018/pdf/6024.pdf. https://onlinelibrary.wiley.com/doi/epdf/10.1111/maps.13146 - a whole issue. An oral presentation. WOS: 000438212400115. (indexed in WoS).

106. Ipatov S.I. Migration of interplanetary dust particles to the Earth and the Moon. *81st Annual Meeting of the Meteoritical Society* (July 22-27, 2018, Moscow, Russia). LPI Contrib. No. 2067. *Meteoritics and Planetary Science*. 2018. V. 53. Issue S1, #6075, p. 115. https://www.hou.usra.edu/meetings/metsoc2018/pdf/6075.pdf. https://onlinelibrary.wiley.com/doi/epdf/10.1111/maps.13146 - a whole issue. A poster. WOS: 000438212400116. (indexed in WoS).

107. Ipatov S.I., Elenin L.V. Suggested models for calculation of the probabilities of detection of near-Earth objects in different sky regions. *81st Annual Meeting of the Meteoritical Society* (July 22-27, 2018, Moscow, Russia). LPI Contrib. No. 2067. *Meteoritics and Planetary Science*. 2018. V. 53. Issue S1, p. 116. https://www.hou.usra.edu/meetings/metsoc2018/pdf/6249.pdf. https://onlinelibrary.wiley.com/doi/epdf/10.1111/maps.13146 - a whole issue. An oral presentation. WOS: 000438212400117. (indexed in WoS).

108. Marov M.Ya., Ipatov S.I. Water and volatiles inventory from beyond Jupiter's orbit to the terrestrial planets and the Moon. *81st Annual Meeting of the Meteoritical Society* (July 22-27, 2018, Moscow, Russia). LPI Contrib. No. 2067. *Meteoritics and Planetary Science*. 2018. V. 53. Issue S1, #6144,

p. 187. https://www.hou.usra.edu/meetings/metsoc2018/pdf/6144.pdf. https://onlinelibrary.wiley.com/doi/epdf/10.1111/maps.13146 - a whole issue. An oral presentation. WOS: 000438212400188. (indexed in WoS).

109. Ipatov S.I. Accumulation of planetesimals by forming terrestrial planets from different regions of their feeding zone. *82nd Annual Meeting of the Meteoritical Society* (July 7-12, 2019, Sapporo, Hokkaido, Japan). https://www.hou.usra.edu/meetings/metsoc2019/pdf/6147.pdf. *Meteoritics and Planetary Science*. 2019. V. 54. Issue S2, #6147. P. 173. LPI Contribution No. 2157, id. 6147. https://onlinelibrary.wiley.com/toc/19455100/2019/54/S2. https://www.hou.usra.edu/meetings/metsoc2019/eposter/6147.pdf - e-poster. WOS:000472113600174. (indexed in WoS).

110. Ipatov S.I. Probabilities of collisions with the Earth and the Moon of planetesimals migrated from outside the orbit of Mars. *82nd Annual Meeting of the Meteoritical Society* (July 7-12, 2019, Sapporo, Hokkaido, Japan). https://www.hou.usra.edu/meetings/metsoc2019/pdf/6290.pdf. *Meteoritics and Planetary Science*. 2019. V. 54. Issue S2, #6290, p. 174. LPI Contribution No. 2157, id.6290. https://www.hou.usra.edu/meetings/metsoc2019/eposter/6290.pdf - e-poster. https://onlinelibrary.wiley.com/toc/19455100/2019/54/S2. WOS:000472113600175. (indexed in WoS).

111. Ipatov S.I. Formation of the Earth-Moon system. *Proc. IAU Symp. No. 345 "Origins: from the Protosun to the First Steps of Life*", ed. by Bruce G. Elmegreen, L. Viktor Tóth, Manuel Gudel, Proceedings of the International Astronomical Union, Vol. 14. Symposium S345, Cambridge University Press, 2020, 148-151. doi:10.1017/S1743921319001455. https://www.cambridge.org/core/journals/proceedings-of-the-international-astronomical-union/article/formation-of-the-earthmoon-system/2D426A2F9E04167ADD5F22AB653E8083, https://www.researchgate.net/publication/338568709, https://www.academia.edu/44453178. https://elibrary.ru/item.asp?id=43225106 (indexed in Scopus).

112. Marov M.Ya., **Ipatov S.I**. Water inventory from beyond the Jupiter orbit to the terrestrial planets and the Moon. *Proc. IAU Symp. No. 345 "Origins: from the Protosun to the First Steps of Life*", ed. by Bruce G. Elmegreen, L. Viktor Tóth, Manuel Gudel, Proceedings of the International Astronomical Union, Vol. 14. Symposium S345, Cambridge University Press, 2020, p. 164-167. doi:10.1017/S1743921319001479. https://www.cambridge.org/core/journals/proceedings-of-the-international-astronomical-union/article/water-inventory-from-beyond-the-jupiters-orbit-to-the-terrestrial-planets-and-the-moon/916B963BACB9DAAB9519EE71AE7EA059. https://www.researchgate.net/publication/338567896. https://www.academia.edu/44453190. https://www.elibrary.ru/item.asp?id=43232574. (indexed in Scopus).

113. Ipatov S.I., Feoktistova E.A., Svetsov V.V. Near-Earth object population and formation of lunar craters during the last billion of years. *Proc. IAU Symp. No. 345 "Origins: from the Protosun to the First Steps of Life"*, ed. by Bruce G. Elmegreen, L. Viktor Tóth, Manuel Gudel, Proceedings of the International Astronomical Union, Vol. 14. Symposium S345, Cambridge University Press, 2020, 299-300. doi:10.1017/S1743921319001467. https://www.cambridge.org/core/journals/proceedings-of-the-international-astronomical-union/article/nearearth-object-population-and-formation-of-lunar-craters-during-the-last-billion-of-years/E26746F910B82DD3B77E3ED5AA1033D3. https://www.researchgate.net/publication/336738161, https://www.academia.edu/44453199. https://elibrary.ru/item.asp?id=43235365 (indexed in Scopus).

114. Ipatov S.I. Delivery of bodies to the Earth and the Moon from the zone of the outer asteroid belt. *84th Annual Meeting of the Meteoritical Society* 2021 (LPI Contrib. No. 2609). *Meteoritics and Planetary Science*. 2021. V. 56. Issue S1, #6040, p. 113. LPI Contribution No. 2609, id.6040. a poster; https://onlinelibrary.wiley.com/doi/pdf/10.1111/maps.13727, https://www.webofscience.com/wos/woscc/full-record/WOS:000684014300114. (indexed in WoS).

115. Ipatov S.I. Delivery of icy planetesimals to inner planets in the Proxima Centauri system. *84th Annual Meeting of the Meteoritical Society* on August 15–21, 2021 (Chicago, Illinois), #6042, https://www.hou.usra.edu/meetings/metsoc2021/pdf/6040.pdf a poster. *Meteoritics and Planetary Science*. 2021. V. 56. Issue S1, #6042, p. 114. LPI Contribution No. 2609, id.6042. https://onlinelibrary.wiley.com/doi/pdf/10.1111/maps.13727, p. 114, https://doi.org/10.1111/maps.13727 – one DOI for all abstracts. https://www.webofscience.com/wos/woscc/full-record/WOS:000684014300115 (indexed in WoS).

116. Ipatov S.I. Mixing of planetesimals in the TRAPPIST-1 exoplanetary system. Abstracts of *85th Annual Meeting of The Meteoritical Society* (August 14–19, 2022, Glasgow, Scotland). - *Meteoritics and Planetary Science*. 2022. V. 57. Issue S1. P. A208.

https://www.hou.usra.edu/meetings/metsoc2022/pdf/6059.pdf. https://www.webofscience.com/wos/woscc/full-record/WOS:000834630400210. (indexed in WoS). A poster.

118. Ipatov S.I. Probabilities of collisions of bodies ejected from the Earth with planets and the Moon. Abstracts of *85th Annual Meeting of The Meteoritical Society* (August 14–19, 2022, Glasgow, Scotland). https://www.hou.usra.edu/meetings/metsoc2022/pdf/6104.pdf. *Meteoritics and Planetary Science*. 2022. Volume 57, Issue S1. P. A209. https://www.webofscience.com/wos/woscc/full-record/WOS:000834630400211. (indexed in WoS). A poster.

119. Ipatov S.I. Migration of dust from the orbit of planet Proxima Centauri c. Abstracts of *86th Annual Meeting of The Meteoritical Society* (August 13–18, 2023, Los Angeles, California, USA). - *Meteoritics and Planetary Science*. 2023. V. 58. Issue S1. P. A131. https://www.hou.usra.edu/meetings/metsoc2023/pdf/6078.pdf. https://www.webofscience.com/wos/woscc/full-record/WOS:001048626300132 (indexed in WoS).

120. Ipatov S.I. Scattering of planetesimals from the feeding zone of planet Proxima Centauri c. Abstracts of *86th Annual Meeting of The Meteoritical Society* (August 13–18, 2023, Los Angeles, California, USA). - *Meteoritics and Planetary Science*. 2023. V. 58. Issue S1. P. A132. https://www.hou.usra.edu/meetings/metsoc2023/pdf/6077.pdf . https://www.webofscience.com/wos/woscc/full-record/WOS:001048626300133 (indexed in WoS). A poster.

121. Ipatov S.I. Exchange of meteorites between the terrestrial planets. *86th Annual Meeting of the Meteoritical Society*. Jul 28-Aug 2, 2024. Brussels, Belgium. LPI Contribution No. 3036, 2024, id.6025. https://www.hou.usra.edu/meetings/metsoc2024/pdf/6025.pdf - abstract. *Meteoritics and Planetary Science*. 2024. V. 59. Supplement 1. P. A204-A204. WOS:001317679600205. https://www.academia.edu/126487408 (indexed in WoS).

122. Ipatov S.I. Migration of bodies to the Earth from different distances from the Sun // Proc. *IAU Symp. No. 374 "Astronomical Hazards for Life on Earth*" (the symposium was on August 9-11, 2022, Busan, Korea), ed. by Tancredi, G. Proceedings of the International Astronomical Union, Vol. 19. Symposium S374, Cambridge University Press, 2023. p. 49-54, DOI: https://doi.org/10.1017/S1743921324000735. http://arxiv.org/abs/2411.06777, https://www.researchgate.net/publication/398827571. https://www.academia.edu/145480012. (indexed in Scopus).

123. Ipatov S.I. Migration of bodies in the Proxima Centauri and Trappist-1 planetary systems. *Proc. IAU Symposium 393 "Planetary Science and Exoplanets in the Era of the James Webb Space Telescope"* (August 13-15, 2024, Cape Town, South Africa). Proceedings of the International Astronomical Union. V. 20, Symposium S393. 2024, pp. 29-33. https://doi.org/10.1017/S1743921324001546. https://www.researchgate.net/publication/399617826. http://arxiv.org/abs/2411.05954. https://www.academia.edu/145875624 (indexed in Scopus).

124. Ipatov S.I. Migration of bodies ejected from the Earth and the Moon. *Proc. IAU Symposium 393 "Planetary Science and Exoplanets in the Era of the James Webb Space Telescope"* (August 13-15, 2024, Cape Town, South Africa). Proceedings of the International Astronomical Union. Volume 20, Symposium S393. 2024. pp. 9-13. DOI: https://doi.org/10.1017/S1743921324001911. https://www.researchgate.net/publication/399618742. https://www.academia.edu/145875450. http://arxiv.org/abs/2411.04218 (indexed in Scopus).

**Non-journal publications in scientific collections and periodicals not indexed in WoS or Scopus**

125. Ipatov S.I. Axial rotation accumulating planets. "*O. Yu. Schmidt and Soviet Geophysics of the 1980s*". Moscow "Science". Ed. by M.A. Sadovsky, 1984. pp. 239-243. https://search.rsl.ru/ru/record/01001194033. In Russian.

126. Ipatov S.I. Numerical studies of planetary accumulation. "*Planetary cosmogony and Earth sciences*". Moscow: Nauka. Ed. by V.A. Magnitsky. 1989. Pp. 89-105. https://search.rsl.ru/ru/record/01001469164/. https://rusneb.ru/catalog/000200_000018_rc_339040/. In Russian.

127. Ipatov S.I. Computer Simulation of the Planetary Accumulation Process. *Origin of the Solar System. Kinetic and Thermodynamic Aspects*. Moscow. "Science". Ed. by A.V. Vityazev. 1993. Pp. 73-87. https://bik.sfu-kras.ru/elib/view?id=BOOK1-%D0%91%D0%91%D0%9A22.6/%D0%9F%20802-112348461. https://spblib.ru/ru/catalog/-/books/11198903-proiskhozhdeniye-solnechnoy-sistemy. In Russian.

128. Ipatov S.I. Evolution times for disks of planetesimals, "*The origin of the Solar System: Soviet*

*Research 1925-1991*". Ed. by A.E. Levin and S.G. Brush. American Inst. of Physics Press, through Oxford University Press. 1995. pp. 243-249. https://www.amazon.com/Origin-Solar-System-Research-1925-1991/dp/1563962810.

129. Marov M.Ya., **Ipatov S.I**. Volatile inventory and early evolution of planetary atmospheres, *Collisional processes in the solar system*, eds. M.Ya. Marov and H. Rickman, *Astrophysics and space science library*, Vol. 261, Dordrecht: Kluwer Academic Publishers, 2001, pp. 223-247. http://books.google.com/books?id=ICEFIUZmoSwC&pg=PA223&lpg=PA223&dq=ipatov&source=web&ots=tl62C5KVCi&sig=yxJPIsB74zjJtZbZfMfMVbQkWEM&output=html

130. Ipatov S.I. Formation and migration of trans-Neptunian bodies to the terrestrial planets. In the collection: "*Modern Problems of Space Mechanics and Physics*" (eds. V.S. Avduevsky and A.V. Kolesnichenko), Fizmatlit, Nauka, 2003, pp. 58-82. https://1drv.ms/b/c/c67d93a65f0a2a17/ERcqCl-mk30ggMawDAAAAAABaCGFa0F24wBCRD5-A-j6bA?e=yVAGgS/. https://elibrary.ru/item.asp?id=24056518. In Russian.

131. Cooray A. et al., including **Ipatov S.** (53 co-authors). A New Era in Extragalactic Background Light Measurements: The Cosmic History of Accretion, Nucleosynthesis and Reionization, the science white paper for the *US Astro2010-2020 Decadal Survey Committee*. http://arxiv.org/abs/0902.2372 , http://www8.nationalacademies.org/astro2010/publicview.aspx. https://www.researchgate.net/publication/24008955.

132. Ipatov S.I. September 23, 2024 marks the 100th anniversary of the birth of Academician of the Russian Academy of Sciences Timur Magometovich Eneev. On the website of the journal "*Science and Life*". https://www.nkj.ru/info/51055. 2024. In Russian.

133. Ipatov S.I. The Moon is moving away from us. On a new model of the Moon's formation. *Delphis* Magazine. 2024. No. 3. Pp. 39-42. http://www.delphis.ru/journal/article/soderzhanie-zhurnala-delfis-3-119-2024. https://www.academia.edu/126459146. https://www.researchgate.net/publication/388874304. In Russian.

134. Ipatov S.I. T.M. Eneev as a scientific supervisor and a person. *Delphis*. 2024, no. 4. pp. 50-54. http://www.delphis.ru/journal/article/soderzhanie-zhurnala-delfis-4-120-2024. https://www.academia.edu/127567432. https://www.researchgate.net/publication/388870640. In Russian.

**Publications in materials of scientific meetings (not less than 3 pages)**

135. Ipatov S.I. Asteroid-type orbit evolution near the 5:2 resonance, Proc. Intern. Conference "*Asteroids, comets, meteors 1991*" (June 24-28, 1991, Flagstaff). Ed. by A. Harris and E. Bowell, 1992, pp. 245-248. https://1drv.ms/b/c/c67d93a65f0a2a17/ERcqCl-mk30ggMb9BwAAAAAB0L7bG2_caS0wt5z10qaORg?e=HzrCSF or http://ntrs.nasa.gov/archive/nasa/casi.ntrs.nasa.gov/19930009982_1993009982.pdf; http://adsabs.harvard.edu/abs/1992acm..proc..245I - starting from this webpage you can get a free file with the paper; http://articles.adsabs.harvard.edu/full/1992acm..proc..245I - see each page separately.

136. Ipatov S.I. Migration of bodies to the Earth's orbit from the Kirkwood gaps and from the zones of giant planets. Proceedings of the All-Union Conference (with international participation) " *Asteroid Hazard* " (October 10-11, 1991, St. Petersburg). St. Petersburg. ed. A. G. Sokolsky. 1992. P. 121-125. In Russian.

137. Ipatov S.I. Migration of small bodies in the Solar system, *Worlds in interaction*: *Small bodies and planets of the Solar System*. Kluwer Academic Publishers. Dordrecht/Boston/London. (Proc. of the Meeting "*Small bodies in the Solar System and their interactions with the planets*" held in Mariehamn, Finland, August 8-12, 1994). Ed. by H. Rickman and M.J. Valtonen, 1996, pp. 211-214. http://adsabs.harvard.edu/full/1996EM%26P...72..211I.

138. Ipatov S.I. Migration of celestial bodies in the solar system. Proceedings of the International Scientific and Practical Conference "*Analysis of Systems on the Threshold of the 21st Century: Theory and Practice*" (February 27-29, 1996, Moscow). Intellect Publishing House. Moscow, 1997. Vol. 3, pp. 293-305. https://www.elibrary.ru/item.asp?edn=tggcvf.

139. Ipatov S.I. Migration of celestial bodies in the forming solar system, Proc. of 9th Rencontres de Blois (June 22-28, 1997) "*Planetary systems: the long view*", Editions Frontieres, Gif. sur Yvette Cedex, Ed. by L.M. Celnikier and J. Tran Thanh Van, 1998, pp. 93-94. https://books.google.ru/books?id=mBTs2a7hEIIC&hl=ru&source=gbs_navlinks_s.

140. Ipatov S.I. Migration of Kuiper-belt objects inside the solar system, Proc. of 9th Rencontres de Blois (June 22-28, 1997) "*Planetary systems: the long view*", Editions Frontieres, Gif. sur Yvette Cedex, Ed. by L.M. Celnikier and J. Tran Thanh Van, 1998, pp. 157-160.

https://1drv.ms/b/c/c67d93a65f0a2a17/ERcqCl-mk30ggMb8BwAAAAABSXUkCzFYsztGs5SzBJLA9g?e=pc34oh.
141. Ipatov S.I., Hahn G.J. Orbital evolution of the objects P/1996 R2 and P/1996 N2, Proc. of 9th Rencontres de Blois (June 22-28, 1997) "*Planetary systems: the long view*", Editions Frontieres, Gif. sur Yvette Cedex, Ed. by L.M. Celnikier and J. Tran Thanh Van, 1998, pp. 179-180. https://books.google.ru/books?id=mBTs2a7hEIIC&hl=ru&source=gbs_navlinks_s.
142. Ipatov S.I. Migration of trans-Neptunian objects to the Earth, Proc. of the IAU Colloquium 172 "*The impact of modern dynamics in astronomy*" (July 6-11, 1998, Namur, Belgium), Kluwer Academic Publishers, Dordrecht/Boston/London, Ed. by J. Henrard and S. Ferraz-Mello, 1999, pp. 107-116. http://adsabs.harvard.edu/full/1999imda.coll..107I. http://books.google.com/books?id=YQlrxy7u_JwC&pg=PA107&lpg=PA107&dq=ipatov&source=web&ots=eBP5m7qZ7a&sig=tR72OH8hMNq-aYFLqVh0R5a1e5A&output=html. https://www.cambridge.org/core/services/aop-cambridge-core/content/view/2855D111D053F887836683B40D3A185F/S0252921100072468a.pdf. DOI: 10.1007/978-94-011-4527-5_11, https://www.academia.edu/44448423/.
143. Ipatov S.I. Migration of trans-Neptunian bodies to the Earth. Proceedings of the Conf. " *Near-Earth Astronomy and Problems of Studying Small Bodies of the Solar System* " (October 25-29, 1999, Obninsk), Moscow, 2000, 151-162. https://search.rsl.ru/ru/record/01000662025. In Russian.
144. Ipatov S.I. Formation of trans-Neptunian objects and their migration to the Earth. Proc. of the international conference "*Comets, asteroids, meteors, meteorites, astroblems, craters*" "CAMMAC-99" (September 26 - October 1, 1999, Vinnitsa, Ukraine), 2000, pp. 41-46. https://1drv.ms/b/c/c67d93a65f0a2a17/ERcqCl-mk30ggMb7BwAAAAABg5DLCCZYaXUoP_Tf_6bgRQ?e=igUID0. https://ui.adsabs.harvard.edu/abs/2000camm.conf...41I/abstract.
145. Ipatov S.I. Formation of trans-Neptunian bodies and their migration to the Earth. Proceedings of the conference "*Near-Earth Astronomy of the XXI century (scientific and practical aspects)*" (May 21-25, 2001, Zvenigorod), Moscow, INASAN, 2001, pp. 388-400. https://1drv.ms/w/c/c67d93a65f0a2a17/ERcqCl-mk30ggMb_BwAAAAAByALHlh71dx1sDP7Xf4ABsA?e=60b9cJ. In Russian.
146. Ipatov S.I. Ozernoy L.M. Formation and evolution of the trans-Neptunian belt and dust, Proc. of International Conference "*Kazan Astronomy 2001*" (September 24-29, 2001, Kazan, Russia), 2001, pp. 137-142, http://arXiv.org/format/astro-ph/0107591. https://1drv.ms/w/c/c67d93a65f0a2a17/ERcqCl-mk30ggMb6BwAAAAABY9lp3VhC61nVd41DYljiBg?e=4zOZrT.
147. Ipatov S.I. Migration of Tunguska-type objects from the trans-Neptunian belt to the Earth, *Proc. of the Third International Aerospace Congress* (August 23-27, 2000, Moscow), both Russian and English texts, 2003. https://1drv.ms/w/c/c67d93a65f0a2a17/ERcqCl-mk30ggMb1BwAAAAABNCSLKQ0waFeeay06o4oZ_w?e=Y9P3k6
148. Ipatov S.I., Mather J.C. Migration of Jupiter-family comets and resonant asteroids to near-Earth space, Proc. of the international conference "*New trends in astrodynamics and applications*" (20-22 January 2003, University of Maryland, College Park), 2003, CD-ROM http://arXiv.org/format/astro-ph/0303219.
149. Ipatov S.I., Mather J.C., Taylor P.A. Migration of asteroidal dust particles, Proc. of the international conference "*New trends in astrodynamics and applications*" (20-22 January 2003, University of Maryland, College Park), CD-ROM, 2003, http://arXiv.org/format/astro-ph/0303398.
150. Ipatov S.I. and Mather J.C., Migration of trans-Neptunian objects to the terrestrial planets, in EKO [Proceedings of the international scientific workshop on the "*First Decadal Review of the Edgeworth-Kuiper Belt - Towards New Frontiers*" (11-14 March 2003, Antofagasta, Chile)], p. 89-98. Kluwer, reprinted from Earth, Moon, and Planets. DOI: 10.1007/978-94-017-3321-2_8, https://doi.org/10.1007/978-94-017-3321-2_8, https://link.springer.com/chapter/10.1007/978-94-017-3321-2_8, https://www.semanticscholar.org/paper/Migration-of-Trans-Neptunian-Objects-to-the-Planets-Ipatov-Mather/426d16d232eaae19b4410bbcbfdc82ad8834fd51
151. Ipatov S.I. Migration of comets to near-Earth space, Proc. of the Joint International Scientific Conference "*New Geometry of Nature: Mathematics, Mechanics, Geophysics, Astronomy & Biology*" (August 25 – September 5, 2003, the Kazan University, Kazan, Russia), V. 3 "Astronomy", Kazan University, 2003, pp. 94-101. https://1drv.ms/w/c/c67d93a65f0a2a17/ERcqCl-mk30ggMb3BwAAAAABhDFEvGrsJtYqcpI-idariw?e=cQgoxI.
152. Ipatov S.I. Migration of interplanetary dust, Proc. of the international conference "*Near-Earth Astronomy – 2003*" (NEA-2003) (8-13 September, 2003, Terskol, Russia), vol. 1, Saint-Petersburg,

Institute of chemistry, 2003, p. 81-86. https://1drv.ms/w/c/c67d93a65f0a2a17/ERcqCl-mk30ggMb5BwAAAAABmUGxgw2q8IC48EhzXHs-uA?e=9lKmcA.

153. Ipatov S.I. Migration of trans-Neptunian objects to the terrestrial planets, Proc. of the international conference "*Near-Earth Astronomy – 2003*" (NEA-2003) (8-13 September, 2003, Terskol, Russia), vol. 1, Saint-Petersburg, Institute of chemistry, 2003, p. 87-94. https://1drv.ms/w/c/c67d93a65f0a2a17/ERcqCl-mk30ggMb5BwAAAAABmUGxgw2q8IC48EhzXHs-uA?e=RKWUGH.

154. Ipatov S.I. Formation and migration of trans-Neptunian objects, *Proceedings of the Second TPF/Darwin International Conference* (San Diego, CA, July 26-29, 2004), 2004. https://1drv.ms/w/c/c67d93a65f0a2a17/ERcqCl-mk30ggMb0BwAAAAABw2oZK95o5broCatCBkY7pA?e=PJuSKc

155. Ipatov S.I. Sources of zodiacal dust, Proceedings of the conference "*Near-Earth astronomy-2007*" (Terskol, Russia, 3-7 September 2007), ed. by L.V. Rykhlova & V.K. Tarady, Nalchik, 2008, p. 132-138. http://arxiv.org/abs/0712.2624. https://1drv.ms/b/c/c67d93a65f0a2a17/ERcqCl-mk30ggManDAAAAAABg2XSiZqQTvCtA2bYIFtXzg?e=pCC96r.

156. Ipatov S.I. Formation of small-body binaries at the stage of rarefied preplanetesimals, Proceedings of the conference *"Near-Earth astronomy - 2009"* (Kazan, Russia, 23-27 August, 2009). Moscow: GEOS. Ed. by L.V. Rykhlova, V.V. Emelyanenko, E.S. Bakanas. 2010, pp. 203-208. https://1drv.ms/w/c/c67d93a65f0a2a17/ERcqCl-mk30ggMamDAAAAAABB_9c3xveoFBGk-eIHaFM2A?e=AJq7qZ. http://agora.guru.ru/display.php?conf=oza-2009&page=program&PHPSESSID=6ifgl49ih14machq21if7f1q54 – program.

157. Ipatov S.I. Formation of embryos of the Earth-Moon system as a result of a collision of two rarefied condensations, "*The Fifth Moscow Solar System Symposium 5M-S3*" (Space Research Institute, Moscow, Russia, October 13-18, 2014), 5MS3-MN-01, book of abstracts, CD-ROM, http://ms2014.cosmos.ru/sites/ms2014.cosmos.ru/files/5m-s3_abstract_book.pdf, 2014, pp. 30ab-32ab, http://arxiv.org/abs/1412.8453, an **oral** presentation.

158. Ipatov S.I. The Earth-Moon system as a typical binary in the Solar System, "*SPACEKAZAN-IAPS-2015*", ed. by. M.Ya. Marov, The International Space Forum "SPACEKAZAN-IAPS-2015" (June 1-7, 2015, Kazan. Russia), Publishing house of Kazan University, 2015, pp. 97-105. http://arxiv.org/abs/1607.07037. https://1drv.ms/b/c/c67d93a65f0a2a17/ERcqCl-mk30ggMYeCAAAAAAB1f5IeBKRMHuOFECfCykccw?e=bnhc7l.

159. Ipatov S.I., Marov M.Ya. Migration of small bodies and dust to the terrestrial planets. Proc. of the int. conference " *Near-Earth Astronomy-2015* " (August 31 – September 5, 2015, Terskol), eds. B.M. Shustov, L.V. Rykhlova, E.S. Bakanas, A.P. Kartashova. Moscow Janus-K. 2015. Pp. 6-11. Oral presentation. https://1drv.ms/w/c/c67d93a65f0a2a17/ERcqCl-mk30ggMaTDAAAAAABZC1D4pPudvT9uprtQzgQ9A?e=dqJJF3. In Russian.

160. Ipatov S.I. Origin of orbits of secondaries in discovered trans-Neptunian binaries, "*The Sixth Moscow Solar System Symposium 6M-S3*" (Space Research Institute, Moscow, Russia, October 5-9, 2015), 6MS3-SB-10, CD-ROM, http://ms2015.cosmos.ru/sites/ms2015.cosmos.ru/files/6m-s3_abstract_book.pdf , 2015, pp. ab-74 - ab-76 (3 pages), or CD-ROM, an **oral** presentation.

161. Marov M.Ya., **Ipatov S.I.** Mechanics of space processes of migration of small bodies in the Solar system: Nature and models. *Proceedings of the XI International Conference on Nonequilibrium Processes in Nozzles and Jets* (NPNJ'2016, May 25-31, 2016, Alushta, Russia). Moscow: MAI, 2016. Pp. 359-362. Plenary presentation. http://www.npnj.ru/files/npnj2016_web.pdf , https://elibrary.ru/download/elibrary_27460266_61827676.pdf. In Russian.

162. Marov M.Ya., **Ipatov S.I.** Water and volatiles on the Earth: the problem of origin. Seventeenth International Conference " *Physicochemical and Petrophysical Research in Geosciences* " (Moscow, September 26-28, Borok, September 30, 2016). Conference Proceedings. Moscow. IGEM RAS. 2016. P. 223-226. https://elibrary.ru/item.asp?id=26657334. http://www.igem.ru/petromeeting_XVII/docs/sbornik_2016.pdf. In Russian.

163. Ipatov S.I., Marov M.Ya. Delivery of water and planetesimals from the feeding zone of Jupiter and Saturn to forming terrestrial planets, "*The Seventh Moscow Solar System Symposium 7M-S3*" (Space Research Institute, Moscow, Russia, October 10-14, 2016), 7MS3-AB-05, 115-ab – 117-ab (3 pages), an **oral** presentation. http://ms2016.cosmos.ru/sites/ms2016.cosmos.ru/files/7ms3-2016_abstract_book_www.pdf

164. Ipatov S.I. Formation of the Earth and Moon embryos from a common condensation and their subsequent growth. Eighteenth International Conference " *Physicochemical and Petrophysical Research in*

*Earth Sciences* " (Moscow, October 2-4, Borok, October 6, 2017). Conference Proceedings. Moscow. IGEM RAS. 2017. P. 122-125. http://www.igem.ru/petromeeting_XVIII/tbgdocs/sbornik_2017.pdf. https://elibrary.ru/item.asp?id=26657334. In Russian.

165. Marov M.Ya., **Ipatov S.I**. Heterogeneous accretion: some results of the computer modeling, "*The Eighth Moscow Solar System Symposium 8M-S3*" (Space Research Institute, Moscow, Russia, October 9-13, 2017), 2017, 8MS3-PA-01, p. 1-3, https://ms2017.cosmos.ru/docs/8m-s3_abstract_book_2.pdf . An oral presentation.

166. Ipatov S.I. Formation and growth of embryos of the Earth-Moon system, "*The Eighth Moscow Solar System Symposium 8M-S3*" (Space Research Institute, Moscow, Russia, October 9-13, 2017), 2017, 8MS3-PS-36, p. 287-289, https://ms2017.cosmos.ru/docs/8m-s3_abstract_book_2.pdf . A poster.

167. Ipatov S.I. Sources of objects falling onto the Earth. *Geography and geoecology in the service of science and innovative education:* Proceedings of the XIII All-Russian scientific and practical conference with international participation dedicated to the 70th anniversary of the Museum of Geology and Geosciences of KSPU named after V.P. Astafiev, the 110th anniversary of Mikhail Vasilyevich Kirillov's birth, and the 110th anniversary of the Tunguska phenomenon. Krasnoyarsk, April 20, 2018. / ed. M.V. Prokhorchuk, ed. staff; Krasnoyarsk State Pedagogical University named after V.P. Astafiev. – Krasnoyarsk, 2018. – Issue. 13. pp. 114-116. ISBN 978-5-00102-205-3. Indexed in the RSCI. https://elibrary.ru/item.asp?id=32854839. In Russian.

168. Marov M.Ya., **Ipatov S.I**. Calculations of migration of small bodies in the Solar system. *Proceedings of the XII International Conference on Applied Mathematics and Mechanics in the Aerospace Industry* (NPNJ '2018) (May 24-31, 2018, Alushta, Crimea). Publisher: MAI Publishing House, Moscow. http://www.npnj.ru/files/npnj2018_web.pdf. P. 414-416. https://elibrary.ru/item.asp?id=35651552 . ISBN 978-5-4316-0491-1. Oral presentation. In Russian.

169. Ipatov S.I. Probabilities of collisions of bodies from the feeding zone of the terrestrial planets with the planets and the Moon. Nineteenth International Conference " *Physicochemical and Petrophysical Research in Earth Sciences* " (Moscow, September 24-30, 2018). Conference Proceedings. Moscow: IGEM RAS. 2018. Pp. 134-137. Poster presentation. ISBN 978-5-88918-053-1. https://elibrary.ru/item.asp?id=36060214. In Russian.

170. Ipatov S.I. Migration of small bodies to the Earth and the Moon from different distances from the Sun. Nineteenth International Conference " *Physicochemical and Petrophysical Research in Earth Sciences* " (Moscow, September 24-30, 2018). Conference Proceedings. Moscow: IGEM RAS. 2018. Pp. 138-141. Oral presentation. Indexed in the Russian Science Citation Index. https://elibrary.ru/item.asp?id=36060221. ISBN 978-5-88918-053-1. In Russian.

171. Ipatov S.I. Migration of dust particles to the Earth and the Moon from the zone of Jupiter and Saturn. Nineteenth International Conference " *Physicochemical and Petrophysical Research in Earth Sciences* " (Moscow, September 24-30, 2018). Conference Proceedings. Moscow: IGEM RAS. 2018. Pp. 142-145. Poster presentation. https://elibrary.ru/item.asp?id=36060225. ISBN 978-5-88918-053-1. In Russian.

172. Ipatov S.I. Migration of bodies to the Earth and the Moon from different distances from the Sun. *The Ninth Moscow Solar System Symposium 9M-S3* (Space Research Institute, Moscow, Russia, October 8-12, 2018). https://ms2018.cosmos.ru/, 2018, # 9MS3-SB-11, p. 104-106. an oral presentation. ISBN 978-5-00015-008-5. https://elibrary.ru/download/elibrary_37178225_95329110.pdf, https://www.academia.edu/44453329.

173. Ipatov S.I. Migration of interplanetary dust particles to the Earth and the Moon. *The Ninth Moscow Solar System Symposium 9M-S3* (Space Research Institute, Moscow, Russia, October 8-12, 2018). https://ms2018.cosmos.ru/, # 9MS3-DP-01, p. 144-146, 3 pages, an oral presentation. ISBN 978-5-00015-008-5. https://elibrary.ru/download/elibrary_37178447_89380013.pdf, https://www.academia.edu/44453504.

174. Busarev V.V., **Ipatov S.I**. Observational evidences and possible dynamical reasons of sublimation activity of primitive asteroids in the main-belt. *The Ninth Moscow Solar System Symposium 9M-S3* (Space Research Institute, Moscow, Russia, October 8-12, 2018). https://ms2018.cosmos.ru/, 2018, 9MS3-PS-48, p. 293-295, 3 pages, a poster. Стендовый доклад. ISBN 978-5-00015-008-5. https://elibrary.ru/download/elibrary_37178457_93734719.pdf.

175. Ipatov S.I., Feoktistova E.A., Svetsov V.V. Variation of near-Earth object population based on analysis of diameters of lunar craters. *The Ninth Moscow Solar System Symposium 9M-S3* (Space Research Institute, Moscow, Russia, October 8-12, 2018). https://ms2018.cosmos.ru/, 2018, 9MS3-PS-77. P. 349-

351, 3 pages, a poster. ISBN 978-5-00015-008-5. https://elibrary.ru/download/elibrary_37178333_94707009.pdf

176. Feoktistova E.A., **Ipatov S.I.**, Svetsov V.V. Lunar craters formed by encounters of satellite systems of near-Earth objects with the Moon. *The Ninth Moscow Solar System Symposium 9M-S3* (Space Research Institute, Moscow, Russia, October 8-12, 2018). https://ms2018.cosmos.ru, 2018, 9MS3-PS-79, p. 352-355, 4 pages, a poster. ISBN 978-5-00015-008-5. https://elibrary.ru/download/elibrary_37178336_58137643.pdf

177. Ipatov S.I. Probabilities of collisions of planetesimals from different regions of the feeding zone of the terrestrial planets with the forming planets and the Moon. *Proceedings of the All-Russian Annual Seminar on Experimental Mineralogy, Petrology and Geochemistry*. (VESEMPG-2019). April 16-17, 2019. Moscow. http://intranet.geokhi.ru/rasempg/Shared%20Documents/2019/%D0%A2%D1%80%D1%83%D0%B4%D1%8B%20%D0%92%D0%95%D0%A1%D0%AD%D0%9C%D0%9F%D0%93-2019.pdf. pp. 285-288. https://elibrary.ru/item.asp?id=41404886 (these are proceedings, the status of the presentation is given next to the abstracts). In Russian.

178. Ipatov S.I. Probabilities of collisions of planetesimals migrated from beyond the orbit of Mars with the Earth and the Moon. *Proceedings of the All-Russian Annual Seminar on Experimental Mineralogy, Petrology and Geochemistry*. (VESEMPG-2019). April 16-17, 2019, Moscow. http://intranet.geokhi.ru/rasempg/Shared%20Documents/2019/%D0%A2%D1%80%D1%83%D0%B4%D1%8B%20%D0%92%D0%95%D0%A1%D0%AD%D0%9C%D0%9F%D0%93-2019.pdf . pp. 289-292. https://elibrary.ru/item.asp?id=41404887 (these are works, the status of the presentation is given next to the abstracts). In Russian.

179. Ipatov S.I., Feoktistova E.A., Svettsov V.V. The number of near-Earth objects and the formation of lunar craters over the last billion years. *Proceedings of the All-Russian Annual Seminar on Experimental Mineralogy, Petrology and Geochemistry.* (VESEMPG-2019). April 16-17, 2019. Moscow. http://intranet.geokhi.ru/rasempg/Shared%20Documents/2019/%D0%A2%D1%80%D1%83%D0%B4%D1%8B%20%D0%92%D0%95%D0%A1%D0%AD%D0%9C%D0%9F%D0%93-2019.pdf. pp. 293-296. https://elibrary.ru/item.asp?id=41404888. https://www.elibrary.ru/download/elibrary_41404888_17746407.pdf. In Russian.

180. Ipatov S.I. Mixing of planetesimals in the feeding zone of the terrestrial planets. *Twentieth International Conference "Physicochemical and Petrophysical Research in Earth Sciences"* (Moscow, September 23-25, 2019). Conference Proceedings. Moscow: IGEM RAS. 2019. Pp. 122-125. 4 p. Poster presentation. ISBN 978-5-88918-056-2. http://www.igem.ru/petromeeting_XX/tbgdocs/sbornik_2019.pdf, https://elibrary.ru/download/elibrary_40938725_91893136.pdf. https://elibrary.ru/item.asp?id=40938725. In Russian.

181. Ipatov S.I., Feoktistova E.A., Svettsov V.V. Changes in the number of near-Earth objects and lunar craters over the last billion years. *Twentieth International Conference "Physicochemical and Petrophysical Research in Earth Sciences* " (Moscow, September 23-25, 2019). Conference Proceedings. Moscow: IGEM RAS. 2019. pp. 126-129. ISBN 978-5-88918-056-2. Oral presentation. http://www.igem.ru/petromeeting_XX/tbgdocs/sbornik_2019.pdf. https://elibrary.ru/item.asp?id=40938726, https://elibrary.ru/download/elibrary_40938726_66245553.pdf. In Russian.

182. Marov M.Ya., **Ipatov S.I.** Migration of planetesimals to the terrestrial planets and the Moon from beyond the orbit of Mars. *Twentieth International Conference "Physicochemical and Petrophysical Research in Earth Sciences* " (Moscow, September 23-25, 2019). Conference Proceedings. Moscow: IGEM RAS. 2019. Pp. 226-229. Oral presentation. ISBN 978-5-88918-056-2. http://www.igem.ru/petromeeting_XX/tbgdocs/sbornik_2019.pdf, https://elibrary.ru/item.asp?id=40938778. In Russian.

183. Marov M.Ya., Ipatov S.I. Migration of planetesimals from different distances outside Mars' orbit to the terrestrial planets and the Moon. *The Tenth Moscow Solar System Symposium 10M-S3* (Space Research Institute, Moscow, Russia, October 7-11, 2019). https://ms2019.cosmos.ru/docs/10m-s3-abstract-book.pdf, # 10MS3-SB-02, 3 стр. Vol. 1, p. 171-173, ISBN-978-5-00015-028-3. https://ms2019.cosmos.ru/docs/10m-s3-abstract-book.pdf, p. 189-191. an oral presentation. https://elibrary.ru/item.asp?id=41829253. https://elibrary.ru/download/elibrary_41829253_49027230.pdf.

184. Ipatov S.I. Probabilities of collisions of planetesimals from different parts of the feeding zone of the terrestrial planets with the forming planets, the Moon, and their embryos. *The Tenth Moscow Solar System Symposium 10M-S3* (Space Research Institute, Moscow, Russia, October 7-11, 2019). # 10MS3-

SB-04, Vol. 1, p. 177-179, 3 стр. ISBN-978-5-00015-028-3, https://ms2019.cosmos.ru/docs/10m-s3-abstract-book.pdf, p. 195-197. an oral presentation. https://elibrary.ru/item.asp?id=41829255. https://elibrary.ru/download/elibrary_41829255_76116769.pdf.

185. Ipatov S.I. Angular momenta of colliding rarefied condensations and formation of the Earth-Moon system. *The Tenth Moscow Solar System Symposium 10M-S3* (Space Research Institute, Moscow, Russia, October 7-11, 2019). # 10MS3-PS-56, 3 стр., vol. 2, p. 126-128. ISBN-978-5-00015-028-3. a poster presentation. https://ms2019.cosmos.ru/docs/10m-s3-abstract-book.pdf, p. 386-388. https://elibrary.ru/item.asp?id=41829343.

186. Busarev V.V., **Ipatov S.I**. Estimation of the fraction of ice material delivered to the Main asteroid belt in the early Solar system. *The Tenth Moscow Solar System Symposium* 10M-S3 (Space Research Institute, Moscow, Russia, October 7-11, 2019). # 10MS3-PS-75, 2019. Vol. 2, p. 436-438, 3 стр., ISBN-978-5-00015-028-3, a poster presentation. https://ms2019.cosmos.ru/docs/10m-s3-abstract-book.pdf, https://elibrary.ru/item.asp?id=41829362. https://www.researchgate.net/publication/337869943.

187. Ipatov S.I. The contribution of T.M. Eneev to planetary cosmogony. *XLIV Academic readings on cosmonautics* (January 28-31, 2020, Moscow). Publishing house of Bauman Moscow State Technical University. ISBN 978-5-7038-5343-6 (vol. 1). 2020. Vol. 1. P. 251-254. http://www.korolevspace.ru/sites/default/files/uploads/Abstracts_44_2020_Vol1.pdf. https://elibrary.ru/item.asp?id=42702351. Invited presentation. In Russian.

188. Ipatov S.I. "Migration of exoplanetesimals in system Proxima Centauri". Twenty-first International Conference "*Physicochemical and Petrophysical Research in Earth Sciences*" (Moscow, September 21-23, 25, 2020). Conference Proceedings. Moscow: IGEM RAS. 2020. pp. 101-104. ISBN 978-5-88918-059-3, ISSN 2686-8938. Poster presentation. https://elibrary.ru/item.asp?id=44139396, https://elibrary.ru/download/elibrary_44139396_54723641.pdf - conference. https://elibrary.ru/download/elibrary_44835907_68393380.pdf – my abstracts. In Russian.

189. Ipatov S.I., Marov M.Ya. Migration of planetesimals to the Earth from the outer asteroid belt zone. *Twenty-first international conference "Physicochemical and petrophysical studies in Earth sciences*" (Moscow, September 21-24, 2020). Conference proceedings. Moscow: IGEM RAS. 2020. 105-108. ISBN 978-5-88918-059-3, ISSN 2686-8938. Oral presentation. https://www.elibrary.ru/item.asp?id=44835908 – the entire conference. https://elibrary.ru/download/elibrary_44835908_21183755.pdf – my abstracts. In Russian.

190. Ipatov S.I., Feoktistova E.A., Svetsov V.V. Estimates of the number of near-Earth objects based on the number of lunar craters formed during the last billion years. *11th Moscow International Solar System Symposium*, 5-9 October 2020, IKI, Moscow, Russia, 11MS3-MN-PS-06, (p. 206-208, https://ms2020.cosmos.ru/docs/Abstract_book_full_version_05.pdf), (p. 191-193, a hard copy), (p. 211-213, https://www.elibrary.ru/item.asp?id=45672827&pff=1, https://elibrary.ru/download/elibrary_45672827_29294412.pdf ) 3 pages, (DOI: 0.21046/11MS3-2020 – for a whole book), a poster. https://www.youtube.com/watch?v=DlSfYoy-tlA – day 3 of the conference, my presentation was at the end (without sounds).

191. Feoktistova E.A., **Ipatov S.I**. Depths of the lunar Copernicans craters located on lunar maria and highlands. *11th Moscow International Solar System Symposium*, 5-9 October 2020, IKI, Moscow, Russia, 11MS3-MN-PS-09, (p. 214-216, https://ms2020.cosmos.ru/docs/Abstract_book_full_version_05.pdf), (p. 199-201, a hard copy), (p. 219-221, https://www.elibrary.ru/item.asp?id=45672830&pff=1, https://elibrary.ru/download/elibrary_45672830_45675027.pdf), (DOI: 0.21046/11MS3-2020 – for a whole book), 3 pages, a poster.

192. Marov M.Ya., **Ipatov S.I**. Migration of planetesimals to the Earth from the zone of the outer asteroid belt. *11th Moscow International Solar System Symposium*, 5-9 October 2020, IKI, Moscow, Russia, 11MS3-SM-11, (p. 277-279, https://ms2020.cosmos.ru/docs/Abstract_book_full_version_05.pdf), (p. 262-264, a hard copy), 3 pages, (p. 282-284, https://www.elibrary.ru/item.asp?id=45672856&pff=1, https://elibrary.ru/download/elibrary_45672856_87819955.pdf) (DOI: 0.21046/11MS3-2020 – for a whole book), an oral presentation. https://www.youtube.com/watch?v=nHHS0lX6qtw - day 4 of the conference, my presentation from 4h34 to 4h47. Oral presentation.

193. Ipatov S.I. Probabilities of collisions of exoplanetesimals with exoplanets in the Proxima Centauri planetary system. *11th Moscow International Solar System Symposium*, 5-9 October 2020, IKI, Moscow, Russia, 11MS3-EP-14, (p. 353-355, https://ms2020.cosmos.ru/docs/Abstract_book_full_version_05.pdf), (p. 338-340, a hard copy), (p. 358-360, https://www.elibrary.ru/item.asp?id=45672890&pff=1,

https://elibrary.ru/download/elibrary_45672890_53369576.pdf), (DOI: 0.21046/11MS3-2020 – for a whole book), https://www.academia.edu/44447360/. 3 pages, an oral presentation. https://www.youtube.com/watch?v=z6aqnzU41OY - day 5 of the conference, my presentation from 3h 34 to 3.42 with questions. Oral presentation.

194. Ipatov S.I. Probabilities of collisions of near-Earth objects with the Earth. *Proceedings of the All-Russian annual seminar on experimental mineralogy, petrology and geochemistry*. Moscow. October 20-21, 2020. / Responsible. editor O.A. Lukanin, - M: GEOKHI RAS, 2020. ISBN 978-5-905049-24-8. pp. 242-246. https://www.elibrary.ru/item.asp?id=45676471&pff=1, http://www.geokhi.ru/rasempg/Shared%20Documents/2020/ТрудыХит%202020_НА%20ПЕЧАТЬ_ФИНАЛ.pdf. Oral presentation. In Russian.

195. Ipatov S.I. Delivery of water and volatiles to planets in the habitable zone in the Proxima Centauri system. Abstracts of the *AASTCS Habitable Worlds 2021 Workshop* (22-26 February 2021, a virtual conference). Open Engagement Abstracts, https://aas.org/sites/default/files/2021-02/Abstracts__Open_Engagement_HW2_final.pdf, # 123, p. 63-66. https://kistorm.com/0MqpKkBD0J8Cw8V6y3Gg/Y58cURvfngDM20wlHkSt - an abstract. https://drive.google.com/file/d/1nn_Mu3bjeDdMS8QUS7sVxwClHoAd20uY/view - a poster. https://ui.adsabs.harvard.edu/abs/2021BAAS...53c1126I/abstract. AASTCS8, Habitable Worlds 2021, id. 1126. *Bulletin of the American Astronomical Society*, 2021, Vol. 53, No. 3 e-id 2021n3i1126, 5 pages. https://baas.aas.org/pub/2021n3i1126/release/2. Poster presentation.

196. Ipatov S.I. Migration of planetesimals in the exoplanetary system Trappist and in the zone of terrestrial planets. *Twenty-second international conference "Physicochemical and Petrophysical Research in Earth Sciences* " (September 27 - October 3, 2021, Moscow). Conference Proceedings. Moscow: IGEM RAS. 2021. pp. 114-117. 4 p. http://www.igem.ru/petromeeting_XXII/tbgdocs/sbornik_2021.pdf. Ihttps://elibrary.ru/item.asp?id=47336179. https://elibrary.ru/download/elibrary_47336179_93660561.pdf/. SBN 978-5-88918-064-7. ISSN 2686-8938. Oral presentation. In Russian.

197. Ipatov S.I. Migration of bodies in the Proxima Centauri planetary system. *12th Moscow International Solar System Symposium*, 11-15 October 2021, IKI, Moscow, Russia, https://ms2021.cosmos.ru/docs/2021/12ms3_book_5.pdf, 12MS3-EP-11, p. 199-201, 3 pages. https://elibrary.ru/item.asp?id=48368710. https://elibrary.ru/download/elibrary_48368710_77849862.pdf. An oral presentation.

198. Ipatov S.I., Marov M.Ya. Collisions of planetesimals with the Earth and the Moon, *12th Moscow International Solar System Symposium*, 11-15 October 2021, IKI, Moscow, Russia, https://ms2021.cosmos.ru/docs/2021/12ms3_book_5.pdf. 12MS3-SB-11, p. 293-295, 3 pages. https://elibrary.ru/item.asp?id=48368762. https://elibrary.ru/download/elibrary_48368762_60937783.pdf. An oral presentation.

199. Ipatov S.I. Probabilities of fallout on the Earth and the Moon of bodies ejected from the Earth during fallout of impactor bodies. Abstracts of the Scientific and Practical Conference with International Participation " *Near-Earth Astronomy–2022* " (April 18-21, 2022, Moscow). http://www.inasan.ru/wp-content/uploads/2022/04/Abstract-Book-2022.pdf. 2022. pp. 120-123. Oral presentation. In Russian.

200. Ipatov S.I., Marov M.Ya. Migration of bodies to the Earth and the Moon from different distances from the Sun. Abstracts of the Scientific and Practical Conference with International Participation "*Near-Earth Astronomy–2022*" (April 18-21, 2022, Moscow). http://www.inasan.ru/wp-content/uploads/2022/04/Abstract-Book-2022.pdf. 2022. 162-164. Poster presentation. In Russian.

201. Ipatov S.I. Probabilities of collisions of bodies ejected from the Earth and the Moon with the terrestrial planets and the Moon. Proceedings of the XXIII International Conference " *Physicochemical and Petrophysical Research in Earth Sciences* " (September 26 - October 2, 2022, Moscow). ISBN 978-5-88918-069-2. ISSN 2686-8938. 2022. P. 104-107. Oral presentation. http://www.igem.ru/petromeeting_XXIII/tbgdocs/sbornik_2022.pdf - abstracts book. In Russian.

202. Ipatov S.I. Migration of icy planetesimals to the inner planets in the Proxima Centauri system. Proceedings of the XXIII International Conference " *Physicochemical and Petrophysical Research in Earth Sciences* " (September 26 - October 2, 2022, Moscow, IGEM RAS). 2022. P. 108-111. ISBN 978-5-88918-069-2 / ISSN 2686-8938. Oral presentation. http://www.igem.ru/petromeeting_XXIII/tbgdocs/sbornik_2022.pdf - abstracts book. In Russian.

203. Ipatov S.I. Probabilities of collisions of bodies ejected from the Earth with the terrestrial planets and the Moon. *Thirteenth Moscow Solar System Symposium* (13M-S3) (October 10-14, 2022, Moscow, the Space Research Institute). ISBN: 978-5-00015-057-3. DOI: 10.21046/13MS3-2022. 2022. 13MS3-SB-11.

P. 264-266. https://ms2022.cosmos.ru/docs/2022/13-MS3_BOOK_final.pdf. https://elibrary.ru/item.asp?id=50086655. https://elibrary.ru/download/elibrary_50086655_43391244.pdf. Oral presentation.

204. Ipatov S.I. Scattering of planetesimals from the feeding zone of Proxima Centauri c. *Thirteenth Moscow Solar System Symposium* (13M-S3) (October 10-14, 2022, Moscow, the Space Research Institute). 2022. 13MS3-EP-08. P. 372-374. ISBN: 978-5-00015-057-3. DOI: 10.21046/13MS3-2022. https://elibrary.ru/item.asp?id=50086723. https://elibrary.ru/download/elibrary_50086655_43391244.pdf. Oral presentation.

205. Ipatov S.I. Mixing of planetesimals in the TRAPPIST-1 exoplanetary system. *Thirteenth Moscow Solar System Symposium* (13M-S3) (October 10-14, 2022, Moscow, the Space Research Institute). 2022. 13MS3-EP-PS-02. P. 378-380. URL: https://ms2022.cosmos.ru/docs/2022/13-MS3_BOOK_final.pdf. ISBN: 978-5-00015-057-3. DOI: 10.21046/13MS3-2022. https://elibrary.ru/item.asp?id=50086725. https://elibrary.ru/download/elibrary_50086725_88823830.pdf. Poster presentation.

206. Marov M.Ya., **Ipatov S.I.** The role of celestial bodies migration in the evolution of the Earth and planets. Proceedings of *the XXIII International Conference on Computational Mechanics and Modern Applied Software Systems* (VMSPS'2023), September 4–10, 2023, Divnomorskoye, Krasnodar Krai. - M.: MAI Publishing House, 2023. P. 493–495. http://www.cmmass.ru/files/cmmass2023_web.pdf - conference materials. http://www.cmmass.ru/files/cmmass2023program.pdf - conference program. https://elibrary.ru/download/elibrary_65489375_69967172.pdf . Poster presentation. In Russian.

207. Ipatov S.I. Migration of bodies from the feeding zone of the planet Proxima Centauri c. // Proceedings of the XIV International Conference " *Physicochemical and Petrophysical Research in Earth Sciences* ". (September 25 – September 27, 2023, Moscow; September 29, Borok). Moscow: IGEM RAS, 2023. Pp. 115-118. ISBN 978-5-88918-072-2 / ISSN 2686-8938. Oral presentation. https://elibrary.ru/download/elibrary_59367516_20494363.pdf. https://elibrary.ru/item.asp?id=59367516&pff=1 – my abstracts. http://www.igem.ru/petromeeting_XXIV/tbgdocs/sbornik_2023.pdf - book of abstracts. http://www.igem.ru/petromeeting_XXIV/tbgdocs/programm_2023.pdf – program of the conference. https://www.youtube.com/watch?v=eWI_n9T8H6s – presentations. In Russian.

208. Ipatov S.I. Migration of bodies ejected from the Earth and the Moon. *Fourteenth Moscow Solar System Symposium* (14M-S3) (October 9-13, 2023, Moscow, the Space Research Institute), https://ms2023.cosmos.ru/docs/2023/14ms3_ABSTRACT-BOOK-2023-10-06.pdf. ISBN: 978-5-00015-061-0. DOI: 10.21046/14MS3-2023. P. 274-276. An oral presentation. https://ms2023.cosmos.ru/docs/2023/14M-S3_PROGRAM_3B-0926.pdf - program of the conference. https://elibrary.ru/item.asp?id=65489334. https://elibrary.ru/download/elibrary_65489334_81821420.pdf

209. Ipatov SI Mixing of planetesimals in the Glisse 581 planetary system. *Fourteenth Moscow Solar System Symposium* (14M-S3) (October 9-13, 2023, Moscow, the Space Research Institute), https://ms2023.cosmos.ru/docs/2023/14ms3_ABSTRACT-BOOK-2023-10-06.pdf. ISBN: 978-5-00015-061-0. DOI: 10.21046/14MS3-2023. P. 333-335. An oral presentation. https://ms2023.cosmos.ru/docs/2023/14M-S3_PROGRAM_3B-0926.pdf - program of the conference. https://elibrary.ru/download/elibrary_65489334_81821420.pdf .

210. Ipatov S. I. Exchange of meteorites between the terrestrial planets. // Proceedings of the XXV International Conference " *Physicochemical and Petrophysical Studies in Earth Sciences* " (September 30 - October 4, 2024, GEOKHI RAS, IGEM RAS, IPE RAS and GO "Borok" IPE RAS). Moscow. IGEM RAS. 2024. ISBN 978-5-88918-075-3. ISSN 2686-8938. Pp. 87-90. Oral presentation. https://www.youtube.com/live/FL2IlkBuAL4 - broadcast of my presentation on September 30 at the conference from the 11th minute to the 25th minute of the broadcast. In Russian.

211. Ipatov S.I. Migration of bodies ejected from Mars. The *Fifteenth Moscow Solar System Symposium* (15M-S3). Moscow, Russia. October 21-25, 2024. 15MS3-SB-06. P. 236-238. http://ms2024.cosmos.ru, an oral presentation. https://ms2024.cosmos.ru/docs/2024/15M-S3_Program_07.10.pdf – conference program. https://ms2024.cosmos.ru/docs/2024/15-MS3_Abstract_Book-10-18.pdf. https://elibrary.ru/download/elibrary_80856144_39006326.pdf . https://youtube.com/playlist?list=PLGyFz6xpxOYlfhol5ORap9INvTVQargy-&si=Tvkk325LEhPOzqwy - My presentation on October 24 from 12 to 12-10 Moscow time (approximately between 2 hours 9 minutes and 2 hours 20 minutes of the recording). Oral presentation.

212. Ipatov S.I. Migration of planetesimals in the TRAPPIST-1 and GLISSE 581 exoplanetary systems. *The Fifteenth Moscow Solar System Symposium* (15M-S3). Moscow, Russia. October 21-25, 2024.

http://ms2024.cosmos.ru, an oral presentation. https://ms2024.cosmos.ru/docs/2024/15M-S3_Program_07.10.pdf – conference program. https://ms2024.cosmos.ru/docs/2024/15-MS3_Abstract_Book-10-18.pdf. 15MS3-EP-01. p. 320-322. https://www.academia.edu/126487193/. https://youtube.com/playlist?list=PLGyFz6xpxOYlfhol5ORap9INvTVQargy-&si=Tvkk325LEhPOzqwy - My presentation on October 25 from 10 to 10-20 Moscow time (approximately between 11 and 29 minutes of the recording). Oral presentation.

213. Ipatov S.I. Migration of planetesimals and dust particles in the Proxima Centauri exoplanetary system. *Modern astronomy: from the Early Universe to exoplanets and black holes*. Federal State Budgetary Institution of Science the Special Astrophysical Observatory of the Russian Academy of Sciences. 2024. C. 845-851. DOI: 10.26119/VAK2024.134. https://sao.editorum.ru/ru/nauka/conference_article/11858/view. https://doi.org/10.26119/VAK2024.134. https://www.researchgate.net/publication/387473114. https://www.academia.edu/126685835 .

214. Ipatov S.I. Migration of planetesimals in the TRAPPIST-1 exoplanetary system. *Modern astronomy: from the Early Universe to exoplanets and black holes.* Federal State Budgetary Institution of Science the Special Astrophysical Observatory of the Russian Academy of Sciences. 2024. C. 852-855. DOI: 10.26119/VAK2024.135 https://sao.editorum.ru/ru/nauka/conference_article/11832/view . https://doi.org/10.26119/VAK2024.135. https://www.researchgate.net/publication/387472279. https://www.academia.edu/126684939.

215. Ipatov S.I. Exchange of meteorites between the terrestrial planets and the Moon. *Modern astronomy: from the Early Universe to exoplanets and black holes.* Federal State Budgetary Institution of Science the Special Astrophysical Observatory of the Russian Academy of Sciences. 2024. C. 904-909. DOI: 10.26119/VAK2024.143. https://sao.editorum.ru/ru/nauka/conference_article/11887/view. https://doi.org/10.26119/VAK2024.143. https://www.researchgate.net/publication/387471384. https://www.academia.edu/126685916. https://arxiv.org/abs/2501.00134

***Preprints of Institute of Applied Mathematics (in Russian)***

https://library.keldysh.ru/author_page.asp?aid=1077 .

IPM preprints were sent to central libraries as mandatory mailing lists.

Abstracts of my preprints in English are available at https://ui.adsabs.harvard.edu/ (search for Ipatov, S. I.)

216. Ipatov S.I. Evolution of a flat ring of gravitating material points. Preprint In. appl. mat. AN USSR. 1978. N 2. 60 p. https://rusneb.ru/catalog/000199_000009_007779447/. https://ui.adsabs.harvard.edu/abs/1978iam..reptQ...1I/abstract .

217. Ipatov S.I. Evolution of a flat ring of gravitating bodies merging during collisions. Preprint In. appl. mat. AN USSR. 1978. N 101. 66 p. https://rusneb.ru/catalog/000199_000009_007779448/. https://ui.adsabs.harvard.edu/abs/1978iam..rept....1I/abstract.

218. Ipatov S.I. Mutual gravitational influence of two protoplanets in the planar three-body problem with initially circular orbits. Preprint In. appl. mat. AN SSSR. 1979. N 183. 32 p. https://rusneb.ru/catalog/000199_000009_007748475/. https://ui.adsabs.harvard.edu/abs/1979iam..reptQ...1I/abstract .

219. Ipatov S.I. The Three-Body Problem and Interaction of Protoplanets in a Protoplanetary Cloud. Preprint In. appl. mat. AN SSSR. 1979. N 192. 28 p. https://rusneb.ru/catalog/000199_000009_007748476/ . https://ui.adsabs.harvard.edu/abs/1979iam..reptR...1I/abstract .

220. Ipatov S.I. Evolution of resonant asteroid orbits in the planar three-body problem: Sun-Jupiter-asteroid. Preprint In. appl. mat. AN SSSR. 1980. N 30. 32 p. https://rusneb.ru/catalog/000200_000018_rc_1141265/. https://ui.adsabs.harvard.edu/abs/1980iam..reptQ...1I/abstract .

221. Ipatov S.I. On an approximate method for studying the mutual gravitational influence of bodies in a protoplanetary cloud. On the evolution of Pluto's orbit. Preprint In. appl. mat. AN USSR. 1980. N 43. 33 p. https://rusneb.ru/catalog/000200_000018_rc_1141264/. https://ui.adsabs.harvard.edu/abs/1980iam..reptR...1I/abstract .

222. Ipatov S.I. Numerical studies of spins of accumulating bodies. Preprint In. appl. mat. AN USSR. 1981. N 101. 28 p. https://rusneb.ru/catalog/000199_000009_001061162/. https://ui.adsabs.harvard.edu/abs/1981iam..reptQ...1I/abstract.

223. Ipatov S.I. Some problems of formation of axial rotations of planets. Preprint In. appl. mat. AN USSR. 1981. N 102. 28 p. https://rusneb.ru/catalog/000199_000009_001061163/. https://ui.adsabs.harvard.edu/abs/1981iam..reptR...1I/abstract .

224. Ipatov S.I. Numerical studies of the accumulation of terrestrial planets. Preprint In. appl. mat. AN SSSR. 1982. N 144. 28 p. https://rusneb.ru/catalog/000200_000018_rc_1105978/. https://ui.adsabs.harvard.edu/abs/1982iam..reptR...1I/abstract .

225. Ipatov S.I. Estimates of the evolution of spatial rings of gravitating bodies merging during collisions. Preprint In. appl. mat. AN USSR. 1982. N 186. 28 p. https://rusneb.ru/catalog/000199_000009_001146726/. https://ui.adsabs.harvard.edu/abs/1982iam..reptQ...1I/abstract .

226. Ipatov S.I. Studies of accumulation of cores of the giant planets based on computer simulations of plane disks. Preprint In. appl. mat. AN USSR. 1983. N 117. 28 p. https://rusneb.ru/catalog/000199_000009_001177306/. https://ui.adsabs.harvard.edu/abs/1983iam..rept....1I/abstract .

227. Ipatov S.I. Numerical studies of the evolution of spatial rings of gravitating bodies corresponding to the feeding zones of giant planets. Preprint In. appl. mat. AN USSR. 1984. N 1. 28 p. https://rusneb.ru/catalog/000199_000009_001195896/. https://ui.adsabs.harvard.edu/abs/1984iam..rept....1I/abstract .

228. Ipatov S.I. Evolution of orbital eccentricities at the initial stage of solid-state accumulation of planets. Preprint In. appl. mat. AN SSSR. 1985. N 4. 28 p. https://rusneb.ru/catalog/000199_000009_010820805/. https://ui.adsabs.harvard.edu/abs/1985iam..rept....1I/abstract .

229. Ipatov S.I. Numerical studies of the evolution of the orbits of asteroids and planetesimals in the framework of the plane three-body problem. Preprint. In. appl. mat. AN SSSR. 1988. N 62. 23 p. https://rusneb.ru/catalog/000199_000009_001411809/. https://ui.adsabs.harvard.edu/abs/1988iam..rept....1I/abstract .

230. Ipatov S.I. Origin of the Kirkwood Gap 5:2. Preprint In. appl. mat. AN USSR. 1989. N 67. 28 p. https://rusneb.ru/catalog/000199_000009_001490552/. https://ui.adsabs.harvard.edu/abs/1989iam..rept....1I/abstract.

231. Kozlov N.N., **Ipatov S.I.**, Toroptseva V.N. Minimization of the number of transitions in the routing of two-layer microcircuits. Preprint In. appl. mat. USSR Academy of Sciences. 1990. N 118. 14 p.

231b. Ipatov S.I. Evolution of asteroid-type orbits at 5:2 resonance. Preprint In. appl. mat. AN USSR. 1991. N 125. 48 p. https://ui.adsabs.harvard.edu/abs/1991iam..rept....1I/abstract .

232. Ipatov S.I. Gravitational interaction of planetesimals moving in close orbits. Preprint In. appl. mat. RAS. 1994. N 23. 60 p. https://rusneb.ru/catalog/000199_000009_001689071/ - information about the preprint in the Russian State Library. https://ui.adsabs.harvard.edu/abs/1994iam..reptQ...1I/abstract .

233. Ipatov S.I. Migration of small bodies to the Earth. Preprint In. appl. mat. RAS. 1994. N 105. 48 p. https://rusneb.ru/catalog/000199_000009_001700550/ . https://cat.gpntb.ru/index.php?id=EC/ShowFull&bid=a4a227fc039ec8dae04ee652c51e681d&irbDb=ESVODT – order in GPNTB (other preprints can also be found in central libraries). https://ui.adsabs.harvard.edu/abs/1994iam..reptR...1I/abstract .

**Abstracts of conferences and other small publications (1-2 pages)**

234. Ipatov S.I. Numerical studies of the evolution of disks of gravitating solid bodies moving around a massive central body. Abstracts of presentations at *the VI All-Union Congress on Theoretical and Applied Mechanics* (Tashkent, September 1986). 1986. P. 314. In Russian.

235. Ipatov S.I. Evolution of asteroid-type orbits in the vicinity of the 2:5 resonance. Abstracts of presentations at the All-Union Conference "*Methods of Studying the Motion, Physics, and Dynamics of Small Bodies of the Solar System*." Dushanbe. August 22-26, 1989. P. 33. https://rusneb.ru/catalog/000199_000009_001496710/?ysclid=m7xilbwul2504403932/. In Russian.

236. Ipatov S.I. Computer Simulation of the Evolution of Disks of Gravitating Bodies Moving Around the Sun. Abstracts of Papers at the All-Union Conference "*Methods of Computer Construction of Models of Classical and Celestial Mechanics-89* " Section 1. Leningrad. November 21-23, 1989. P. 41-42. https://cat.gpntb.ru/index.php?id=EC/ShowFull&bid=bc9c4d045a07bf40589bf6738a7bb370&irbDb=ESVODT. In Russian.

237. Ipatov S.I., Computer modelling of the process of solar system formation, Abstr. Intern. IMACS Conference "*Mathematical modelling and applied mathematics*" (June 18-23, 1990, Moscow), 1990, pp. 200-201.

238. Ipatov S.I. Methods for selecting pairs of contacting objects in an evolving disk. Abstracts of presentations at the All-Union Conference " *Algorithms and Programs of Celestial Mechanics* " Leningrad. November 20-22, 1990. P. 31-32. In Russian.

239. Ipatov S.I. Possible migration of the giant planets' embryos, Abstracts of *22nd Lunar and Planetary Science Conference* (March 18-22, 1991, Houston), 1991, pp. 607-608. https://1drv.ms/b/c/c67d93a65f0a2a17/ERcqCl-mk30ggMYQCAAAAAABPW80jdwKLqebQaBXBjijMA?e=2lfL2w, https://articles.adsabs.harvard.edu/pdf/1991LPI....22..607I

240. Ipatov S.I. Asteroid-type orbit evolution near the 5:2 resonance, Abstr. Intern. Conference "*Asteroids, comets, meteors 1991*" (June 24-28, 1991, Flagstaff, Arizona), LPI Contribution No 765, 1991, p. 98. http://articles.adsabs.harvard.edu/full/1992LPI....23..567I.

241. Ipatov S.I. Numerical studies of the migration of bodies in the forming Solar System. Abstracts of presentations at *the VII All-Union Congress on Theoretical and Applied Mechanics* (August 15-21, 1991, Moscow). 1991. P. 172-173. In Russian.

242. Ipatov SI Computer simulation of the bodies migration in the forming solar system, Abstr. Intern. Conference " *Origin and evolution of the solar system* " (August 27-31, 1991, Moscow), 1991, p. 31.

243. Ipatov S.I. Migration of bodies to the Earth's orbit from the Kirkwood gaps and from the zones of giant planets. Proceedings of the All-Union Conference (with international participation) "*Asteroid Hazard*" (St. Petersburg, October 10-11, 1991. Ed. A.G. Sokolsky. St. Petersburg. ITA. 1992. Pp. 121-125. In Russian.

244. Ipatov S.I. Migration of bodies during the accumulation of terrestrial planets, Abstracts of *23nd Lunar and Planetary Science Conference* (March 16-20, 1992, Houston), 1992, pp. 567-568. https://1drv.ms/b/c/c67d93a65f0a2a17/ERcqCl-mk30ggMYRCAAAAAABL7U2JmZYjgaaoB4IvSEqvQ?e=DcgTNM. http://adsabs.harvard.edu/abs/1992LPI....23..567I - starting from this webpage you can get a free file with the abstract; http://articles.adsabs.harvard.edu/full/1992LPI....23..567I.

245. Ipatov S.I. Migration of planetesimals and planets in the forming solar system, Abstracts of *24th Meeting of Division for Planetary Sciences of the American Astronomical Society* (Munich, October 12-16, 1992). *Bulletin of the American Astronomical Society*, 1992, 25-14P, N 3, p. 984. https://1drv.ms/b/c/c67d93a65f0a2a17/ERcqCl-mk30ggMarDAAAAAABG6PujDfolItGWjZv9Nn62Q?e=zRgeTd, http://adsabs.harvard.edu/full/1992DPS....24.2516P or http://articles.adsabs.harvard.edu/full/1992DPS....24.2514I .

246. Ipatov S.I. Migration of bodies to the orbit of Earth, Abstracts of the Conference "*Hazards due to comets and asteroids*" (January 5-9, 1993, Tucson, Arizona), 1993.

247. Ivashkin V.V., **Ipatov S.I.**, Smirnov V.V. Ballistical analysis of the problem of the asteroid hazard mitigation. Abstr. of the conference (with international participation) "*Asteroid hazard-93*" (May 25-27, 1993, St. Petersburg), ITA, St. Petersburg, 1993, p. 94.

248. Ipatov S.I. Migration of bodies to Earth, Abstr. Intern. Conference "*Asteroids, comets, meteors 1993*" (June 14-18, 1993, Belgirate, Italy), 1993, p. 138. http://articles.adsabs.harvard.edu/full/1993LPICo.810..138I

249. Ipatov S.I. Interactions of two bodies moving around the Sun, Abstr. of the Conference with international participation "*Theoretical, applied, and computer celestial mechanics*" (12-14 October 1993, St. Petersburg, Russia), 1993, pp. 48-49.

250. Ipatov S.I. Dynamics of two interacting objects orbiting the Sun, Abstracts of *25th Lunar and Planetary Science Conference* (March 14-18, 1994, Houston, USA), 1994, pp. 593-594. http://articles.adsabs.harvard.edu/pdf/1994LPI....25..593I. https://1drv.ms/b/c/c67d93a65f0a2a17/ERcqCl-mk30ggMYSCAAAAAAByBknNYFwjYu6tn8DX0JKtg?e=dNki9M

251. Ipatov S.I. Migration of bodies in the accumulation of planets, Abstracts of the Intern. *Conference on comparative planetology* (June 6-8, 1994, Pasadena, USA), 1994. https://link.springer.com/article/10.1007/BF00613304

252. Ipatov S.I. Migration of bodies in the forming Solar system. Abstracts of presentations at the International Conference "*Modern Problems of Theoretical Astronomy*" (June 20-24, 1994, St. Petersburg) / Ed. A.G. Sokolsky. 1994. P. 33-35. In Russian.

253. Ipatov S.I. Migration of small bodies in the Solar system, Abstracts of the Intern. Conference "*Small bodies in the Solar system and their interactions with the planets*" (August 8-12, 1994, Mariehamn, Finland), 1994, p. 69.

254. Ipatov S.I. Migration of celestial bodies and formation of planets, Abstracts of the *20th Intern.*

*Meeting on comparative planetology* (October 10-12, 1994, Moscow), 1994, pp. 31-32.
255. Ipatov S.I. Migration of small bodies to the Earth, Abstracts of *26th Lunar and Planetary Science Conference* (March 13-17, 1995, Houston, USA), 1995, pp. 655-656. http://adsabs.harvard.edu/full/1995LPI....26..655I, https://1drv.ms/b/c/c67d93a65f0a2a17/ERcqCl-mk30ggMYTCAAAAAABTaFWEv6QULhXVvi8W2vyhA?e=hDgcB7
256. Ipatov S.I. Migration of icy bodies in the Solar System, Abstr. of the Symposium "*Solar System Ices*" (March 27-30, 1995, Toulouse, France), 1995, p. 58.
257. Ipatov S.I., Migration of small bodies to the Earth, Abstracts of the *XX General Assembly of the European Geophysical Society* (April 3-7, 1995, Hamburg, Germany). *Annales Geophysicae*, Supplement of vol. 13. 1995.
258. Ipatov S.I., Ivashkin V.V. Possible sources and mechanisms of replenishment of the family of celestial bodies approaching the Earth. Abstracts of presentations at the Complex Conference with International Participation " *Asteroid Hazard-95* " (May 23-25, 1995, St. Petersburg). 1995. Vol. 2. P. 55-56. In press.
259. Ipatov S.I. Migration of celestial bodies in the Solar System, Abstr. of IAU Symposium 172 "*Dynamics, ephemerides and astrometry in the Solar System*" (July 3-8, 1995, Paris, France), 1995, p. 21-22. ftp://ftp.imcce.fr/pub/colloquia/IAU95/abstracts/Ipatov.tex. http://www.imcce.fr/fr/publications/colloque_sympo/SYMP172/abs.php.
260. Ipatov S.I. Sources of near-Earth objects, Abstracts of the conference “*Tunguska-95*” (July 18-24, 1995, Moscow-Tomsk-Vanavara, USSR), 1995.
261. Ipatov S.I. Migration of small bodies to Earth from the beyond-Neptune and asteroid belts, Abstr. of Intern. Conference "*Asteroids, comets, meteors 1996*" (July 8-12, 1996, Versailles, France), 1996, p. 19.
262. Ipatov S.I., Henrard J. Orbital evolution at the 2:3 resonance with Neptune, Abstr. of Intern. Conference "*Asteroids, comets, meteors 1996*" (July 8-12, 1996, Versailles, France), 1996, p. 88.
263. Ipatov S.I. Migration of small bodies to Earth from the Kuiper belt, Abstr. of Intern. Workshop "*Tunguska96*" (July 15-17, 1996, Bologna, Italy), 1996, p. 30.
264. Ipatov S.I. Migration of small bodies to Earth from the beyond-Neptune belt, Abstr. of Intern. Conference "*Asteroid hazard-96*" (July15-19, 1996, St. Petersburg, Russia), 1996, pp. 59-61.
265. Ipatov S. I. Migration of celestial bodies in the Solar system. Abstracts of presentations at *the Second Symposium on Classical and Celestial Mechanics* (August 23-27, 1996, Velikiye Luki). 1996. P. 35-37.
266. Ipatov S.I. Migration of small bodies to the Earth's orbit from the Kuiper belt, Abstracts of *28th Lunar and Planetary Science Conference* (March 17-21, 1997, Houston, USA), 1997, pp. 615-616. http://www.lpi.usra.edu/meetings/lpsc97/pdf/1106.PDF.
267. Ipatov S.I. Mutual gravitational influence of beyond-Neptune bodies, Abstracts of *28th Lunar and Planetary Science Conference* (March 17-21, 1997, Houston, USA), 1997, pp. 617-618. http://www.lpi.usra.edu/meetings/lpsc97/pdf/1588.PDF.
268. Ipatov S.I., Hahn G.J. Evolution of the orbits of the objects P/1996 R2 (Lagerkvist) and P/1996 N2 (Elst-Pizarro), Abstracts of *28th Lunar and Planetary Science Conference* (March 17-21, 1997, Houston, USA), 1997, pp. 619-620. http://www.lpi.usra.edu/meetings/lpsc97/pdf/1590.PDF. (Here: .PDF not .pdf).
269. Ipatov S.I., Henrard J. Evolution of orbits at the 2:3 resonance with Neptune, Abstracts of *28th Lunar and Planetary Science Conference* (March 17-21, 1997, Houston, USA), 1997, p. 621. http://www.lpi.usra.edu/meetings/lpsc97/pdf/1186.PDF.
270. Ipatov S.I., Hahn G.J. Evolution of the orbits of objects P/1996 R2 and P/1996 N2. Abstracts of the All-Russian conference (with international participation) "*Problems of Celestial Mechanics*" (St. Petersburg, June 3-6, 1997). ITA RAS. St. Petersburg. 1997. P. 92-95.
271. Ipatov S.I., Hahn G.J. Migration of Kuiper-belt objects inside the solar system, Abstracts of *9th Rencontres de Blois* (June 22-28, 1997) “*Planetary systems: the long view*”, 1997, pp. 35-36.
272. Ipatov S.I. Migration of celestial bodies in the forming solar system, Abstracts of *9th Rencontres de Blois* (June 22-28, 1997) “*Planetary systems: the long view*”, 1997, p. 50.
273. Ipatov S.I., Hahn G.J. Orbital evolution of the objects P/1996 R2 (Lagerkvist) and P/1996 N2 (Elst-Pizarro), Abstracts of *9th Rencontres de Blois* (June 22-28, 1997) “*Planetary systems: the long view*”, 1997, pp. 53-54.
274. Ipatov S.I. Migration of objects to the Earth from the Edgeworth-Kuiper belt, Abstracts of *Joint European and National Astronomical Meeting for 1997 JENAM-97* (2-5 July 1997, Thessaloniki, Greece), 1997, p. 10. https://ui.adsabs.harvard.edu/abs/1997jena.confE..10I/abstract.

275. Ipatov S.I. Migration of small bodies from the Kuiper belt, Abstracts of the *23th General Assembly of the IAU*, Joint Discussian 6 (Interaction between planets and small bodies), 22-23 August 1997, Kyoto, Japan, 1997. https://ui.adsabs.harvard.edu/abs/1997IAUJD...6E..15I/abstract (only in electronic form).

276. Ipatov S.I. Migration of small bodies to Earth, Abstracts of *Joint discussion on comets and minor planets with Japanese amateur astronomers* (August 29-31, 1997, Lake Biwa-Ko, Japan), 1997.

277. Ipatov S.I. Orbital evolution of planets, Abstracts of the international conference " *Results and perspectives of investigations of planets: new models and informational technologies* " (November 10-14, 1997, Ulyanovsk, Russia), pp. 16-17. In Russian.

278. Ipatov S.I. Mutual gravitational influence of bodies in the trans-Neptune belt. Abstracts of the All-Russian scientific conference "*New theoretical results and practical problems of celestial mechanics*" (Moscow, December 2-4, 1997). 1997. P. 45-46. In Russian.

279. Ipatov S.I. Migration of bodies to the Earth from the asteroid and Edgeworth-Kuiper belts, Abstracts of the *XXIII General Assembly of European Geophysical Society* (April 20-24, 1998, Nice, France), *Annales Geophysicae*, Part 3. Space and Planetary Sciences. 1998, Supplement 3 to vol. 16. p. 1043.

280. Ipatov S. I. Migration of small bodies to the Earth from the trans-Neptunian belt. Abstracts of the Third international conference " *90 years of Tunguska problem* " (June 30 - July 2, 1998, Krasnoyarsk-Vanavara), p. 25 (1998). http://omzg.sscc.ru/TUNGUSKA/en/newse/abstracts/ipat.htm. In Russian.

281. Ipatov S.I. Dynamics of trans-Neptunian objects. Abstracts of the IAU colloquium N 172 "*The impact of modern dynamics in astronomy*" (July 6-11, 1998, Namur, Belgium), 1998, pp. 56-57. https://ui.adsabs.harvard.edu/abs/1999imda.coll..107I/abstract.

282. Ipatov S.I. Formation and evolution of the trans-Neptunian belt. Abstracts *of the third international symposium on classical and celestial mechanics* (August 23-28, 1998, Velikiye Luki). 1998. P. 75-77. In Russian.

283. Ipatov S.I. Migration of Edgeworth-Kuiper belt objects to the Earth. Abstracts of *ESO workshop on minor bodies in the outer Solar System* (November 2-5, 1998, Garching, Germany), www.eso.org/gen-fac/meetings/mboss98/abstract.ps. 1998, p. 21.

284. Ipatov S.I., Mardon A.A. Delivery of meteorites to the Earth from the Edgeworth-Kuiper belt. Abstracts of *30th Lunar and Planetary Science Conference* (March 15-19, 1999, Houston, USA), CD-ROM. 1999. http://www.lpi.usra.edu/meetings/LPSC99/pdf/1147.pdf.

285. Ipatov S.I. Formation of the Edgeworth-Kuiper belt and migration of trans-Neptunian objects to the Earth. Abstracts of the conference "*Asteroids, Comets, Meteors*" (July 26-30, 1999, Cornell University, USA), 1999, pp. 132-133.

286. Ipatov S.I. Formation of trans-Neptunian objects and their migration to the Earth. Abstracts of the international conference "*Comets, asteroids, meteors, meteorites, astroblems, craters*" "CAMMAC-99" (September 26 - October 1, 1999, Vinnitsa, Ukraine), ed. Churuymov, 1999, p. 21.

287. Elst E.W., **Ipatov S**, Ticha J., Tichy M., Blythe M., Shelly F., Bezpalko M., Elowitz M., Huber R., Stuart J., Viggh H., Sayer R., Durig D.T., Marsden B.G. Comet C/1999 T3 (LINEAR). *Minor Planet Electronic Circ.*, 1999, 1999-U27. October 1999. http://cfa-www.harvard.edu/cfa/ps/mpec/J99/J99U27.html.

288. Elst E.W., **Ipatov S**, Ticha J., Tichy M., Durig D. Comet C/1999 T3 (LINEAR). *IAU Circ.*, 1999, 7289, 1. http://cfa-www.harvard.edu/iauc/07200/07289.html#Item1. https://ui.adsabs.harvard.edu/abs/1999IAUC.7289....1E/abstract.

289. Ipatov S.I. Migration of trans-Neptunian objects to the Earth. Abstracts of the conference "*Near-Earth astronomy and problems of investigations* of *small bodies in the Solar System" (*October 25-29, *1999,* Obninsk), 1999, p. 27. In Russian.

290. Ipatov S.I. Migration of matter from the Edgeworth-Kuiper belt to the Earth. Abstracts of IAU Colloquium No 181 and COSPAR Colloquium No. 11 "*Dust in the solar system and other planetary systems*" (April 10-14, 2000, Canterbury, UK), 2000, p. 68.

291. Ipatov S.I. Migration of trans-Neptunian bodies to Earth. Abstracts of the *XXV General Assembly of European Geophysical Society* (April 25-29, 2000, Nice, France), CD-ROM, 2000.

292. Ipatov S.I. Safronov's mass of a protoplanet cloud. Abstracts of the *XXV General Assembly of European Geophysical Society* (April 25-29, 2000, Nice, France), CD-ROM, 2000.

293. Ipatov S.I. Migration of small bodies from the main asteroid and Edgeworth-Kuiper belts, Abstracts of *JENAM-2000* (May 29-June 3, 2000, Moscow), 2000, p. 162.

294. Ipatov S.I. Migration of celestial bodies to the Earth from the trans-Neptunian and asteroid belts. Abstracts of the conference "*Astrometry, geodynamics and celestial mechanics at the turn of the century*" (June 19-24, 2000, St. Petersburg), 2000. P. 292-293. In Russian.

295. Ipatov S.I. Migration of trans-Neptunian bodies to Earth, Abstr. of *US- European celestial mechanics workshop* (July 3-7, 2000, Poznan, Poland), http://www.astro.amu.edu.pl/Science/Conference/pdf/ipatov_r.pdf, 2000.

296. Ipatov S.I. Comet hazard to the Earth, Abstracts of *33rd COSPAR Scientific Assembly* (July 16-23, 2000, Warsawa, Poland), CD-ROM, 2000.

297. Ipatov S.I. Evolution of the Edgeworth-Kuiper belt, Abstracts of the *24th General Assembly of the IAU* (August 7-18, 2000, Manchester, UK), http://www.jb.man.ac.uk/~abstract/abstracts/JD4P.html#abstract8p, 2000, p. 156. https://ui.adsabs.harvard.edu/abs/2000IAUJD...4E..28I/abstract.

298. Ipatov S.I. Migration of bodies during planet formation, Abstracts of the *24th General Assembly of the IAU* (August 7-18, 2000, Manchester, UK), http://ast.star.rl.ac.uk/symp202/posters/ipatov.html, 2000, p. 45.

299. Ipatov S.I. Migration of Tunguska-type objects from the trans-Neptunian belt to the Earth. Abstracts of the *Symposium on small satellites* (August 23-27, 2000, *Third International Aerospace Congress*, Moscow), floppy disk (pp. 299-300 for Russian text). In Russian.

300. Ipatov S.I. Evolution of the Edgeworth-Kuiper belt and migration of its bodies inside the Solar System, Abstracts of the international conference "*The Forth Vsekhsvyatsky Readings. Modern problems of physics and dynamics of the Solar System*" (October 5-9, 2000, Kyiv, Ukraine), ed. K.I. Churyumov, 2000, pp. 20-21.

301. Ipatov S.I. Migration of trans-Neptunian bodies to the Earth. Abstracts of the conference " *New results of analytical and qualitative celestial mechanics* " (December 5-6, 2000, SAI, Moscow), 2000, p. 42. http://www.sai.msu.ru/neb/rw/confer/ipatov.htm. In Russian.

302. Ipatov S.I., Mardon A.A. Delivery of trans-Neptunian objects to the Earth. Abstracts of *Near-Earth asteroid sample return workshop* (December 11-12, 2000, LPI, Houston, USA), 2000, p. 21, http://www.lpi.usra.edu/meetings/astreroid2000/pdf/8004.pdf. https://ui.adsabs.harvard.edu/abs/2000neas.work...21I/abstract.

303. Mardon A.A., **Ipatov S.I**. Impact threat from the source of Edgeworth-Kuiper belt, *Teaching of Astronomy in Asian-Pacific Region*. Bulletin No. 17. Mitaka Tokyo Japan, 2001, 2001.01.30, pp. 61-62.

304. Ipatov S.I. Formation of trans-Neptunian objects. Abstracts of *32th Lunar and Planetary Science Conference* (March 12-16, 2001, Houston, USA), 2001, #1165, http://www.lpi.usra.edu/meetings/lpsc2001/pdf/1165.pdf.

305. Ipatov S.I. Formation of trans-Neptunian objects. Abstracts of the *XXVI General Assembly of European Geophysical Society* (March 25-30, 2001, Nice, France), CD-ROM, http://www.copernicus.org/EGS/egsga/nice01/programme/abstracts/aai4551.pdf, 2001.

306. Ipatov S.I. Migration of trans-Neptunian bodies to the Earth, Abstracts of the conference "*Near-Earth Astronomy of the 21st Century"* (scientific and practical aspects). (May 21-25, 2001, Zvenigorod, 2001. P. 79-80. In Russian.

307. Marov M.Ya., Ipatov S.I. Asteroids migration: Role in the inner planets evolution. Abstracts of the conference "*Asteroids 2001: From Piazzi to the 3rd Millennium*" (June 11-16, 2001, Santa Flavia, Palermo, Italy), 2001, p. 215. (IV.13). www.astropa.unipa.it/Asteroids2001/Abstracts/Posters/marov.doc.

308. Ipatov S.I. Formation of trans-Neptunian objects and asteroids. Abstracts of the conference "*Asteroids 2001: From Piazzi to the 3rd Millennium*" (June 11-16, 2001, Santa Flavia, Palermo, Italy), 2001, pp. 307-308 (VIII.7). www.astropa.unipa.it/Asteroids2001/Abstracts/Posters/ipatov.doc.

309. Ipatov S.I. Formation and evolution of the trans-Neptunian belt. Abstracts *of the All-Russian Astronomical Conference*, St. Petersburg, August 6-12, 2001. P. 77-78. In Russian.

310. Ipatov S.I. Migration of bodies from the trans-Neptunian belt. Abstracts of *the Forth international conference on classical and celestial mechanics* (August 15-20, 2001, V. Luki, Russia), 2001, p. 75 ( English text), pp. 73-74 (Russian text).

311. Ipatov S.I. Migration of celestial bodies to the Earth from the trans-Neptunian and asteroid belts, Abstracts *of the VIII All-Russian Congress on Theoretical and Applied Mechanics* (August 23-29, 2001, Perm), Ekaterinburg. Ural Branch of the Russian Academy of Sciences, 2001. P. 296. In Russian.

312. Ozernoy L.M., **Ipatov S.I**. Evolution of the Edgeworth-Kuiper belt and kuiperoidal dust, Abstracts of *JENAM-2001* (September 10-15, 2001, Munich, Germany), *Astronomische Gesellschaft*, Abstract Series 18, 2001, abstract #P40, p. 155. www.ari.uni-heidelberg.de/AG/aga18.ps.

https://ui.adsabs.harvard.edu/abs/2001AGM....18..P40O/abstract.

313. Ipatov S.I., Mardon A.A. Delivery of material from a trans-Neptunian region to the Earth, abstracts of *64th Annual meeting of meteoritical society* (Vatican, September 10-14, 2001), *Meteoritics & Planetary Science*, 2001, vol. 36, Supplement, p. A86. http://adsabs.harvard.edu/full/2001M%26PSA..36S..86I. https://www.lpi.usra.edu/meetings/metsoc2001/pdf/5209.pdf.

314. Ozernoy L.M., **Ipatov S.I.** Origins of water on Mars, Earth, and Venus: Evaluating the supply from the outer Solar system, Abstracts of the *199th Meeting of the American Astronomical Society* (6-10 January 2002, Washington, DC, USA), The *Bulletin of the American Astronomical Society*, 2001, v. 33, N 4, pp. 1352-1353. http://www.aas.org/publications/baas/v33n4/aas199/261.htm. https://ui.adsabs.harvard.edu/abs/2001AAS...199.2806O/abstract.

315. Ipatov S.I., Ozernoy L.M. Characteristic times elapsed up to collisions of minor bodies with planets, Abstracts of the *199th Meeting of the American Astronomical Society* (6-10 January 2002, Washington, DC, USA), *The Bulletin of the American Astronomical Society*, 2001, v. 33, N 4, p. 1352. http://www.aas.org/publications/baas/v33n4/aas199/262.htm. https://ui.adsabs.harvard.edu/abs/2001AAS...199.2803I/abstract.

316. Ipatov S.I., Mather J.C. Migration of trans-Neptunian objects to a near-Earth space. Abstracts of workshop "*From here to Pluto-Charon: The new horizons Pluto-Kuiper Belt mission*" (20-21 May, 2002, Boulder, Colorado, USA), 2002.

317. Emelyanenko V.V., **Ipatov S.I.** Migration of celestial bodies from different regions of the Solar system into near-Earth space. Abstracts of the conference "*International Cooperation in Astronomy*" within the framework of the VI Congress of Astronomy (May 25 – June 2, 2002, Moscow), 2002. P. 26-27. In Russian.

318. Ipatov S.I., Mather J.C. Orbital evolution of Jupiter-family comets. Abstracts of the *200th Meeting of the American Astronomical Society* (2-6 June 2002, Albuquerque, NM, USA), *The Bulletin of the American Astronomical Society*, 2002, v. 34, n 2, p. 783. http://www.aas.org/publications/baas/v34n2/aas200/296.htm.

319. Ipatov S.I. Formation and migration of trans-Neptunian objects. Abstracts of international conference "*Scientific Frontiers in Research of Extrasolar Planets*" (18-21 June 2002, Washington D.C., USA), 2002, p. 44.

320. Ipatov S.I. Formation and migration of trans-Neptunian objects and asteroids. Abstracts of the conference "*Asteroids, comets, meteors, 2002*" (July 29 - August 2, 2002, Berlin), 2002, p. 58, #05-26p, http://berlinadmin.dlr.de/SGF/acm2002.

321. Ipatov S.I. Orbital evolution of Jupiter-crossers and asteroids, Abstracts of the conference "*Current status and perspectives of international researches in observational astronomy, ecology, and extreme physiology in the Elbrus region* (ASTROECO-2002)" (12-16 August, 2002, Terskol, Russia), 2002.

322. Ipatov S.I., Mather J.C. Decoupling comets from Jupiter, Abstracts of *JENAM 2002 "The Unsolved Universe: Challenges for the Future*" (2-7 September 2002, Porto, Portugal), 2002.

323. Ipatov S.I. Formation and migration of trans-Neptunian objects. *IAA Transactions*. No 8. Celestial Mechanics. – St. Petersburg: Inst. Appl. Astron. of Russian Acad. of Sciences, Abstracts of the Conference "*Celestial Mechanics -2002: Results and Prospects*" (10-14 September 2002, St. Petersburg, Russia, Institute of Applied Astronomy of Russian Academy of Sciences), 2002, pp. 86-87, http://arXiv.org/format/astro-ph/0205307. http://iaaras.ru/library/paper/332.

324. Ipatov S.I. Migration of asteroids from the 3/1 and 5/2 resonances with Jupiter to the Earth. *IAA Transactions*. No 8. Celestial Mechanics. – St. Petersburg: Inst. Appl. Astron. of Russian Acad. of Sciences, Abstracts of the Conference "*Celestial Mechanics -2002: Results and Prospects*" (10-14 September 2002, St. Petersburg, Russia, Institute of Applied Astronomy of Russian Academy of Sciences, 2002, pp. 88-89, http://arXiv.org/format/astro-ph/0205309. http://iaaras.ru/library/paper/333.

325. Ipatov S.I., Mather J.C. Migration of comets to near-Earth orbits, Abstracts of the 34th annual meeting of the *Division of Planetary Sciences of AAS* (6-11 October 2002, Birmingham, USA). ID 188. *The Bulletin of the American Astronomical Society*, 2002, v. 34, N 3, pp. 887-888, #27.10, http://www.aas.org/publications/baas/v34n3/dps2002/188.htm. https://ui.adsabs.harvard.edu/abs/2002DPS....34.2710I/abstract.

326. Marov M.Ya., Ipatov S.I. Volatiles inventory from planetesimals and trans-Neptunian objects: An Estimate, Abstracts of the *34th annual meeting of the Division of Planetary Sciences of AAS* (6-11 October 2002, Birmingam, USA), *The Bulletin of the American Astronomical Society*, 2002, v. 34, N 3, p.

891, #28.12, http://www.aas.org/publications/baas/v34n3/dps2002/164.htm. https://ui.adsabs.harvard.edu/abs/2002DPS....34.2812M/abstract.

327. Ipatov S.I., Mather J.C. Comet and asteroid hazard to the terrestrial planets. Abstracts of *34th Scientific Assembly of the Committee on Space Research (COSPAR) at the 2nd World Space Congress* (10-19 October 2002, Houston, Texas, USA) (ID Nr: COSPAR02-A-00845), CD-ROM, http://www.cosis.net/abstracts/COSPAR02/00845/COSPAR02-A-00845.pdf. 2002.

328. Ipatov S.I., Mather J.C. Migration of trans-Neptunian objects to the terrestrial planets, Abstracts of the international scientific workshop on the "*First Decadal Review of the Edgeworth-Kuiper Belt - Towards New Frontiers*" (11-14 March 2003, Antofagasta, Chile, 2003.

329. Ipatov S.I. and Mather J.C. Decoupling of Jupiter-family comets, Abstracts of AAS Division on Dynamical Astronomy (4-7 May 2003, Cornell University, Ithaca, NY, USA), The Bulletin of the American Astronomical Society, v. 35, N 4, p. 1034, #2.05, 2003. http://www.aas.org/publications/baas/v35n4/dda2003/27.htm. https://ui.adsabs.harvard.edu/abs/2003DDA....34.0205I/abstract.

330. Ipatov S.I., Mather J.C., Taylor P. Migration of asteroidal dust particles, Abstracts of *AAS Division on Dynamical Astronomy* (4-7 May 2003, Cornell University, Ithaca, NY, USA), *The Bulletin of the American Astronomical Society*, 2003, v. 35, N 4, p. 1037, #6.06. (http://www.aas.org/publications/baas/v35n4/dda2003/38.htm. https://ui.adsabs.harvard.edu/abs/2003DDA....34.0606I/abstract.

331. Ipatov S.I., Mather J.C., Taylor P. Migration of dust in the solar system, Abstracts of the *Astrophysics of Dust Symposium* (26-30 May 2003, Estes Park, Colorado, USA), 2003, P.2.23. id 101. http://www.physics.utoledo.edu/~aod03/abs.html. https://ui.adsabs.harvard.edu/abs/2003asdu.confE.101I/abstract.

332. Ipatov S.I. Migration of trans-Neptunian objects to the terrestrial planets, Abstracts of the *Fourth International Aerospace Congress* (18-23 August, 2003, Moscow), Symposium on small satellites, Russia. Moscow: SIP RIA, 2003, 473-474 (in English), 471-472 (in Russian).

333. Ipatov S.I., Mather J.C. Extinct comets in near-Earth object orbits, Abstracts of the *35th annual meeting of the Division of Planetary Sciences of AAS* (2-6 September 2003, Monterey, CA, USA). *The Bulletin of the American Astronomical Society,* 2003, v. 35, N 4, p. 974-975, #33.04. http://www.aas.org/publications/baas/v35n4/dps2003/54.htm. https://ui.adsabs.harvard.edu/abs/2003DPS....35.3304I/abstract.

334. Ipatov S.I. Former comets in near-Earth object orbits, Abstracts of the Conference "*Astrometry, Geodynamics and Solar System Dynamics: from milliarcseconds to microarcseconds*" (September 22-25, 2003, Institute of applied astronomy, Saint-Petersburg, Russia), 2003, p. 31.

335. Ipatov S.I. Migration of interplanetary dust, Abstracts of the Conference "Astrometry, *Geodynamics and Solar System Dynamics: from milliarcseconds to microarcseconds*" (September 22-25, 2003, Institute of applied astronomy, Saint-Petersburg, Russia), 2003, p. 32.

336. Ipatov S.I. Formation of trans-Neptunian objects, Abstracts of the *203th Meeting of the American Astronomical Society* (4-8 January 2004, Atlanta, USA), *The Bulletin of the American Astronomical Society*, 2003, v. 35, N 5, #15.06. http://www.aas.org/publications/baas/v35n5/aas203/818.htm. https://ui.adsabs.harvard.edu/abs/2003AAS...203.1506I/abstract.

337. Marov M.Ya., **Ipatov S.I**. Volatiles inventory to the inner planets due to small bodies migration, Abstracts of *34th Lunar and Planetary Science Conference* (March 17-21, 2003, League City, TX, USA), CD-ROM, 2003, #1099, 2 pages, http://www.lpi.usra.edu/meetings/lpsc2003/pdf/1099.pdf. .

338. Ipatov S.I., Mather J.C., Taylor P. Migration of asteroidal dust, Abstracts of *34th Lunar and Planetary Science Conference* (March 17-21, 2003, League City, TX, USA), CD-ROM, 2003, #1501, 2 pages, http://www.lpi.usra.edu/meetings/lpsc2003/pdf/1501.pdf.

339. Ipatov S.I., Mather J.C. Migration of dust particles from Comet 2P Encke, Abstracts of *workshop on cometary dust in astrophysics* (Crystal Mountain, Washington, August 10-15, 2003), 2003, N 6039. https://www.lpi.usra.edu/meetings/stardust2003/pdf/6039.pdf.

340. Ipatov S.I. Formation and migration of trans-Neptunian objects and asteroids, Abstracts of mini-symposium "*Dynamics of formation, evolution and stability of planetary systems" at JENAM-2003* (25-30 August 2003, Budapest, Hungary), 2003.

341. Ipatov S.I. Migration of trans-Neptunian objects to the Earth, Abstracts of the international conference "*Near-Earth Astronomy – 2003*" (NEA-2003) (9-14 September, 2003, Terskol, Russia), 2003, p. 17, http://www.inasan.rssi.ru/rus/conferences/OZA_2003/.

342. Ipatov S.I. Migration of dust to the near-Earth space, Abstracts of the international conference "*Near-Earth Astronomy – 2003*" (NEA-2003) (9-14 September, 2003, Terskol, Russia), 2003, p. 19, http://www.inasan.rssi.ru/rus/conferences/OZA_2003/.

343. Marov M.Ya., **Ipatov S.I**. Migration processes and volatiles inventory to the inner planets, Abstracts of *35th Lunar and Planetary Science Conference* (March 15-19, 2004, League City, TX, USA), 2004, #1410, 2 pages, http://www.lpi.usra.edu/meetings/lpsc2004/pdf/1410.pdf.

344. Ipatov S.I., Mather J.C., Guillory J.U. Migration of dust particles and their collisions with the terrestrial planets, Abstracts of *35th Lunar and Planetary Science Conference* (March 15-19, 2004, League City, TX, USA), 2004, #1446, 2 pages, http://www.lpi.usra.edu/meetings/lpsc2004/pdf/1446.pdf.

345. Ipatov S.I., Mather J.C., Guillory J.U. Migration of interplanetary dust particles, Abstracts of *35th Meeting of the AAS Division on Dynamical Astronomy* (April 20-23, 2004, Cannes, France), *The Bulletin of the American Astronomical Society*, 2004, v. 36, N 2, #05.06. p. 856. Also pp. 15-16 in the abstracts of the conference. http://www.aas.org/publications/baas/v36n2/dda04/50.htm. https://ui.adsabs.harvard.edu/abs/2004DDA....35.0506I/abstract.

346. Ipatov S.I., Mather J.C. Orbital evolution of Jupiter-family comets, Abstracts of *35th Meeting of the AAS Division on Dynamical Astronomy* (April 20-23, 2004, Cannes, France), The *Bulletin of the American Astronomical Society*, 2004, v. 36, N 2, #7.07, p. 860. Also p. 25 in the abstracts of the conference. http://www.aas.org/publications/baas/v36n2/dda04/65.htm. https://ui.adsabs.harvard.edu/abs/2004DDA....35.0707I/abstract.

347. Ipatov S.I., Mather J.C. Migration of comets and asteroids to near-Earth space, Abstracts of *1st General Assembly of European Geosciences Union* (Nice, France, 25-30 April 2004), 2004, ID-NR: EGU04-A-02392.

348. Ipatov S.I., Mather J.C. Migration of small bodies and dust to near-Earth space, Abstracts of *35th Scientific Assembly of the Committee on Space Research* (COSPAR) (Paris, France, July 18-25 2004), 2004, Abstract-Nr. COSPAR04-A-01012 , CD-ROM, http://www.cosis.net/abstracts/COSPAR04/01012/COSPAR04-A-01012.pdf. http://crdlx1.yerphi.am/CONFERENCES/CONFERENCEPAPERS/documents/data/B0.2_C02.3-0002-04.html. https://ui.adsabs.harvard.edu/abs/2004cosp...35.1012I/abstract.

349. Ipatov S.I. Formation and migration of trans-Neptunian objects, Abstracts of the *second TPF/Darwin international conference "Dust disks and the formation, evolution and detection of habitable planets*" (Mission Bay, San Diego, California, July 26-29, 2004). 2004. https://ui.adsabs.harvard.edu/abs/2004tpf..conf...45I/abstract. https://arxiv.org/pdf/astro-ph/0401279.

350. Ipatov S.I., Mather J.C. Migration of small bodies and dust to terrestrial planets, Abstracts of the IAU Colloq. N 197 "*Dynamics of populations of planetary systems*" (Belgrade, Serbia and Montenegro, 31 August – 4 September, 2004), 2004, p. 48.

351. Ipatov S.I., Kutyrev A., Madsen G.J., Mather J.C., Moseley S.H., Reynolds R.J. Dynamical Zodiacal cloud models constrained by high resolution spectroscopy of the Zodiacal light, Abstracts of the *36th annual meeting of the Division of Planetary Sciences of AAS* (8-12 November 2004, Lousville, Kentucky, USA), *The Bulletin of the American Astronomical Society,* 2004, v. 36, N 4, #34.14. p. 1150. http://www.aas.org/publications/baas/v36n4/dps2004/82.htm. https://ui.adsabs.harvard.edu/abs/2004DPS....36.3414I/abstract.

352. Ipatov S.I., Kutyrev A., Madsen G.J., Mather J.C., Moseley S.H., Reynolds R.J. Dynamical Zodiacal cloud models constrained by high resolution spectroscopy of the Zodiacal light, Abstracts of *36th Lunar and Planetary Science Conference* (March 14-18, 2005, League City, TX, USA), 2005, #1266, 2 pages, http://www.lpi.usra.edu/meetings/lpsc2005/pdf/1266.pdf.

353. Marov M.Ya., **Ipatov S.I.** Migration of dust particles and volatiles delivery to the inner planets, Abstracts of *36th Lunar and Planetary Science Conference* (March 14-18, 2005, League City, TX, USA), 2005, #1268, 2 pages, http://www.lpi.usra.edu/meetings/lpsc2005/pdf/1268.pdf.

354. Ipatov S.I., Kutyrev A., Madsen G.J., Mather J.C., Moseley S.H., Reynolds R.J. Velocities of zodiacal dust particles, late abstracts of *AAS 206 Meeting* (29 May - 2 June 2005, Minneapolis, Minnesota USA), *The Bulletin of the American Astronomical Society,* 2005, v. 37, #34.08, p. 791. #449. https://ui.adsabs.harvard.edu/abs/2005AAS...206.3408I/abstract.

355. Ipatov S.I., Mather J.C. Migration of interplanetary dust and small bodies, Abstracts of the IAU Symposium 229 "*Asteroids, Comets, Meteors – 2005*" (7-12 August 2005, Buzios, Rio de Janeiro, Brazil), 2005, p. 25. http://www.on.br/acm2005/visualiza-abstract.php?acao=Migration+of+interplanetary+dust+and+small+bodies.

356. Ipatov S.I., A'Hearn M.F. Automatic recognition of cosmic rays at Deep Impact CCDs, Abstracts of the IAU Symposium 229 "*Asteroids, Comets, Meteors – 2005*" (7-12 August 2005, Buzios, Rio de Janeiro, Brazil), 2005, p. 11-12. http://www.on.br/acm2005/visualiza-abstract.php?acao=Automatic+recognition+of+cosmic+rays+at+Deep+Impact+CCDs.

357. Ipatov S.I., Mather J.C. Dynamics and distribution of interplanetary dust, abstracts of 37th DPS Meeting and 31st HAD Meeting (September 4-9, 2005, Cambridge, UK), *The Bulletin of the American Astronomical Society*, 2005, v. 37, N 3, #17.04, p. 649. #262. http://www.aas.org/publications/baas/v37n3/dps2005/262.htm. https://ui.adsabs.harvard.edu/abs/2005DPS....37.1704I/abstract.

358. Ipatov S.I., Mather J.C. Migration of dust particles to the terrestrial planets, Abstracts of the conference "*Dust in Planetary Systems*" (September 26-30, 2005, Kaua'i, Hawaii, USA), 2005, p. 71-72, http://www.lpi.usra/meetings/dust2005/pdf/4049.pdf/.

359. Madsen G.J., Reynolds R.J., **Ipatov S.I.**, Kutyrev A., Mather J.C., Moseley S.H. New observations of the kinematics of the zodiacal dust cloud, Abstracts of the conference "*Dust in Planetary Systems*" (September 26-30, 2005, Kaua'i, Hawaii, USA), 2005, p. 111-112, http://www.lpi.usra.edu/meetings/dust2005/pdf/4072.pdf.

360. Ipatov S.I. Formation of trans-Neptunian objects, Abstracts of the conference "*Protostars and planets V*" (October 24-28, 2005, Hilton Waikoloa Village, The Big Island, Hawaii, USA), CD-ROM, Contribution No. 1286, 2005, p. 8054, http://www.lpi.usra.edu/meetings/ppv2005/pdf/8054.pdf. .

361. Ipatov S.I., A'Hearn M.F. Automatic recognition of cosmic rays at Deep Impact CCDs, Abstracts of the *AAS 207 Meeting* (8-12 January 2006, Washington, DC, USA), *The Bulletin of the American Astronomical Society*, v. 37 (#867), #154.08, 2005, p. 1415. http://www.aas.org/publications/baas/v37n4/aas207/867.htm. https://ui.adsabs.harvard.edu/abs/2005AAS...20715408I/abstract.

362. Ipatov S.I., A'Hearn M.F. Velocities of material ejected from comet Tempel 1, Abstracts of *37th Lunar and Planetary Science Conference* (March 13-17, 2006, League City, TX, USA), 2006, #1462, 2 pages, http://www.lpi.usra.edu/meetings/lpsc2006/pdf/1462.pdf.

363. Ipatov S.I., Kutyrev A., Madsen G.J., Mather J.C., Moseley S.H., Reynolds R.J. Dynamical zodiacal cloud models, Abstracts of *37th Lunar and Planetary Science Conference* (March 13-17, 2006, League City, TX, USA), 2006, #1471, 2 pages, http://www.lpi.usra.edu/meetings/lpsc2006/pdf/1471.pdf.

364. Ipatov S.I., Marov M.Ya., Mather J.C. Delivery of volatiles to the terrestrial planets, Abstracts of the *Astrobiology Science Conference* 2006 (March 26-30, 2006, Washington, D.C.), *Astrobiology*, 2006, v. 6, N 1, p. 241; http://abscicon2006.arc.nasa.gov/abstract/id/421.doc.

365. Ipatov S.I., Mather J.C. Probabilities of collisions of Jupiter-family comets with the terrestrial planets, *First International Conference on Impact Cratering in the Solar System* (ESTEC, Noordwijk, The Netherlands, May 08-12, 2006), 2006. http://www.rssd.esa.int/SYS/docs/ll_transfers/295234_ipatov.pdf.

366. Ipatov S.I. [misprinted as Sergei, I.I.], Mather J.C., Marov M.Ya. Migration of icy bodies to the terrestrial planets, *2006 Joint Assembly* (a partnership between AGU, GS, MAS, MSA, SEG, and UGM; May 23-26, 2006, Baltimore, Maryland, USA), 2006, http://www.agu.org/meetings/ja06/ [use SERGEI as author] (2006). Eos Trans. AGU, 87(36), Jt. Assem. Suppl., Abstract P32A-07.

367. Ipatov S.I., Mather J.C. Dynamic of interplanetary dust and comets, Abstracts of *36th Scientific Assembly of the Committee on Space Research* (COSPAR) (July 16-23, 2006, Beijing, China), 2006, (ID Nr: COSPAR2006-A-00458) http://www.cosis.net/abstracts/COSPAR2006/00458/COSPAR2006-A-00458-1.pdf. https://ui.adsabs.harvard.edu/abs/2006cosp...36..458I/abstract.

368. Ipatov S.I. A'Hearn M.F., Klaasen K.P., Desnoyer M., Lindler D. Automatic recognition and removal of cosmic rays at Deep Impact CCDs, Abstracts of *36th Scientific Assembly of the Committee on Space Research* (COSPAR) (July 16-23, 2006, Beijing, China) 2006, (ID Nr: COSPAR2006-A-00459) http://www.cosis.net/abstracts/COSPAR2006/00459/COSPAR2006-A-00459-3.pdf. https://ui.adsabs.harvard.edu/abs/2006cosp...36..459I/abstract.

369. Ipatov S.I., A'Hearn M.F. Velocities of material ejected from comet Tempel 1, Abstracts of *36th Scientific Assembly of the Committee on Space Research* (COSPAR) (July 16-23, 2006, Beijing, China), 2006, (ID Nr: COSPAR2006-A-00460) http://www.cosis.net/abstracts/COSPAR2006/00460/COSPAR2006-A-00460-1.pdf. https://ui.adsabs.harvard.edu/abs/2006cosp...36..460I/abstract.

370. Ipatov S.I., Mather J.C. Migration of comets into NEO orbits and probabilities of their collisions with the terrestrial planets, Abstracts of *26th IAU General Assembly* (August 14-25, 2006, Prague, Czech Republic; session S236), 2006, #0361; Abstract No.: S236-56, p. 90.

371. Ipatov S.I., A'Hearn M.F. Velocities of dust particles ejected from comet Tempel 1, Abstracts of *26th IAU General Assembly* (August 14-25, 2006, Prague, Czech Republic; JD10) #0362; 2006, Abstract No.: JD10-19, p. 348. http://www.astronomy2006.com/list-of-registered-abstracts.php?event=jd10.

372. Ipatov S.I. A'Hearn M.F., Klaasen K.P., Desnoyer M., Lindler D. Automatic Removal of Cosmic Rays on Deep Impact CCDs, Abstracts of *26th IAU General Assembly* (August 14-25, 2006, Prague, Czech Republic; JD10) #0364; 2006, Abstract No.: JD10-16, p. 348. http://www.astronomy2006.com/list-of-registered-abstracts.php?event=jd10, http://www.ppt2txt.com/r/1dc4448b/.

373. Marov M.Ya., Ipatov S.I. Volatiles Delivery to the Terrestrial Planets, Abstracts of *26th IAU General Assembly* (August 14-25, 2006, Prague, Czech Republic; JD11) #0365, 2006, Abstract No.: JD11-24, p. 359. http://www.astronomy2006.com/list-of-registered-abstracts.php?event=jd11.

374. Ipatov S.I., Mather J.C. Migration of Interplanetary Dust and Comets, Abstracts of the *European Planetary Science Congress 2006* (Berlin, Germany, September 18–22, 2006), 2006, p. 197, CD-ROM, EPSC2006-A-00197, http://www.cosis.net/abstracts/EPSC2006/00197/EPSC2006-A-00197-1.pdf or start from http://meetings.copernicus.org/epsc2006/).

375. Ipatov S.I., Kutyrev A., Madsen G.J., Mather J.C., Moseley S.H., Reynolds R.J., Computer simulations of spectrum of the zodiacal light. Abstracts of *38th DPS Meeting* (October 8-13, 2006, Pasadena, CA), *The Bulletin of the American Astronomical Society*, 2006, v. 38, N 3, p. 558. #41.02. https://ui.adsabs.harvard.edu/abs/2006DPS....38.4102I/abstract . https://1drv.ms/b/c/c67d93a65f0a2a17/ERcqCl-mk30ggMarDAAAAAABG6PujDfolItGWjZv9Nn62Q?e=kZ72qt. https://ui.adsabs.harvard.edu/abs/2006DPS....38.4102I/abstract.

376. Ipatov S.I. Growth of eccentricities and inclinations of planetesimals due to their mutual gravitational influence, Abstracts of *38th Lunar and Planetary Science Conference* (March 12-16, 2007, League City, TX, USA), 2007, #1260, 2 pages, http://www.lpi.usra.edu/meetings/lpsc2007/pdf/1260.pdf.

377. Ipatov S.I., Boss A.P., Myhill E.A. Triggering presolar cloud collapse and injection of short-lived radioisotopes by a supernova shock wave: adaptive mesh refinement calculations with the FLASH code, Abstracts of *38th Lunar and Planetary Science Conference* (March 12-16, 2007, League City, TX, USA), 2007, #1018, 2 pages, http://www.lpi.usra.edu/meetings/lpsc2007/pdf/1018.pdf.

378. Ipatov S.I., Mather J.C. Sources of zodiacal dust particles, Abstracts of the *European Planetary Science Congress 2007* (Potsdam, Germany, 19-24 August 2007), 2007, ID-NR: EPSC2007-A-00209. http://www.cosis.net/abstracts/EPSC2007/00209/EPSC2007-J-00209.pdf.

379. Ipatov S.I. Sources of zodiacal dust, Abstracts of the conference "*Near-Earth astronomy-2007*" (Terskol, Russia, 3-7 September 2007), 2007, p. 20. http://www.inasan.rssi.ru/rus/conferences/OZA_2007/Abstracts_sections.pdf.

380. Ipatov S.I., A'Hearn M.F. Velocities and relative amount of material ejected after the Deep Impact collision, abstracts of *39th DPS Meeting* (October 7-12, 2007, Orlando, FL), 2007, id.21.01; *Bulletin of the American Astronomical Society*, Vol. 39, p. 449. https://ui.adsabs.harvard.edu/abs/2007DPS....39.2101I/abstract

381. Marov M.Ya., **Ipatov S.I**. Volatiles and biogenic matter delivered by small bodies and dust particles, Abstracts of the II International Conference "*Biosphere origin and evolution*" (October 28 – November 2, 2007, Loutraki, Greece). 2007, KN-4. CD-ROM, P. 31-32. http://catalysis.ru/resources/institute/Publishing/Report/2007/015-2007-abstracts-BOE-2_Greece.pdf.

382. Boss A.P., **Ipatov S.I**., Myhill E.A. Triggering presolar cloud collapse and injection of short-lived radioisotopes with an outflow from a massive star, Abstracts of *Workshop on The Chronology of Meteorites and the Early Solar System* (November 5-7, 2007, Kauai, HI, USA), LPI Contribution No. 1374, 2007, pp. 34-35, http://www.lpi.usra.edu/meetings/metchron2007/pdf/4004.pdf.

383. Ipatov S.I., A'Hearn M.F. Velocities of material ejected after the Deep Impact collision, Abstracts of *39th Lunar and Planetary Science Conference* (March 10-14, 2008, League City, Texas), 2008, #1024, 2 pages, http://www.lpi.usra.edu/meetings/lpsc2008/pdf/1024.pdf.

384. Ipatov S.I., Cho J.Y-K. Synthetic spectra from a GCM simulation of a model exo-Earth, Abstracts of *39th Lunar and Planetary Science Conference* (March 10-14, 2008, League City, Texas), 2008, # 2554, 2 pages, http://www.lpi.usra.edu/meetings/lpsc2008/pdf/2554.pdf.

385. Cho J., **Ipatov S**. Circulation, cloud, and spectra: a first look at a 3-D model exo-Earth, Abstracts of the *General Assembly of European Geosciences Union* (April 13-18, 2008, Vienna, Austria), session PS9: Extrasolar planets and planet formation session, Geophysical Research Abstracts, 2008, v. 10, ID-No.: EGU2008-A-11813. http://www.cosis.net/abstracts/EGU2008/11813/EGU2008-A-11813.pdf.

386. Ipatov S.I. Migration of comets into NEO orbits and probabilities of their collisions with the Earth, Abstracts of the international conference "*100 years since Tunguska phenomenon: Past, present and future*" (Moscow, Russia, June 26-28, 2008), 2008, p. 125 (English), p. 124 (Russian), http://smerdyachee.ucoz.ru/_ld/0/8_abstract_all.pdf. http://tunguska.sai.msu.ru/content/abstract_all.pdf.

387. Ipatov S.I., A'Hearn M.F. Velocities of material ejected from a comet at its collision with a celestial body, Abstracts of the conference "*100 years since Tunguska phenomenon: Past, present and future*" (Moscow, Russia, June 26-28, 2008), 2008, p. 127 (English), p. 126 (Russian). http://smerdyachee.ucoz.ru/_ld/0/8_abstract_all.pdf. http://tunguska.sai.msu.ru/content/abstract_all.pdf.

388. Ipatov S.I., A'Hearn M.F. Velocities and relative amounts of material ejected after the collision of DI impactor with comet 9P/Tempel 1, Abstracts of *10th conference "Asteroids, Comets, Meteors*" (Baltimore, USA, July 14-18, 2008), 2008, paper id. 8102. http://www.lpi.usra.edu/meetings/acm2008/pdf/8102.pdf.

389. Ipatov S.I. Probabilities of collisions of migrating small bodies and dust particles with planets, Abstracts of *10th conference "Asteroids, Comets, Meteors*" (Baltimore, USA, July 14-18, 2008), 2008, paper id. 8103. http://www.lpi.usra.edu/meetings/acm2008/pdf/8103.pdf.

390. Ipatov S.I., A'Hearn M.F. Velocities and relative amount of material ejected from comet Tempel 1 after the Deep Impact collision, Abstracts of *37th Scientific Assembly of the Committee on Space Research* (COSPAR) (Montréal, Canada, 13 - 20 July 2008), abstract ID is 2960, Paper number: B04-0052-08, 2008, p. 1323. http://www.cospar-assembly.org/user/download.php?id=2960&type=abstract. http://adsabs.harvard.edu/abs/2008cosp...37.1323I

391. Ipatov S.I., A'Hearn M.F. Rate of ejection and velocities of material ejected from comet Tempel 1 after the Deep Impact collision, abstracts of *40th DPS Meeting* (October 10-15, 2008, Ithaca, NY), *The Bulletin of the American Astronomical Society*, 2008, v. 40, #3, p. 388-389. https://ui.adsabs.harvard.edu/abs/2008DPS....40.0208I/abstract.

392. Boss A.P., **Ipatov S.I.,** Keiser S.A., Myhill E.A., Vanhala H.A.T. Simultaneous triggered presolar cloud collapse and injection of short-lived radioisotopes by a supernova shock wave. Abstracts of *40th Lunar and Planetary Science Conference* (March 23-27, 2009, The Woodlands, Texas), 2009, #1002, 2 pages, http://www.lpi.usra.edu/meetings/lpsc2009/pdf/1002.pdf.

393. Ipatov S.I. Formation of binaries at the stage of rarefied preplanetesimals, Abstracts of *40th Lunar and Planetary Science Conference* (March 23-27, 2009, The Woodlands, Texas), 2009, # 1021, 2 pages. http://www.lpi.usra.edu/meetings/lpsc2009/pdf/1021.pdf.

394. Ipatov S.I., A'Hearn M.F. Deep Impact ejection from Comet Tempel 1 as a triggered outburst, Abstracts of *40th Lunar and Planetary Science Conference* (March 23-27, 2009, The Woodlands, Texas), 2009, #1022, 2 pages. http://www.lpi.usra.edu/meetings/lpsc2009/pdf/1022.pdf.

395. Ipatov S.I. Formation of binaries at the stage of rarefied preplanetesimals, Abstracts of the *General Assembly of European Geosciences Union* (April 19-24, 2009, Vienna, Austria), session PS8: Extrasolar planets and planet formation session, *Geophysical Research Abstracts,* 2009, ID-No.: EGU2009-10080. http://meetingorganizer.copernicus.org/EGU2009/EGU2009-10080.pdf. https://ui.adsabs.harvard.edu/abs/2009EGUGA..1110080I/abstract, a poster.

396. Ipatov S.I., A'Hearn M.F. Ejection of material after the Deep Impact collision with Comet Tempel 1, abstracts of *40th DDA Meeting* (May 2-5, 2009, Virginia Beach, VA, USA). *The Bulletin of the American Astronomical Society*, session 13, 2009, vol. 41, 13.01, p. 906. https://ui.adsabs.harvard.edu/abs/2009DDA....40.1301I/abstract, an oral presentation, session 13.

397. Ipatov S.I. Formation of binaries at the stage of rarefied preplanetesimals, abstracts of *40th DDA Meeting* (May 2-5, 2009, Virginia Beach, VA, USA). *The Bulletin of the American Astronomical Society*, 2009, vol. 41, 6.01, p. 898. https://ui.adsabs.harvard.edu/abs/2009DDA....40.0601I/abstract, a poster.

398. Ipatov S.I., A'Hearn M.F. Deep Impact ejection from Comet 9P/Tempel 1 as a triggered outburst, Abstracts of *27th IAU General Assembly* (August 3-14, 2009, Rio de Janeiro, Brazil; Symp. 263), 2009, #0495, p. 52, a poster.

399. Ipatov S.I. Angular momentum of two collided rarefied preplanetesimals and formation of binaries, Abstracts of *27th IAU General Assembly* (August 3-14, 2009, Rio de Janeiro, Brazil; Symp. 263), 2009, #0496, p. 52, a poster.

400. Ipatov S.I. Collision probabilities of migrating small bodies and dust particles with planets, Abstracts of *27th IAU General Assembly* (August 3-14, 2009, Rio de Janeiro, Brazil; Symp. 263), 2009, #0498, p. 53, a poster.

401. Ipatov S.I. Formation of small-body binaries at the stage of rarefied preplanetesimals, Abstracts of "*Near-Earth astronomy 2009*" (August 23-27, 2009, Kazan, Russia), 2009, a poster.

402. Ipatov S.I., A'Hearn M.F. Models of the ejection of material from Comet Tempel 1 caused by the Deep Impact experiment, Abstracts of *41st DPS Meeting* (October 4-9, 2009, Fajardo, Puerto Rico), *The Bulletin of the American Astronomical Society*, 2009, vol. 41, No 3, 20.06, p. 1028-1029. https://ui.adsabs.harvard.edu/abs/2009DPS....41.2006I/abstract, an oral presentation.

403. Ipatov S.I. Probabilities of collisions of migrating small bodies and dust particles with planets, Abstracts of *the 215th Meeting of the American Astronomical Society* (January 3-7, 2010, Washington, D.C.), 344.01. *The Bulletin of the American Astronomical Society*, 2010, v. 42, *N* 1, p. 450. https://ui.adsabs.harvard.edu/abs/2010AAS...21534401I/abstract, an oral presentation.

404. Ipatov S.I. Delivery of dust particles and small bodies to planets, Abstracts *of 41th Lunar and Planetary Science Conference* (March 1-5, 2010, The Woodlands, Texas), 2010, #1267, 2 pages, http://www.lpi.usra.edu/meetings/lpsc2010/pdf/1267.pdf.

405. Ipatov S.I. Formation of trans-Neptunian binaries at the stage of rarefied preplanetesimals, Abstracts of the conference "*TNO 2010: Dynamical and Physical properties of Trans-Neptunian Objects*" (Philadelphia, June 27 - July 1, 2010), 2010, a poster.

406. Ipatov S.I., A'Hearn M.F. Triggered outburst after the collision of the Deep Impact impactor with Comet Tempel 1, Abstracts of *38th Scientific Assembly of the Committee on Space Research* (COSPAR) (Bremen, Germany, July 18-25, 2010), 2010, abstract ID is 5702, Paper number: B04-0056-10. https://www.cospar-assembly.org/abstractcd/COSPAR-10/abstracts/data/pdf/abstracts/B04-0056-10.pdf, https://ui.adsabs.harvard.edu/abs/2010cosp...38..659I/abstract, a poster.

407. Ipatov S.I. The triggered outburst after the Deep Impact collision with Comet Tempel 1 and the composition of the comet, Abstracts of *42nd DPS Meeting* (October 4-8, 2010, Pasadena, California). 2010, 28.35. *The Bulletin of the American Astronomical Society*, 2010, v. 42, *N* 4. https://ui.adsabs.harvard.edu/abs/2010DPS....42.2835I/abstract, a poster.

408. Ipatov S.I. The outburst triggered by the collision of the Deep Impact module with Comet Tempel 1, and cavities in comets, Abstracts of *42nd Lunar and Planetary Science Conference* (March 7-11, 2011, The Woodlands, Texas), 2011, #1317, 2 pages, http://www.lpi.usra.edu/meetings/lpsc2011/pdf/1317.pdf.

409. Ipatov S.I. Location of the upper borders of the cavities excavated after the Deep Impact collision, Abstracts of *43rd Lunar and Planetary Science Conference* (March 19-23, 2012, The Woodlands, Texas), 2012, #1318, 2 pages, http://www.lpi.usra.edu/meetings/lpsc2012/pdf/1318.pdf.

410. Ipatov S.I. Cavities as a source of outbursts from comets, Abstracts of the conference *JENAM-2011, European week of astronomy and space science* (St. Petersburg, Russia, July 4-8, 2011), 2011, S2-9, a poster.

411. Ipatov S.I. Collision probabilities of small bodies and dust particles with planets, Abstracts of the international conference "*Near-Earth astronomy*" (Krasnoyarsk, September 5-10, 2011), 2011, session 13.

412. Ipatov S.I. Cavities as a source of outbursts from comets, Abstracts of *DPS-EPSC 2011 joint meeting*, (October 2–7, 2011, Nantes, France), EPSC abstracts, 2011, Vol. 6, EPSC-DPS2011-586. http://meetingorganizer.copernicus.org/EPSC-DPS2011/EPSC-DPS2011-586.pdf, a poster.

413. Ipatov S.I. Angular momenta of rarefied preplanetesimals and formation of small-body binaries, Abstracts of the conference "*Signposts of planets*" (NASA Goddard Space Flight Center, Greenbelt, MD, October 18-20, 2011). 2011, http://science.gsfc.nasa.gov/667/conferences/posters/signposts/gsfc2011oq-binaries.pdf, a poster.

414. Ipatov S.I. Sources of the zodiacal dust cloud, Abstracts of the conference "*Signposts of planets*" (NASA Goddard Space Flight Center, Greenbelt, MD, October 18-20, 2011), 2011, http://science.gsfc.nasa.gov/667/conferences/posters/signposts/gsfc2011oq-zodi.pdf, a poster.

415. Ipatov S.I. Cavities in Comet Tempel 1 excavated after the Deep Impact collision, Abstracts of *39th Scientific Assembly of the Committee on Space Research* (COSPAR) (Mysore, India, July 14-22, 2012), 2012, abstract ID is 12188, Paper number: B0.4-0046-12, a poster.

416. Ipatov S.I. Angular momenta of collided rarefied preplanetesimals, Abstracts of *28th IAU General Assembly* (August 20-31, 2012, Beijing, China), IAUS 293. 2012, #5517, p. 547-548.

417. Ipatov S.I., Horne, K., Alsubai K., Bramich D., Dominik M., Hundertmark M., Liebig C., Snodgrass C., Street R., Tsapras Y. Simulator for microlens planet surveys, Abstracts of *28th IAU General Assembly* (August 20-31, 2012, Beijing, China), 2012, #5753, IAUS 293. p. 583-584, a poster.

418. Ipatov S.I. Location of the upper border of the main cavity excavated after the Deep Impact collision, Abstracts of *28th IAU General Assembly* (August 20-31, 2012, Beijing, China), 2012, #5494, JD5, p. 868-869, an oral presentation.

419. Ipatov S.I. Outbursts and cavities in comets, Abstracts of the *Joint Qatar Foundation Annual Research Forum and Arab Expatriate Scientists Network 2012* (21-23 October 2012, Doha, Qatar), *Qatar Foundation Annual Research Forum Proceedings*: 2012, Vol. 2012, EEP15. http://www.qscience.com/doi/abs/10.5339/qfarf.2012.EEP15, a poster.

420. Ipatov S.I. Angular momenta of collided rarefied preplanetesimals, Abstracts of *44th Lunar and Planetary Science Conference* (March 18-22, 2013, The Woodlands, Texas), 2013, #1488, 2 pages, http://www.lpi.usra.edu/meetings/lpsc2013/pdf/1488.pdf.

421. Ipatov S.I. Angular momenta of collided rarefied preplanetesimals, Abstracts of *European Planetary Science Congress 2013* (September 08 – 13, 2013, London, UK), Session PD3/EX4, 2013. EPSC Abstracts. 2013. V. 8. EPSC2013-290-3. http://meetingorganizer.copernicus.org/EPSC2013/EPSC2013-290-3.pdf, 2 pages. Presentation: http://presentations.copernicus.org/EPSC2013-290_presentation.pdf. A poster.

422. Ipatov S.I. Cavities in comets, Abstracts of *European Planetary Science Congress 2013* (September 08 – 13, 2013, London, UK). EPSC Abstracts. 2013. V. 8. EPSC2013-335-1. http://meetingorganizer.copernicus.org/EPSC2013/EPSC2013-335-1.pdf. 2 pages. Presentation: http://presentations.copernicus.org/EPSC2013-335_presentation.pdf. An oral presentation.

423. Ipatov S.I., Horne K., Alsubai K., Bramich D., Dominik M., Hundertmark M., Liebig C., Snodgrass C., Street R., Tsapras Y. Exoplanet detection capability of microlensing observations, Abstracts of *European Planetary Science Congress 2013* (September 08 – 13, 2013, London, UK), Program EX5, 2013. EPSC Abstracts. 2013. V. 8. EPSC2013-331-2. http://meetingorganizer.copernicus.org/EPSC2013/EPSC2013-331-2.pdf. 2 pages. Presentation: http://presentations.copernicus.org/EPSC2013-331_presentation.pdf. A poster.

424. Ipatov S.I. Cavities in comets, Abstracts of International conference "*Near-Earth astronomy 2013*" (Krasnodar, Russia, October 7-11, 2013). 2013. P. 37-38. An oral presentation.

425. Ipatov S.I. The role of collisions of rarefied preplanetesimals in formation of satellites of small bodies, Abstracts of International conference "*Near-Earth astronomy 2013*" (Krasnodar, Russia, October 7-11, 2013). 2013. P. 87-88, a poster.

426. Ipatov S.I. Outbursts and cavities in comets, Abstracts of "*The Fourth Moscow Solar System Symposium 4M-S3*" (Space Research Institute, Moscow, Russia, October 14-18, 2013), 2013, 4MS3-PS-38, CD-ROM, http://ms2013.cosmos.ru/sites/ms2013.cosmos.ru/files/4m-s3_abstract_book.pdf. 2013. P. 203.

427. Ipatov S.I. The angular momentum of colliding rarefied preplanetesimals allows the formation of binaries, Abstracts of "*The Fourth Moscow Solar System Symposium 4M-S3*" (Space Research Institute, Moscow, Russia, October 14-18, 2013), 4MS3-PS-39, CD-ROM, http://ms2013.cosmos.ru/sites/ms2013.cosmos.ru/files/4m-s3_abstract_book.pdf. 2013. P. 204.

428. Ipatov S.I. Delivery of water and volatiles to the terrestrial planets. Abstracts of the 1st COSPAR Symposium "*Planetary systems of our Sun and other stars, and the future of space astronomy*" (Bangkok, Thailand, November 11-15, 2013), session 4, 2013, p. 88, an oral presentation at the session 4.

429. Ipatov S.I., Cho J.Y-K. Synthetic spectra from a general circulation simulation of a model extrasolar Earth, Abstracts of the *1st COSPAR Symposium "Planetary systems of our Sun and other stars, and the future of space astronomy*" (Bangkok, Thailand, November 11-15, 2013), session 4, 2013, p. 101, a poster at the session 4.

430. Ipatov S.I., Keith H., Alsubai K., Bramich D., Dominik M., Hundertmark M., Liebig C., Snodgrass C., Street R., Tsapras Y. Exoplanet detection capability of microlensing observations, Abstracts of the *1st COSPAR Symposium "Planetary systems of our Sun and other stars, and the future of space astronomy*" (Bangkok, Thailand, November 11-15, 2013), session 5, 2013, p. 116.

431. Ipatov S.I. Formation and migration of planetesimals, Abstracts of the *1st COSPAR Symposium "Planetary systems of our Sun and other stars, and the future of space astronomy*" (Bangkok, Thailand, November 11-15, 2013), session 5, 2013, p. 118, an oral presentation at the session 5.

432. Ipatov S.I. Formation of planetesimals from rarefied preplanetesimals. *Modern understanding of the Solar system and open questions:* All-Russian scientific Internet conference with international participation: conf. (Kazan, December 10, 2013) / Pax Grid virtual conference service; compiled by Sinyaev D.N. - Kazan: IP Sinyaev D.N., 2014. ISBN 978-5-906217-39-4, http://www.paxgrid.ru/conference/new_conf_pages.php?p=proceedings&c=solarem2013. P. 43. http://www.paxgrid.ru/proceedings_solarem2013.pdf#page=43. Oral presentation. In Russian.

433. Ipatov SI, Horne K. Models of sky brightness, Abstracts of *the 45th Lunar and Planetary Science Conference* (March 17-21, 2014, The Woodlands, Texas), 2014, #1390, 2 pages,

http://www.hou.usra.edu/meetings/lpsc2014/pdf/1390.pdf.
http://www.hou.usra.edu/meetings/lpsc2014/eposter/1390.pdf – eposter.

434. Ipatov S.I. Outbursts from cavities in comets, Abstracts of the *5th Bredikhin International conference* (Zavolzsk, Russia, 12-16 May, 2014), 2014, p. 17. http://www.inasan.ru/rus/conferences/bredikhin2014/presentations/2_Comets/Ipatov.pdf. Устный доклад.

435. Ipatov S.I., Cho Y-K. Spectra of Earth-like exoplanets with different rotation periods. Abstracts of the seminar *"Exoplanet Studies"* (June 3-4, 2014, Space Research Institute of the Russian Academy of Sciences, Moscow), 2014, p. 24. http://exo2014.cosmos.ru/sites/exo2014.cosmos.ru/files/exo2014-program-and-abstracts-4.pdf. Poster presentation. In Russian.

436. Ipatov S.I., Horne K. Efficiency of exoplanet search by microlensing method using different telescopes. Abstracts of the seminar *"Exoplanet Studies* " (June 3-4, 2014, Space Research Institute of the Russian Academy of Sciences, Moscow), 2014, pp. 8-9. http://exo2014.cosmos.ru/sites/exo2014.cosmos.ru/files/exo2014-program-and-abstracts-4.pdf. Oral presentation. In Russian.

437. Ipatov S.I., Marov M.Ya. Migration of small bodies and dust to the terrestrial planets. Abstracts of the int. conference “ *Modern Problems of Computational Mathematics and Mathematical Physics* ” (June 16-17, 2014, Moscow), MakS Press. 2014, pp. 101-103. http://vm.cs.msu.su/samarski2014/Abstracts_Samarskii_2014.pdf. Oral presentation. In Russian.

438. Ipatov S.I. The Earth-Moon system as a typical binary in the Solar System, Abstracts of the *Asteroids, Comets, Meteors conference* (June 30–July 4, 2014, Helsinki, Finland), Karri Muinonen, Antti Penttilä, Mikael Granvik, Anne Virkki, Grigori Fedorets, Olli Wilkman, and Tomas Kohout (eds.), a USB flash drive and website, http://www.helsinki.fi/acm2014/pdf-material/ACM2014.pdf. 2014, p. 248, an oral presentation.

439. Ipatov S.I. Outbursts from cavities in comets, Abstracts of the *Asteroids, Comets, Meteors conference* (June 30–July 4, 2014, Helsinki, Finland), Karri Muinonen, Antti Penttilä, Mikael Granvik, Anne Virkki, Grigori Fedorets, Olli Wilkman, and Tomas Kohout (eds.), a USB flash drive and website, in http://www.helsinki.fi/acm2014/pdf-material/ACM2014.pdf, 2014, p. 249, an oral presentation.

440. Ipatov S.I. Outbursts and cavities in comets, Abstracts of *40th Scientific Assembly of the Committee on Space Research* (COSPAR) (Moscow, Russia, August 02-10, 2014), a USB flash drive and website, http://adsabs.harvard.edu/abs/2014cosp...40E1284I, 2014, abstract ID is 12736, B0.4-0039-14, an oral presentation.

441. Ipatov S.I. Formation of satellites of trans-Neptunian objects at the stage of rarefied preplanetesimals, Abstracts of *40th Scientific Assembly of the Committee on Space Research* (COSPAR) (Moscow, Russia, August 02-10, 2014), a USB flash drive and website, http://adsabs.harvard.edu/abs/2014cosp...40E1283I, 2014, abstract ID is 12600, B0.3-0029-14, an oral presentation.

442. Ipatov S.I. The role of collisions of rarefied condensations in formation of embryos of the Earth and the Moon, Abstracts of *46th Lunar and Planetary Science Conference* (March 16-20, 2015, The Woodlands, Texas), 2015, #1355, 2 pages, http://www.hou.usra.edu/meetings/lpsc2015/pdf/1355.pdf. http://www.lpi.usra.edu/meetings/lpsc2015/eposter/1355.pdf - eposter.

443. Ipatov S.I. Origin of orbits of secondaries in discovered trans-Neptunian binaries, Abstracts of *46th Lunar and Planetary Science Conference* (March 16-20, 2015, The Woodlands, Texas), 2015, # 1512, 2 pages, http://www.hou.usra.edu/meetings/lpsc2015/pdf/1512.pdf. http://www.lpi.usra.edu/meetings/lpsc2015/eposter/1512.pdf - e-poster

444 . Ipatov S.I. Formation of satellite systems of trans-Neptunian objects and the Earth-Moon system. Scientific Conference " *Astronomy from Near Space to Cosmological Distants* " (May 25-30, 2015, Moscow). http://www.sai.msu.ru/EAAS/rus/confs/abstr_book2.pdf. 2015, pp . 57-58. Oral presentation. In Russian.

445. Ipatov S.I., Marov M.Ya. Migration of small bodies and dust to the terrestrial planets, Abstracts of *IX international conference “Near-Earth astronomy*” (August 31 – September 5, 2015, Terskol, Russia). http://www.terskol.com/conferences/abstracts.pdf.
http://www.inasan.ru/rus/conferences/OZA_2015/Absrtract_book.pdf. 2015, pp. 6-7. Oral presentation.

446. Ipatov S.I. Formation of embryos of the Earth-Moon system at the stage of rarefied condensations, Session EX3, Abstracts of *European Planetary Science Congress 2015* (27 September – 02 October 2015, Nantes, France), EPSC Abstracts, 2015, Vol. 10, EPSC2015-310, 2 pages, http://meetingorganizer.copernicus.org/EPSC2015/EPSC2015-310.pdf, a poster presentation.

447. Ipatov S.I., Marov M.Ya. Migration of planetesimals to forming terrestrial planets from the feeding zone of Jupiter and Saturn. Abstracts of *47th Lunar and Planetary Science Conference* (March 21-25, 2016, The Woodlands, Texas), 2016, #1458, 2 pages, http://www.hou.usra.edu/meetings/lpsc2016/pdf/1458.pdf – abstract, http://www.hou.usra.edu/meetings/lpsc2016/eposter/1458.pdf - e-poster.

448. Ipatov S.I., Marov M.Ya. Migration of icy planetesimals to forming terrestrial planets, Abstracts of *41st Scientific Assembly of the Committee on Space Research* (COSPAR) (Istanbul, Turkey, 30 July - 7 August 2016), 2016, presentation number B0.5-0011-16, selected for an oral presentation. COSPAR-2016 has been cancelled. https://www.cospar-assembly.org/abstractcd/COSPAR-16/abstracts/B0.5-0011-16.pdf. https://ui.adsabs.harvard.edu/abs/2016cosp...41E.867I/abstract.

449. Ipatov S.I. Formation and growth of embryos of the Earth-Moon system, Abstracts of *41st Scientific Assembly of the Committee on Space Research* (COSPAR) (Istanbul, Turkey, 30 July - 7 August 2016), 2016, presentation number B0.5-0017-16, selected for a poster presentation. COSPAR-2016 has been cancelled. https://www.cospar-assembly.org/abstractcd/COSPAR-16/abstracts/B0.5-0017-16.pdf. https://ui.adsabs.harvard.edu/abs/2016cosp...41E.868I/abstract.

450. Ipatov S.I. Angular momenta of collided rarefied preplanetesimals needed for formation of trans-neptunian satellite systems. Abstracts of *48th Lunar and Planetary Science Conference* (March 20-24, 2017, The Woodlands, Texas), 2017, #1554, 2 pages, http://www.hou.usra.edu/meetings/lpsc2017/pdf/1554.pdf – abstract. http://www.hou.usra.edu/meetings/lpsc2017/eposter/1554.pdf - e-poster.

451. Ipatov S.I., Marov M.Ya. Migration of small bodies to the terrestrial planets. Abstracts of the *6th Bredikhin International conference* (Zavolzsk, Russia, September 4-8, 2017), 2017, p. 26, an oral presentation. Website of the conference: http://agora.guru.ru/display.php?conf=bredikhin2017.

452. Ipatov S.I. Formation of satellite systems of small bodies and the embryos of the Moon and the Earth. Abstracts of the *6th Bredikhin International conference* (Zavolzsk, Russia, September 4-8, 2017), 2017, p. 27, an oral presentation. Website of the conference: http://agora.guru.ru/display.php?conf=bredikhin2017.

453. Ipatov S.I., Marov M.Ya. Migration of icy objects to forming terrestrial planets. Abstracts of *European Planetary Science Congress 2017* (September 17-22, 2017, Riga, Latvia), 2017, EPSC2017-211, http://meetingorganizer.copernicus.org/EPSC2017/EPSC2017-211.pdf, 2 pages, an oral presentation - https://presentations.copernicus.org/EPSC2017-211_presentation.pdf.

454. Ipatov S.I. Formation of trans-Neptunian satellite systems at the stage of rarefied condensations, Abstracts of *European Planetary Science Congress 2017* (September 17-22, 2017, Riga, Latvia), 2017, EPSC2017-225, 2 pages, http://meetingorganizer.copernicus.org/EPSC2017/EPSC2017-225.pdf, an oral presentation - https://presentations.copernicus.org/EPSC2017-225_presentation.pdf.

455. Ipatov S.I. Formation and growth of embryos of the Earth and the Moon. Abstracts of *European Planetary Science Congress 2017* (September 17-22, 2017, Riga, Latvia), 2017, EPSC2017-355, http://meetingorganizer.copernicus.org/EPSC2017/EPSC2017-355.pdf, 2 pages, a poster presentation - https://presentations.copernicus.org/EPSC2017-355_presentation.pdf.

456. Ipatov S.I. Formation of satellite systems of small bodies and the embryos of the Moon and the Earth, Abstracts of "The X international conference «*Near-Earth Astronomy-2017*" (Agoy, Krasnodar region, Russia, October 2-6, 2017). An oral presentation. Website of the conference: http://agora.guru.ru/display.php?conf=oza-2017.
https://docviewer.yandex.ru/view/0/?*=hA2OT0drleOHLXXG6roDUco%2BLfh7InVybCI6InlhLWRpc2stcHVibGljOi8vU09DMUlSVnRrZGZndHBnUEFFNDBORkRtamF6L3pLRXpxbGFMR3dBeVZ1WT0iLCJ0aXRsZSI6Ik9aQS0yMDE3IEFic3RhY3QgYm9vay5wZGYiLCJ1aWQiOiIwIiwieXUiOiI4MjY5MzA3MzQxNTE1NTIxMzUwIiwibm9pZnJhbWUiOmZhbHNlLCJ0cyI6MTUxNzYwNzIzNTIwOX0%3D – book of abstracts.

457. Ipatov S.I., Elenin L.V. Suggested models of the probabilities of discovery of near-Earth objects in different regions of the sky based on studies of migration of celestial bodies, Abstracts of "The X international conference «*Near-Earth Astronomy-2017*" (Agoy, Krasnodar region, Russia, October 2-6, 2017). C. 23. An oral presentation. Website of the conference: http://agora.guru.ru/display.php?conf=oza-2017.
https://docviewer.yandex.ru/view/0/?*=hA2OT0drleOHLXXG6roDUco%2BLfh7InVybCI6InlhLWRpc2stcHVibGljOi8vU09DMUlSVnRrZGZndHBnUEFFNDBORkRtamF6L3pLRXpxbGFMR3dBeVZ1WT0iLCJ0aXRsZSI6Ik9aQS0yMDE3IEFic3RhY3QgYm9vay5wZGYiLCJ1aWQiOiIwIiwieXUiOiI4MjY5MzA3MzQxNTE1NTIxMzUwIiwibm9pZnJhbWUiOmZhbHNlLCJ0cyI6MTUxNzYwNzIzNTIwOX0%3D

– book of abstracts.

458. Ipatov S.I. Formation and growth of embryos of the Earth and the Moon. Abstracts of *49th Lunar and Planetary Science Conference* (March 19-23, 2018, The Woodlands, Texas), (LPI Contrib. No. 2083), 2018, #1602, 2 pages, https://www.hou.usra.edu/meetings/lpsc2018/pdf/1602.pdf - abstract (2018). https://www.hou.usra.edu/meetings/lpsc2018/eposter/1602.pdf - e-poster.

459. Ipatov S.I. Formation of satellite systems of small bodies and the embryos of the Moon and the Earth. Abstracts of *42nd COSPAR Scientific Assembly* (Pasadena, California, USA, July 14-22, 2018), presentation number B1.3-0049-18, http://cospar2018.org/wp-content/uploads/2018/07/COSPAR-2018-Abstract-Book_July21-2018-UPDATE.pdf, 2018, p. 326. http://adsabs.harvard.edu/abs/2018cosp...42E1558I.

460. Ipatov S.I., Marov M.Ya. Migration of bodies from beyond Jupiter's orbit to the Earth and the Moon. Abstracts of *42nd COSPAR Scientific Assembly* (Pasadena, California, USA, July 14-22, 2018), 2018, presentation number B1.3-0050-18. http://cospar2018.org/wp-content/uploads/2018/07/COSPAR-2018-Abstract-Book_July21-2018-UPDATE.pdf, 2018, p. 326-327. http://adsabs.harvard.edu/abs/2018cosp...42E1557I.

461. Marov M.Ya., **Ipatov S.I**. Water inventory from beyond the Jupiter orbit to the terrestrial planets and the Moon. Abstracts of XXXth General Assembly of the International Astronomical Union (Vienna, Austria, August 20-31, 2018), an oral presentation. https://astronomy2018.univie.ac.at/abstractsiaus345/#iaus345abstr15 (only the first author is in the web file).

462. Ipatov S.I. Formation of the Earth-Moon system. Abstracts of *XXXth General Assembly of the International Astronomical Union* (Vienna, Austria, August 20-31, 2018), 2018. https://astronomy2018.univie.ac.at/abstractsiaus345/#iaus345abstr16. an oral presentation.

463. Ipatov S.I., Feoktistova E.A., Svetsov V.V. Near-Earth object population and formation of lunar craters during the last billion of years. Abstracts of *XXXth General Assembly of the International Astronomical Union* (Vienna, Austria, August 20-31, 2018), 2018, https://astronomy2018.univie.ac.at/PosterAbstracts/posteriaus345 (only the first author is in the web file), a poster.

464. Ipatov S.I. Formation of trans-Neptunian satellite systems at the stage of rarefied condensations. Abstracts of *XXXth General Assembly of the International Astronomical Union* (Vienna, Austria, August 20-31, 2018), 2018, https://astronomy2018.univie.ac.at/PosterAbstracts/posterDivF, a poster.

465. Ipatov S.I. Migration of bodies to the Earth from different distances from the Sun. EPSC Abstracts Vol. 12, EPSC2018-516, 2018. *European Planetary Science Congress 2018* (Technische Universitat Berlin, September 16–21, 2018, Berlin, Germany), a poster P36 at session: Exo-2, 2 pages. https://meetingorganizer.copernicus.org/EPSC2018/EPSC2018-516.pdf.

466. Ipatov S.I. Probabilities of collisions of bodies from the feeding zone of the terrestrial planets with the planets, the Moon, and their embryos. Abstracts of *50th Lunar and Planetary Science Conference* (March 18-22, 2019, The Woodlands, Texas), 2019. #2289, 2 pages. https://www.hou.usra.edu/meetings/lpsc2019/pdf/2289.pdf. e-poster: https://www.hou.usra.edu/meetings/lpsc2019/eposter/2289.pdf.

467. Ipatov S.I. Migration of planetesimals to the Earth and the Moon from different distances from the Sun. Abstracts of *50th Lunar and Planetary Science Conference* (March 18-22, 2019, The Woodlands, Texas), 2019. #2594, 2 pages. https://www.hou.usra.edu/meetings/lpsc2019/pdf/2594.pdf. e-poster: https://www.hou.usra.edu/meetings/lpsc2019/eposter/2594.pdf

468. Feoktistova E.A., **Ipatov S.I.**, Svetsov V.V. Tripples of lunar craters formed by encounters of satellite systems of near-Earth objects with the Moon. Abstracts of *50th Lunar and Planetary Science Conference* (March 18-22, 2019, The Woodlands, Texas), 2019. #1946, 2 pages. https://www.hou.usra.edu/meetings/lpsc2019/pdf/1946.pdf. e-poster: https://www.hou.usra.edu/meetings/lpsc2019/eposter/1946.pdf

469. Ipatov S.I. Probabilities of collisions of planetesimals from different regions of the feeding zone of the terrestrial planets with the forming planets and the Moon. Abstracts *of the All-Russian Annual Seminar on Experimental Mineralogy, Petrology and Geochemistry*. (VESEMPG-2019). April 16-17, 2019. Moscow. http://intranet.geokhi.ru/rasempg, 2019. #063. Oral presentation. In Russian.

470. Ipatov S.I. Probabilities of collisions of planetesimals migrated from beyond the orbit of Mars with the Earth and the Moon. Abstracts *of the All-Russian Annual Seminar on Experimental Mineralogy, Petrology and Geochemistry*. (VESEMPG-2019). April 16-17, 2019. Moscow. http://intranet.geokhi.ru/rasempg, 2019. #064. Poster presentation. In Russian.

471. Ipatov S.I., Feoktistova E.A., Svettsov V.V. The number of near-Earth objects and the formation of lunar craters during the last billion years. Abstracts *of the All-Russian Annual Seminar on Experimental Mineralogy, Petrology and Geochemistry*. (VESEMPG-2019). April 16-17, 2019. Moscow. http://intranet.geokhi.ru/rasempg, 2019. #065. Poster presentation. In Russian.

472. Marov M.Ya., Ipatov S.I. Delivery of water from beyond Jupiter's orbit to the terrestrial planets and the Moon. Abstracts of *EPSC-DPS Joint Meeting 2019* (September 15-20, 2019, Geneva, Switzerland), Vol. 13, EPSC-DPS2019-458-1, 2019. https://meetingorganizer.copernicus.org/EPSC-DPS2019/EPSC-DPS2019-458-1.pdf, a poster.

473. Ipatov S.I. The Earth-Moon system as a typical satellite system in the Solar System. Abstracts of *EPSC-DPS Joint Meeting 2019* (September 15-20, 2019, Geneva, Switzerland), Vol. 13, EPSC-DPS2019-733, 2019. https://meetingorganizer.copernicus.org/EPSC-DPS2019/EPSC-DPS2019-733-1.pdf, a poster.

474. Ipatov S.I. Initial distances from the Sun for planetesimals that collided with forming terrestrial planets and the Moon. Abstracts of *EPSC-DPS Joint Meeting 2019* (September 15-20, 2019, Geneva, Switzerland), Vol. 13, EPSC-DPS2019-462-1, 2019. https://meetingorganizer.copernicus.org/EPSC-DPS2019/EPSC-DPS2019-462-1.pdf, a poster.

475. Ipatov S.I. Angular momenta of colliding rarefied condensations Abstracts of *EPSC-DPS Joint Meeting 2019* (September 15–20, 2019, Geneva, Switzerland), Vol. 13, EPSC-DPS2019-2034, 2019. https://meetingorganizer.copernicus.org/EPSC-DPS2019/EPSC-DPS2019-2034.pdf, a poster.

476. Marov M.Ya., **Ipatov S.I.** Migration of planetesimals from different distances from the feeding zone of the giant planets to the terrestrial planets and the Moon. Abstracts of the XI International scientific conference "*Near-Earth astronomy and space heritage*" (September 30 – October 4, 2019, Kazan, Russia). http://agora.guru.ru/display.php?conf=oza2019, https://yadi.sk/i/ifmloGZ-X-w8cg, pp. 65-66. Oral presentation. In Russian.

477. Ipatov S.I. Probabilities of collisions of planetesimals from different parts of the feeding zone of the terrestrial planets with the forming planets, the Moon, and their embryos, Abstracts of the XI International scientific conference "*Near-Earth astronomy and space heritage*" (September 30 – October 4, 2019, Kazan, Russia). http://agora.guru.ru/display.php?conf=oza2019, https://yadi.sk/i/ifmloGZ-X-w8cg, pp. 60-61. Poster presentation. In Russian.

478. Ipatov S.I., Feoktistova E.A., Svettsov V.V. Estimates of variations in the number of near-Earth objects based on the ages of lunar craters over the past billion years. Abstracts of the XI International scientific conference "*Near-Earth astronomy and space heritage*" (September 30 – October 4, 2019, Kazan, Russia), http://agora.guru.ru/display.php?conf=oza2019, https://yadi.sk/i/ifmloGZ-X-w8cg, pp. 46-47. An oral presentation. In Russian.

479. Ipatov S.I., Feoktistova E.A., Svetsov V.V. Variations in the number of near-Earth objects and lunar craters during the last billion years. Abstracts of *51st Lunar and Planetary Science Conference* (The Woodlands, Texas - March 16-20, 2020), LPI Contribution No. 2326, 2020, id.1910, 2 pages. https://www.hou.usra.edu/meetings/lpsc2020/pdf/1910.pdf, https://www.hou.usra.edu/meetings/lpsc2020/eposter/1910.pdf - e-poster.

480. Ipatov S.I. Probabilities of collisions of near-Earth objects with the Earth. Abstracts *of the All-Russian annual seminar on experimental mineralogy, petrology and geochemistry.* (VESEMPG-2020). April 14-15, 2020, Moscow. http://www.geokhi.ru/rasempg/Shared%20Documents/2020/Тезисы%20ВЕСЭМПГ%202020.pdf, 2020. #047. P. 49. Oral presentation. In Russian.

481. Ipatov S.I. Formation of the terrestrial planets and the Moon. Abstracts of Virtual American Astronomical Society meeting #236 (1–3 June 2020), id. 113.01. *Bulletin of the American Astronomical Society*, Vol. 52, No. 3. https://baas.aas.org/pub/aas236-113p01-ipatov/release/1, https://ui.adsabs.harvard.edu/abs/2020AAS...23611301I/abstract. iPoster Plus (poster+oral) – https://aas236-aas.ipostersessions.com/default.aspx?s=EA-3E-83-9E-0A-34-58-51-1C-4A-6D-28-4B-9E-9D-36 – iPoster; https://my.aas.org/services/AAS236/AAS236Recordings.aspx (Presenter Name Contains: Ipatov, - my oral presentation is at the end of session 113).

482. Ipatov S.I., Feoktistova E.A., Svetsov V.V. Estimates of the number of near-Earth objects based on the number of lunar craters formed during the last billion years. *11th Planetary Crater Consortium Meeting* (August 5-7, 2020). #2038. https://www.hou.usra.edu/meetings/crater2020/pdf/2038.pdf, 2 pages. An oral presentation.

483. Feoktistova E.A. **Ipatov S.I.** Depths of the lunar Copernicans craters. *11th Planetary Crater Consortium Meeting* (August 5-7, 2020). #2021. https://www.hou.usra.edu/meetings/crater2020/pdf/2021.pdf. Print only, 2 pages.

484. Ipatov S.I. Migration of planetesimals from beyond Mars' orbit to the Earth. *14th Europlanet Science Congress 2020*, online, 21 September – 9 Oct 2020, EPSC2020-71, 2020. https://meetingorganizer.copernicus.org/EPSC2020/EPSC2020-71.html, https://doi.org/10.5194/epsc2020-71, 5 pages, https://presentations.copernicus.org/EPSC2020/EPSC2020_71_presentation.mp4 - an oral presentation.

485. Ipatov S.I. Probabilities of collisions of planetesimals with exoplanets in the Proxima Centauri planetary system. *14th Europlanet Science Congress 2020*, online, 21 September–9 Oct 2020, EPSC2020-416, 2020. https://meetingorganizer.copernicus.org/EPSC2020/EPSC2020-416.html, https://doi.org/10.5194/epsc2020-416. 2 pages, a poster.

486. Ipatov S.I. Formation of the terrestrial planets from planetesimals originated at different distances from the Sun. Abstracts of *43rd COSPAR Scientific Assembly* (Sydney, Australia, 28 January – 4 February, 2021). B0.1-0036-21. Id. 148. https://ui.adsabs.harvard.edu/abs/2021cosp...43E.148I/abstract, https://www.cospar-assembly.org/user/download.php?id=26592&type=abstract&section=congressbrowser - an abstract; https://assets.entegy.com.au/documents/c85a7e61-cbb7-464a-ba9f-3248c1181880.pdf - a poster; https://www.youtube.com/watch?v=9W6ZfZHns_w – a video presentation.

487. Ipatov S.I., Feoktistova E.A., Svetsov V.V. Estimates of variations in the number of near-Earth objects based on the ages of lunar craters over the past billion years. Abstracts, *43rd COSPAR Scientific Assembly* (Sydney, Australia, 28 January – 4 February, 2021). B3.1-0027-21. Id. 362. https://ui.adsabs.harvard.edu/abs/2021cosp...43E.362I/abstract, https://www.cospar-assembly.org/user/download.php?id=26645&type=abstract&section=congressbrowser – an abstract; https://assets.entegy.com.au/documents/a9244ee6-770d-41ba-bc90-8181a687b100.pdf - a poster; https://www.youtube.com/watch?v=woK-x4R3ZKQ - a video presentation.

488. Ipatov S.I. Migration of bodies from the zone of the outer asteroid belt to the Earth. Abstracts of *52nd Lunar and Planetary Science Conference* (March 15–19, 2021, a virtual conference), #1618, 2 pages, https://www.hou.usra.edu/meetings/lpsc2021/pdf/1618.pdf, https://lpsc2021.ipostersessions.com/?s=EA-59-31-3B-A2-7B-45-F1-E5-68-07-AA-0B-FC-FC-05 - an iposter.

489. Ipatov S.I. Exocometary bombardment of planets in the habitable zone in the Proxima Centauri planetary system. Abstracts of *international conference "VII Bredikhin conference*" (May 24-28, 2021, Zavolzsk, Russia), pp. 18-19. http://www.inasan.ru/wp-content/uploads/2021/06/Abstracts_2021_V2.pdf. Oral presentation.

490. Ipatov S.I., Marov M.Ya. Collisions of small bodies with the forming Earth and Moon. *International Conference " VII Bredikhin readings* "(May 24-28, 2021, Zavolzhsk). p. 20-22. http://www.agora.guru.ru/display.php?conf=bredikhin2020, http://www.inasan.ru/wp-content/uploads/2021/06/Abstracts_2021_V2.pdf. Invited presentation.

491. Ipatov S.I. Migration of planetesimals from the outer planets to the inner planets in the Proxima Centauri system and in the solar system. Abstracts of *the All-Russian Annual Seminar on Experimental Mineralogy, Petrology and Geochemistry* (VESEMPG-2021, May 25-26, 2021, Moscow), VESEMPG – 050, p. 52, oral presentation. http://www.geokhi.ru/rasempg/Shared%20Documents/Forms/AllItems.aspx?RootFolder=%2Frasempg%2FShared%20Documents%2F2021. http://www.geokhi.ru/rasempg/Shared%20Documents/2021/Краткие%20тезисы%20ВЕСЭМПГ2021-31052021.pdf. In Russian.

492. Ipatov S.I. Migration of planetesimals to planets in the habitable zone in the Solar System and the Proxima Centauri system. Abstracts. *All-Russian Astronomical Conference VAK-2021 "Astronomy in the Era of Multichannel Research"* (August 23-28, 2021, Sternberg Astronomical Institute of Moscow State University, Moscow, Russia, http://www.vak2021.ru), 2021, p. 225. Oral presentation. https://www.vak2021.ru/wp-content/uploads/2021/08/vak2021_abstracts.pdf. In Russian.

493. Ipatov S.I. Formation of the terrestrial planets and the Moon. Abstracts. *All-Russian Astronomical Conference VAK-2021 "Astronomy in the Era of Multichannel Research"* (August 23-28, 2021, Sternberg Astronomical Institute of Moscow State University named after M.V. Lomonosov, Moscow, Russia, http://www.vak2021.ru), 2021, p. 226. Oral presentation. https://www.vak2021.ru/wp-content/uploads/2021/08/vak2021_abstracts.pdf. In Russian.

494. Ipatov S.I. Formation of the terrestrial planets and the Moon. *Astronomy at the epoch of multimessenger studies*. Proceedings of the *VAK-2021 conference* (Aug 23–28, 2021). Ed. by A.M. Cherepashchuk et al. - Moscow, Janus-K. 2022. P. 236-237. DOI: 10.51194/VAK2021.2022.1.1.083. https://www.vak2021.ru/proceedings/. https://www.vak2021.ru/wp-content/uploads/2022/03/VAK_2021_proceedings.pdf, http://www.inasan.ru/wp-content/uploads/2022/02/VAK_2021_.pdf#page=237. https://www.academia.edu/99225166. https://elibrary.ru/item.asp?id=67932703.

495. Ipatov S.I. Migration of planetesimals to planets located in habitable zones in the Solar System and in the Proxima Centauri system. *Astronomy at the epoch of multimessenger studies.* Proceedings of the *VAK-2021 conference* (Aug 23–28, 2021). Ed. by A.M. Cherepashchuk et al. - Moscow, Janus-K. 2022. P. 238-239. DOI: 10.51194/VAK2021.2022.1.1.084. https://www.vak2021.ru/proceedings/. http://www.inasan.ru/wp-content/uploads/2022/02/VAK_2021_.pdf#page=239. https://elibrary.ru/item.asp?id=67932703.

496. Ipatov S.I. Migration of planetesimals to the Earth and the Moon from the region of the outer asteroid belt, *European Planetary Science Congress 2021*, online, September 13-24, 2021, EPSC2021-100, https://doi.org/10.5194/epsc2021-100. https://www.researchgate.net/publication/369585726. https://meetingorganizer.copernicus.org/EPSC2021/EPSC2021-100.html, an oral presentation.

497. Ipatov S.I. Migration of planetesimals to forming terrestrial planets, Abstracts of *IAU Symposium 364 "Multi-scale (time and mass) dynamics of space objects*" (October 18-22, 2021, Iași, Romania), The book of abstracts is available at: https://www.math.uaic.ro/~IAU_S364/adm/IAUS_364_book_of_abstracts.pdf, p. 31, an oral presentation.

498. Ipatov S.I. Migration of planetesimals in the TRAPPIST-1 exoplanetary system. Abstracts of *53nd Lunar and Planetary Science Conference* (The Woodlands, Texas - March 7–11, 2022), #1182, 2 pages, https://www.hou.usra.edu/meetings/lpsc2022/pdf/1182.pdf, https://lpsc2022.ipostersessions.com/?s=3B-6A-91-5C-9C-63-DE-66-1F-99-E9-7B-E0-7C-B5-CA - an iposter.

499. Ipatov S.I. Migration of bodies ejected from the Earth and collided with the Earth and the Moon. Abstracts of *53nd Lunar and Planetary Science Conference* (The Woodlands, Texas - March 7–11, 2022), #1298, 2 pages, https://www.hou.usra.edu/meetings/lpsc2022/pdf/1298.pdf, https://lpsc2022.ipostersessions.com/?s=9B-5E-DB-47-82-D4-92-21-8B-2F-D9-53-5D-DF-8A-B3 - an iposter.

500. Ipatov S.I. Probabilities of fallout of bodies ejected from the Earth during fallout of impactor bodies onto the Earth onto the planets and the Moon. Abstracts of *the All-Russian Annual Seminar on Experimental Mineralogy, Petrology and Geochemistry* (VESEMPG-2022, April 19-20, 2022). VESEMPG 2022. #045. P. 47. 1 p. - Oral presentation. http://www.geokhi.ru/rasempg/SitePages/%D0%94%D0%BE%D0%BC%D0%B0%D1%88%D0%BD%D1%8F%D1%8F.aspx?RootFolder=%2Frasempg%2FShared%20Documents%2F2022&FolderCTID=0x01200077A1861A98FF9D4FB359CBBC44A5DE86&View=%7B4A98432A%2DB52E%2D452B%2D95D6%2DACAAEC61D655%7D. In Russian.

501. Ipatov S.I. Migration of bodies to the Earth from different distances from the Sun. Abstracts of *IAU General Assembly* (Busan, South Korea, Aug. 2-11, 2022). 2022. No 3066. P. 708. https://www.youtube.com/watch?v=R1Ao1Tc-5nA - e-talk.

502. Ipatov S.I. Migration of bodies ejected from the Earth into heliocentric orbits. Abstracts of *IAU General Assembly* (Busan, South Korea, Aug. 2-11, 2022). 2022. No 3107. P. 1398. e-poster.

503. Ipatov S.I. Motion of planetesimals in the Proxima Centauri planetary system. Abstracts of *IAU General Assembly* (Busan, South Korea, Aug. 2-11, 2022). 2022. No 3070. P. 1404. e-talk. https://www.iauga2022.org/program/poster_view.asp?A_s_code=gRazgRa2gJqzgZgzGBKPgBczGRc=&pttype=Mj2kmlYV&abs_key=gRazgRa2gReygBgyGBOPGJi1GZa=, https://www.youtube.com/watch?v=ZsZ645ofdlY.

504. Ipatov S.I. Motion of planetesimals from the feeding zone of Proxima Centauri c. Abstracts of *54nd Lunar and Planetary Science Conference* (The Woodlands, Texas - March 13–17, 2023), #1309, 2 pages. https://www.hou.usra.edu/meetings/lpsc2023/pdf/1309.pdf, https://lpsc2023.ipostersessions.com/?s=55-BB-36-8D-89-E0-55-68-EA-39-AF-2A-4C-B8-63-8C - a poster. https://www.hou.usra.edu/meetings/lpsc2023/program/ - program.

505. Ipatov S.I. Collisions of bodies ejected from several places on the Earth and the Moon with the terrestrial planets and the Moon. Abstracts of *54nd Lunar and Planetary Science Conference* (The Woodlands, Texas - March 13–17, 2023), #1508.

https://www.hou.usra.edu/meetings/lpsc2023/pdf/1508.pdf , https://lpsc2023.ipostersessions.com/?s=48-B7-96-FD-BD-8D-42-97-9C-5D-C2-4D-89-78-9E-54. a poster. https://www.hou.usra.edu/meetings/lpsc2023/program/ - program.

506. Ipatov S.I. Growth of exoplanets due to planetesimals, initially located at different distances from the star, in the TRAPPIST-1 system. Abstracts of *the All-Russian Annual Seminar on Experimental Mineralogy, Petrology and Geochemistry* (VESEMPG-2023, April 11-12, 2023, Moscow), VESEMPG 2023. 048, p. 50, https://new.ras.ru/upload/iblock/fff/%D0%A2%D0%B5%D0%B7%D0%B8%D1%81%D1%8B%20%D0%92%D0%95%D0%A1%D0%AD%D0%9F%D0%9C%D0%932023.pdf. poster presentation. http://www.geokhi.ru/rasempg. In Russian.

507. Ipatov S.I. Dimensions of the feeding zone of the planet Proxima Centauri. Abstracts of *the All-Russian Annual Seminar on Experimental Mineralogy, Petrology and Geochemistry* (VESEMPG-2023, April 11-12, 2023, Moscow), VESEMPG 2023 - 049, p. 51, https://new.ras.ru/upload/iblock/fff/%D0%A2%D0%B5%D0%B7%D0%B8%D1%81%D1%8B%20%D0%92%D0%95%D0%A1%D0%AD%D0%9F%D0%9C%D0%932023.pdf. poster presentation. http://www.geokhi.ru/rasempg. In Russian.

508. Ipatov S.I. Migration of exocomets in the Proxima Centauri and Trappist-1 planetary systems. Abstracts of *international conference "VIII Bredikhin conference*, September 4-8, 2023, Zavolzsk, Russia). http://www.inasan.ru/wp-content/uploads/2023/08/Abstract-book_1.pdf. P. 11-12. Oral presentation.

509. Ipatov S.I. Exchange of meteorites between the Earth and the Moon. Abstracts of *International Conference "VII Bredikhin Readings"* (September 4-8, 2023, Zavolzhsk). http://www.inasan.ru/wp-content/uploads/2023/08/Abstract-book_1.pdf. P. 32-33. Oral Presentation. http://www.agora.guru.ru/bredikhin2023 – conference website. http://www.inasan.ru/wp-content/uploads/2023/08/programma_3.pdf - conference program.

510. Ipatov S.I. Migration of planetesimals from the orbit of planet Proxima Centauri c. *DPS-EPSC 2023* (San Antonio, Texas, USA, 1-6 October 2023). https://aas.org/meetings/dps55. Program Number: 222.01. https://baas.aas.org/pub/2023n8i222p01/release/1, https://submissions.mirasmart.com/DPS55/ViewSubmissionFile.aspx?sbmID=95&mode=html&validate=false - abstract. https://submissions.mirasmart.com/Verify/DPS55/Submission/Temp/radlzf5t4qg.pdf - a poster.

511. Ipatov S.I. Motion of planetesimals in the Proxima Centauri, Trappist-1 and Gliese 581 planetary systems. Abstracts of *55th Lunar and Planetary Science Conference* (The Woodlands, Texas - March 11–15, 2024), #1054, 2 pages. https://www.hou.usra.edu/meetings/lpsc2024/pdf/1054.pdf .https://lpsc2024.ipostersessions.com/?s=7E-78-4F-7C-57-01-9A-CE-60-A9-E6-04-08-1F-19-00 – iposter. https://www.hou.usra.edu/meetings/lpsc2024/program - program. https://www.academia.edu/126487537.

512. Ipatov S.I. Probabilities of collisions of bodies ejected from the Earth with the terrestrial planets. Abstracts of *55th Lunar and Planetary Science Conference* (The Woodlands, Texas - March 11–15, 2024), #1231, 2 pages. https://www.hou.usra.edu/meetings/lpsc2024/pdf/1231.pdf . https://lpsc2024.ipostersessions.com/?s=4A-4B-CD-60-CD-FB-C7-D6-69-03-3B-F2-98-87-34-75 – iposter. https://www.hou.usra.edu/meetings/lpsc2024/program - program. https://www.academia.edu/126487489.

513. Ipatov S.I. Migration of bodies in the exoplanetary systems Proxima Centauri and Trappist-1. Abstracts. *All-Russian annual seminar on experimental mineralogy, petrology and geochemistry VESEMPG-2024*. Moscow. April 16-17, 2024. http://portal.geokhi.ru/Newsdata/%D0%A2%D0%B5%D0%B7%D0%B8%D1%81%D1%8B_%D0%B4%D0%BE%D0%BA%D0%BB%D0%B0%D0%B4%D0%BE%D0%B2_%D0%92%D0%95%D0%A1%D0%AD%D0%9C%D0%9F%D0%93_2024.pdf. VESEPMG 2024 – 049. Page. 51. Poster presentation. In Russian.

514. Ipatov S.I. Migration of bodies ejected from the Earth and the Moon. Abstracts. *All-Russian Annual Seminar on Experimental Mineralogy, Petrology, and Geochemistry* VESEMPG-2024. Moscow. April 16-17, 2024. http://portal.geokhi.ru/Newsdata/%D0%A2%D0%B5%D0%B7%D0%B8%D1%81%D1%8B_%D0%B4%D0%BE%D0%BA%D0%BB%D0%B0%D0%B4%D0%BE%D0%B2_%D0%92%D0%95%D0%A1%D0%AD%D0%9C%D0%9F%D0%93_2024.pdf . VESEPMG 2024 - 050. Page. 52. Poster presentation. In Russian.

515. Ipatov S.I. Delivery of material to the terrestrial planets and the Moon at the late stages of their formation. *Earth and Planets Origin and Evolution Workshop* 2024. Paris – May 13-17. 2024. Abstracts. https://drive.google.com/file/d/1Chmidh88sgUkmH2eftI1GWAhfJFXvRfs/view. A poster.

516. Ipatov S.I. Probabilities of collisions of bodies ejected from the Earth with the terrestrial planets and the Moon. *45th COSPAR Scientific Assembly*. July 13-21, 2024. A poster no TWT-032. https://app.cospar-assembly.org/2024/browser/presentation/34076 - abstracts. https://www.cospar-assembly.org/admin/session_cospar.php?session=1184 – the session COSPAR-2024-B0.1: PLANETARY SCIENCE HIGHLIGHTS.

517. Ipatov S.I. Migration of bodies in the Proxima Centauri and Trappist-1 planetary systems. *IAU Symposium 393: Planetary Science and Exoplanets in the Era of the James Webb Space Telescope* (August 13-15, 2024, Cape Town, South Africa). Abstract no 288. https://ui.adsabs.harvard.edu/abs/2024IAUGA..32P.288I/abstract. https://astronomy2024.org/wp-content/uploads/2024/07/Abstracts-book_July30.pdf - abstracts book, p. 2378. https://ui.adsabs.harvard.edu/abs/2024IAUGA..32P.288I/abstract. https://astronomy2024.org/wp-content/uploads/posters/288.pdf - a poster. https://iaus393.uca.ma/program/. https://astronomy2024.org/. Poster session on Aug 14 from 15:00, screen number 71. https://iaus393.uca.ma.

518. Ipatov S.I. Migration of bodies ejected from the Earth and the Moon. *IAU Symposium 393: Planetary Science and Exoplanets in the Era of the James Webb Space Telescope* (August 13-15, 2024, Cape Town, South Africa). Abstracts. A poster. Abstract no 289. https://ui.adsabs.harvard.edu/abs/2024IAUGA..32P.289I/abstract. https://astronomy2024.org/wp-content/uploads/2024/07/Abstracts-book_July30.pdf - abstracts book, p. 2379. https://astronomy2024.org/wp-content/uploads/posters/289.pdf - a poster. In https://astronomy2024.org/poster-schedule/ make a search for Ipatov. Poster session on Aug 15 from 10:00, screen number 73. https://iaus393.uca.ma.

519. Ipatov S.I. Exchange of meteorites between the terrestrial planets and the Moon. *All-Russian Conference (with international participation) "Modern Astronomy: from the early Universe to exoplanets and black holes* ". August 25-31, 2024. Nizhny Arkhyz, Karachay-Cherkess Republic. Oral presentation. https://vak2024.ru/ru/prog.php. https://vak2024.ru/doc/CAO_PAH_сборник_тезисов.pdf – collection of abstracts. P. 223. https://vak2024.ru/ru/lk/pdf.php?a1=224 – abstract. https://vak2024.ru/doc/reportFiles/T9_O224_%D0%98%D0%BF%D0%B0%D1%82%D0%BE%D0%B2_%D0%A1%D0%98.pdf - presentation slides. In Russian.

520. Ipatov S.I. Migration of planetesimals and dust particles in the Proxima Centauri exoplanetary system. *All-Russian conference (with international participation)* "*Modern Astronomy: from the early Universe to exoplanets and black holes*". August 25-31, 2024. Nizhny Arkhyz, Karachay-Cherkess Republic. Oral presentation. https://vak2024.ru/ru/prog.php. https://vak2024.ru/doc/SAO_RAN_collection_of_theses.pdf - book of abstracts. P. 205. https://vak2024.ru/ru/lk/pdf.php?a1=225 - abstract. https://vak2024.ru/doc/reportFiles/T8_O225_%D0%98%D0%BF%D0%B0%D1%82%D0%BE%D0%B2_%D0%A1%D0%98.pdf - presentation slides. In Russian.

521. Ipatov S.I. Migration of planetesimals in the TRAPPIST-1 exoplanetary system. *All-Russian Conference (with international participation) "Modern Astronomy: from the early Universe to exoplanets and black holes*". August 25-31, 2024. Nizhny Arkhyz, Karachay-Cherkess Republic. https://vak2024.ru/ru/prog.php. https://vak2024.ru/doc/CAO_PAH_сборник_тезисов.pdf - book of abstracts. P. 206. https://vak2024.ru/ru/lk/pdf.php?a1=252– abstract. https://vak2024.ru/doc/reportFiles/T8_OP252_%D0%98%D0%BF%D0%B0%D1%82%D0%BE%D0%B2_%D0%A1%D0%98.pdf – presentation slides. https://vak2024.ru/doc/posterFiles/T8_P252_%D0%98%D0%BF%D0%B0%D1%82%D0%BE%D0%B2_%D0%A1%D0%98.pdf, poster presentation. In Russian.

522. Ipatov S.I. Motion of planetesimals and dust particles in the Proxima Centauri planetary system.

*Europlanet Science Congress 2024*. Berlin, Germany. 8–13 September 2024. Abstract EPSC2024-68. A poster. Virtual presentation in EXOA4 – Astrobiology and Origins. EPSC Abstracts. Vol. 17, EPSC2024-68, 2024. https://doi.org/10.5194/epsc2024-68, 2024. https://meetingorganizer.copernicus.org/EPSC2024/EPSC2024-68.html - abstract. https://ui.adsabs.harvard.edu/abs/2024epsc.conf...68I/abstract.
https://meetingorganizer.copernicus.org/EPSC2024/sessionprogramme. Poster presentation.

523. Ipatov S.I. Migration of bodies ejected from the terrestrial planets. *Europlanet Science Congress 2024*. Berlin, Germany. 8–13 September 2024. Abstract EPSC2024-51. A poster. Virtual presentation in TP9 – Impact Processes in the Solar System. EPSC Abstracts. Vol. 17, EPSC2024-51, 2024. https://doi.org/10.5194/epsc2024-51, https://meetingorganizer.copernicus.org/EPSC2024/EPSC2024-51.html - abstract. https://ui.adsabs.harvard.edu/abs/2024epsc.conf...51I/abstract. Poster presentation.

524. Ipatov S.I. Evolution of orbits of bodies in exoplanetary systems Proxima Centauri, TRAPPIST-1, and GLISSE 581. Scientific and practical seminar "Study of *the planets of the Solar System and exoplanets"* (December 16-21, 2024, Special Astrophysical Observatory of the Russian Academy of Sciences, Nizhny Arkhyz, Karachay-Cherkess Republic). Oral online presentation. https://planetaexo.wixsite.com/planets2024 - seminar program. In Russian.

525. Ipatov S.I. Probabilities of collisions of bodies ejected from the terrestrial planets with planets. Abstracts of *56th Lunar and Planetary Science Conference* (The Woodlands, Texas - March 10–14, 2025), #1593, 2 pages, https://www.hou.usra.edu/meetings/lpsc2025/pdf/1593.pdf. https://lpsc2025.ipostersessions.com/Default.aspx?s=62-F1-C5-A2-68-4E-73-52-70-52-16-55-61-AD-68-7B – iposter. https://www.hou.usra.edu/meetings/lpsc2025/program/.

526. Ipatov S.I. Probabilities of collisions of bodies ejected from the terrestrial planets with these planets // *Abstracts of the All-Russian annual seminar on experimental mineralogy, petrology and geochemistry* (VESEMPG-2025) (Moscow, April 15-16, 2025). VESEMPG 2025 - 044, p. 24. http://www.geokhi.ru/rasempg/Shared%20Documents/2025/Тезисы%20докладов%20ВЕСЭМПГ%202025.pdf. Oral presentation. on the website https://www.youtube.com/playlist?list=PLF5HdevRKsXLgmqCR31AmMDERjFnfo1or April 16, 2025. My presentation at 1:11:40-1:21:50 (10 minutes). http://www.geokhi.ru/rasempg/Shared%20Documents/2025/Програм%202025.pdf. In Russian.

527. Ipatov S.I. Exchange of matter between the terrestrial planets during the ejection of matter from the planets // *Abstracts of the conference "Near-Earth Astronomy–2025"* (Moscow, Space Research Institute of the Russian Academy of Sciences, June 2–4, 2025). P. 45–47. https://disk.yandex.ru/i/TnerA0kaOHgGBg- book of abstracts. Oral presentation. June 2, 2025. 15:30–15:45 – according to the program. Conference broadcast on YouTube: https://www.youtube.com/playlist?app=desktop&list=PLGyFz6xpxOYn1xRlQ518DXr8WfvL-HV0p . https://oza2025.tilda.ws/full_program - program. In Russian.

528. Ipatov S.I. Migration of bodies ejected from Mars, Mercury, and Venus // *Abstracts of 87th Annual Meeting of the Meteoritical Society* (Perth, Western Australia, July 14–18, 2025). ID. 5019. https://www.hou.usra.edu/meetings/metsoc2025/pdf/5019.pdf. A poster. Poster Session 2: Meteorite Delivery to Earth and Recovery. *Meteoritics and Planetary Science*. 2025. V. 60. Supplement S1. #5019. https://onlinelibrary.wiley.com/doi/epdf/10.1111/maps.70033 - p. 117 in this file. https://metsoc2025.au/program/ .

529. Ipatov S.I. Exchange of ejected material between the terrestrial planets and the Moon. *Abstracts of EPSC-DPS Joint Meeting 2025*. (Finland, Helsinki, 7-12 September 2025). EPSC-DPS2025-11. SB15 – Computational and experimental astrophysics of small bodies and planets. https://doi.org/10.5194/epsc-dps2025-11. Slides of the presentation are in Supplementary materials. https://meetingorganizer.copernicus.org/EPSC-DPS2025/session/55076#Orals_WED-OB6 – program of session SB15. An oral presentation. Wednesday, 10 Sep, 16:30–16:42 (EEST). Room Mercury.

530. Ipatov S.I. Probabilities of collisions of planetesimals with planets in the Proxima Centauri, Trappist-1 and Gliese 581 exoplanetary systems. *Abstracts of EPSC-DPS Joint Meeting 2025*. (Finland, Helsinki, 7-12 September 2025). EPSC-DPS2025-15. EXOA17 – Dynamics and stability of extrasolar systems. doi: https://doi.org/10.5194/epsc-dps2025-15. Slides of the presentation are in Supplementary materials. https://meetingorganizer.copernicus.org/EPSC-DPS2025/session/55193#Posters – program EXOA17, posters. A poster. Thursday, 11 Sep, 18:00–19:30 (EEST), Display time Thursday, 11 Sep, 08:30–19:30| Finlandia Hall foyer, F230.

531. Ipatov S.I. Migration of bodies in the Proxima Centauri and Trappist-1 planetary systems // *Abstracts of the conference "Solar System in Context 2025"* (29 September - 2 October 2025, Tucson,

Arizona), https://noirlab.edu/science/events/websites/solar-system-in-context-2025/schedule (this website includes also abstracts). p. 30-31. Virtual poster. 29.09.2025.

532. Ipatov S.I. Exchange of ejected material between the Moon and the terrestrial planets // *The Sixteenth Moscow Solar System Symposium (16M-S3)*. Moscow, Russia. October 20- 24, 2025, 16MS3-MN-PS-11. https://ms2025.cosmos.ru/sites/default/files/2025/abstracts/pdf/16M-S3-Ipatov-from-Moon_0.pdf - abstracts, DOI: 10.21046/16MS3-2025. https://ms2025.cosmos.ru/docs/2025/16-MS3_ABSTRACT-BOOK_1.pdf, p. 146-148. https://ms2025.cosmos.ru/docs/2025/16M-S3_PROGRAM_25.09.25.pdf - program. A poster. 21.10.2025.

533. Ipatov S.I. Migration of bodies ejected from Mercury and Venus // *The Sixteenth Moscow Solar System Symposium (16M-S3)*. Moscow, Russia. October 20- 24, 2025, 16MS3-SB-25. https://ms2025.cosmos.ru/sites/default/files/2025/abstracts/pdf/16M-S3-Ipatov-from-Mercury-and-Venus_1.pdf - abstracts, DOI: 10.21046/16MS3-2025. https://ms2025.cosmos.ru/docs/2025/16-MS3_ABSTRACT-BOOK_1.pdf, p. 243-245. https://ms2025.cosmos.ru/docs/2025/16M-S3_PROGRAM_25.09.25.pdf - program. An oral presentation. 23.10.2025

**Articles and a book submitted in 2025**

534. Ipatov S.I, Cho Y-K. Spectra of Earth-like exoplanets with different rotation periods // *Solar System Research*, 2025. V. 59, N 7. id. 83 (12 p.), DOI: 10.1134/S0038094625600234, https://doi.org/10.1134/S0038094625600234. https://doi.org/10.1134/S0038094625600234. https://rdcu.be/eEdC7. https://www.researchgate.net/publication/395293250. https://www.academia.edu/143801710. http://arxiv.org/abs/2509.19174. https://www.webofscience.com/wos/woscc/full-record/WOS:001564023500004.

535. Ipatov S.I. Models of the night-sky brightness and the efficiency of searching for exoplanets with the microlensing method // *Solar System Research*, 2025. V. 59, N 7. id. 82 (14 p.), DOI: 10.1134/S003809462560026X, https://doi.org/10.1134/S003809462560026X, https://rdcu.be/eEdDa. https://www.researchgate.net/publication/395291951. https://www.academia.edu/143801624. http://arxiv.org/abs/2509.19134. https://www.webofscience.com/wos/woscc/full-record/WOS:001564023500002.

536. Ipatov S.I. Exchange of ejected bodies between the terrestrial planets and the Moon // *Moscow University Physics Bulletin*, 2025, v. 80, Suppl. 1, pp. S423–S427. DOI: 10.3103/S0027134925701279. https://doi.org/10.3103/S0027134925701279. https://rdcu.be/fcPx4. https://www.researchgate.net/publication/403719268. https://www.academia.edu/168356256/.

537. Ipatov S.I. *Several notes about the history of the USSR and Russia*. 2025. 52 p. https://www.litres.ru/72597838, https://www.litres.ru/author/sergeyivanovich-ipatov/ - ISBN 978-5-532-89332-0, or https://1drv.ms/b/c/c67d93a65f0a2a17/EYqt02TO9zNBn4S_4ZKJhL8Bf6yi6ZZLPUDFF2tfAjhmJA?e=p0HE76 or https://cloud.mail.ru/public/sZ7x/irEEtRHsW - ISBN 978-5-6055159-0-6. (pdf file on my website). https://1drv.ms/w/c/c67d93a65f0a2a17/EYATE3DUbPZDnIXGziD2PmQBhUzyQbtzGrixZofpvTrtOw?e=gXow8u is docx file. https://www.academia.edu/145085402. https://cloud.mail.ru/public/mFnh/tBePebKC2 - English translation of the book.

**Posters and oral presentations without abstracts**

Ipatov S.I., Horne K., Alsubai K., Bramich D., Dominik M., Hundertmark M., Liebig C., Snodgrass C., Street R., Tsapras Y., Simulator for Microlens Planet Surveys, a **poster** at "*Gravitational microlensing. 101 years from theory to practice*" (10-13 February 2013, Doha, Qatar, http://www.astrodoha2013.org/).

Ipatov S.I., Horne K., Alsubai K., Bramich D., Dominik M., Hundertmark M., Liebig C., Snodgrass C., Street R., Tsapras Y., Exoplanet detection capability of microlensing observations, a poster at *18th International Conference on Microlensing* (January 19-24, 2014, Santa Barbara, USA, http://lcogt.net/microlensing18).

Ipatov S.I., Formation and growth of embryos of the Earth-Moon system. *The 1st IUGG Symposium on Planetary Sciences* (IUGG-PS 2017) (July 3-5, 2017, Berlin, Germany), http://www.dlr.de/iugg-ps2017, a **poster** presentation.

Ipatov S.I., Formation of trans-Neptunian satellite systems at the stage of rarefied condensations. *The 1st IUGG Symposium on Planetary Sciences* (IUGG-PS 2017) (July 3-5, 2017, Berlin, Germany), http://www.dlr.de/iugg-ps2017, a **poster** presentation.

Ipatov S.I., Marov M.Ya. Migration of icy planetesimals from the feeding zones of Jupiter and Saturn to forming terrestrial planets. *The 1st IUGG Symposium on Planetary Sciences* (IUGG-PS 2017) (July 3-5, 2017, Berlin, Germany), http://www.dlr.de/iugg-ps2017, an **oral** presentation.

Marov M.Ya., Ipatov S.I. Oceans of the Earth and the terrestrial planets: hypotheses of origin. *XXII International Scientific Conference (School) on Marine Geology* (Moscow, November 20-24, 2017). **Plenary presentation**. Conference website - http://geoschool.ocean.ru/. In Russian.

Ipatov S.I., Formation of the terrestrial planets, the Moon, and satellite systems of small bodies. *Conference* “*Star Formation and Planet Formation*” (November 12–13, 2019, Moscow, FIAN). Oral presentation. In Russian.

Ipatov S.I. Migration of planetesimals in the exoplanetary systems Proxima Centauri and TRAPPIST-1. *All-Russian conference* "*Study of stars with exoplanets-2021*" Simeiz Observatory of the Institute of Astronomy of the Russian Academy of Sciences (INASAN) October 24-29, 2021. Oral presentation without abstracts. http://www.inasan.ru/scientific_activities/conferences/simeiz-2021/. In Russian.

Ipatov S.I. Presentation of GEOKHI RAS under grant No. 075-15-2020-780. *All-Russian conference* "*Study of stars with exoplanets*" (November 23-27, 2022, Suzdal). http://www.inasan.ru/wp-content/uploads/2023/01/program_2022_exo.pdf . Oral presentation. In Russian.

Ipatov S.I. "Migration processes in the Solar system and in some exoplanetary systems and their role in the evolution of planets." *Meeting-discussion* " *Astronomical problems of the origin and development of life. The young Sun and the Earth* " (GAISH MSU, Moscow. March 19-20, 2024). http://www.sai.msu.su/EAAS/images/marov.pdf - Meeting program. Oral 30-minute invited presentation (without abstracts). https://www.youtube.com/watch?v=7i9YFPrma0U - recording of the first day of the meeting-discussion. In Russian.

Ipatov S.I. Exchange of ejected bodies between the terrestrial planets and the Moon. // *Modern Astronomy: Science and Education 2025.* On the 270th anniversary of Moscow University. Moscow. GAISH. June 23-27, 2025. https://modast2025.sai.msu.ru. Oral sectional presentation June 27, 2025. In Russian.

**Grant Reports. In Russian**

Ipatov S.I. Numerical modeling of the processes of formation and evolution of the solar system. Research: grant No. 93-02-17035. Russian Foundation for Basic Research. 1993. https://www.elibrary.ru/item.asp?id=50542320.

Ipatov S.I. Symposium No. 172 of the International Astronomical Union "Dynamics, Ephemeris, and Astrometry in the Solar System. Research: grant No. 95-02-87020. Russian Foundation for Basic Research. 1995. https://www.elibrary.ru/item.asp?id=50568449.

Ipatov S.I. Participation in the Sixth International Conference on Asteroids, Comets, Meteors (ACM96). Research: grant No. 96-02-26747. Russian Foundation for Basic Research. 1996. https://www.elibrary.ru/item.asp?id=50569204.

Ipatov S.I. Migration of celestial bodies to the Earth from the trans-Neptune and asteroid belts. Research: grant No. 96-02-17892. Russian Foundation for Basic Research. 1996. https://www.elibrary.ru/item.asp?id=50545230.

Ipatov S.I. Migration of celestial bodies to Earth from the trans-Neptune and asteroid belts. Research Report No. 96-02-17892. Russian Foundation for Basic Research. 1996. https://www.elibrary.ru/item.asp?id=227105.

Ipatov S.I. Participation in a joint discussion on comets and minor planets with Japanese amateur astronomers. This "discussion" is a merger of the 27th Japanese Cometary Conference and the 13th Meeting on Minor Planets. Participation in the 23rd General Assembly of the International Astronomical Union. Participation in the international seminar "Planet Formation". Research: grant No. 97-02-27124. Russian Foundation for Basic Research. 1997. https://www.elibrary.ru/item.asp?id=50571647.

Ipatov S.I. Participation in the conference "The impact of modern dynamics in astronomy". Research: grant No. 98-02-26832. Russian Foundation for Basic Research. 1998. https://www.elibrary.ru/item.asp?id=50574106.

Ipatov S.I. Migration of celestial bodies in the solar system. Research: grant No. 98-02-30069. Russian Foundation for Basic Research. 2000. Publication of a monograph. https://www.elibrary.ru/item.asp?id=50530335.

Ipatov S.I. Participation in the IAU Colloquium No. 181 and COSPAR No. 11 'Dust in the solar system and other planetary systems'. Research: grant No. 00-02-26614. Russian Foundation for Basic Research. 2000. https://www.elibrary.ru/item.asp?id=50576439.

Ipatov S.I. Construction of a computer model of the origin and evolution of the trans-Neptunian and asteroid belts and the migration of the bodies that comprised them to the Earth. Research: grant No. 01-02-17540. Russian Foundation for Basic Research. 2001. https://www.elibrary.ru/item.asp?id=50580936.

Final (January 23, 2008 – January 22, 2010) report on NASA grant NNX08AG25G "Velocities and amount of material ejected at different times after the Deep Impact collision" (PI – Sergei Ipatov). https://1drv.ms/b/c/c67d93a65f0a2a17/ERcqCl-mk30ggMasDAAAAAABV2Zu7RS8vVQdwtRL9jfE_Q?e=6lMloL.

Report at https://rosrid.ru. Registration number (state registration number of НИОКТР) 122113000124-9. Studies of the formation of exoplanetary disks and extrasolar planetary systems in 2022. Participants: M.Ya. Marov, S.I. Ipatov, A.V. Rusol, et al.

Ipatov was the head of the Moscow group of scientists in the INTAS grant in 2001-2004. He was a participant in grants in the USA in 2001-2007 and National Research Council / National Academy of Sciences Senior Research Associate (May 2002 – April 2003), and was also a participant in the Qatar grant for the search for exoplanets using microlensing (Qatar National Research Fund proposal NPRP 09-476-1-078) in 2011-2013. Since 2014, Ipatov has been an executor in various reports at GEOKHI RAS (annual reports and grant reports).

**IPM reports and dissertations. In Russian**

Ipatov S.I. Evolution of the orbits of gravitating particles and the problem of accumulation of planets of the solar system. Candidate's dissertation, IPM USSR Academy of Sciences, 264 p., 1981, Abstract, 18 p., 1982. https://search.rsl.ru/ru/record/01008805386, https://rusneb.ru/catalog/000200_000018_rc_1105979/. https://rusneb.ru/catalog/000199_000009_008805386/ - dissertation.

Ipatov S.I. An algorithm for simulating, on the PS-2000 multiprocessor computer, the evolution of a disk of gravitating bodies orbiting a massive central body. Research report, Keldysh Institute of Applied Mathematics (USSR Academy of Sciences), No. O-1211, 1985, 182 pp.

Ipatov S.I. and Gonzalez-Menendez E.A. Computational experiments on PS-2000 for problems of planetary cosmogony. Hand. Report of the Institute of Problems of Mathematics of the USSR Academy of Sciences and IPU N O-3385, 1985, 109 pages.

Ipatov S.I. Formation of the Kirkwood gap 5:2. Hand. Report of the Institute of Applied Mathematics of the USSR Academy of Sciences, N 181-90, 1990, 47 pages.

Eneev T.M., Ipatov S.I., Toroptseva V.N. Numerical studies of the process of formation and evolution of the solar system. Hand. Report of the Institute of Applied Mathematics of the USSR Academy of Sciences, N 256-90, 1990, 8 pp.

Eneev T.M., Ipatov S.I., Toroptseva V.N. Numerical studies of the process of formation and evolution of the solar system. Hand. Report of the Institute of Problems of Applied Mathematics of the Russian Academy of Sciences, No. 5-6-92, 1992, 4 pp.

Eneev T.M., Ipatov S.I., Toroptseva V.N. Numerical studies of the process of formation and evolution of the solar system. Hand. Report of the Keldysh Institute of Applied Mathematics of the Russian Academy of Sciences, N 5-16-93, 1993, 9 p.

Eneev T.M., Ipatov S.I., Toroptseva V.N. Numerical studies of the process of formation and evolution of the solar system. Hand. Report of the Keldysh Institute of Applied Mathematics of the Russian Academy of Sciences, N 5-2-94, 1994, 7 p.

Ipatov S.I. Modeling of celestial bodies migration in the solar system. Handbook. Doctoral dissertation, Keldysh Institute of Applied Mathematics, Russian Academy of Sciences, 1996, 387 p., abstract 48 p. http://www.dissercat.com/content/modelirovanie-migratsii-nebesnykh-tel-v-solnechnoi-sisteme. http://fizmathim.com/modelirovanie-migratsii-nebesnyh-tel-v-solnechnoy-sisteme. http://search.rsl.ru/ru/record/01000293573, https://dlib.rsl.ru/viewer/01000293573#?page=1. https://1drv.ms/b/c/c67d93a65f0a2a17/ERcqCl-mk30ggMYZDwAAAAAB5cPSkV4aIDOT_NdWXRWsQQ?e=3gt2C4. https://elibrary.ru/item.asp?id=30198870. https://rusneb.ru/catalog/000199_000009_000293573/. https://rusneb.ru/catalog/000199_000009_000167189/ - dissertation.

**Slides of presentations on websites (except conference websites). In Russian.**

https://ppt-online.org/659923 and https://theslide.ru/uncategorized/nekotorye-voprosy-formirovaniya-evolyutsii-i-poiska - Some issues of formation, evolution and search for planetary systems. S.I. Ipatov. 71 slides. Report at GEOKHI in March 2014.

https://astronomer.ru/data/0238/7-ipatov-elenin-ppt.ppt and https://present5.com/predlagaemye-modeli-veroyatnosti-obnaruzheniya-v-razlichnyx-oblastyax-neba - Proposed models of the probability of detecting near-Earth objects in different regions of the sky, based on the study of celestial body migration. S.I. Ipatov, L.V. Elenin. 21 slides. Oral report at the X International Conference "Near-Earth Astronomy-2017" (Agoy village, Krasnodar Krai, October 2-6, 2017).

**Press releases. In Russian.**

2023

According to the article in the UFN (Marov, Ipatov, 2023): The origin of water on Earth, there were a few press releases: https://www.minobrnauki.gov.ru/press-center/news/main/65417/ (Ministry of Education and Science website), https://nauka.tass.ru/nauka/17261341 (TASS) and https://new.ras.ru/activities/news/kommentariy-akademika-ran-mikhaila-marova-i-vedushchego-nauchnogo-sotrudnika-geokhi-ran-sergeya-ipat/ (RAS). The origin of water on Earth http://portal.geokhi.ru/SitePages/News082.aspx/. This article was also mentioned at https://t.me/geokhi/897, http://portal.geokhi.ru/SitePages/News107.aspx, https://t.me/geokhi/771, https://t.me/geokhi/729. https://t.me/geokhi/734.

The plausibility of existing scientific theories of the origin of the Moon. http://portal.geokhi.ru/SitePages/News107.aspx. https://www.kommersant.ru/doc/5900105 (Kommersant newspaper).

Longread. RIA. News on exoplanets: https://ria.ru/20231030/ekzoplanety-1905699613.html.

Based on an article in “Solar System Research”, No. 3. Stable orbits in the feeding zone of the planet Proxima Centauri c. http://portal.geokhi.ru/SitePages/News_200.aspx. https://new.ras.ru/activities/news/ustoychivye-orbity-v-zone-pitaniya-planety-proksima-tsentavra-s/.

According to an article in Meteoritics and Planetary Science: http://portal.geokhi.ru/SitePages/News_206.aspx - Delivery of icy bodies to the inner planets of the Proxima Centauri system. https://nauka.tass.ru/nauka/19184129 - Scientists believe that one of the planets in the Proxima Centauri system may be habitable.

For section 2 in the article in the GEOKHI-75 collection: on the RAS website - https://new.ras.ru/activities/news/issledovaniya-v-oblasti-zvyezdno-planetnoy-kosmogonii-metodami-matematicheskogo-modelirovaniya, on the GEOKHI website - http://portal.geokhi.ru/SitePages/News_226.aspx, in Telegram - https://t.me/geokhi/1514.

According to the article in No. 6 of the “Solar System Research” in 2023: - https://t.me/geokhi/1630, http://portal.geokhi.ru/SitePages/News_260.aspx, https://new.ras.ru/activities/news/dvizhenie-planetezimaley-v-sfere-khilla-zvezdy-proksima-tsentavra/.

2024

https://www.mk.ru/science/2024/05/08/uchenyy-sergey-ipatov-predlozhil-svoyu-versiyu-rosta-luny.html - Moskovsky Komsomolets.

https://russian.rt.com/science/article/1268428-luna-mars-voda-intervyu-kolonizaciya. "The presence of water is mandatory": a researcher at the Russian Academy of Sciences' Institute of Geochemistry and Analytical Chemistry (GEOKHI) on the chances of human colonization of the Moon and Mars.

Interview with Russia Today. February 7, 2024. http://xn--c1aejx9a.xn--p1ai/SitePages/News_265.aspx. https://t.me/geokhi/1935 http://portal.geokhi.ru/SitePages/News_338.aspx .

https://www.mk.ru/science/2024/05/08/uchenyy-sergey-ipatov-predlozhil-svoyu-versiyu-rosta-luny.html - interview with the newspaper Moskovsky Komsomolets. https://t.me/geokhi/1936. https://t.me/geokhi/1938 .

https://www.kommersant.ru/doc/6743452 June 2, 2024, 12:00 PM. "The Earth and the Moon will perish together." Kommersant. Interview with mathematician Sergei Ipatov about a new model for the formation of our satellite.

https://poisknews.ru/astronomiya/evolyucziya-orbit-i-vodnye-resursy-zahvatyvayushhee-issledovanie-o-planetah/ 04.12.2024. Orbital evolution and water resources: a fascinating study about planets. This is a note in the Poisk newspaper about my article in the journal Solar System Research. The note provides a link to a paid journal. The same text can be downloaded for free from one of the websites below: https://rdcu.be/d04Q0, http://arxiv.org/abs/2411.05436, https://www.researchgate.net/publication/385627526 or https://www.academia.edu/125363162/. https://t.me/geokhi/2411.https://minobrnauki.gov.ru/press-center/news/nauka/92482/, - The news about my article was published on December 4 in the Poisk newspaper and on December 9, 2024 on the website of the Ministry of Education and Science of Russia.

2025

https://www.mk.ru/science/2025/01/04/zemlya-atakovala-mars-podschitano-vozmozhnoe-chislo-zemnykh-meteoritov-na-krasnoy-planete.html - Earth attacked Mars: the possible number of terrestrial meteorites on the red planet has been calculated. - interview in the Moskovsky Komsomolets newspaper on January 4, 2025, based on an article in the Icarus journal. https://t.me/geokhi/2455 - Press release on an article in the Icarus journal. https://t.me/frnved/2745.

https://minobrnauki.gov.ru/press-center/news/nauka/93451/ Scientists have calculated the probability of collisions between Earth fragments and other planets in the solar system. January 10, 2025

https://www.kommersant.ru/doc/8159218 - Spectrum of distant Earth. What can be learned about a planet from its radiation. October 28, 2025.

https://ria.ru/20251030/earth-2051544273.html - A discovery deep within the Earth reveals a secret about our planet. RIA Novosti, October 30, 2025.

https://minobrnauki.gov.ru/press-center/news/nauka/98938/ - Scientists have studied the spectra of Earth-like planets. Material on the Ministry of Education and Science website. October 31, 2025.

**Information about S.I. Ipatov on the GEOKHI website and Telegram channel in 2024-2025. In Russian.**

http://portal.geokhi.ru/SitePages/News_251.aspx On this site, among the outstanding results of Section No. 15 “Planetary Research”, the results of M. Ya. Marov and S. I. Ipatov are noted in first place.

https://t.me/geokhi/1613 - The most important achievements in astronomy in 2023.

http://portal.geokhi.ru/SitePages/News_259.aspx - Asteroids named in favour of scientists from GEOKHI RAS. - https://t.me/geokhi/1627 .

http://portal.geokhi.ru/SitePages/News_260.aspx - Motion of planetesimals in the Hill sphere of the star Proxima Centauri.

http://portal.geokhi.ru/SitePages/News_265.aspx - Interview with leading researcher of GEOKHI RAS S.I. Ipatov about the possible colonization of the Moon and Mars (Russia Today).

http://portal.geokhi.ru/SitePages/News_338.aspx - Growth of the Moon due to bodies ejected from the Earth.

http://portal.geokhi.ru/SitePages/News_351.aspx - "The Earth and the Moon will perish together": interview with leading researcher of the GEOKHI RAS Sergei Ipatov to Kommersant.

http://portal.geokhi.ru/SitePages/News_455.aspx - Migration of celestial bodies in the solar system and in exoplanetary systems. 12/10/2024.

http://portal.geokhi.ru/SitePages/News_474.aspx - Probabilities of collisions of bodies ejected from Earth with planets. 01/06/2025.

https://t.me/geokhi/2563 - information about my articles about T.M. Eneev on the GEOKHI telegram channel. March 4, 2025.

https://t.me/geokhi/2815, https://t.me/geokhi/2818 - information about the conference "Modern Astronomy: Science and Education" (June 23-27, 2025, SAI MSU) https://modast2025.sai.msu.ru), in which S.I. Ipatov participated.

https://t.me/geokhi/3062 - information about the international conference EPSC-DPS2025 (September 8-12, 2025, Helsinki, Finland, https://www.epsc-dps2025.eu/), in which S.I. Ipatov participated. 09.21.2025.

https://t.me/geokhi/3109 - information about the international conference "Solar System in Context» (September 29 to October 2, 2025, Tucson, USA, https://noirlab.edu/science/events/websites/solar-system-in-context-2025/schedule), dedicated to the solar system and exoplanetary systems, in which S.I. Ipatov participated. September 30, 2025.

https://t.me/geokhi/3134, http://portal.geokhi.ru/SitePages/News_631.aspx. - the effectiveness of the search for exoplanets using the microlensing method (press release on Ipatov's article in Solar System Research). 10.10.2025.

http://portal.geokhi.ru/SitePages/News_351.aspx - information about the presentations at the Sixteenth Moscow International Symposium on Solar System Research (16M-S3). https://t.me/geokhi/3182. 10.19.2025.

https://t.me/geokhi/3212, http://portal.geokhi.ru/SitePages/News_644.aspx - 31.10.2025. https://www.kommersant.ru/doc/8159218. - Spectra of Earth-like planets with different periods of axial rotation. (press release on Ipatov's article in Solar System Research). October 30, 2025.

https://t.me/geokhi/3235 - comment on the article (https://www.nature.com/articles/s41561-025-01811-3) in the journal "Nature Geosciences ", 01.11.2025.
http://portal.geokhi.ru/SitePages/News_648.aspx - similar comment on the GEOKHI website 10.31.2025.

**Publications published after this Russian book has been printed (Addition to the Russian book)**

Ipatov S.I. *Formation and evolution of planetary systems* (in Russian). ISBN 978-5-6055159-1-3. Moscow. Onebook.ru. 2025. 132 p. DOI: 10.17513/np.649. https://dx.doi.org/10.17513/np.649. https://cloud.mail.ru/public/Yeoy/cDwaFQjVV – English translation (this file).

Ipatov S.I. *My meetings with astronomers and other people* (in Russian). Moscow. 2026. 372 p. ISBN 978-5-6055159-2-0 for printed book. https://dx.doi.org/10.17513/np.668. The pdf file with the book with photos is also free on https://cloud.mail.ru/public/b9E2/fNj2PJHQ3. https://cloud.mail.ru/public/DWT8/S6wWoeETJ - English translation of the book with photos.

Ipatov S.I. Migration of bodies in the Proxima Centauri and Trappist-1 planetary systems // Proc. IAU Symposium 393 "*Planetary Science and Exoplanets in the Era of the James Webb Space Telescope*" (August 13-15, 2024, Cape Town, South Africa). 5 pages, http://arxiv.org/abs/2411.05954, Online ISBN: 9781009546157. Proceedings of the International Astronomical Union, Volume 20, Symposium S393. 2026, pp. 29–33. https://doi.org/10.1017/S1743921324001546. https://www.researchgate.net/publication/399617826. https://www.academia.edu/145875624.

Ipatov S.I. Migration of bodies ejected from the Earth and the Moon // Proc. IAU Symposium 393 "*Planetary Science and Exoplanets in the Era of the James Webb Space Telescope*" (August 13-15, 2024, Cape Town, South Africa). 5 pages, http://arxiv.org/abs/2411.04218, Online ISBN: 9781009546157. Proceedings of the International Astronomical Union, Volume 20, Symposium S393. 2026. pp. 9-13. DOI: https://doi.org/10.1017/S1743921324001911. https://www.researchgate.net/publication/399618742. https://www.academia.edu/145875450.

Ipatov S.I. Migration of bodies from the zone of the giant planets to the terrestrial planets and the Moon // *The Seventeenth Moscow Solar System Symposium* (17M-S3). Moscow, Russia. October 12- 16, 2026. 17MS3-SB-16, an oral presentation. https://ms2026.cosmos.ru/docs/2026/MS3-17_BOOK-1.pdf, p. 257-260.

Ipatov S.I. Migration of bodies from the zone of the giant planets to the terrestrial planets and the Moon // Proceedings of the Twenty-seventh international conference "*Physical-chemical and petrophysical researches in the Earth's sciences*" (28 September – October 2, 2026, Moscow; Borok), Moscow. IGEM, 2026, ISBN 978-5-88918-085-2. ISSN 2686-8938. P. 109-112. https://igem.ru/petromeeting_XXVII/tbgdocs/sbornik_2026.pdf. https://rutube.ru/video/04282c883b47bda51b09c83b6c40e04b/ or https://youtu.be/My3RDcdKRHQ - my oral presentation is the first lecture. In Russian.

Ipatov S.I. Migration of dust particles ejected from the Earth and Mars // Proceedings of the Twenty-seventh international conference "*Physical-chemical and petrophysical researches in the Earth's sciences*" (28 September – October 2, 2026, Moscow; Borok), Moscow. IGEM, 2026, ISBN 978-5-88918-085-2. ISSN 2686-8938. C. 105-108. https://igem.ru/petromeeting_XXVII/tbgdocs/sbornik_2026.pdf. In Russian.

Ipatov S.I. Migration of dust particles ejected from the Earth. Abstracts of *88th Annual Meeting of the Meteoritical Society* (Frankfurt am Main, Germany - August 9–14, 2026). ID 5015, a poster. https://www.hou.usra.edu/meetings/metsoc2026/pdf/5015.pdf.

Ipatov S.I. Exchange of ejected bodies between the Moon and the terrestrial planets // Abstracts of *57th Lunar and Planetary Science Conference* (The Woodlands, Texas - March 16–20, 2026), #1110, 2 pages. https://www.hou.usra.edu/meetings/lpsc2026/pdf/1110.pdf, a poster.

Ipatov S.I. Migration of dust particles ejected from the Earth and Mars. *Europlanet Science Congress 2026* (The Hague, The Netherlands. 6–11 September 2026). SB12 – Dynamics of Small Bodies. Identification number EPSC2026-20. A virtual poster. https://doi.org/10.5194/epsc2026-20. https://meetingorganizer.copernicus.org/EPSC2026/EPSC2026-20.html.

Ipatov S.I. Exchange of ejected material between the terrestrial planets and the Moon // Abstracts of the *All-Russian annual seminar on experimental mineralogy, petrology and geochemistry* (VESEMPG-2026) (Moscow, April 14-15, 2026). A poster. VESEMPG 2026 - 051, p. 53. http://www.geokhi.ru/rasempg/Shared%20Documents/2026/Тезисы%20докладов%20ВЕСЭМПГ%202026.pdf. http://www.geokhi.ru/rasempg/. In Russian.

Ipatov S.I. Lunar impacts of bodies ejected from Mars, Mercury, and Venus // *Solar System Research*, 2027. V. 61, N 1, in press

**Press releases in 2026** (in Russian)

http://portal.geokhi.ru/SitePages/News_697.aspx - Press release about my book "Formation and evolution of planetary systems". 23.01.2026.

https://iz.ru/export/google/amp/2034172 - Water data: a new model of the young Solar System will help in the search for life on Mars. Izvestia newspaper. 02.02.2026.

https://iz.ru/2035360/2026-02-02/uchenye-smodelirovali-usloviia-poiavleniia-organizmov-na-marse-v-drevnosti - Scientists have modelled the conditions for the emergence of organisms on Mars in ancient times. Izvestia newspaper. 02.02.2026.

https://minobrnauki.gov.ru/press-center/news/nauka/99669/ - Russian scientists have built a model of the migration of bodies in the early Solar System. Ministry of Science and Higher Education of Russia website. 03.02.2026.

https://vkvideo.ru/video-20484756_456239038 or https://rutube.ru/video/75a67383f501a9d3ef30cb8c203a69d3/ - video by S.I. Ipatov "Formation and evolution of planetary systems". 05.03.2026.

http://portal.geokhi.ru/SitePages/News_796.aspx - Press release about exchange of bodies between the terrestrial planets and the Moon on the paper in Moscow University Physics Bulletin.

https://www.kp.ru/daily/277819.5/5308432/ - 28.09.2026. Exchange of material between planets

**Information about Ipatov on the website or telegram of GEOKHI in 2026** (In Russian)

https://max.ru/id7729081339_gos/AaDsc6_xGfs . https://t.me/geokhi/4020. - 29.09.2026.
https://max.ru/id7729081339_gos/AaCO_ijQIkE . https://t.me/geokhi/3964 11.09.2026
https://max.ru/id7729081339_gos/AZ8H4Nj6cKE . https://t.me/geokhi/3799 - 27.06.2026
https://max.ru/id7729081339_gos/AZ7PA_-UdUo . https://t.me/geokhi/3781 - 16.06.2026
https://t.me/geokhi/3591 13.03.2026.
https://t.me/geokhi/3496 - 04.02.2026.
https://t.me/geokhi/3461 - 23.01.2026
http://portal.geokhi.ru/SitePages/News_767.aspx - 24.06.2026
http://portal.geokhi.ru/SitePages/News_759.aspx 16.06.2026
http://portal.geokhi.ru/SitePages/News_697.aspx - 12.03.2026.

**Information about the author**

Sergei Ivanovich Ipatov is a leading researcher at the Vernadsky Institute of Geochemistry and Analytical Chemistry of the Russian Academy of Sciences, Doctor of Physical and Mathematical Sciences, laureate of the F.A. Bredikhin Prize of the Russian Academy of Sciences in Astronomy, academician of the Russian Academy of Natural Sciences, specialist in the field of dynamic astronomy, author of more than 500 scientific publications (including about 80 articles in peer-reviewed scientific journals, more than 100 other articles, a monograph and more than 300 conference abstracts). The h-index according to WoS and Scopus is 19. More than 1500 citations in Scopus. The main scientific interests of S.I. Ipatov are related to the study of the origin and evolution of the Solar System and some exoplanetary systems. Ipatov is a co-discoverer of 8 asteroids. Asteroid 14360 is named Ipatov in honor of S.I. Ipatov. More information about the author can be found on the websites https://sites.google.com/view/siipatov/, http://siipatov.webnode.ru (rarely upgraded), https://ru.wikipedia.org/wiki/Ипатов,_Сергей_Иванович and https://en.wikipedia.org/wiki/Sergei_Ipatov. In particular, on the websites https://siipatov.webnode.ru/publications/ and https://sites.google.com/view/siipatov/publications-in-russian you can find a list of publications by S.I. Ipatov (with indication of the websites where these publications are located), and on the websites https://siipatov.webnode.ru/link-to-albums-with-photos and https://sites.google.com/view/siipatov/links-to-photos there are internet links to a large number of photo albums of astronomers and places where conferences were held in which Ipatov participated.

RESEARCH EXPERIENCE (excluding payment from grants via some institutions):

● V.I. Vernadsky Institute of Geochemistry and Analytical Chemistry of Russian Academy of Sciences (Moscow, Russia), Leading scientist. December 2013 – present.
● Space Research Institute (Moscow, Russia), Leading scientist (part time). May 2011 – March 2017.
● Alsubai Establishment for Scientific Studies (Doha, Qatar). Researcher. August 2011 – August 2013.
● Studies of astronomical problems. February 2010 – July 2011.
● Catholic University of America (Washington, DC). Research associate. April 2008 – January 2010.
● Department of Terrestrial Magnetism of Carnegie Institution for Science (Washington, DC, USA). Research scientist. September 2006 – March 2008.
● University of Maryland (College Park, MD, USA). Research associate. January 2005 – August 2006.
● Catholic University of America (Washington, DC). Research associate. May 2004 – October 2004.
● George Mason University (VA, USA). Visiting senior research associate. May 2003 – April 2004.
● NASA/Goddard Space Flight Center (MD, USA). National Research Council/National Academy of Sciences Senior Research Associate. May 2002 – April 2003.
● George Mason University (VA, USA). Visiting researcher. July 2001 – April 2002.
● Institute of Applied Mathematics of Russian Academy of Sciences (Moscow, Russia). Leading scientist, 1997-2001 (formally until December 2003). Senior scientist, 1990-1997. Scientist, 1987-1990. Junior

scientist, 1977-1987. Probationer-investigator, 1975-1977.

**Visits to scientific institutions:** For one or two months, Ipatov visited Institute for Theoretical Physics of the University of California in Santa Barbara (USA) in 1992, Notre-Dame University in Namur (Belgium) in 1995, Berlin Institute of Planetary Exploration in 1996, Royal Observatory of Belgium (Brussels) in 1998, Dresden Technical University in 2001, Inst. d'Astrophysique de Paris in 2007, St-Andrews University (UK) in 2011. During 5 months' visit to the Royal observatory of Belgium in 1999, he found (with E. Elst and T. Pauwels) several new asteroids. Eight of the asteroids have gotten numbers. He made shorter trips to the Department of Terrestrial Magnetism (1990), University of London (1995), Armagh Observatory (Northern Ireland, 1995, 2000), Nice Observatory (1999), etc.

**Web sites:** https://sites.google.com/view/siipatov/, https://siipatov.webnode.ru/ (rarely upgraded) https://en.wikipedia.org/wiki/Sergei_Ipatov.

https://siipatov.webnode.ru/publications/ - links to publications,

https://siipatov.webnode.ru/link-to-albums-with-photos/ - links to albums with photos.

**review of S.I. Ipatov's book "Formation and Evolution of Planetary Systems" by O.L. Kuskov**

The book-essay "Formation and Evolution of Planetary Systems" by Sergei Ivanovich Ipatov is a unique and multifaceted work that bridges the gap between fundamental scientific research and popular science literature. It offers readers a broad panorama of studies concerning the origin and development of planetary systems, with a particular focus on our own Solar System.

The book is based on the synthesized results of the author's own long-term research. The scope of the work is impressive, covering topics ranging from classic problems - such as the collapse of a protoplanetary cloud and the accretion of terrestrial planets - to contemporary issues like the migration of celestial bodies, the dynamics of the asteroid belt, and the formation of craters on the Moon. Particularly noteworthy are the sections dedicated to modeling the evolution of exoplanetary systems, as well as those addressing the practical aspects of astronomy.

The defining feature of this work lies in its concept. It is neither a traditional academic monograph nor a popular-science retelling of well-established facts. Instead, the author describes the book as a "guide to publications" - a key to understanding both its structure and its intended audience. In this essay-style navigator, the author shares his personal perspective on the subject, drawing upon arguments and examples from his own body of work. The 130-page volume contains 76 pages of text, while the remainder is devoted to a meticulously numbered list of 537 publications. This comprehensive bibliography encompasses the author's entire output, including journal articles, conference abstracts, various shorter publications, and both poster and oral presentations - some of which were delivered without accompanying abstracts. The author has deliberately limited the number of illustrations and references to the work of other scientists to keep the volume concise, opting instead to provide the reader with direct, unprecedented access to the primary sources of his own research. A key merit of this book is its openness. The author generously provides links to his own work, most of which is freely accessible on various scientific platforms. S. I. Ipatov's book, "Formation and Evolution of Planetary Systems", is recommended for publication; it will be a valuable resource for astronomy enthusiasts, as well as for undergraduate and graduate students in the fields of physics and astronomy, for whom it may serve as a starting point for their own research.

Chief Researcher at the Vernadsky Institute of Geochemistry and Analytical Chemistry (GEOKHI) of the Russian Academy of Sciences (RAS), Doctor of Chemical Sciences,

Corresponding Member of the RAS /O. L. Kuskov/

I hereby certify the signature of O. L. Kuskov, Corresponding Member of the RAS.

Scientific Secretary of GEOKHI RAS, Candidate of Geological and Mineralogical Sciences /L. I. Kolmykova/

**Contents**